\documentclass[a4paper,12pt,twoside]{article}

\usepackage{amsmath}
\usepackage{amssymb}
\usepackage{fullpage}

\usepackage{dsfont} 

\usepackage{verbatim}

\makeatletter
\def\cleardoublepage{\clearpage\if@twoside \ifodd\c@page\else
    \hbox{}
    \thispagestyle{empty}
    \newpage
    \if@twocolumn\hbox{}\newpage\fi\fi\fi}
\makeatother \clearpage{\pagestyle{empty}\cleardoublepage}

\usepackage[T1,OT1]{fontenc}

\usepackage{graphicx}
\graphicspath{{figures/}}

\usepackage[hang]{footmisc}
\usepackage[compress,square,numbers]{natbib}

\usepackage[colorlinks,linktocpage,linkcolor=blue,citecolor=blue,urlcolor=blue]{hyperref}

\numberwithin{equation}{section}
\numberwithin{figure}{section}
\numberwithin{table}{section}

\allowdisplaybreaks[2]

\renewcommand{\d}{\mathrm{d}}
\newcommand{\e}{\mathrm{e}}

\newcommand{\w}{\wedge}

\newcommand{\de}[2]{\frac{\delta #1}{\delta #2}}
\newcommand{\diff}[2]{\frac{\d #1}{\d #2}}

\newcommand{\nl}{\notag \\ &\quad\,}
\newcommand{\nll}{\notag \\ &}

\newcommand{\dsix}{$\overline{\textrm{D}6}$}
\newcommand{\dsixb}{$\overline{\textrm{D}\boldsymbol{6}}$}
\newcommand{\dthree}{$\overline{\textrm{D}3}$}

\begin{document}

\pagestyle{empty}

\begin{center}

\vspace*{5.8cm}

{\huge {\textsf{Backreaction of Localised Sources}}}

\vspace{0.7cm}

{\huge {\textsf{in String Compactifications}}}

\vspace{4cm}

{Von der Fakult\"{a}t f\"{u}r Mathematik und Physik\\der Gottfried Wilhelm Leibniz Universit\"{a}t Hannover\\zur Erlangung des Grades}

\vspace{0.7cm}

{\large Doktor der Naturwissenschaften (Dr. rer. nat.)}

\vspace{0.7cm}

{genehmigte Dissertation\\von\\}

\vspace{0.7cm}

{\large\bf Dipl.-Phys. Daniel Junghans\\}

\vspace{0.4cm}

{\large geboren am 10. Januar 1984 in Zittau\\}

\vspace{4cm}

{2013\\}

\end{center}

\newpage

\pagestyle{empty}

\vspace*{20.5cm}

\begin{center}
\begin{tabular*}{1 \textwidth}{@{} l l @{\extracolsep{\fill} } l }

Referent: & Prof. Dr. Marco Zagermann & \\[0.3cm]
Korreferenten: & Prof. Dr. Olaf Lechtenfeld & \\
 & Prof. Dr. Ulf Danielsson & \\[0.3cm]
Tag der Promotion: & 14. Mai 2013 & \\

\end{tabular*}
\end{center}

\newpage

\pagestyle{plain}

\begin{center}
{\bf Zusammenfassung\\}
\end{center}

\noindent Lokalisierte Quellen wie D-Branen oder Orientifold-Fl{\"{a}}chen spielen eine wichtige Rolle in vielen ph{\"{a}}nomenologisch relevanten Stringkompaktifizierungen. Die Anwesenheit dieser Objekte f{\"{u}}hrt typischerweise zu einer komplizierten Dynamik in den kompakten Dimensionen, so dass eine volle L{\"{o}}sung der zehndimensionalen Bewegungsgleichungen oft nicht gefunden werden kann. Um dennoch Aussagen {\"{u}}ber die vierdimensionale effektive Theorie treffen zu k{\"{o}}nnen, die im Limes niedriger Energien aus einer Stringkompaktifizierung hervorgeht, werden {\"{u}}blicherweise nur die integrierten Bewegungsgleichungen gel{\"{o}}st, w{\"{a}}hrend die R{\"{u}}ckwirkung der lokalisierten Quellen auf die internen Felder vernachl{\"{a}}ssigt wird. Diese Vereinfachung wird oft als Schmieren bezeichnet. In dieser Arbeit untersuchen wir, inwieweit das Schmieren lokalisierter Quellen Observablen der effektiven Niederenergietheorie beeinflusst und ob es zu falschen L{\"{o}}sungen f{\"{u}}hren kann, die bei voller Ber{\"{u}}cksichtigung der R{\"{u}}ckwirkung nicht konsistent w{\"{a}}ren. Wir analysieren verschiedene Beispiele, f{\"{u}}r die geschmierte L{\"{o}}sungen existieren, und stellen fest, dass die Vertrauensw{\"{u}}rdigkeit der geschmierten N{\"{a}}herung davon abzuh{\"{a}}ngen scheint, ob die je-weiligen L{\"{o}}sungen eine BPS-Schranke saturieren. Zudem untersuchen wir den Vorschlag, \dthree-Branen an der Spitze halsartiger Geometrien wie der Klebanov-Strassler-L{\"{o}}sung zu platzieren, um metastabile De-Sitter-Vakua in der Stringtheorie zu konstruieren. Wir zeigen unter wenigen allgemeinen Annahmen, dass die R{\"{u}}ckwirkung der vollst{\"{a}}ndig loka-lisierten Anti-Branen zu einer Singularit{\"{a}}t in Feldern f{\"{u}}hrt, die nicht direkt an die Anti-Branen koppeln, und f{\"{u}}hren ein Argument an, welches nahelegt, dass die Singularit{\"{a}}t nicht durch eine Polarisierung der Branen behoben wird. Schlie\ss lich demonstrieren wir, dass in vielen Stringkompaktifizierungen mit nichttrivialen Hintergrundfl{\"{u}}ssen die effektive niederdimensionale kosmologische Konstante durch die Wirkung der lokalisierten Quellen bestimmt ist, die in der jeweiligen L{\"{o}}sung auftreten. Dies macht es oft m{\"{o}}glich, die kosmologische Konstante zu berechnen, ohne auf potentiell unzuverl{\"{a}}ssige N{\"{a}}herungen wie das Schmieren zur{\"{u}}ckgreifen zu m{\"{u}}ssen.
\\

{\noindent \bf Schl{\"{u}}sselw{\"{o}}rter:} Stringtheorie, Kompaktifizierung, D-Branen und Orientifold-Fl{\"{a}}chen

\cleardoublepage

\begin{center}
{\bf Abstract\\}
\end{center}

\noindent Localised sources such as D-branes or orientifold planes play an important role in many string compactifications that are relevant for phenomenology. The presence of these objects typically induces complicated dynamics in the compact dimensions such that a full solution to the ten-dimensional equations of motion is often out of reach. In order to still be able to make statements about the four-dimensional effective theory arising in the low-energy limit, the equations of motion are usually only solved in an integrated sense, while the backreaction of the localised sources on the internal fields is neglected. This simplification is often referred to as smearing. In this work, we investigate to what extent smearing may affect observables in the effective low-energy theory and whether it may lead to fake solutions that would cease to exist once the backreaction is properly taken into account. We analyse explicit examples for which smeared solutions exist and find that the reliability of the smeared approximation appears to depend on whether or not these solutions saturate a BPS bound. We also address the proposal of placing \dthree-branes at the tip of a warped throat geometry like the Klebanov-Strassler solution in order to construct meta-stable de Sitter vacua in string theory. We show that, under a few general assumptions, the backreaction of the fully localised anti-branes yields a singularity in fields that do not directly couple to them, and we give an argument suggesting that the singularity is not resolved by brane polarisation. We furthermore demonstrate that, in many string compactifications involving non-trivial background fluxes, the effective lower-dimensional cosmological constant is determined by the on-shell action of the localised sources present in the solution. This often allows to compute the cosmological constant without resorting to potentially unreliable approximations like smearing.
\\

{\noindent \bf Keywords:} string theory, compactification, D-branes and orientifold planes

\cleardoublepage

\tableofcontents

\cleardoublepage
\thispagestyle{plain}

\section{Introduction}
\label{ch:intro}

General relativity and quantum field theory are arguably the two major achievements of 20th century theoretical physics. Both theories have been confirmed by experiments with great accuracy, and there is in fact not a single known experiment that is in conflict with the predictions of either of the two. While the large scale structure of the universe is successfully described by general relativity, its microscopic behaviour is known to be governed by a quantum field theory, the Standard Model of particle physics, which---if extended by non-zero neutrino masses---agrees with the observed particle spectrum up to presently accessible energy scales. Nonetheless, a more fundamental theory of nature is believed to be ultimately required in order to answer open questions such as the origin of dark matter and dark energy or to solve the hierarchy problem. Moreover, it is known that general relativity and quantum field theory lead to mutually incompatible results in a regime where both gravity and quantum effects become significant, as it is the case in the vicinity of blackhole singularities or in the early universe shortly after the big bang. In order to explore these extreme situations, a yet to be found theory of quantum gravity appears to be necessary.

String theory (see e.\ g.\ \cite{Green:1987sp, Green:1987mn, Polchinski:1998rq, Polchinski:1998rr, Zwiebach:2004tj, Becker:2007zj} and references therein) is a candidate for such a theory of quantum gravity. Since gravity is perturbatively non-renormalisable, its naive quantisation leads to divergent amplitudes unless some unexpected cancellation takes place.\noindent\footnote{This is only conjectured to happen for highly symmetric theories such as $\mathcal{N}=8$ supergravity (see \mbox{e.\ g.} \cite{Kallosh:2011dp, Banks:2012dp} as well as references therein for recent arguments in favour of and against this possibility).} Amplitudes in string theory, on the other hand, are believed to be finite. This can intuitively be understood from the extended nature of the strings, which provides a UV cutoff at the string scale. More precisely, the finiteness of string amplitudes was argued to follow from a property called modular invariance. Furthermore, a spin-two particle, the graviton, necessarily appears when string theory is quantised, and so gravity in fact arises quite naturally in this theory. At the same time, string theory provides a natural framework for a unification of the different particles and forces in nature, which all arise as different vibrational modes of the fundamental string. Thus, string theory is not only a candidate for a theory of quantum gravity but in fact a candidate for a ``theory of everything''.

It is also remarkable that all string theories with both bosonic and fermionic degrees of freedom necessarily have spacetime supersymmetry. These theories are anomaly-free only in ten dimensions and thus have the peculiar property of ``predicting'' the number of spacetime dimensions they live in. There are five such superstring theories---type I string theory, type IIA and IIB string theory, as well as two heterotic string theories with gauge groups $\mathrm{E}_8 \times \mathrm{E}_8$ and $\mathrm{SO}(32)$. In the limit of low energies and small string coupling, they effectively reduce to classical supergravity theories, with corrections due to higher derivative terms and perturbative and non-perturbative quantum effects. Furthermore, it is known that all superstring theories are related to one another and to the eleven-dimensional M-theory by a web of dualities and should therefore be understood as different limits of one unique underlying theory.
\\

Despite these and many other intriguing features of string theory that have been discovered in the past decades, it is fair to say that the theory also faces several problems. One of the biggest challenges for string theory is to understand how it can make contact with the real world and whether it can reproduce the observed properties of our universe in its low-energy limit. That this is a problem is closely related to the extra dimensions predicted by string theory. Since superstring theories are only consistent in ten dimensions, it has to be explained why only four of them are observed macroscopically. One way to do this is to assume that the extra dimensions are compactified on an internal manifold and thus only become observable above a certain energy scale, which is determined by the size of the compact dimensions. In the low-energy limit, one then obtains a four-dimensional effective theory, whose properties are determined by the specifics of the compactification. For phenomenological reasons, this effective theory is often engineered to be an $\mathcal{N}=1$ supergravity, and supersymmetry is supposed to be broken at a scale much lower than the compactification scale.

While the procedure of compactification explains why we do not observe a ten-dimensional spacetime in everyday life, it also comes with a number of technical and conceptual difficulties, not all of which have been solved so far. A problem of many compactification scenarios is the problem of moduli stabilisation. Moduli fields are massless scalar fields in the four-dimensional effective action, whose vacuum expectation values (vevs) parametrise certain properties of the compactification such as the volume of the internal manifold. Since the moduli fields are massless, their effective potential is flat, and their vevs and the associated physical parameters remain completely undetermined, which is certainly not desirable in a realistic compactification. Furthermore, massless scalars mediate long-range forces in the non-compact dimensions, which is not consistent with observations. In many models, the appearance of such moduli fields is quite generic. In particular, it is a serious drawback of compactifications on Calabi-Yau manifolds, which were traditionally popular due to their attractive property of preserving an $\mathcal{N}=1$ supersymmetry in the four-dimensional effective low-energy theory.

A way out of the moduli problem is provided by flux compactifications \cite{Grana:2005jc, Blumenhagen:2006ci, Douglas:2006es, Denef:2007pq}, where a potential for the moduli is generated by turning on background fluxes threading non-trivial cycles of the internal manifold. This procedure generically stabilises at least some of the moduli, although this is not true for all of them. Therefore, additional effects such as quantum corrections to the effective potential must often be taken into account, which can be hard to control explicitly. Moreover, the potential generated by the fluxes and other effects can sometimes destabilise rather than stabilise some of the moduli such that tachyons appear in the four-dimensional theory. For these and other reasons---such as the additional complication of flux quantisation conditions---it is technically challenging to realise explicit constructions with stabilised moduli, in particular if an absence of tachyons is not guaranteed by special properties such as supersymmetry or the saturation of a BPS bound.

Apart from these rather technical difficulties, there is another, possibly more fundamental obstacle to deriving the familiar low-energy physics from string theory. Due to the many possible choices for compactification details such as the internal geometry or the number of flux quanta, the vacuum of the four-dimensional effective theory seems to be far from unique. Instead, the effective potential yields a vast number of meta-stable vacua with different physical properties (which was estimated to be roughly of order $\sim\! 10^{500}$ in \cite{Ashok:2003gk}), the so-called string landscape \cite{Bousso:2000xa,Susskind:2003kw, Kachru:2003aw}. It is therefore not obvious how four-dimensional theories that resemble our universe are selected in string theory out of the huge number of possibilities at its disposal. Although there have been attempts to investigate the distribution of vacua statistically \cite{Ashok:2003gk, Douglas:2003um, Denef:2004ze, Douglas:2004zg, Denef:2004cf, Acharya:2005ez}, the exact properties of the landscape and its implications on the predictive power of string theory have not been fully understood so far. It also remains to be seen whether the landscape is truly a feature of string theory in a yet to be found complete formulation or just an artifact of our limited current understanding of it.
\\

A further source of complication is that many compactifications require the presence of higher-dimensional extended objects such as D-branes and orientifold planes (O-planes). These objects are localised on submanifolds of ten-dimensional spacetime and act as electric and magnetic sources for various $p$-form fields. Moreover, they source the gravitational field since they carry a tension that contributes to the vacuum energy density. Extended sources appear naturally in string theory as non-perturbative excitations, in addition to the fundamental strings, which are visible in the perturbative spectrum. D-branes arise in string theory as dynamical objects on which open strings can end. In the classical supergravity approximation, their dynamics is captured by a number of scalar and vector fields, which live on the brane worldvolume and come from the massless sector of the open string degrees of freedom. While the scalar fields are moduli determining the brane position, the vector fields can give rise to non-abelian gauge interactions. O-planes are special loci on a generalised manifold called orientifold, which is defined by taking the quotient of a manifold by a discrete symmetry of the background fields (such as an isometry of the metric) and an orientation reversal of the string world sheet. O-planes are the fixed points of such a projection, where only unoriented string states survive. Unlike D-branes, they are not dynamical. In addition to D-branes and O-planes, string theory also contains a variety of other extended objects such as NS$5$-branes, whose existence follows from string dualities and other arguments.

Localised sources are essential for several reasons. For one thing, they are often required in order to satisfy Gauss' law in compactifications with non-trivial background fluxes. For another, they have many attractive properties from a phenomenological point of view, such as to break supersymmetry or, in case of D-branes, allow for non-abelian gauge theories. They also induce a non-trivial warping of spacetime, which in their absence is ruled out at the supergravity level \cite{Gibbons:1984kp, deWit:1986xg, Maldacena:2000mw} and was proposed as a solution to the hierarchy problem \cite{Randall:1999ee, Giddings:2001yu}. Furthermore, they play an important role in string cosmology, as we will discuss in more detail below.

The downside is that localised sources significantly complicate the equations of motion such that a full solution is often out of reach, even in the limit of classical supergravity. Therefore, a simplification called smearing is often used in the literature, where the sources are assumed to extend over the whole compact space instead of being localised on a submanifold. Technically, this is done by replacing the delta functions, which determine the positions of the sources in the supergravity equations of motion, by regularised (often constant) functions that integrate to the same value. This has the advantage that computations simplify considerably since the equations of motion are only solved in an integrated sense, while the backreaction of the sources on the compact geometry and the internal fields is neglected. Besides, compactifications with smeared sources often allow to explicitly construct a four-dimensional effective low-energy theory using standard techniques such as Kaluza-Klein reduction or consistent truncations (see e.\ g.\ \cite{Roest:2004pk, Caviezel:2008ik, Caviezel:2009tu, Danielsson:2011au} for a discussion). In the presence of localised sources, on the other hand, this requires a formalism called warped effective field theory, which, although extensively studied in the literature, has not been fully understood so far \cite{DeWolfe:2002nn, Giddings:2005ff, Frey:2006wv, Burgess:2006mn, Douglas:2007tu, Koerber:2007xk, Shiu:2008ry, Douglas:2008jx, Frey:2008xw, Marchesano:2008rg, Martucci:2009sf, Chen:2009zi, Douglas:2009zn, Underwood:2010pm, Marchesano:2010bs, Blaback:2012mu}.

While convenient, the procedure of smearing also raises several doubts. For one thing, neglecting the backreaction of localised sources may affect four-dimensional observables such as moduli vevs or the cosmological constant. Since there is in general no reason to believe that such effects are small, it is not obvious that smeared solutions are good approximations to fully localised ones. Another worry is that smeared solutions do not satisfy the full ten-dimensional equations of motion. In particular, smearing conceals all non-trivial charge and energy profiles generated by the localised sources in compact space and thus potentially eliminates forces that render the solution unstable. It is therefore in general not clear whether the existence of a smeared solution implies the existence of a fully localised one. For these reasons, understanding the backreaction of localised sources is an important problem, in particular in view of the fact that most solutions in the literature are only known in the smeared limit (with a few exceptions being discussed \mbox{e.\ g.\ }in \cite{Giddings:2001yu, Schulz:2004ub, Grana:2006kf, Blaback:2010sj}).
\\

Given the huge number of possible universes distributed over the string landscape, one might expect that at least a few vacua with the right properties should exist in string theory, despite subtleties like moduli stabilisation. However, it has proven to be surprisingly hard to construct explicit string compactifications with features that agree with observations. While there has already been considerable progress in obtaining the (Minimally Supersymmetric) Standard Model from string theory (see e.\ g.\ \cite{Blumenhagen:2006ci, Blumenhagen:2005mu, Ibanez:2012xs} and references therein for reviews in the context of type II string theory and \cite{Braun:2005zv, Braun:2005nv} for some recent work on heterotic string theory), the construction of consistent solutions that are relevant for cosmology appears to be problematic. In particular, solutions with a positive cosmological constant are notoriously hard to obtain within string theory. These are believed to be necessary in order to describe the early universe, which is likely to have undergone an inflationary epoch of exponential expansion \cite{Guth:1980zm, Linde:1981mu, Albrecht:1982wi, Linde:1983gd, Bennett:2012fp, Hinshaw:2012fq}. Recent measurements have furthermore shown that the universe is also at present in a stage of accelerated expansion and approaches an asymptotically de Sitter phase in the distant future \cite{Bennett:2012fp, Hinshaw:2012fq, Riess:1998cb, Perlmutter:1998np}.

That solutions with a positive cosmological constant are so elusive in string theory has several reasons. Partly, it is due to the fact that they are not supersymmetric, which makes it much harder to ensure that all equations of motion are solved than for supersymmetric vacua. Since there is no particular reason for a vacuum to be \mbox{(meta-)} stable in absence of supersymmetry, also moduli stabilisation is an issue. To make matters even worse, de Sitter vacua are excluded for the simplest compactifications at the classical supergravity level \cite{Gibbons:1984kp, deWit:1986xg, Maldacena:2000mw} such that all candidate constructions require a considerable degree of complexity.

There are two main strategies pursued in the literature in order to nevertheless obtain de Sitter vacua in string theory (see also \mbox{\cite{Marsh:2011aa, Chen:2011ac, Sumitomo:2012wa, Bachlechner:2012at, Sumitomo:2012vx, Danielsson:2012by}} for related statistical studies). One idea is to take advantage of corrections to the classical supergravity approximation such as non-perturbative effects and higher derivative terms, which allow for a stabilisation of all moduli in an AdS vacuum of the four-dimensional low-energy theory. The vacuum energy density is then lifted to a positive value either due to spontaneous supersymmetry breaking or by including further ingredients that explicitly break supersymmetry. These constructions are best understood in the context of type IIB string theory. A second approach is to construct de Sitter solutions purely at the level of classical two-derivative supergravity. Such solutions, commonly referred to as classical de Sitter vacua, have been studied in both type IIA and type IIB string theory.
\\

The prime example for constructions of the first type is the KKLT scenario \cite{Kachru:2003aw}. There, the starting point are the type IIB flux compactifications on warped Calabi-Yau orientifolds described in \cite{Giddings:2001yu}, which arise as weak coupling limits of certain F-theory compactifications (a class of non-perturbative solutions of type IIB string theory \cite{Vafa:1996xn, Sen:1997gv}). These models admit a region in compact space with exponentially large but finite warping, which is realised by an embedding of a non-compact solution due to Klebanov and Strassler, the so-called warped deformed conifold \cite{Klebanov:2000hb}. In the four-dimensional low-energy theory, the models yield a Minkowski vacuum and are ``no-scale'', which means that some of the moduli, in particular the volume modulus, are left unstabilised. In the KKLT scenario, this is cured by taking non-perturbative effects into account, which modify the effective potential in such a way that all moduli are stabilised in a supersymmetric AdS vacuum. Such effects were shown to come, for example, from Euclidean D$3$-brane instantons \cite{Witten:1996bn} or gaugino condensation on stacks of D$7$-branes \cite{Kachru:2003aw}. The last step is then to add a small number of \dthree-branes, which explicitly break the supersymmetry of the four-dimensional vacuum. In the probe approximation, it was shown that, for a suitable choice of parameters, this procedure lifts the vacuum energy to a positive value without destabilising the moduli, thus leading to a meta-stable de Sitter vacuum. Since the flux background generates a non-trivial potential for the anti-branes, they are attracted towards the tip of the conifold, where their tension is highly redshifted due to the large warping present in that region \cite{Kachru:2002gs}. The cosmological constant generated by the anti-branes therefore turns out very small, as supported by observations.

A drawback of the KKLT proposal is that the non-perturbative corrections used for moduli stabilisation cannot be computed exactly (see however \cite{Koerber:2007xk, Baumann:2010sx, Heidenreich:2010ad, Dymarsky:2010mf} for some progress), and so it is very difficult to construct explicit models realising the scenario. Furthermore, the \dthree-branes responsible for the uplift are, as mentioned above, only considered in the probe approximation, which is similar to the previously discussed procedure of smearing in that it does not properly take into account the backreaction of the anti-branes. This might be an issue for the construction since its reliability heavily depends on a delicate balance of different contributions to the effective potential, which, lacking an understanding of the effects of backreaction, can only be estimated.

The backreaction of \dthree-branes on the Klebanov-Strassler background has been heavily studied in the literature \cite{DeWolfe:2008zy, McGuirk:2009xx, Bena:2009xk, Bena:2011hz, Bena:2011wh, Dymarsky:2011pm, Blaback:2011nz, Blaback:2011pn, Blaback:2012nf, Massai:2012jn, Bena:2012tx, Bena:2012bk, Bena:2012vz, Bena:2012ek, Gautason:2013, Vanriet:2013}. Increasing evidence suggests that it generates a singularity in the energy densities of the NSNS and RR three-form field strengths $H$ and $F_3$, although they do not directly couple to the anti-branes. Until recently, the presence of this singularity had only been demonstrated in simplified setups using various approximations. In earlier works on the subject, this involved a partial smearing of the anti-branes and a linearisation of the equations of motion around the Klebanov-Strassler background \cite{McGuirk:2009xx, Bena:2009xk, Bena:2011hz, Bena:2011wh}. In \cite{Dymarsky:2011pm}, it was therefore argued that the singularity might just be an artifact of perturbation theory and disappear in the full solution (see however \cite{Massai:2012jn}). Although it could then be shown in \cite{Bena:2012bk} that also the non-linear equations of motion necessarily lead to a singular solution, the analysis still required a partial smearing of the anti-branes. A similar singularity was, however, also observed in a toy model with \dsix-branes in a non-BPS flux background \cite{Blaback:2011nz, Blaback:2011pn, Bena:2012tx}, which is T-dual to a model with partially smeared \dthree-branes on $\mathbb{R}^{1,3}\! \times T^3\! \times \mathbb{R}^3$ \cite{Massai:2012jn}. In this simplified setup, it could be shown that the non-linear backreaction of fully localised \dsix-branes yields a divergent energy density of the $H$ flux, which is not directly sourced by them. Recently, a simple global argument was then presented in \cite{Gautason:2013}, which suggests that the singularity is not an artifact of the various approximations that had been used before but also generated by fully localised \dthree-branes that backreact on the full non-linear equations in the Klebanov-Strassler background. Whether this indicates that the corresponding solutions are unphysical or whether there is a mechanism to resolve the singularity in string theory is still under debate, although there are arguments supporting the first interpretation \cite{Blaback:2012nf, Bena:2012tx, Bena:2012vz, Bena:2012ek, Vanriet:2013}.

Apart from the KKLT scenario, several other constructions of meta-stable de Sitter vacua involving corrections to the classical supergravity equations have been proposed in the literature. One well-known example is the large volume scenario \cite{Balasubramanian:2005zx}, where the moduli are stabilised in a non-supersymmetric AdS vacuum at exponentially large volume by a combination of non-perturbative effects and higher derivative corrections. An uplift to de Sitter can then again be obtained by the inclusion of \dthree-branes, which might involve similar difficulties to the ones described above. Another example, which might evade at least some of the above-mentioned issues, is the recently proposed K{\"{a}}hler uplifting scenario, where both moduli stabilisation and the uplift to de Sitter are realised by an interplay of non-perturbative effects and higher derivative corrections to the K{\"{a}}hler potential of the four-dimensional effective supergravity theory \cite{Balasubramanian:2004uy, Westphal:2006tn, Rummel:2011cd, Louis:2012nb}.
\\

In contrast to the KKLT scenario and similar proposals, classical de Sitter vacua have the advantage of only including ingredients whose properties (such as their dependence on the supergravity fields) are well understood. They therefore allow, at least in principle, for the construction of fully explicit models. The search for such models is, however, complicated by a number of recently discovered no-go theorems, which are more restrictive than the older theorems found in \cite{Gibbons:1984kp, deWit:1986xg, Maldacena:2000mw}. These no-go theorems exclude de Sitter vacua and, more generally, also solutions admitting slow-roll inflation for a large class of compactifications in type IIA and type IIB supergravity \cite{Caviezel:2008ik, Caviezel:2009tu, Hertzberg:2007wc, Haque:2008jz, Steinhardt:2008nk, Caviezel:2008tf, Flauger:2008ad, Danielsson:2009ff, Wrase:2010ew} (see also \cite{VanRiet:2011yc} for an analysis of higher-dimensional de Sitter vacua). In order to keep the models as simple as possible, it is reasonable to consider orientifold compactifications in absence of additional ingredients such as branes or non-geometric fluxes. A minimal requirement to evade the no-go theorems is then an internal space with negative scalar curvature, which is induced by the negative tension of O-planes and can uplift the four-dimensional vacuum to de Sitter. Compactifications involving such negatively curved internal spaces have been considered in a variety of papers. However, all vacua that have been found so far either fail to satisfy the full ten-dimensional equations of motion \cite{Haque:2008jz, Silverstein:2007ac} or contain at least one tachyon in the spectrum of the four-dimensional low-energy theory \cite{Caviezel:2009tu, Danielsson:2011au, Caviezel:2008tf, Flauger:2008ad, Danielsson:2009ff, Danielsson:2010bc}.\footnote{Stable solutions have been argued to exist in setups involving non-geometric fluxes \cite{deCarlos:2009fq, deCarlos:2009qm, Dibitetto:2010rg, Danielsson:2013} (see also \cite{Danielsson:2012by}), but it is not clear whether the supergravity approximation is reliable there. Furthermore, there are solutions where supersymmetry is broken at the compactification scale \cite{Saltman:2004jh, Andriot:2010ju, Dong:2010pm}. In such setups, however, it may not be self-consistent to neglect Kaluza-Klein modes.} The origin of these tachyons is not explained by presently known no-go theorems against stability \cite{Covi:2008ea, Shiu:2011zt}. Whether they are just coincidental or a generic feature of classical de Sitter vacua is therefore not fully understood, although some progress on that issue has recently been made \cite{Danielsson:2012et}.

Since classical de Sitter solutions have so far only been obtained in the smeared limit, also backreaction is a potential issue. This is illustrated by a problem raised by Douglas and Kallosh in \cite{Douglas:2010rt}, where they pointed out that an internal manifold with an everywhere negative scalar curvature gives a positive contribution to the cosmological constant. They then investigated the effects of various ingredients of string compactifications on the internal manifold and found that, in the absence of large higher derivative corrections, negative contributions to its scalar curvature can only be generated by O-planes and warping. If the O-planes are smeared as in the classical de Sitter solutions described above, their negative energy density can produce a negative scalar curvature everywhere on the compact space. Fully localised O-planes, on the other hand, can only contribute a negative energy density on lower-dimensional submanifolds, which implies that large warping must be present everywhere else. Whether this is realised in a hypothetical localised version of classical de Sitter solutions is not obvious.
\\

We have seen from the above discussion that string compactifications, in particular those that are relevant for phenomenology, face a series of difficult problems that presently prevent us from fully understanding the low-energy physics of string theory. Many of these problems are, at least partially, related to our limited understanding of the backreaction of localised sources. The aim of this thesis is therefore to investigate different aspects of backreaction and thus shed light on some of the issues raised above, such as the Douglas-Kallosh problem, the problem of singularities that arise from \dthree-brane uplifts, and the problem of how backreaction modifies moduli values and the cosmological constant. This thesis is based on a series of articles that were published previously in \cite{Blaback:2010sj, Blaback:2011nz, Blaback:2011pn, Bena:2012tx, Gautason:2013}. It is organised as follows.

In Chapter \ref{ch:smeared-vs-loc}, we present a class of flux compactifications of type II string theory, both in the smeared limit and with fully localised sources. These solutions are T-dual to the solutions described in \cite{Giddings:2001yu} and have the special property that they saturate a BPS bound. By comparing the smeared and the localised solutions, we are able to explicitly study the effects of backreaction in these solutions. In particular, we show that their BPS property guarantees that the moduli vevs do not shift in the localised solutions as compared to the smeared ones. Since some of our solutions have an everywhere negatively curved internal space in the smeared limit, we can also explicitly study the Douglas-Kallosh problem. We argue that it is resolved in the corresponding localised solutions by large warping that is induced in all regions of compact space where no O-planes are present. Furthermore, we discuss an example for a non-BPS solution and find that the simple localisation prescription that worked for the BPS case does not lead to consistent localised solutions in the non-BPS case.

In Chapter \ref{ch:simple-non-bps}, we elaborate on the effects of backreaction in non-BPS solutions using the example of a simple toy model on AdS$_7 \times S^3$ with D$6$-branes or \dsix-branes that are not mutually BPS with the background fluxes. We first show that a solution with smeared sources exists for this setup and prove that there is a consistent truncation that is stable in the closed string sector. We then consider the most general ansatz for a localisation of the solution that is compatible with the symmetries of the setup. For partially localised sources with different types of regularised source profiles, we show that the only solution allowed by the equations of motion is the smeared solution. For fully localised sources, we find that only two boundary conditions for the supergravity fields are locally allowed in the near-brane region. One of the boundary conditions is globally excluded due to a topological argument. The second boundary condition is consistent both locally and globally but yields a singularity in the energy density of the NSNS $H$ flux, which is reminiscent of the singularity found in the Klebanov-Strassler solution perturbed by \dthree-branes. We then discuss several possibilities for a resolution of the singularity in our model and explicitly analyse whether the D$6$/\dsix-branes polarise into a D$8$-brane due to the Myers effect. In order to make contact with the KKLT scenario, we first consider a non-compact version of our model, which exhibits the same flux singularity and was argued to be T-dual to a setup that approximates the near-tip physics of partially smeared \dthree-branes in the Klebanov-Strassler background \cite{Massai:2012jn}. In the non-compact model, we find that the singularity is not resolved by a polarisation of the branes, suggesting that the same may be true for the setup of \dthree-branes in the Klebanov-Strassler background. We also comment on the compact version of our model and find that the polarisation potential is modified there due to curvature terms such that, contrary to the non-compact case, a polarisation is not excluded.

In Chapter \ref{ch:scaling-symmetries}, we exploit two global scaling symmetries of the classical type II supergravity action to show that the classical cosmological constant in type II flux compactifications can be written as a sum of terms from the action of localised sources and a contribution due to non-trivial background fluxes. We furthermore argue that the symmetries often allow to set the flux contribution to zero by a convenient gauge choice such that the cosmological constant is fully determined by the classical boundary conditions of the supergravity fields in the near-source region. We discuss several examples of well-understood compactifications and show that our arguments are consistent with all expectations. We then apply our result to the KKLT scenario and give a simple global argument indicating that, under a few assumptions that we discuss in detail, the backreaction of fully localised \dthree-branes in this setup yields a singularity in the energy densities of the NSNS and RR three-form field strengths $H$ and $F_3$.

We conclude in Chapter \ref{ch:conclusions} with a discussion of our results and an outlook on interesting future directions of research. In Appendix \ref{app:conventions}, we state our notation and conventions and collect a number of useful formulae. Appendices \ref{app:computations.curvature} to \ref{app:ccandsources} contain several lengthy calculations that were separated from the main text for readability.

\cleardoublepage
\thispagestyle{plain}

\section{Smeared and Localised Sources in BPS Compactifications}
\label{ch:smeared-vs-loc}

In this chapter, we study backreaction effects in a class of solutions of type II supergravity, which we construct both in the smeared limit and with fully localised sources from the point of view of the ten-dimensional equations of motion \cite{Blaback:2010sj}. The solutions we consider are orientifold compactifications to $d$-dimensional Minkowski space with an internal space that, in the smeared limit, is either Ricci-flat or an everywhere negatively curved twisted torus. The solutions with Ricci-flat internal space are $d$-dimensional generalisations of the compactifications to four dimensions discussed in \cite{Giddings:2001yu}, which we will refer to as the GKP solution. The solutions on twisted tori are $d$-dimensional generalisations of solutions that were obtained in the smeared limit in \cite{Gurrieri:2002wz, Kachru:2002sk} and discussed with localised sources in \cite{Schulz:2004ub, Grana:2006kf}. Furthermore, all solutions are related to one another and to the GKP solution by a chain of T-dualities.

The solutions we discuss are not necessarily supersymmetric. However, they have the special property of saturating a BPS bound. This is analogous to the GKP solution, which is also BPS but not necessarily supersymmetric.\footnote{The GKP solution is only supersymmetric if the (imaginary self-dual) $G$ flux is of complexity type $\left({2,1}\right)$. See also \cite{Lust:2008zd} for an analysis of BPS but non-supersymmetric solutions in the language of generalised geometry.} We will argue in this chapter that BPSness plays a crucial role in whether or not smearing is a reasonable approximation. In particular, we will find that, due to their property of being BPS, our solutions can be localised using a relatively simple localisation prescription. This property also guarantees that the individual localisation corrections cancel out in the effective potential for the moduli such that their vevs do not get shifted in the localised solutions as compared to the smeared ones. In non-BPS solutions, on the other hand, it is less obvious that smearing is justified, and we will show for an example that the simple localisation procedure that is succesful in the BPS case does not yield a consistent localised solution there. Even a more general localisation ansatz can lead to problems in non-BPS solutions, as will be discussed in Chapter \ref{ch:simple-non-bps} in more detail. Since some of our BPS solutions have an everywhere negatively curved internal space in the smeared limit, we can also explicitly address the Douglas-Kallosh problem. We find that the problem is evaded in these solutions since their localisation induces large warping everywhere on the compact space.

This chapter is based on the results presented in \cite{Blaback:2010sj} and organised as follows. In Sections \ref{ch:smeared-vs-loc.bps-ricciflat} and \ref{ch:smeared-vs-loc.bps-twisted}, we consider BPS compactifications to $d$-dimensional Minkowski space that either have a Ricci-flat or an everywhere negatively curved internal space in the smeared limit. We then show how the solutions can be promoted to solutions with fully localised sources. In Section \ref{ch:smeared-vs-loc.non-bps}, we discuss a simple example for a smeared non-BPS solution and find that the localisation procedure that was succesful for the BPS solutions does not lead to a consistent solution in the non-BPS case. In Section \ref{ch:smeared-vs-loc.discussion}, we comment on localisation in BPS and non-BPS solutions and discuss how the Douglas-Kallosh problem is evaded in the localisation of the BPS solutions that have an everywhere negatively curved internal space in the smeared limit.
\\

\subsection{BPS Solutions with Ricci-flat Internal Space}
\label{ch:smeared-vs-loc.bps-ricciflat}

In this section, we present a class of flux compactifications to $(p+1)$-dimensional Minkow-ski space with spacetime-filling O$p$-planes, which have a Ricci-flat internal manifold in the smeared limit. We consider solutions with $p = 1, \ldots, 6$, where the special case $p=3$ corresponds to the well-known GKP solution \cite{Giddings:2001yu}.\footnote{More precisely, the sources in our solutions correspond to O$p$-planes for $p=2,3,6$ and $\overline{\textrm{O}p}$-planes for $p=1,4,5$. By flipping the signs of the RR fields, however, all solutions can also be obtained with the corresponding sources of the opposite charge such that this subtlety is not relevant here. For $p=6$, our solutions are also related to the solution of \cite{Janssen:1999sa}, where the same setup is considered in a non-compact setting with D$6$-branes instead of O$6$-planes. This suggests that a non-compact version, where the O$p$-planes are replaced by D$p$-branes, exists for all of our solutions with general $p$.} We first construct the solutions in the smeared limit and then show how they can be promoted to solutions with fully localised sources. Our notation and conventions as well as the equations of motion for type II supergravity used throughout this thesis are stated in Appendix \ref{app:conventions}.

\subsubsection{Smeared Solutions}

Our ansatz in the smeared limit is as follows. We assume the non-zero fields
\begin{equation}
\phi = \phi_0, \qquad H, \qquad F_{6-p}, \label{eq:bps1_ansatz_smeared}
\end{equation}
where the dilaton is constant in compact space and all RR fields other than $F_{6-p}$ are identically zero. Furthermore, we assume that the metric has the form of a direct product,
\begin{equation}
\d s^2 = \d s_{p+1}^2 + \d s_{9-p}^2,
\end{equation}
where
\begin{equation}
\d s_{p+1}^2 = g_{\mu\nu} \d x^\mu \d x^\nu, \qquad \d s_{9-p}^2 = g_{ik} \d x^i \d x^k
\end{equation}
and the external spacetime is Minkowski. The O$p$-planes enter the Einstein equation, the dilaton equation and the Bianchi identity for $F_{8-p}$. In order to smear the sources, we set the delta functions determining their location on the internal manifold equal to one in the Einstein and dilaton equations,
\begin{equation}
\delta(\Sigma) \to 1.
\end{equation}
This implies that the delta form $\delta_{9-p}$ in the Bianchi identity is replaced by the internal volume form $\epsilon_{9-p}$.\footnote{Since the delta function contains a factor $\sqrt{g_{9-p}}\,^{-1}$, this is only consistent if we set the volume of the internal manifold to $1$. That this can always be done is guaranteed by scaling symmetries of the classical supergravity action, which are discussed in Chapter \ref{ch:scaling-symmetries} in more detail.}

The external Einstein equation then reads
\begin{equation}
0 = R_{\mu\nu} = -\frac{1}{8} g_{\mu\nu} \e^{-\phi_0} |H|^2 - \frac{5-p}{16} g_{\mu\nu} \e^{\tfrac{p-1}{2}\phi_0} |F_{6-p}|^2 + \frac{7-p}{16} g_{\mu\nu} \e^{\tfrac{p-3}{4}\phi_0} \mu_p, \label{eq:bps1_ext_einst_smeared}
\end{equation}
and the Bianchi identity for $F_{8-p}$ is
\begin{equation}
0 = H \w F_{6-p} - \mu_p \, \epsilon_{9-p}. \label{eq:bps1_bianchi1_smeared}
\end{equation}
Rewrititing \eqref{eq:bps1_ext_einst_smeared} in form notation (e.\ g.\ using the identities collected in Appendix \ref{app:conventions.tensors}), it can be combined with \eqref{eq:bps1_bianchi1_smeared} such that the source terms cancel out. This yields the equation
\begin{equation}
0 = H \wedge F_{6-p} - \frac{2}{7-p} \e^{-\tfrac{p+1}{4}\phi_0} \star_{9-p} H \wedge H - \frac{5-p}{7-p} \e^{\tfrac{p+1}{4}\phi_0} \star_{9-p} F_{6-p} \wedge F_{6-p}.
\end{equation}
The ansatz
\begin{equation}
F_{6-p} = \left({-1}\right)^p \e^{-\tfrac{p+1}{4}\phi_0} \kappa \star_{9-p} H \label{eq:bps1-ansatz-f}
\end{equation}
then leads to a quadratic equation for $\kappa$ with the solutions $\kappa = 1$ and $\kappa =\frac{2}{5-p}$. For $p=5$, one finds a linear equation with the single solution $\kappa = 1$.

For $p \neq 5$, one of the solutions for $\kappa$ can be discarded due to the dilaton equation,
\begin{equation}
0 = \nabla^2 \phi_0 = -\frac{1}{2} \e^{-\phi_0} |H|^2 + \frac{p-1}{4} \e^{\tfrac{p-1}{2}\phi_0} |F_{6-p}|^2 - \frac{p-3}{4} \e^{\tfrac{p-3}{4}\phi_0} \mu_p. \label{eq:bps1_eomdil}
\end{equation}
Again rewriting the equation in form notation, we can substitute \eqref{eq:bps1_bianchi1_smeared} and \eqref{eq:bps1-ansatz-f} in order to eliminate the source terms. This yields another quadratic equation for $\kappa$ with the solutions $\kappa = 1$ and $\kappa = \frac{2}{1-p}$. For the case $p=1$, we only obtain a linear equation with the solution $\kappa = 1$. Comparing this with our previous results for $\kappa$, we conclude that $\kappa = 1$ for all $p$. We thus find
\begin{equation}
F_{6-p} = \left({-1}\right)^p \e^{-\tfrac{p+1}{4}\phi_0} \star_{9-p} H. \label{eq:bps1_bps_cond}
\end{equation}
As will become clear below, this condition is responsible for the BPS bound saturated by the solutions. We will therefore refer to it as the BPS condition. For $p=3$, it is equivalent to the ISD condition on the $G$ flux in the language of \cite{Giddings:2001yu}.\footnote{It is sometimes convenient to combine $F_3$ and $H$ into a complex three-form flux $G = F_3 - i \e^{-\phi} H$. Imaginary self-dual (ISD) flux then satisfies $\star_6 G = i G$, while imaginary anti-self-dual (IASD) flux satisfies $\star_6 G = - i G$.} For $p=1$, there is a subtlety due to the self-duality of $F_5$ that has to be taken into account in the derivation. The result is that the ten-dimensional Hodge dual has to be added to the right-hand side of \eqref{eq:bps1_bps_cond}.

Substituting \eqref{eq:bps1_bps_cond} in \eqref{eq:bps1_bianchi1_smeared}, we find that the fluxes are related to the charge of the O$p$-planes,
\begin{equation}
\mu_p = \e^{\tfrac{p+1}{4}\phi_0} |F_{6-p}|^2 = \e^{-\tfrac{p+1}{4}\phi_0} |H|^2. \label{eq:bps1_cond2}
\end{equation}
The internal Einstein equation is
\begin{align}
R_{ik} &= \frac{1}{2} \e^{-\phi_0} |H|^2_{ik} + \frac{1}{2} \e^{\tfrac{p-1}{2}\phi_0} |F_{6-p}|^2_{ik} - \frac{1}{8} g_{ik} \e^{-\phi_0} |H|^2 - \frac{5-p}{16} g_{ik} \e^{\tfrac{p-1}{2}\phi_0} |F_{6-p}|^2 \nl - \frac{p+1}{16} g_{ik} \e^{\tfrac{p-3}{4}\phi_0} \mu_p
\end{align}
and, using \eqref{eq:bps1_bps_cond} and \eqref{eq:bps1_cond2}, yields the condition $R_{ik} = 0$. Thus, it follows from our ansatz that the internal space is Ricci-flat.
Furthermore, we have to impose that the Bianchi identities for $H$ and $F_{6-p}$ are satisfied,
\begin{equation}
\d H = 0, \qquad \d F_{6-p} = 0. \label{eq:bps1_bianchis}
\end{equation}
All other equations of motion are then satisfied trivially.
\\

{\noindent\bf Summary of the solution.} The a priori non-trivial fields in our ansatz for solutions with smeared O$p$-planes (with $p=1,\ldots,6$) are
\begin{equation}
\phi=\phi_0, \qquad H, \qquad F_{6-p}, \qquad R_{ik}.
\end{equation}
This leads to $(p+1)$-dimensional Minkowski solutions if the following conditions are satisfied:
\begin{itemize}
\item The RR and NSNS fluxes are related by the BPS condition,
\begin{equation}
F_{6-p} = \left({-1}\right)^p \e^{-\tfrac{p+1}{4}\phi_0} \star_{9-p} H.
\end{equation}
\item The amount of RR and NSNS flux is fixed in terms of the O$p$-plane charge,
\begin{equation}
\mu_p = \e^{\tfrac{p+1}{4}\phi_0} |F_{6-p}|^2 = \e^{-\tfrac{p+1}{4}\phi_0} |H|^2.
\end{equation}
\item The internal space is Ricci-flat,
\begin{equation}
R_{ik} = 0.
\end{equation}
\item The Bianchi identities for $H$ and $F_{6-p}$ are satisfied,
\begin{equation}
\d H = 0, \qquad \d F_{6-p} = 0.
\end{equation}
\end{itemize}
For $p=1$, the Hodge dual needs to be added to the expression for $F_5$.

\subsubsection{Localised Solutions}
\label{ch:smeared-vs-loc.bps-ricciflat.loc}

Since O$p$-planes couple to the metric, the dilaton and the RR potential $C_{p+1}$, we expect that the localised solutions have non-trivial warping, a dilaton that varies over the internal space and a non-zero $F_{8-p}$ field strength. We therefore assume that the fields
\begin{equation}
\phi, \qquad H, \qquad F_{6-p}, \qquad F_{8-p}
\end{equation}
are non-zero, whereas all other RR fields vanish. We furthermore assume a warped metric of the form
\begin{equation}
\d s^2 = \e^{2aA} \d \tilde s_{p+1}^2 + \e^{2bA} \d \tilde s_{9-p}^2 \label{eq:bp1-warpedmetric}
\end{equation}
with
\begin{equation}
\d \tilde s_{p+1}^2 = \tilde g_{\mu\nu} \d x^\mu \d x^\nu, \qquad \d \tilde s_{9-p}^2 = \tilde g_{ik} \d x^i \d x^k,
\end{equation}
where $a$ and $b$ are numbers that are determined below and $A$ is a function of the internal coordinates called the warp factor. In the following, we always put tildes on the metric if the warp factor dependence is taken out. We will also use tildes to indicate that objects such as covariant derivatives, Hodge operators or contractions of tensors are constructed with the unwarped metric rather than the warped one.

Since we can always absorb powers of the warp factor into the internal metric $\tilde g_{ik}$, the number $b$ is a gauge choice. However, it obtains an absolute meaning if we take $\tilde g_{ik}$ to be the internal metric before localisation. Moreover, we can always normalise $A$ such that $a=1$. The equations of motion then also fix $b$, which yields
\begin{equation}
a = 1, \qquad b=\frac{p+1}{p-7}.
\end{equation}
As derived in Appendix \ref{app:computations.curvature.block}, the non-trivial components of the Ricci tensor for the metric \eqref{eq:bp1-warpedmetric} are then
\begin{align}
R_{\mu\nu} &= - \e^{\tfrac{16}{7-p}A} \tilde g_{\mu\nu} \tilde \nabla^2 A + \tilde R_{\mu\nu}, \label{eq:bps1-ricci1} \\
R_{ik} &= - \frac{p+1}{p-7} \tilde g_{ik} \tilde \nabla^2 A + 8 \frac{p+1}{p-7} (\partial_i A)(\partial_k A) + \tilde R_{ik}, \label{eq:bps1-ricci2}
\end{align}
whereas all mixed components are zero.

Our ansatz for $F_{8-p}$ is
\begin{equation}
F_{8-p} = - \e^{-2\left({p+1}\right)A - \tfrac{p-3}{2}\phi} \tilde \star_{9-p} \d \alpha, \label{eq:bps1_ansatz_f8-p}
\end{equation}
where $\alpha$ is a function of the internal coordinates. For $p=3$, the Hodge dual has to be added to this expression due to the self-duality of $F_5$. The ansatz then coincides with the one in \cite{Giddings:2001yu}. A subtlety also occurs for $p=2$, where $F_6$ and $F_4$ are dual to one another. The correct expression for $F_4$ is then obtained by adding the dual of \eqref{eq:bps1_ansatz_f8-p} to the expression for $F_{6-p}$ that is stated in \eqref{eq:bps1_bps_cond_loc} for general $p$. For later convenience, we also note that
\begin{align}
\d F_{8-p} &= \left[{\e^{\tfrac{\left({p-5}\right)\left({p+1}\right)}{7-p}A - \tfrac{p-3}{2}\phi} \tilde \nabla^2 \alpha + \e^{\tfrac{\left({9-p}\right)\left({p+1}\right)}{7-p}A} \tilde g^{ik}  \Big(\partial_i \e^{-2\left({p+1}\right)A - \tfrac{p-3}{2}\phi}\Big) \left({\partial_k \alpha}\right)}\right] \nl \cdot \left({-1}\right)^{p+1} \epsilon_{9-p} \label{eq:bps1_f8-p_and_alpha1}
\end{align}
and
\begin{align}
& |F_{8-p}|^2 = \e^{2\tfrac{\left({p+1}\right)\left({p-6}\right)}{7-p}A-\left({p-3}\right)\phi} (\partial \alpha)^2, \label{eq:bps1_f8-p_and_alpha2} \\
& |F_{8-p}|^2_{ik} = \e^{-2\left({p+1}\right)A-\left({p-3}\right)\phi} \left[{ \tilde g_{ik} (\partial \alpha)^2 - (\partial_i \alpha)(\partial_k \alpha) }\right], \label{eq:bps1_f8-p_and_alpha3}
\end{align}
which can be verified using the identities of Appendix \ref{app:conventions.tensors}.

In order to show that the localised solutions satisfy a BPS condition like the smeared ones, we have to combine three different equations. These are the dilaton equation,
\begin{align}
\e^{2\tfrac{p+1}{7-p}A} \tilde \nabla^2 \phi &= -\frac{1}{2} \e^{-\phi} |H|^2 + \frac{p-1}{4} \e^{\tfrac{p-1}{2}\phi} |F_{6-p}|^2 + \frac{p-3}{4} \e^{\tfrac{p-3}{2}\phi} |F_{8-p}|^2 \nl - \frac{p-3}{4} \e^{\tfrac{p-3}{4}\phi} \mu_p \delta(\Sigma), \label{eq:bps1_dilaton_loc}
\end{align}
the external Einstein equation,
\begin{align}
R_{\mu\nu} &= - \e^{\tfrac{16}{7-p}A} \tilde g_{\mu\nu} \tilde \nabla^2 A + \tilde R_{\mu\nu} \nll =
-\frac{1}{8} \tilde g_{\mu\nu} \e^{2A -\phi} |H|^2 - \frac{5-p}{16} \tilde g_{\mu\nu} \e^{2A + \tfrac{p-1}{2}\phi} |F_{6-p}|^2 - \frac{7-p}{16} \tilde g_{\mu\nu} \e^{2A + \tfrac{p-3}{2}\phi} |F_{8-p}|^2 \nl + \frac{7-p}{16} \tilde g_{\mu\nu} \e^{2A + \tfrac{p-3}{4}\phi} \mu_p \delta(\Sigma), \label{eq:bps1_ext_einst_loc}
\end{align}
and the Bianchi identity for $F_{8-p}$,
\begin{equation}
\d F_{8-p} = H \wedge F_{6-p} - \mu_p \delta(\Sigma)\, \epsilon_{9-p}. \label{eq:bps1_bianchi_loc}
\end{equation}
Using \eqref{eq:bps1_f8-p_and_alpha1} and \eqref{eq:bps1_f8-p_and_alpha2} in these equations, they can be combined into
\begin{align}
\tilde \nabla^2\! \left[{\e^{\left({p+1}\right)A+\tfrac{p-3}{4}\phi} + \left({-1}\right)^p \alpha}\right] &= \e^{\tfrac{\left({p+1}\right)\left({9-p}\right)}{p-7}A-\tfrac{p-3}{4}\phi} \left[{\partial\Big( \e^{\left({p+1}\right)A+\tfrac{p-3}{4}\phi} + \left({-1}\right)^p \alpha \Big)}\right]^2 \nl + \frac{1}{2} \e^{\tfrac{\left({p+1}\right)\left({p-5}\right)}{p-7}A+\tfrac{3p-5}{4}\phi} \left|{F_{6-p} - \left({-1}\right)^p \e^{-\tfrac{p+1}{4}\phi} \star_{9-p} H}\right|^2 \nl + \e^{\tfrac{\left({p-3}\right)^2}{p-7}A+\tfrac{p-3}{4}\phi} \tilde R_{p+1},
\end{align}
where the two squares on the right-hand side are contracted with warped metrics. Since the left-hand side of the equation integrates to zero on a compact space, we find $\tilde R_{p+1} \le 0$. For Minkowski solutions with $\tilde R_{p+1}=0$, both squares on the right-hand side need to vanish, which yields
\begin{equation}
\alpha = - \left({-1}\right)^p \e^{\left({p+1}\right)A+\tfrac{p-3}{4}\phi} + \textrm{const.} \label{eq:bps1_cond_alpha}
\end{equation}
and the BPS condition
\begin{equation}
F_{6-p} = \left( {-1}\right)^p \e^{-\tfrac{p+1}{4}\phi} \star_{9-p} H. \label{eq:bps1_bps_cond_loc}
\end{equation}

The remaining equations of motion are solved as follows. Combining \eqref{eq:bps1_dilaton_loc} and \eqref{eq:bps1_ext_einst_loc} to eliminate the source terms and using \eqref{eq:bps1_bps_cond_loc}, we find
\begin{equation}
\tilde \nabla^2\! \left({4\frac{p-3}{7-p}A - \phi}\right) = 0.
\end{equation}
Since harmonic functions are constant on a compact space, this implies
\begin{equation}
\phi = 4\frac{p-3}{7-p}A + \phi_0. \label{eq:bps1_dilaton_expr}
\end{equation}
Substituting this into \eqref{eq:bps1_bps_cond_loc} and pulling out the warp factor that is hidden in the Hodge operator, we obtain
\begin{equation}
F_{6-p} = \left( {-1}\right)^p \e^{-\tfrac{p+1}{4}\phi_0} \tilde \star_{9-p} H
\end{equation}
and thus recover \eqref{eq:bps1_bps_cond}. The BPS condition has therefore not changed as compared to the smeared limit. Using \eqref{eq:bps1_cond_alpha}, \eqref{eq:bps1_bps_cond_loc} and \eqref{eq:bps1_dilaton_expr} in \eqref{eq:bps1_bianchi_loc}, we furthermore obtain the condition
\begin{equation}
\tilde \nabla^2 \e^{\tfrac{16}{p-7}A} = -\e^{-\phi_0} |\tilde H|^2 + \e^{\tfrac{p-3}{4}\phi_0} \mu_p \tilde \delta(\Sigma), \label{bps1_cond2_loc}
\end{equation}
where we have pulled out the warp factor from the inverse metric determinant $\sqrt{g_{9-p}}\,^{-1}$ in $\delta(\Sigma)$. This condition relates the fluxes to the O$p$-plane charge as in the smeared limit and fixes the warp factor in terms of the delta function. Integrating it over the compact space, we find that \eqref{eq:bps1_cond2} is still valid in the localised solution. Moreover, the condition implies that warping is comparable to the fluxes everywhere on the compact space where the O$p$-planes do not contribute.

The internal Einstein equation reads
\begin{align}
R_{ik} &= - \frac{p+1}{p-7} \tilde g_{ik} \tilde \nabla^2 A + 8 \frac{p+1}{p-7} (\partial_i A) (\partial_k A) + \tilde R_{ik} \nll = \frac{1}{2} (\partial_i \phi) (\partial_k \phi) + \frac{1}{2} \e^{\tfrac{p-1}{2}\phi} |F_{6-p}|^2_{ik} - \frac{5-p}{16} g_{ik} \e^{\tfrac{p-1}{2}\phi} |F_{6-p}|^2 + \frac{1}{2} \e^{\tfrac{p-3}{2}\phi} |F_{8-p}|^2_{ik} \nl - \frac{7-p}{16} g_{ik} \e^{\tfrac{p-3}{2}\phi} |F_{8-p}|^2 + \frac{1}{2} \e^{-\phi} |H|^2_{ik} - \frac{1}{8} g_{ik} \e^{-\phi} |H|^2 \nl - \frac{p+1}{16} g_{ik} \e^{\tfrac{p-3}{4}\phi} \mu_p \delta(\Sigma)
\end{align}
and, using above results and the identities \eqref{eq:bps1_f8-p_and_alpha2} and \eqref{eq:bps1_f8-p_and_alpha3}, reduces to
\begin{equation}
\tilde R_{ik} = 0. \label{bps1_conf_ricciflat}
\end{equation}
Assuming that $H$ and $F_{6-p}$ satisfy the Bianchi identities
\begin{equation}
\d H = 0, \qquad \d F_{6-p} = 0
\end{equation}
as in the smeared limit, one can then check that all other equations of motion are also satisfied.
\\

{\noindent\bf Summary of the solution.} We have shown that the solutions obtained in the smeared limit can be localised by introducing warping, allowing the dilaton to vary over the compact space and adding a non-zero $F_{8-p}$ field strength. For $p=2$, there is a subtlety since $F_6$ and $F_4$ are dual to one another, which leads to
\begin{equation}
F_4 = \e^{-\tfrac{3}{4}\phi} \star_7 H + \e^{-6A+\tfrac{1}{2}\phi} \star_{10} \tilde \star_7 \d \alpha.
\end{equation}
The dilaton and $F_{8-p}$ can be written in terms of the warp factor using \eqref{eq:bps1_ansatz_f8-p}, \eqref{eq:bps1_cond_alpha} and \eqref{eq:bps1_dilaton_expr}, while the warp factor itself is fixed in terms of the O$p$-plane charge by \eqref{bps1_cond2_loc}. The BPS condition \eqref{eq:bps1_bps_cond} has not changed after localisation due to a cancellation of warp and dilaton factors in \eqref{eq:bps1_bps_cond_loc}. Integrating \eqref{bps1_cond2_loc} over the compact space, we find that also \eqref{eq:bps1_cond2} is still valid in the localised solution. Furthermore, the Bianchi identities \eqref{eq:bps1_bianchis} remain satisfied. The internal space changes from Ricci-flat to conformally Ricci-flat, as follows from \eqref{bps1_conf_ricciflat}.
\\

\subsection{BPS Solutions on Negatively Curved Twisted Tori}
\label{ch:smeared-vs-loc.bps-twisted}

In this section, we consider flux compactifications to $p$-dimensional Minkowski space for $p=1,\ldots,6$ with O$p$-planes that are spacetime-filling and wrap a one-cycle on the internal manifold. In the smeared limit, the solutions have an everywhere negatively curved internal space and can be obtained from the solutions discussed above by performing a T-duality \cite{Buscher:1987sk, Hassan:1999mm} along one of the directions of the $H$ flux. Starting, for example, with a solution that has a torus as internal space, this leads to a solution on a negatively curved twisted torus.

Let us assume that $H$ has one leg in the $9$-direction, i.\ e.\ $H = H_{ik9}\, \d x^i \w \d x^k \w \d x^9$, and that the internal space has a $\mathrm{U}(1)$ isometry corresponding to shifts of $x^9$ such that we can perform a T-duality along this direction. It is then convenient to write the T-dual solutions in terms of the basis forms
\begin{equation}
\d x^\mu, \qquad \d x^i, \qquad e^9 = \d x^9 + \frac{1}{2} f_{ik}^9 x^i \d x^k,
\end{equation}
where $\mu = 0, \ldots, p-1$ and $i = p, \ldots, 8$ \cite{Shelton:2005cf}. In this basis, $e^9$ is not closed but satisfies
\begin{equation}
\d e^9 = \frac{1}{2} f_{ik}^9 \d x^i \w \d x^k
\end{equation}
with non-trivial structure constants $f_{ik}^9 = H_{ik9}$, which give a contribution to the Ricci curvature. The T-dual solutions thus have vanishing $H$ flux but non-vanishing $f_{ik}^9$, which is often referred to as ``metric flux''.

The T-duality relations in the smeared limit suggest that our setup should also allow for localised solutions, where the O$p$-planes extend along the directions $\mu = 0,\ldots,p-1$ and the $9$-direction. Our ansatz for the metric is then\footnote{Unlike in a usual vielbein basis, $e^9$ does not have unit norm here since $g_{99}$ is in general not equal to one.}
\begin{equation}
\d s^2 = \e^{2A} \tilde g_{\mu\nu} \d x^\mu \d x^\nu + \e^{2\tfrac{p+1}{p-7}A} \tilde g_{ik} \d x^i \d x^k + \e^{2A} \tilde g_{99} e^9 e^9. \label{eq:bp2-warpedmetric}
\end{equation}
Since $e^9$ is not closed, the Ricci tensor is more complicated for this metric than for the one used in Section \ref{ch:smeared-vs-loc.bps-ricciflat}. As derived in Appendix \ref{app:computations.curvature.twisted}, it gets contributions from the unwarped metric $\tilde g_{MN}$, the metric fluxes $f_{ik}^9$ and the warp factor $A$. This yields
\begin{align}
R_{\mu\nu} &= - \e^{\tfrac{16}{7-p}A} \tilde g_{\mu\nu} \tilde \nabla^2 A + \tilde R_{\mu\nu}, \label{ricci-twistedtori0} \\
R_{ik} &= - \frac{p+1}{p-7} \tilde g_{ik} \tilde \nabla^2 A + 8 \frac{p+1}{p-7} (\partial_i A) (\partial_k A) - \frac{1}{2} \e^{\tfrac{16}{7-p}A} \tilde g_{99} |\d \tilde e^9|^2_{ik} + \tilde R_{ik}, \label{ricci-twistedtori1} \\
R_{9k} &= - \frac{8}{p-7} \e^{\tfrac{16}{7-p}A} \tilde g_{99} |e^9 \wedge \d A \cdot \d \tilde e^9|_{9k} + \tilde R_{9k}, \\
R_{99} &= - \e^{\tfrac{16}{7-p}A} \tilde g_{99} \tilde \nabla^2 A + \frac{1}{2} \e^{\tfrac{32}{7-p}A} \tilde g_{99} \tilde g_{99} |\d \tilde e^9|^2 + \tilde R_{99}, \label{ricci-twistedtori2}
\end{align}
where we again use tildes for unwarped metrics and objects constructed from unwarped metrics such as contractions of tensors. Furthermore, we will assume $\tilde R_{99} = 0$ in the following.

\subsubsection{Smeared Solutions}

Our ansatz in the smeared limit is as follows. The a priori non-zero fields are
\begin{equation}
\phi = \phi_0, \qquad F_{8-p} = m_{7-p} \wedge e^9, \qquad R_{\mu\nu}, \qquad R_{ik}, \qquad R_{9k}, \qquad R_{99}, \label{eq:bps2_ansatz_smeared}
\end{equation}
where the dilaton is assumed to be constant and $m_{7-p}$ is a closed $\left({7-p}\right)$-form. All other RR fields and $H$ vanish. As in Section \ref{ch:smeared-vs-loc.bps-ricciflat}, we set the delta functions in the equations of motion equal to one in order to smear the O$p$-planes,
\begin{equation}
\delta(\Sigma) \to 1.
\end{equation}
Since smeared sources do not induce any warping, we furthermore take the metric to be of the form \eqref{eq:bp2-warpedmetric} with $A=0$. For $p=3$, the self-duality of $F_5$ requires us to add the Hodge dual to above ansatz. Some of the equations in this section then have to be modified accordingly.

Substituting the dilaton equation
\begin{equation}
0 = \nabla^2 \phi_0 = \frac{p-3}{4} \e^{\tfrac{p-3}{2}\phi_0} |F_{8-p}|^2 - \frac{p-3}{4} \e^{\tfrac{p-3}{4}\phi_0} \mu_p \label{eq:bps2_eomdil}
\end{equation}
into the external Einstein equation
\begin{equation}
R_{\mu\nu} = -\frac{7-p}{16} g_{\mu\nu} \e^{\tfrac{p-3}{2}\phi_0} |F_{8-p}|^2 + \frac{7-p}{16} g_{\mu\nu} \e^{\tfrac{p-3}{4}\phi_0} \mu_p, \label{eq:bps2_exteinstein}
\end{equation}
we find $R_{\mu\nu} = 0$. Our setup therefore only allows for $p$-dimensional Minkowski solutions. An exception is the case $p=3$, where \eqref{eq:bps2_eomdil} is solved trivially and we take $R_{\mu\nu} = 0$ as an assumption.

The Bianchi identity for $F_{8-p}$ is
\begin{equation}
\d F_{8-p} = - \left({-1}\right)^p m_{7-p} \wedge \d e^9 = - \mu_p \, \epsilon_{9-p}.
\end{equation}
Using \eqref{eq:bps2_exteinstein} to eliminate the source term, we obtain
\begin{equation}
\d F_{8-p} = - \e^{\tfrac{p-3}{4}\phi_0} |F_{8-p}|^2 \epsilon_{9-p},
\end{equation}
which can be rewritten with \eqref{eq:bps2_ansatz_smeared} to find
\begin{equation}
\left({-1}\right)^p m_{7-p} \wedge \d e^9 = g^{99} \e^{\tfrac{p-3}{4}\phi_0} \star_{9-p} m_{7-p} \wedge m_{7-p}. \label{eq:bps2_bianchi_rewr}
\end{equation}
The Einstein equation in $(99)$-direction reads
\begin{align}
R_{99} &= \frac{1}{2} g_{99}g_{99} |\d e^9|^2 \nll =
\frac{1}{2} \e^{\tfrac{p-3}{2}\phi_0} |F_{8-p}|^2_{99} - \frac{7-p}{16} g_{99} \e^{\tfrac{p-3}{2}\phi_0} |F_{8-p}|^2 + \frac{7-p}{16} g_{99} \e^{\tfrac{p-3}{4}\phi_0} \mu_p \nll =
\frac{1}{2} \e^{\tfrac{p-3}{2}\phi_0} |F_{8-p}|^2_{99},
\end{align}
where we substituted \eqref{eq:bps2_exteinstein} in the last line. Since $|F_{8-p}|^2_{99} = g_{99} |F_{8-p}|^2$, it follows
\begin{equation}
g_{99} |\d e^9|^2 = \e^{\tfrac{p-3}{2}\phi_0} |F_{8-p}|^2
\end{equation}
and, using \eqref{eq:bps2_ansatz_smeared},
\begin{equation}
g_{99} \star_{9-p} \d e^9 \wedge \d e^9 = g^{99} \e^{\tfrac{p-3}{2}\phi_0} \star_{9-p} m_{7-p} \wedge m_{7-p}.
\end{equation}
Together with \eqref{eq:bps2_bianchi_rewr}, this equation implies the condition
\begin{equation}
\d e^9 = \left({-1}\right)^p g^{99} \e^{\tfrac{p-3}{4}\phi_0} \star_{9-p} m_{7-p}, \label{eq:bps2_bps_cond}
\end{equation}
which relates the RR flux to the metric flux and is the T-dual version of the BPS condition \eqref{eq:bps1_bps_cond}. Furthermore, we find from \eqref{eq:bps2_exteinstein} that the RR and metric fluxes are related to the charge of the O$p$-planes via
\begin{equation}
\mu_p = \e^{\tfrac{p-3}{4}\phi_0} |F_{8-p}|^2 = \e^{-\tfrac{p-3}{4}\phi_0} g_{99} |\d e^9|^2, \label{eq:bps2_cond2}
\end{equation}
which is the T-dual analogue of \eqref{eq:bps1_cond2}.

The Einstein equations in the directions transverse to the O$p$-planes are
\begin{align}
R_{ik} &= -\frac{1}{2} g_{99} |\d e^9|^2_{ik} + \tilde R_{ik} \nll =
\frac{1}{2} \e^{\tfrac{p-3}{2}\phi_0} |F_{8-p}|^2_{ik} - \frac{7-p}{16} g_{ik} \e^{\tfrac{p-3}{2}\phi_0} |F_{8-p}|^2 - \frac{p+1}{16} g_{ik} \e^{\tfrac{p-3}{4}\phi_0} \mu_p \nll =
\frac{1}{2} \e^{\tfrac{p-3}{2}\phi_0} |F_{8-p}|^2_{ik} - \frac{1}{2} g_{ik} \e^{\tfrac{p-3}{2}\phi_0} |F_{8-p}|^2,
\end{align}
where we substituted \eqref{eq:bps2_exteinstein} in the last line. This can be rewritten with \eqref{eq:bps2_bps_cond} as
\begin{equation}
\tilde R_{ik} = \frac{1}{2} g^{99} \e^{\tfrac{p-3}{2}\phi_0} \left({|\star_{9-p} m_{7-p}|^2_{ik} + |m_{7-p}|^2_{ik} - g_{ik} |m_{7-p}|^2}\right) = 0,
\end{equation}
where it is useful to take advantage of the identities stated in Appendix \ref{app:conventions.tensors}. Note that $\tilde R_{ik} = 0$ does not mean that there is no curvature since the contribution of the metric fluxes yields $R_{ik} \neq 0$. The remaining Einstein equation is
\begin{equation}
R_{9k} = \tilde R_{9k} = \frac{1}{2} \e^{\tfrac{p-3}{2}\phi_0} |F_{8-p}|^2_{9k} = 0,
\end{equation}
where the last equality follows from our ansatz \eqref{eq:bps2_ansatz_smeared}. All other equations of motions are trivially satisfied.
\\

{\noindent\bf Summary of the solution.} The a priori non-trivial fields in our ansatz for solutions with smeared O$p$-planes (with $p=1,\ldots,6$) are
\begin{equation}
\phi = \phi_0, \qquad F_{8-p} = m_{7-p} \wedge e^9, \qquad R_{\mu\nu}, \qquad R_{ik}, \qquad R_{9k}, \qquad R_{99}.
\end{equation}
This leads to $p$-dimensional Minkowski solutions if the following conditions are satisfied:
\begin{itemize}
\item The RR and metric fluxes are related by the BPS condition,
\begin{equation}
\d e^9 = \left({-1}\right)^p g^{99} \e^{\tfrac{p-3}{4}\phi_0} \star_{9-p} m_{7-p}.
\end{equation}
\item The amount of RR and metric flux is fixed in terms of the O$p$-plane charge,
\begin{equation}
\mu_p = \e^{\tfrac{p-3}{4}\phi_0} |F_{8-p}|^2 = \e^{-\tfrac{p-3}{4}\phi_0} g_{99} |\d e^9|^2.
\end{equation}
\item The internal Ricci curvature is only due to the metric flux such that
\begin{equation}
\tilde R_{ik} = 0, \qquad \tilde R_{9k} = 0, \qquad \tilde R_{99} = 0.
\end{equation}
\end{itemize}
For $p=3$, some expressions are modified due to the self-duality of $F_5$, and $R_{\mu\nu} = 0$ has to be imposed instead of being implied by the equations of motion.

\subsubsection{Localised Solutions}

In order to find localised solutions, we consider the metric \eqref{eq:bp2-warpedmetric} with a non-trivial warp factor $A$, which we allow to depend on the coordinates transverse to the O$p$-planes. We also allow the dilaton to vary over the transverse coordinates and introduce a new term in $F_{8-p}$. Our ansatz for $F_{8-p}$ is
\begin{equation}
F_{8-p} = \hat F_{8-p} - \e^{-2\left({p+1}\right)A-\tfrac{p-3}{2}\phi} \tilde \star_{9-p} \d \alpha, \label{eq:bps2_ansatz_f}
\end{equation}
where $\hat F_{8-p} = m_{7-p} \wedge e^9$ is the $\left({8-p}\right)$-form of \eqref{eq:bps2_ansatz_smeared} and $\alpha$ is a function of the coordinates transverse to the O$p$-planes.

In order to derive the BPS condition, we have to combine three equations. These are the dilaton equation,
\begin{align}
\e^{2\tfrac{p+1}{7-p}A} \tilde \nabla^2 \phi &= \frac{p-3}{4} \e^{\tfrac{p-3}{2}\phi} \left[{|\hat F_{8-p}|^2 + \e^{2\tfrac{\left({p+1}\right)\left({p-6}\right)}{7-p}A - \left({p-3}\right)\phi} (\tilde \partial \alpha)^2}\right] \nl - \frac{p-3}{4} \e^{\tfrac{p-3}{4}\phi} \mu_p \delta(\Sigma), \label{eq:bps2_eom_dilaton_loc}
\end{align}
the trace of the Einstein equations along the O$p$-planes,
\begin{align}
\tilde g^{\mu\nu} R_{\mu\nu} + \tilde g^{99} R_{99} &= \tilde R_{p} - \e^{\tfrac{16}{7-p}A} \left({p+1}\right) \tilde \nabla^2 A + \frac{1}{2} \e^{\tfrac{32}{7-p}A} \tilde g_{99} | \d \tilde e^9|^2 \nll =
-\frac{\left({7-p}\right)\left({p+1}\right)}{16} \e^{2A+\tfrac{p-3}{2}\phi} \left[{|\hat F_{8-p}|^2 + \e^{2\tfrac{\left({p+1}\right)\left({p-6}\right)}{7-p}A - \left({p-3}\right)\phi} (\tilde \partial \alpha)^2}\right] \nl + \frac{1}{2} \e^{\tfrac{p-3}{2}\phi} |\hat F_{8-p}|^2_{99} + \frac{\left({7-p}\right)\left({p+1}\right)}{16} \e^{2A+\tfrac{p-3}{4}\phi} \mu_p \delta(\Sigma), \label{eq:bps2_eom_exteinstein_loc}
\end{align}
and the Bianchi identity for $F_{8-p}$,
\begin{align}
\d F_{8-p} &= \d \hat F_{8-p} - \d \left({\e^{-2\left({p+1}\right)A - \tfrac{p-3}{2}\phi} \tilde \star_{9-p} \d \alpha}\right) \nll =
- \left({-1}\right)^p m_{7-p} \wedge \d e^9 - \left({-1}\right)^p \epsilon_{9-p} \nl \cdot \left[{\e^{\tfrac{\left({p-5}\right)\left({p+1}\right)}{7-p}A - \tfrac{p-3}{2}\phi} \tilde \nabla^2 \alpha + \e^{\tfrac{\left({9-p}\right)\left({p+1}\right)}{7-p}A} \tilde g^{ik}  \Big(\partial_i \e^{-2\left({p+1}\right)A - \tfrac{p-3}{2}\phi}\Big) \left({\partial_k \alpha}\right)}\right] \nll = - \mu_p \delta(\Sigma)\, \epsilon_{9-p}.
\end{align}
These equations can be combined into
\begin{align}
\tilde \nabla^2\! \left[{\e^{\left({p+1}\right)A+\tfrac{p-3}{4}\phi} + \left({-1}\right)^p \alpha}\right] &= \e^{\tfrac{\left({p+1}\right)\left({9-p}\right)}{p-7}A-\tfrac{p-3}{4}\phi} \left[{\partial \Big(\e^{\left({p+1}\right)A+\tfrac{p-3}{4}\phi} + \left({-1}\right)^p \alpha \Big)}\right]^2 \nl \!\!\!\!\! \!\!\!\!\! \!\!\!\!\! \!\!\!\!\! \!\!\!\!\! \!\!\! + \frac{1}{2} \e^{\tfrac{\left({p+1}\right)\left({p-5}\right)}{p-7}A + 3\tfrac{p-3}{4}\phi} \left|{\sqrt{g^{99}} m_{7-p} - \left({-1}\right)^p \e^{-\tfrac{p-3}{4}\phi} \sqrt{g_{99}} \star_{9-p} \d e^9}\right|^2 \nl \!\!\!\!\! \!\!\!\!\! \!\!\!\!\! \!\!\!\!\! \!\!\!\!\! \!\!\! + \e^{\tfrac{\left({p-3}\right)^2}{p-7}A+\tfrac{p-3}{4}\phi} \tilde R_{p},
\end{align}
where the squares on the right-hand side are contracted by warped metrics. Since the left-hand side integrates to zero on a compact space, we find for Minkowski solutions with $\tilde R_{p} = 0$ that the two squares on the right-hand side have to vanish. This yields
\begin{equation}
\alpha = - \left({-1}\right)^p \e^{\left({p+1}\right)A+\tfrac{p-3}{4}\phi} + \textrm{const.} \label{eq:bps2_alpha}
\end{equation}
and the BPS condition
\begin{equation}
\d e^9 = \left({-1}\right)^p g^{99} \e^{\tfrac{p-3}{4}\phi} \star_{9-p} m_{7-p}. \label{eq:bps2_bps_cond_loc}
\end{equation}

The remaining equations of motion are then solved as follows. Substituting \eqref{eq:bps2_eom_dilaton_loc} into \eqref{eq:bps2_eom_exteinstein_loc} and using \eqref{eq:bps2_bps_cond_loc}, we obtain
\begin{equation}
\tilde \nabla^2\! \left({4\frac{p-3}{7-p} A - \phi}\right) = 0.
\end{equation}
Since a harmonic function is constant on a compact space, this implies
\begin{equation}
\phi = 4\frac{p-3}{7-p} A + \phi_0. \label{eq:bps2_dilaton_expr}
\end{equation}
Using this in \eqref{eq:bps2_bps_cond_loc}, we find that the dilaton factor exactly cancels the warp factors hidden in the Hodge operator and the inverse metric,
\begin{equation}
\d e^9 = \left({-1}\right)^p \tilde g^{99} \e^{\tfrac{p-3}{4}\phi_0} \tilde \star_{9-p} m_{7-p},
\end{equation}
such that we reproduce \eqref{eq:bps2_bps_cond}. As in the solutions discussed in Section \ref{ch:smeared-vs-loc.bps-ricciflat}, the BPS condition has therefore not changed after localisation. Furthermore, we can use \eqref{eq:bps2_dilaton_expr} and \eqref{eq:bps2_alpha} in \eqref{eq:bps2_eom_dilaton_loc} to get the condition
\begin{equation}
\tilde \nabla^2 \e^{\tfrac{16}{p-7}A} = -\e^{\tfrac{p-3}{2}\phi_0} |\tilde{\hat F}_{8-p}|^2 + \e^{\tfrac{p-3}{4}\phi_0} \mu_p \tilde \delta(\Sigma), \label{bps2_cond2_loc}
\end{equation}
where we have pulled out the warp factor dependence in $|\tilde{\hat F}_{8-p}|^2$ and $\tilde \delta(\Sigma)$. Thus, the flux is again related to the charge of the O$p$-planes, and the warp factor is fixed in terms of the delta function. Integrating the equation over the compact space, we then find that \eqref{eq:bps2_cond2} remains satisfied in the localisation solutions. Furthermore, \eqref{bps2_cond2_loc} implies that warping is comparable to the fluxes everywhere on the compact space where the O$p$-planes do not contribute. The Douglas-Kallosh problem is thus evaded in our solutions, as we will discuss in more detail in Section \ref{ch:smeared-vs-loc.discussion.dk}.

The Einstein equations transverse to the O$p$-planes read
\begin{align}
R_{ik} &= \tilde R_{ik} - \frac{1}{2} \e^{\tfrac{16}{7-p}A} \tilde g_{99} |\d \tilde e^9|^2_{ik} - \frac{p+1}{p-7} \tilde g_{ik} \tilde \nabla^2 A + 8 \frac{p+1}{p-7} (\partial_i A) (\partial_k A) \nll = \frac{1}{2} (\partial_i \phi) (\partial_k \phi) + \frac{1}{2} \e^{\tfrac{p-3}{2}\phi} |\hat F_{8-p}|^2_{ik} + \frac{1}{2} \e^{-2\left({p+1}\right)A - \tfrac{p-3}{2}\phi} \left[{\tilde g_{ik} (\tilde \partial \alpha)^2 - (\partial_i \alpha)(\partial_k \alpha)}\right] \nl - \frac{7-p}{16} \tilde g_{ik} \e^{2\tfrac{p+1}{p-7}A + \tfrac{p-3}{2}\phi} |\hat F_{8-p}|^2 - \frac{7-p}{16} \tilde g_{ik} \e^{-2\left({p+1}\right)A - \tfrac{p-3}{2}\phi} (\tilde \partial \alpha)^2 \nl - \frac{p+1}{16} \tilde g_{ik} \e^{2\tfrac{p+1}{p-7}A+\tfrac{p-3}{4}\phi} \mu_p \delta(\Sigma)
\end{align} \label{eq:bps2_rik}
and, using above results, reduce to $\tilde R_{ik} = 0$. The last non-trivial Einstein equation is
\begin{equation}
R_{9k} = - \frac{8}{p-7} \e^{\tfrac{16}{7-p}A} \tilde g_{99} |e^9 \wedge \d A \cdot \d \tilde e^9|_{9k} + \tilde R_{9k} = \frac{1}{2} \e^{\tfrac{p-3}{2}\phi} |F_{8-p}|^2_{9k}. \label{eq:bps2_r9k}
\end{equation}
The right-hand side of the equation can be rewritten using \eqref{eq:bps2_ansatz_f}, \eqref{eq:bps2_alpha} and the identities of Appendix \ref{app:conventions.tensors}, which yields
\begin{equation}
\frac{1}{2} \e^{\tfrac{p-3}{2}\phi} |F_{8-p}|^2_{9k} = \frac{8}{7-p} \left({-1}\right)^p \e^{\tfrac{16}{7-p}A+\tfrac{p-3}{4}\phi_0} |e^9 \wedge \d A \cdot \tilde \star_{9-p} \tilde m_{7-p}|_{9k}.
\end{equation}
It then follows from \eqref{eq:bps2_bps_cond} that \eqref{eq:bps2_r9k} is satisfied if $\tilde R_{9k} = 0$. All other equations of motion are then also satisfied.
\\

{\noindent\bf Summary of the solution.} We have seen that the solutions obtained in the smeared limit can be localised by introducing warping, allowing the dilaton to vary over the compact space and adding a new term to $F_{8-p}$. The dilaton and the new term in $F_{8-p}$ can be written in terms of the warp factor using \eqref{eq:bps2_ansatz_f}, \eqref{eq:bps2_alpha} and \eqref{eq:bps2_dilaton_expr}, while the warp factor itself is fixed in terms of the O$p$-plane charge by \eqref{bps2_cond2_loc}. The BPS condition \eqref{eq:bps2_bps_cond} has not changed after localisation since the dilaton factor exactly cancels the warp factors in \eqref{eq:bps2_bps_cond_loc}. Integrating \eqref{bps2_cond2_loc} over the compact space, we find that also \eqref{eq:bps2_cond2} is still valid in the localised solution. This is analogous to what we found in Section \ref{ch:smeared-vs-loc.bps-ricciflat} for the solutions that have a Ricci-flat internal space in the smeared limit.

\subsubsection{A Simple Example}
\label{ch:smeared-vs-loc.bps-twisted.simpleo5}

A simple example for Minkowski solutions of the type discussed above are solutions on a twisted torus where the metric fluxes satisfy the Heisenberg algebra. Such a space can easily be compactified and is T-dual to a $3$-torus with $H$ flux \cite{Douglas:2006es}. In order to obtain an explicit example, we consider the case $p=5$, which yields a setup with $F_3$ flux and O$5$-planes wrapping the directions $0,\ldots,4$ and $9$. We furthermore assume $f^9_{78} = - f^9_{87} = n$, which implies
\begin{equation}
\d e^9 = n\, \d x^7 \wedge \d x^8,
\end{equation}
and take the unwarped metric to be $\tilde g_{\mu\nu} = \eta_{\mu\nu}$, $\tilde g_{ik} = \delta_{ik}$ and $\tilde g_{99} = 1$. In the smeared limit, all equations of motion are then satisfied for
\begin{gather}
R_{77} = R_{88} = - R_{99}= - \frac{1}{2}n^2, \qquad F_3 = m\, \d x^5 \wedge \d x^6 \wedge e^9, \qquad \phi=\phi_0=\textrm{const.}, \notag \\ n = - m\, \e^{\tfrac{1}{2}\phi_0}, \qquad \mu_5 = \e^{-\tfrac{1}{2}\phi_0} n^2.
\end{gather}
For the case of localised sources, we find
\begin{gather}
F_3 = m\, \d x^5 \wedge \d x^6 \wedge e^9 - \e^{-12A-\phi} \tilde \star_4 \d \alpha, \qquad \alpha = \e^{6A + \tfrac{1}{2}\phi} + \textrm{const.}, \qquad \phi = 4A + \phi_0, \notag \\ n = - m\, \e^{-2A + \tfrac{1}{2}\phi} = - m\, \e^{\tfrac{1}{2}\phi_0}, \qquad \tilde \nabla^2 \e^{-8A} = - n^2 + \e^{\tfrac{1}{2}\phi_0} \mu_5 \tilde \delta(\Sigma). \label{simpleo5}
\end{gather}
\\[-0.7cm]

\subsection{Non-BPS Solutions}
\label{ch:smeared-vs-loc.non-bps}

In Sections \ref{ch:smeared-vs-loc.bps-ricciflat} and \ref{ch:smeared-vs-loc.bps-twisted}, we discussed Minkowski solutions of type II supergravity that satisfy a BPS condition and found that they could be localised using a simple localisation procedure, which involved adding non-trivial warping, a varying dilaton and a term in the $F_{8-p}$ field strength. In this section, we consider smeared non-BPS solutions and show for a simple example that this localisation procedure does not lead to a consistent localised solution there.

\subsubsection{Smeared Solutions}
\label{ch:smeared-vs-loc.non-bps.smeared}

Our ansatz for non-BPS solutions with smeared sources is as follows. We consider compactifications to $(p+1)$ dimensions for $p=2,\ldots,6$ and assume the non-zero fields
\begin{equation}
\phi=\phi_0, \qquad F_{6-p}, \qquad H,
\end{equation}
where the dilaton is constant and the remaining RR fields are identically zero. We furthermore assume that the metric has the form of a direct product,
\begin{equation}
\d s^2 = \d s_{p+1}^2 + \d s_{9-p}^2.
\end{equation}
For later convenience, we pull out the part of $F_{6-p}$ along the $\star_{9-p} H$ direction,
\begin{equation}
F_{6-p} = \bar F_{6-p} + \left({-1}\right)^p \e^{-\tfrac{p+1}{4}\phi_0} \kappa \star_{9-p} H, \label{eq:nonbps-f6-p}
\end{equation}
where $\kappa$ is a constant determined below and $\bar F_{6-p}$ is a closed and co-closed $(6-p)$-form satisfying $\bar F_{6-p} \w H = 0$.

The dilaton equation then reads
\begin{equation}
0 = \nabla^2 \phi_0 = \left({\frac{p-1}{4}\kappa^2 -\frac{1}{2}}\right) \e^{-\phi_0} |H|^2 + \frac{p-1}{4} \e^{\tfrac{p-1}{2}\phi_0} |\bar F_{6-p}|^2 \mp \frac{p-3}{4} \e^{\tfrac{p-3}{4}\phi_0} \mu_p, \label{eq:nonbps_dilaton-sm}
\end{equation}
where we allow sources that are O$p$-planes (upper sign) or D$p$-branes (lower sign).\footnote{More precisely, the sources correspond to O$p$-planes/D$p$-branes for $p=2,3,6$ and $\overline{\textrm{O}p}$-planes/$\overline{\textrm{D}p}$-branes for $p=1,4,5$. As noted above, it is always possible to flip the signs of the RR fields of a given solution to obtain the corresponding solution with sources of the opposite charge such that we can neglect this subtlety here. Also note that, in order to have consistent conventions throughout this thesis, we take $\mu_p$ to be the absolute value of the charge such that it is a positive number for all sources. Depending on the source type, some of our equations then obtain a minus sign relative to the conventions of \cite{Blaback:2010sj}.} The Bianchi identity for $F_{8-p}$ becomes
\begin{equation}
0 = \e^{-\tfrac{p+1}{4}\phi_0} \kappa |H|^2 \epsilon_{9-p} \mp \mu_p \epsilon_{9-p}, \label{eq:nonbps-bianchi}
\end{equation}
where $|H|^2$ is purely internal and therefore positive. It follows that $\kappa > 0$ for O$p$-planes and $\kappa < 0$ for D$p$-branes. Substituting \eqref{eq:nonbps-bianchi} in \eqref{eq:nonbps_dilaton-sm}, we furthermore find
\begin{equation}
|\bar F_{6-p}|^2 = \left({-\kappa^2 + \frac{p-3}{p-1}\kappa + \frac{2}{p-1}}\right)\e^{-\tfrac{p+1}{2}\phi_0} |H|^2. \label{eq:nonbps-f6}
\end{equation}
Since $|\bar F_{6-p}|^2$ is also internal, both squares in the equation are positive, and we find $-\frac{2}{p-1} \le \kappa \le 1$.

Using \eqref{eq:nonbps-f6-p}, \eqref{eq:nonbps-bianchi} and \eqref{eq:nonbps-f6}, the external Einstein equation simplifies to
\begin{equation}
R_{\mu\nu} = -\frac{1-\kappa}{2(p-1)} g_{\mu\nu} \e^{-\phi_0} |H|^2.
\end{equation}
For $\kappa < 1$, we thus obtain AdS solutions, whereas the special case $\kappa = 1$ leads to the BPS Minkowski solutions discussed in Section \ref{ch:smeared-vs-loc.bps-ricciflat}. The internal Einstein equation yields
\begin{align}
R_{ik} &= -\frac{(1-\kappa)\left[{1+\kappa(p-1)}\right]}{2(p-1)} g_{ik} \e^{-\phi_0} |H|^2 + \frac{1}{2} \e^{\tfrac{p-1}{2}\phi_0} |\bar F_{6-p}|^2_{ik} \nl + \frac{1}{2} \left({1-\kappa^2}\right) \e^{-\phi_0} |H|^2_{ik}, \label{eq:nonbps-internaleinstein}
\end{align}
where we have assumed $|\bar F_{6-p} \cdot H|_{ik} = 0$ and again used \eqref{eq:nonbps-f6-p}, \eqref{eq:nonbps-bianchi} and \eqref{eq:nonbps-f6}. Taking the trace, we then find
\begin{equation}
R_{9-p} = \frac{(1-\kappa)p}{p-1} \e^{-\phi_0} |H|^2 = - \frac{2p}{p+1} R_{p+1}. \label{eq:nonbps-ricciscalars}
\end{equation}
Therefore, if the external space is AdS, the internal space is positively curved.

\subsubsection{A Simple Example}

Let us consider the simple example $\kappa = -\frac{2}{p-1}$, for which \eqref{eq:nonbps-f6} implies $\bar F_{6-p} = 0$. Since $\kappa < 0$, the sources then have a net D$p$-brane charge
\begin{equation}
\mu_p = \frac{2}{p-1} \e^{-\tfrac{p+1}{4}\phi_0} |H|^2. \label{eq:nonbps-simple-1}
\end{equation}
It follows from \eqref{eq:nonbps-ricciscalars} that the external Ricci scalar becomes
\begin{equation}
R_{p+1} = -\frac{(p+1)^2}{2(p-1)^2} \e^{-\phi_0} |H|^2, \label{eq:nonbps-simple-2}
\end{equation}
while we find from \eqref{eq:nonbps-internaleinstein} that the internal Einstein equation reduces to
\begin{equation}
R_{ik} = \frac{p+1}{2(p-1)^2} g_{ik} \e^{-\phi_0} |H|^2 + \frac{(p+1)(p-3)}{2(p-1)^2} \e^{-\phi_0} |H|^2_{ik}. \label{eq:nonbps-internaleinstein-simple}
\end{equation}
In order to write down an explicit solution, we take the internal space to be a product of two spheres,
\begin{equation}
\mathcal{M}_{9-p} = S^3 \times S^{6-p}.
\end{equation}
For $p=3$, this corresponds to the solution described in \cite{Silverstein:2004id}. Evaluating \eqref{eq:nonbps-f6-p} for our setup, we find
\begin{equation}
H = h \epsilon_3, \qquad F_{6-p} = -\frac{2}{p-1} \e^{-\tfrac{p+1}{4}\phi_0} h \epsilon_{6-p},
\end{equation}
where $\epsilon_3$ and $\epsilon_{6-p}$ are the volume forms of the spheres. The internal curvature is then fixed in terms of the fluxes by \eqref{eq:nonbps-internaleinstein-simple},
\begin{equation}
R_{ik}^{(S^3)} = \frac{(p+1)(p-2)}{2(p-1)^2} g_{ik}^{(S^3)} \e^{-\phi_0}  h^2, \qquad R_{ik}^{(S^{6-p})} = \frac{p+1}{2(p-1)^2} g_{ik}^{(S^{6-p})} \e^{-\phi_0}  h^2.
\end{equation}
\\[-1.1cm]

\subsubsection{The Failure to Localise?}

Let us now study the localisation of the non-BPS solutions. For simplicity, we consider $\kappa=-\frac{2}{p-1}$ as in the example discussed above and restrict to the special case $p=3$. We thus have $\kappa=-1$, which, in the smeared limit, yields an AdS solution with D$3$-branes. It is instructive to compare this to the corresponding BPS solution with $\kappa = 1$, which is a Minkowski solution with O$3$-planes, i.\ e.\ the GKP solution \cite{Giddings:2001yu}. Our starting point is the ansatz
\begin{equation}
F_3 = \mp \e^{-\phi} \star_6 H, \label{eq:nonbps-condition}
\end{equation}
where the upper sign corresponds to the BPS case with $\kappa = 1$ and the lower sign to the non-BPS case with $\kappa = - 1$. We thus have ISD flux in the BPS case and IASD flux in the non-BPS case. While this is not the most general ansatz for $F_3$, it is motivated by our success in the BPS case, where it equals the BPS condition \eqref{eq:bps1_bps_cond_loc} and leads to a consistent localised solution. We also assume that the fluxes satisfy the Bianchi identities
\begin{equation}
\d H = \d F_3 = 0,
\end{equation}
and our ansatz for the metric is
\begin{equation}
\d s_{10}^2 = \e^{2A} \d \tilde s_4^2 + \d s_6^2,
\end{equation}
where we keep the internal metric arbitrary. Using \eqref{eq:nonbps-condition} in the dilaton equation
\begin{equation}
\nabla^2 \phi = -\frac{1}{2} \e^{-\phi} |H|^2 + \e^{2\phi} |F_1|^2 + \frac{1}{2} \e^{\phi} |F_3|^2
\end{equation}
and integrating over the compact space, we furthermore find $F_1=0$. The equation then also implies that the dilaton is constant, $\phi = \phi_0$.

From the equations of motion for $H$ and $F_3$,
\begin{equation}
\d \left({\e^{-\phi_0} \star H}\right) + \star F_5 \w F_3 = 0, \qquad \d\left({\e^{\phi_0} \star F_3}\right) + H \w \star F_5 = 0,
\end{equation}
and the self-duality of $F_5$, we find
\begin{equation}
F_5 = -\left({1+\star_{10}}\right)\e^{-4A} \star_6 \d \alpha, \label{eq:nonbps-f5}
\end{equation}
where
\begin{equation}
\alpha = \pm \e^{4A} + \textrm{const.} \label{eq:nonbps-alpha}
\end{equation}
Substituting \eqref{eq:nonbps-condition}, \eqref{eq:nonbps-f5} and \eqref{eq:nonbps-alpha} into the $F_5$ Bianchi identity
\begin{equation}
\d F_5 = H \w F_3 \pm \mu_3 \delta_6
\end{equation}
then yields the equation
\begin{equation}
4 \nabla^2 A = \e^{-\phi_0} |H|^2 - \mu_3 \delta(\Sigma). \label{eq:nonbps-final1}
\end{equation}
The trace of the external Einstein equation reduces to
\begin{equation}
\e^{-2A} \tilde R_4 - 4 \nabla^2 A = -\e^{-\phi_0} |H|^2 \pm \mu_3 \delta(\Sigma), \label{eq:nonbps-final2}
\end{equation}
where we have again used \eqref{eq:nonbps-condition}, \eqref{eq:nonbps-f5} and \eqref{eq:nonbps-alpha}. Combining \eqref{eq:nonbps-final1} and \eqref{eq:nonbps-final2}, we then find
\begin{equation}
\e^{-2A} \tilde R_4 = - \left({1 \mp 1}\right) \mu_3 \delta(\Sigma).
\end{equation}
In the BPS case, we thus recover the Minkowski solution $\tilde R_4 = 0$. In the non-BPS case, however, the right-hand side of the equation is zero away from the D$3$-branes, while its left-hand side is negative for an AdS spacetime. The equation is therefore only satisfied in the smeared limit $\delta(\Sigma) \to 1$ and $A \to 0$, where it is implied by \eqref{eq:nonbps-simple-1} and \eqref{eq:nonbps-simple-2}. Thus, we have shown that the localisation procedure that was successful for the BPS solution does not lead to a localised solution in the non-BPS case. One can verify that similar arguments also hold for the case $p \neq 3$, where the existence of localised non-BPS solutions is excluded for $\kappa=-\frac{2}{p-1}$ and the ansatz
\begin{equation}
F_{6-p} = \left({-1}\right)^p \e^{-\tfrac{p+1}{4}\phi} \kappa \star_{9-p} H.
\end{equation}
\\[-0.7cm]

\subsection{Discussion}
\label{ch:smeared-vs-loc.discussion}

Here, we discuss several implications of the results of this chapter. In particular, we comment on the role of the BPS condition in achieving a successful localisation and show for an explicit example how the Douglas-Kallosh problem is evaded in our BPS solutions on negatively curved twisted tori.

\subsubsection{Localisation in BPS and Non-BPS Solutions}
\label{ch:smeared-vs-loc.discussion.loc}

In Sections \ref{ch:smeared-vs-loc.bps-ricciflat} and \ref{ch:smeared-vs-loc.bps-twisted}, we considered BPS compactifications to Minkowski space that have a Ricci-flat or everywhere negatively curved internal space in the smeared limit and are related by a chain of T-dualities. The T-duality relations can schematically be written as
\begin{equation}
H \propto \star_{9-p} F_{6-p} \quad \begin{array}{c} \xrightarrow[]{\textrm{ along O}p\textrm{ }} \\[-6pt] \xleftarrow[\textrm{along }F_{7-p}]{}
\end{array} \quad H \propto \star_{10-p} F_{7-p} \quad \begin{array}{c} \xrightarrow[]{\textrm{ along }H} \\[-6pt] \xleftarrow[\textrm{ along }e_9]{} \end{array} \quad \d e^9 \propto \star_{9-p} m_{7-p},
\end{equation}
where we have characterised the different solutions by their BPS condition. The solutions with Ricci-flat internal space are thus related to one another by T-dualising in one of the directions of the O$p$-planes or along a cycle with RR flux. A T-duality along a cycle with $H$ flux leads to one of the solutions on negatively curved internal spaces, which are again related to one another by further T-dualities.

The localisation of these solutions introduces warping, a dilaton that varies over the compact space and a term in the $F_{8-p}$ field strength. Although these corrections are large everywhere on the compact space, we found that the localised solutions still preserve many features of the smeared ones. In particular, we found that the BPS conditions \eqref{eq:bps1_bps_cond} and \eqref{eq:bps2_bps_cond} do not change after localisation due to a cancellation between a warp and a dilaton factor. The vevs of the moduli that are fixed by these equations are therefore equal to those in the smeared solutions. From the perspective of the effective potential, this means that its different localisation corrections cancel on-shell, where the BPS condition is satisfied (see also \cite{DeWolfe:2002nn}). The extremum of the potential is therefore not shifted in moduli space, and the moduli stay at their positions.\footnote{Since the BPS condition only guarantees a cancellation in the on-shell effective potential, properties such as moduli masses will in general change after localisation (see e.\ g.\ \cite{Blaback:2012mu} for a discussion and \cite{Douglas:2007tu} for an explicit example).} This can intuitively be understood from the ``no-force'' condition, which states that mutually BPS objects do not exert any force on each other due to an exact cancellation of their gravitational and electromagnetic interactions. The O$p$-planes in our setup are mutually BPS with the flux background, and so it is guaranteed that we can arbitrarily deform their source profile in compact space without destroying the solutions.

For solutions that do not satisfy a BPS condition, this can in general not be expected. Indeed, we showed in Section \ref{ch:smeared-vs-loc.non-bps} for an explicit non-BPS example that the simple ansatz that was successful in the BPS case does not lead to a consistent localised solution. The reason is that the D$3$-branes are not mutually BPS with the IASD flux in this example, and so their localisation can generate non-zero forces. In the smeared limit, these forces are not present since the charge and energy densities of the branes are delocalised over the compact space such that there is no preferred point of attraction. Localised branes, on the other hand, can attract flux and thus, at least for the ansatz tested in this chapter, destabilise the solution. Although our ansatz was not the most general one, its failure already indicates that the localisation of non-BPS solutions is much more involved than in the BPS case. In Chapter \ref{ch:simple-non-bps}, we will find that the localisation of non-BPS solutions can even be problematic using the most general ansatz compatible with the symmetries of the setup. While this shows that BPSness plays an important role in achieving a successful localisation, it does not necessarily mean that all BPS solutions are localisable. In particular, it is not known whether solutions with intersecting sources such as the supersymmetric AdS vacua of \cite{DeWolfe:2005uu} allow for a consistent localisation, and it would be interesting to see if our arguments also extend to those cases.

\subsubsection{The Douglas-Kallosh Problem}
\label{ch:smeared-vs-loc.discussion.dk}

It was shown in \cite{Douglas:2010rt} that, in the absence of large $\alpha^\prime$ corrections, compactifications using an everywhere negative internal curvature to uplift the cosmological constant must have large warping at every point on the internal manifold where no O-planes contribute a negative energy density. This problem can be studied explicitly in the solutions of Section \ref{ch:smeared-vs-loc.bps-twisted}, which are compactified on everywhere negatively curved spaces in the smeared limit. Since these solutions admit a localisation, they evade the problem and must therefore have large warping.\footnote{Since our solutions are compactifications to Minkowski space, there is no real uplift to de Sitter. Still, the negatively curved compact space gives a positive contribution to the cosmological constant in the smeared limit, and so the arguments of \cite{Douglas:2010rt} apply accordingly.} Naively, one might have expected that the warp factor approaches a constant far away from the O-planes such that warping becomes negligible in these regions. This would have ruled out localised solutions in the large volume limit, where we can trust the supergravity approximation. However, \eqref{bps2_cond2_loc} shows that, even for a large internal volume, warping is comparable to the fluxes in our solutions everywhere on the compact space where the contribution of the O-planes vanishes. According to \cite{Douglas:2010rt}, this then allows the ten-dimensional equations of motion to be satisfied in the localised solutions at every point on the compact space.\footnote{By T-duality, the Douglas-Kallosh argument should extend to the solutions with Ricci-flat internal space discussed in Section \ref{ch:smeared-vs-loc.bps-ricciflat}. This is indeed confirmed by \eqref{bps1_cond2_loc}.}

In order to make this more explicit, let us consider the simple example with $F_3$ flux and O$5$-planes discussed in Section \ref{ch:smeared-vs-loc.bps-twisted.simpleo5}.
The discussion of the general case is then completely analogous. Taking the trace of the internal components of the Einstein equation, we find that the internal scalar curvature $R_5 = g^{mn}R_{mn} + g^{99}R_{99}$ is given by
\begin{align}
R_5 &= \underbrace{\frac{7}{8} \e^{14A} n^2 + 56 \e^{6A} (\tilde \partial A)^2}_{F_3} + \underbrace{\vphantom{\frac{1}{8}}8 \e^{6A} (\tilde\partial A)^2}_{\textrm{dilaton}} - \underbrace{\frac{11}{8} \e^{6A} \left[{- 8 \tilde \nabla^2 A + 64 (\tilde\partial A)^2 + \e^{8A} n^2 }\right]}_{\textrm{O}5\textrm{-planes}} \nll = - \frac{1}{2} \e^{14A} n^2 + 11 \e^{6A} \tilde \nabla^2 A - 24 \e^{6A} (\tilde \partial A)^2, \label{r5}
\end{align}
where we have used \eqref{simpleo5} on the right-hand side of the first line to spell out the different contributions of the $F_3$ field strength, the dilaton and the O$5$-planes to the energy-momentum tensor. We can then read off from the second line that the integral of $R_5$ over the compact space is manifestly negative if weighted with the full ten-dimensional metric determinant,
\begin{equation}
\int \d ^5 x \sqrt{g_{10}}\, R_{5} = \int \d ^5 x \sqrt{\tilde g_{10}} \left[{- \frac{1}{2} \e^{8 A} n^2 - 24 (\tilde \partial A)^2}\right] < 0.
\end{equation}
Up to a Weyl rescaling, this expression equals minus the contribution of $R_5$ to the effective potential. As in the smeared limit, the internal scalar curvature therefore yields an overall positive term in the effective potential, and so the Einstein equations are satisfied in an integrated sense.

In contrast to the smeared solutions, however, $R_5$ is not negative \emph{everywhere} on the compact space. Instead, it follows from the first line of \eqref{r5} that this is only true at those points on the internal manifold where the localised O$5$-planes contribute, whereas $R_5$ is positive everywhere else. In order that the ten-dimensional Einstein equations are satisfied pointwise on the internal manifold after localisation, this has to be compensated by large warping. Indeed, one can verify that warping terms, which come from the warped external Ricci scalar and the energy-momentum tensor in the Einstein equations, ``lift'' the cosmological constant to Minkowski space at those points on the compact space where the internal curvature is positive. Thus, warping does not have the effect of preserving an everywhere negative internal curvature in the localised solutions but instead yields additional contributions to the ten-dimensional Einstein equations, which then help to satisfy the constraints of \cite{Douglas:2010rt}. It would be interesting to understand whether similar effects also allow a localisation of smeared non-BPS solutions on negatively curved internal spaces or whether this is inhibited by the problems sketched in Section \ref{ch:smeared-vs-loc.discussion.loc}. This would be an important step towards understanding backreaction in the classical de Sitter vacua of \cite{Caviezel:2009tu, Danielsson:2011au, Caviezel:2008tf, Flauger:2008ad, Danielsson:2009ff, Danielsson:2010bc}.

\cleardoublepage
\thispagestyle{plain}

\section{(Anti-)brane Backreaction in a Simple Non-BPS Setup}
\label{ch:simple-non-bps}

In this chapter, we elaborate on the effects of backreaction in compactifications with sources that are not mutually BPS with the background fluxes. As indicated in the previous chapter, it is not obvious whether smeared non-BPS solutions can always be localised since this can generate non-zero forces that may render the configuration unstable and thus destroy the solution. In order to be able to study this problem in detail, we now consider a simple toy model in type IIA supergravity with spacetime-filling D$6$-branes or \dsix-branes on AdS$_7 \times S^3$ \cite{Blaback:2011nz, Blaback:2011pn, Bena:2012tx}. This setup has a smeared solution, which is perturbatively stable in the left-invariant closed string modes, and is simple enough to allow us to analyse the localised equations of motion without having to rely on non-trivial assumptions in our ansatz for the fields. Moreover, a non-compact version of our model with \dsix-branes is T-dual to a setup with partially smeared \dthree-branes on $\mathbb{R}^{1,3}\! \times T^3\! \times \mathbb{R}^3$ \cite{Massai:2012jn}. This setup approximates the near-tip region of the solution for partially smeared \dthree-branes in the Klebanov-Strassler background. Our model therefore also provides some insight into the issue of backreaction in the KKLT scenario \cite{Kachru:2003aw}.

Although we consider the most general ansatz compatible with the symmetries, localisation in our model turns out problematic. We will first show this for regularised brane profiles, which, if allowed, are useful for studying localisation since they interpolate between the smeared solution and the fully localised one and can approximate a delta function profile with arbitrary precision. Taking the regularised profile to be a smooth function, we find that, up to coordinate transformations, all solutions except for the smeared one are ruled out by the equations of motion. Furthermore, we can use a simple topological argument to also exclude brane profiles in the shape of a step function. This is quite different from what we found for the BPS solutions in Chapter \ref{ch:smeared-vs-loc}, where the no-force condition guaranteed that we can solve the equations of motion for arbitrary source profiles.

The simplicity of our model also allows us to explicitly compute the boundary conditions for the fields in the vicinity of fully localised sources. The computation is rather lengthy but yields a surprisingly simple result: there are only two different boundary conditions that support D$6$/\dsix-branes and are locally consistent with the equations of motion. The first boundary condition is the standard boundary condition for D$6$/\dsix-branes in a flux background \cite{Janssen:1999sa} and yields fluxes that are locally BPS with the sources in their vicinity. Although consistent locally, this boundary condition is excluded globally due to the topological argument that also rules out step function profiles. The second boundary condition is not excluded by this argument but yields an infinite (but integrable) energy density of the $H$ flux. This singularity is unusual since $H$ does not directly couple to the branes. While singularities in directly sourced fields such as the warp factor and the dilaton are expected to appear in the supergravity approximation and are resolved in a string theory completion of the solution, this is less obvious for singularities in fields that do not couple to a localised source. Such singularities can be problematic since they may indicate that a solution has an instability.

Besides the possibility that our model might not have a localised solution, the existence of the flux singularity also suggests a relation to the solution for \dthree-branes in the Klebanov-Strassler background, where a similar unexpected singularity appears. For partially smeared anti-branes, this singularity was noticed in \cite{McGuirk:2009xx, Bena:2009xk} at linear order in a perturbation around the supersymmetric background and recently confirmed in \cite{Bena:2012bk} taking into account the full non-linear backreaction.\footnote{Similar flux singularities also exist in solutions with $\overline{\textrm{M}2}$-branes \cite{Bena:2010gs, Massai:2011vi} and $\overline{\textrm{D}2}$-branes \cite{Giecold:2011gw}.} Moreover, we will argue in Chapter \ref{ch:scaling-symmetries} that the singularity is not an artifact of the partial smearing but also generated by the backreaction of fully localised anti-branes. As stated above, this possibly indicates that the backreaction of the anti-branes causes an instability in one or more of the degrees of freedom of the setup. It was proposed, for example, that the solution is perturbatively unstable in the closed string modes, which leads to an attraction of flux towards the anti-branes and a subsequent decay of the solution via brane-flux annihilation \cite{Blaback:2011pn, Blaback:2012nf, Bena:2012ek, Vanriet:2013}.\footnote{See also \cite{Dvali:2002pe} and references therein for examples of perturbatively unstable systems that develop singularities when forced into a static ansatz.} If true, this would imply that the singular solution is not physical and that there is no meta-stable state with \dthree-branes in the Klebanov-Strassler throat. A second, more optimistic interpretation is that the instability is in the open string degrees of freedom. The singularity might then be resolved by taking into account the non-perturbative dynamics of the anti-branes. According to \cite{Myers:1999ps}, (anti-)branes can expand into a non-commutative configuration in the presence of a background field and thus polarise into a higher-dimensional brane wrapping a trivial cycle. This is the so-called Myers effect, an analogue of the dielectric effect for atoms in an electric field. In \cite{Kachru:2002gs}, it was shown using the probe approximation that \dthree-branes in the Klebanov-Strassler background indeed polarise into an NS$5$-brane. If this is also the case when the backreaction of the anti-branes is taken into account, it might resolve the singularity, similarly to how it happens in the Polchinski-Strassler solution \cite{Polchinski:2000uf}.

In principle, \dthree-branes in the Klebanov-Strassler throat can polarise either into an NS$5$-brane wrapping a two-sphere inside a three-cycle with $F_3$ flux or into a D$5$-brane wrapping a two-sphere inside the orthogonal three-cycle, which is threaded by $H$ flux.\footnote{More generally, there is also the possibility of a polarisation into a $(p,q)$ $5$-brane (a bound state of $p$ D$5$-branes and $q$ NS$5$-branes) wrapping a two-sphere inside a diagonal three-plane \cite{Polchinski:2000uf, Bena:2012vz}.} Since the non-compact version of our toy model is expected to capture the near-tip physics of this setup if the \dthree-branes are smeared along the cycle with $F_3$ flux, we can use it to explicitly analyse whether such a polarisation happens. A subtlety is that the potential for the worldvolume scalars, which determines whether the anti-branes want to polarise, is not visible in the directions of the smearing. We can therefore not directly observe a possible polarisation into an NS$5$-brane but only one into a D$5$-brane. It is known from similar setups such as the Polchinski-Strassler solution, however, that brane polarisation always occurs in two channels.\footnote{This is also true in a solution with M$2$-branes \cite{Bena:2000zb, Bena:2004jw, Lin:2004nb}, which polarise into M$5$-branes in orthogonal planes, and in a solution with D$2$-branes \cite{Bena:2000fz}, which polarise into D$4$-branes and NS$5$-branes in orthogonal planes.} This suggests that \dthree-branes in the Klebanov-Strassler background do either polarise into both NS$5$-branes \emph{and} D$5$-branes or not at all. If this is true, it should be sufficient to analyse one of the two possibilites in order to decide whether polarisation happens.

In our T-dual toy model, the polarisation of the \dthree-branes into a D$5$-brane corresponds to a polarisation of \dsix-branes into a D$8$-brane. While, in the compact model, such a polarisation is not ruled out for the whole parameter range, we will demonstrate below that the non-compact version is much more constrained. Interestingly, we find that the potential for the worldvolume scalars has exactly the terms that appear in solutions with brane polarisation \cite{Polchinski:2000uf, Bena:2000fz, Bena:2001aw}, but the coefficients are such that a polarisation is excluded (see also \cite{Pilch:2000ue} for another example with such a behaviour). Contrary to the compact model, this result is independent of the free parameters of the solution such that a polarisation of the \dsix-branes is excluded universally. This indicates that the same is true for \dthree-branes in the Klebanov-Strassler background, and so brane polarisation does probably not help to resolve the singularity there either. Further evidence supporting this intuition was recently presented in \cite{Bena:2012vz}, where the authors explicitly computed the polarisation potential for partially smeared \dthree-branes in the Klebanov-Strassler background and indeed found that a polarisation into a D$5$-brane does not occur.

This chapter is based on \cite{Blaback:2011nz, Blaback:2011pn, Bena:2012tx} and organised as follows. In Section \ref{ch:simple-non-bps.setup}, we present the supergravity equations of motion in the presence of D$6$/\dsix-branes and show that our setup on AdS$_7 \times S^3$ has a solution in the smeared limit. We then prove in Section \ref{ch:simple-non-bps.stability} that this solution is stable in the sector of the left-invariant closed string modes. In Section \ref{ch:simple-non-bps.localisation}, we consider localised sources with either regularised or fully localised source profiles. For regularised sources, we show that the equations of motion rule out all solutions except for the smeared one. For fully localised sources, we compute the boundary conditions and find that the only consistent one yields a singularity in the energy density of the $H$ flux. In Section \ref{ch:simple-non-bps.myers-effect}, we discuss the non-compact version of our model and explicitly show that the singularity is not resolved by brane polarisation there. We conclude in Section \ref{ch:simple-non-bps.discussion} with a discussion of our results.
\\

\subsection[A Toy Model on AdS$_7 \times S^3$]{A Toy Model on AdS$\boldsymbol{_7 \times S^3}$}
\label{ch:simple-non-bps.setup}

In this section, we state the equations of motion for type IIA supergravity with D$6$/\dsix-branes and present the smeared solution of our simple non-BPS setup on AdS$_7 \times S^3$. This solution corresponds to the special case $p=6$ of the solutions discussed in Chapter \ref{ch:smeared-vs-loc.non-bps.smeared}.

\subsubsection[Type IIA Supergravity with D$6$/\dsix-branes]{Type IIA Supergravity with D$\boldsymbol{6}$/\dsixb-branes}
\label{ch:simple-non-bps.setup.eoms}

Since we compactify to AdS$_7$, we do not consider any field configurations that break maximal symmetry. We can therefore set $F_4=0$ and assume that the other form fields only have internal components. Furthermore, we assume that the dilaton and the form fields only depend on the internal coordinates $x^i$. The equations of motion stated in Appendix \ref{app:conventions.sugra} then simplify as follows. The trace-reversed Einstein equation reads
\begin{align}
R_{MN} &= \frac{1}{2} \partial_M \phi \partial_N \phi + \frac{1}{2}\e^{-\phi} |H|^2_{MN} - \frac{1}{8}\e^{-\phi} g_{MN} |H|^2 + \frac{1}{16} \e^{\tfrac{5}{2}\phi} g_{MN} F_0^2 + \frac{1}{2}\e^{\tfrac{3}{2}\phi} |F_2|^2_{MN} \nl - \frac{1}{16}\e^{\tfrac{3}{2}\phi} g_{MN} |F_2|^2 + \frac{1}{2} \left({T^\textrm{loc}_{MN}-\frac{1}{8} g_{MN} T^\textrm{loc}}\right), \label{nonbps-eom1}
\end{align}
where
\begin{equation}
T^\textrm{loc}_{\mu\nu} = - \e^{\tfrac{3}{4}\phi} \mu_6 g_{\mu\nu} \delta(\Sigma), \qquad T^\textrm{loc}_{mn} = 0.
\end{equation}
$\delta(\Sigma)$ denotes the delta distribution with support on the world-volume(s) of the spacetime-filling D$6$/\dsix-branes. It may implicitly also include a sum over parallel branes that are localised at different points on the internal manifold. The dilaton equation reduces to
\begin{equation}
\nabla^2 \phi = - \frac{1}{2}\e^{-\phi} |H|^2 + \frac{5}{4} \e^{\tfrac{5}{2}\phi} F_0^2 + \frac{3}{4} \e^{\tfrac{3}{2}\phi} |F_2|^2 + \frac{3}{4} \e^{\tfrac{3}{4}\phi} \mu_6 \delta(\Sigma). \label{nonbps-eom2}
\end{equation}
The non-trivial equations of motion for the form fields are
\begin{equation}
\d\left({\e^{-\phi} \star H}\right) = - \e^{\tfrac{3}{2}\phi} \star F_2 F_0, \qquad \d \left({\e^{\tfrac{3}{2}\phi} \star F_2 }\right) = 0, \label{nonbps-eom3}
\end{equation}
and their Bianchi identities read
\begin{equation}
\d H = 0, \qquad \d F_0 = 0, \qquad \d F_2 = H F_0 \mp \mu_6 \delta_3, \label{nonbps-eom4}
\end{equation}
where the upper sign is for D$6$-branes and the lower sign for \dsix-branes.

\subsubsection{Smeared Solution}

In the smeared limit, we replace the delta distributions in the equations of motion by a constant,
\begin{equation}
\delta(\Sigma) \to \frac{1}{V},
\end{equation}
where $V$ is the volume of the space transverse to the branes. We can then set $F_2 = 0$ and assume that all other fields are constant. Furthermore, we take the ten-dimensional spacetime to be the direct product AdS$_7 \times S^3$ and write
\begin{equation}
\d s_{10}^2 = R^2 \d s_7^2 + r^2 \d s_3^2
\end{equation}
for the metric, where $R$ is the AdS radius and $r$ is the radius of the three-sphere. Our ansatz for $H$ is
\begin{equation}
H = \lambda F_0 \e^{\tfrac{7}{4}\phi} \star_3 1,
\end{equation}
with $\lambda$ a constant. The equations of motion \eqref{nonbps-eom1}, \eqref{nonbps-eom2}, \eqref{nonbps-eom3} and \eqref{nonbps-eom4} are then solved for
\begin{equation}
R^2 = 12 r^2, \qquad \lambda = \mp \frac{5}{2}, \qquad F_0^2 = \frac{4}{7} \e^{-\tfrac{5}{2}\phi} r^{-2}, \qquad \mu_6 = \frac{20}{7} \pi^2 \e^{-\tfrac{3}{4}\phi} r, \label{simple-smeared}
\end{equation}
where the upper sign is again for D$6$-branes and the lower sign for \dsix-branes. For large fluxes, one can check that the solution has large internal volume and small string coupling.
\\

\subsection{Perturbative Stability}
\label{ch:simple-non-bps.stability}

In this section, we analyse the perturbative stability of the smeared solution. Since the three-sphere is the group manifold of $\mathrm{SU}(2)$, we can restrict ourselves to the sector of left-invariant modes \cite{Scherk:1979zr}. This is a standard approach to obtain a consistent truncation of the lower-dimensional effective field theory for compactifications on group or coset spaces and gives rise to an effective gauged supergravity (see e.\ g.\ \cite{Roest:2004pk, Caviezel:2008ik, Caviezel:2009tu, Danielsson:2011au} for a discussion and explicit constructions and \cite{AlonsoAlberca:2003jq, Angelantonj:2003rq, Derendinger:2004jn, Cassani:2009ck} for further examples). Furthermore, we only include the closed string modes in our analysis. Instabilities can in principle also arise from the open string modes and may, in particular, lead to a polarisation of the branes into a D$8$-brane due to the Myers effect \cite{Myers:1999ps}. Under certain conditions, such a polarised configuration can be unstable itself and decay perturbatively via brane-flux annihilation \cite{Kachru:2002gs}. We will analyse this possibility further in Section \ref{ch:simple-non-bps.myers-effect} in the context of fully localised sources.

The left-invariant modes are the lower-dimensional fields obtained by expanding the ten-dimensional fields in the Maurer-Cartan forms of $\mathrm{SU}(2)$. These satisfy the equation
\begin{equation}
\d e^i = - \frac{1}{2} f_{kl}^i e^k \w e^l
\end{equation}
with structure constants
\begin{equation}
f_{kl}^i = q\, \delta^{im} \epsilon_{mkl},
\end{equation}
where $q$ is a number related to the curvature of the three-sphere \cite{AlonsoAlberca:2003jq}. The expansion of the fields then yields a total of $14$ scalars in the seven-dimensional effective field theory: six from the metric, three from the $B$ field, three from $C_1$, one from $C_3$ and one from the (ten-dimensional) dilaton.

Let us at first consider the metric scalars. The ten-dimensional metric in Einstein frame can be written as
\begin{equation}
\d s_{10}^2 = \e^{2\alpha v} \d s_7^2 + \e^{2\beta v} \d s_3^2,
\end{equation}
where $v$ is the volume modulus and $\alpha$ and $\beta$ are numbers. Demanding that the seven-dimensional effective field theory is in Einstein frame and the volume modulus is canonically normalised, we find
\begin{equation}
\beta = -\frac{5}{3} \alpha, \qquad \alpha^2 = \frac{3}{80}.
\end{equation}
Since we pulled out the volume modulus, the internal metric is given by a symmetric, positive definite $3 \times 3$ matrix with fixed determinant. We can therefore write
\begin{equation}
\d s_3^2 = M_{ik} e^i e^k,
\end{equation}
where $M = L L^T$ with
\begin{equation}
L = \begin{pmatrix}
  \e^{\tfrac{1}{2}\sigma_1+\tfrac{1}{2\sqrt{3}}\sigma_2} & \e^{-\tfrac{1}{2}\sigma_1+\tfrac{1}{2\sqrt{3}}\sigma_2} \chi_1 & \e^{-\tfrac{1}{\sqrt{3}}\sigma_2} \left({\chi_1\chi_2 + \chi_3}\right) \\
  0 & \e^{-\tfrac{1}{2}\sigma_1+\tfrac{1}{2\sqrt{3}}\sigma_2} &  \e^{-\tfrac{1}{\sqrt{3}}\sigma_2}\chi_2 \\
  0 & 0 &  \e^{-\tfrac{1}{\sqrt{3}}\sigma_2}
 \end{pmatrix}
\end{equation}
and $\det M =1$. One can check that the five scalars $\sigma_1$, $\sigma_2$, $\chi_1$, $\chi_2$ and $\chi_3$ are then canonically normalised. Their vevs are $\sigma_1 = \sigma_2 = \chi_i = 0$ since the smeared solution yields the ``round sphere'' with $M=\mathds{1}$.

The scalars from the form fields are obtained as follows. Since $H$ carries a non-trivial flux, we write
\begin{equation}
H = h \epsilon _3 + \d B.
\end{equation}
The expansion of $B$ then gives rise to three scalar fields $b^i$,
\begin{equation}
B = \frac{1}{2} b^i \epsilon_{ikl} e^k \w e^l.
\end{equation}
The RR fields can be written as
\begin{equation}
C_1 = c_i e^i, \qquad C_3 = a \epsilon_3.
\end{equation}
Thus, $C_1$ yields three scalars $c_i$, and $C_3$ gives one (axionic) scalar $a$. One can check that all scalars are canonically normalised.

The scalar potential of the effective field theory gets contributions from the internal curvature, the $H$ flux, the RR field strengths and the tension of the sources,
\begin{equation}
V = V_R + V_H + V_0 + V_2 + V_{\textrm{D}6}.
\end{equation}
Substituting the expressions for the metric and the form fields, we find
\begin{align}
& V_R = \e^{(7\alpha+\beta)v} q^2 \left[{-\frac{1}{2} \left({\mathrm{Tr}\, M}\right)^2 + \mathrm{Tr} \left({M^2}\right)}\right], \\
& V_H = \frac{1}{2} h^2 \e^{-\phi+(7\alpha-3\beta)v}, \\
& V_0 = \frac{1}{2} \e^{\tfrac{5}{2}\phi+(7\alpha+3\beta)v} F_0^2, \\
& V_2 = \frac{1}{2} \e^{\tfrac{3}{2}\phi + (7\alpha-\beta)v} M_{mn} \left({F_0 b^m-q c_i \delta^{mi}}\right) \left({F_0b^n-q c_k \delta^{nk}}\right), \label{simple-stability} \\
& V_{\textrm{D}6} = \e^{\tfrac{3}{4}\phi+7\alpha v} \mu_6.
\end{align}
In order to determine whether the smeared solution is stable with respect to the $14$ moduli $\varphi_a \in \left\{ \phi, v, \sigma_1, \sigma_2, \chi_i, b^i, c_i, a \right\}$, we now have to evaluate the mass matrix $\partial_{\varphi_a} \partial_{\varphi_b} V$ at an extremum of $V$. Since $F_4$ does not contribute to the scalar potential, $a$ is a massless axion and can therefore be set to zero at the solution. Furthermore, we can read off from \eqref{simple-stability} that the moduli combinations $F_0 b^m - qc_i \delta^{mi}$ are stabilised at $F_0 b^m - qc_i \delta^{mi}=0$, whereas the orthogonal combinations are flat in moduli space. We can therefore also set $b^i = c_i = 0$. The volume modulus $v$ and the dilaton modulus $\phi$ are zero at the solution if we choose
\begin{equation}
q^2 = \frac{28}{25}h^2, \qquad F_0^2 = \frac{4}{25} h^2, \qquad \mu_6 = \frac{2}{5} h^2.
\end{equation}
Finally, we also have $\sigma_1=\sigma_2=\chi_i=0$ at the solution. We therefore have to evaluate the mass matrix at the origin in moduli space. This yields the eigenvalues
\begin{equation}
h^2 \left\{{\frac{56}{25}, \frac{32}{25}, \frac{2}{25}\left({18 \pm \sqrt{79}}\right),0 }\right\},
\end{equation}
where the multiplicity of $\frac{56}{25}h^2$ is $5$, the multiplicity of $\frac{32}{25}h^2$ is $3$, the multiplicity of $0$ is $4$ and the multiplicity of the other eigenvalues is $1$. Since all eigenvalues are positive, we conclude that the smeared solution is stable with respect to the $14$ left-invariant closed string modes.
\\

\subsection{Localisation}
\label{ch:simple-non-bps.localisation}

In this section, we consider the localisation of the smeared non-BPS setup discussed above. We first analyse the equations of motion for regularised brane profiles using a Taylor expansion and find strong constraints on the possible configurations. For smooth brane profiles, we show that the only allowed solution is the smeared one. Furthermore, we formulate a simple topological argument that rules out brane profiles in the shape of a step function since they are not consistent with the required properties of the fluxes away from the branes. For genuine delta function profiles, we compute the boundary conditions that are locally allowed by the equations of motion. This yields two different cases: the first boundary condition does not lead to a globally consistent solution due to our topological argument, while the second one yields a singularity in the energy density of the $H$ flux.

\subsubsection{Ansatz}
\label{ch:simple-non-bps.localisation.ansatz}

In order to study localisation, we assume that all sources are located on the north-pole and/or the south-pole of the three-sphere such that our setup preserves an $\mathrm{SO}(3)$ symmetry of the $\mathrm{SO}(4)$ symmetry exhibited by the setup with smeared sources. We furthermore assume that the symmetries of our setup are not broken in the solution. The most general ansatz for the metric is then
\begin{equation}
\d s_{10}^2 = \e^{2A(\theta)} \d s_7^2 + \e^{2B(\theta)} \left({\d \theta^2 + \e^{2 C(\theta)} \sin^2 \theta\, \d \Omega_2^2}\right),
\end{equation}
where $\d s_7^2$ is the AdS metric with unit radius and $\d \Omega_2^2$ is the metric of the unit two-sphere. The second conformal factor $C$ can be absorbed into $B$ by a suitable coordinate transformation $\theta \to \tilde \theta(\theta)$ that keeps the sources at the poles. One can show that, if the original metric does not have any singularities away from the poles, then the same is true for the new metric for which $C$ is transformed away. We can therefore write
\begin{equation}
\d s_{10}^2 = \e^{2A(\theta)} \d s_7^2 + \e^{2B(\theta)} \left({\d \theta^2 + \sin^2 \theta\, \d \Omega_2^2}\right)
\end{equation}
without loss of generality and assume that $A$ and $B$ are regular away from the poles. Our ansatz for the form fields is
\begin{equation}
H = \lambda F_0 \e^{\tfrac{7}{4}\phi} \star_3 1, \qquad F_2 = \e^{-\tfrac{3}{2}\phi-7A} \star_3 \d \alpha,
\end{equation}
where $\phi$, $\lambda$ and $\alpha$ are functions depending on $\theta$ and $F_0$ is a constant. One can check that this is the most general ansatz compatible with the form equations of motion \eqref{nonbps-eom3} and \eqref{nonbps-eom4} and the symmetries of the setup.

The equations of motion \eqref{nonbps-eom1}, \eqref{nonbps-eom2}, \eqref{nonbps-eom3} and \eqref{nonbps-eom4} then simplify as follows. The $F_2$ Bianchi identity yields
\begin{equation}
\frac{\left({\e^{-\tfrac{3}{2}\phi-7A+B} \sin^2 \theta\, \alpha^\prime}\right)^\prime}{\e^{3B} \sin^2 \theta} = \e^{\tfrac{7}{4}\phi} \lambda F_0^2 + Q \delta(\Sigma), \label{eoms-theta-bianchi}
\end{equation}
where primes are derivatives with respect to $\theta$ and $Q \delta(\Sigma)$ is a sum of delta functions that contains all charge contributions from sources at the north-pole and the south-pole. The $H$ equation implies that $\alpha$ can be eliminated in terms of $\lambda$,
\begin{equation}
\alpha = \e^{\tfrac{3}{4}\phi+7A} \lambda + \alpha_0, \label{lambda-alpha}
\end{equation}
where $\alpha_0$ is an integration constant, which we set to zero in the following. The dilaton equation reads
\begin{equation}
\frac{\left({\e^{7A+B} \sin^2 \theta\, \phi^\prime}\right)^\prime}{\e^{7A+3B} \sin^2 \theta} = \e^{\tfrac{5}{2}\phi} F_0^2 \left({\frac{5}{4} - \frac{\lambda^2}{2}}\right) + \frac{3}{4} \e^{-14A-2B-\tfrac{3}{2}\phi} \left({\alpha^\prime}\right)^2 + \frac{3}{4} \e^{\tfrac{3}{4}\phi} T \delta(\Sigma), \label{eoms-theta-dilaton}
\end{equation}
where $T \delta(\Sigma)$ denotes the sum over all tension contributions from localised sources. Taking the trace of the external Einstein equation, we find
\begin{align}
\frac{16}{7} R_7 &= - 96 \e^{-2A} - 16 \e^{-2B} \left[{7 \left({A^\prime}\right)^2 + A^\prime B^\prime + \frac{\left({\sin^2 \theta\, A^\prime}\right)^\prime}{\sin^2 \theta}}\right] \nll = \e^{\tfrac{5}{2}\phi} F_0^2 \left({1 - 2\lambda^2}\right) - \e^{-14A-2B-\tfrac{3}{2}\phi} \left({\alpha^\prime}\right)^2 - \e^{\tfrac{3}{4}\phi} T \delta(\Sigma). \label{eoms-theta-exteinstein}
\end{align}
The internal Einstein equation in $(\theta\theta)$-direction gives
\begin{align}
0 &= - 2 + \frac{\left({\sin^2 \theta \, B^\prime}\right)^\prime}{\sin^2 \theta} + 7 \left({A^\prime}\right)^2 + B^{\prime\prime} + 7 A^{\prime\prime} - 7 A^\prime B^\prime \nl +\frac{1}{2} \left({\phi^\prime}\right)^2 + \frac{1}{16} \e^{\tfrac{5}{2}\phi+2B} F_0^2 \left({1+6\lambda^2}\right) - \frac{1}{16} \e^{-14A-\tfrac{3}{2}\phi} \left({\alpha^\prime}\right)^2 + \frac{7}{16} \e^{\tfrac{3}{4}\phi+2B}T\delta(\Sigma), \label{eoms-theta-inteinsteintheta}
\end{align}
and the Einstein equations in the remaining directions yield
\begin{align}
0 &= - 2 + \frac{\left({\sin^2 \theta \, B^\prime}\right)^\prime}{\sin^2 \theta} + \left({B^\prime}\right)^2 + \cot \theta \left({7A+B}\right)^\prime + 7 A^\prime B^\prime \nl + \frac{1}{16} \e^{\tfrac{5}{2}\phi+2B} F_0^2 \left({1+6\lambda^2}\right) + \frac{7}{16} \e^{-14A-\tfrac{3}{2}\phi} \left({\alpha^\prime}\right)^2 + \frac{7}{16} \e^{\tfrac{3}{4}\phi+2B}T\delta(\Sigma), \label{eoms-theta-inteinsteintrans}
\end{align}
where we used the results of Appendix \ref{app:computations.curvature.block} to compute the components of the Ricci tensor. All other equations of motion are automatically satisfied with above ansatz for the metric and the form fields. Thus, the problem of finding a localised solution of our setup is reduced to solving the five second-order ordinary differential equations \eqref{eoms-theta-bianchi}, \eqref{eoms-theta-dilaton}, \eqref{eoms-theta-exteinstein}, \eqref{eoms-theta-inteinsteintheta} and \eqref{eoms-theta-inteinsteintrans} for the four unknown functions $A$, $B$, $\phi$ and $\lambda$ (or, equivalently, $A$, $B$, $\phi$ and $\alpha$).

Moreover, the equations of motion yield a simple topological constraint, which can be used to rule out a large class of configurations in our setup without having to know the full solution. At any point away from the sources, we can combine \eqref{eoms-theta-bianchi} and \eqref{lambda-alpha} to find
\begin{equation}
\frac{\left({\e^{-\tfrac{3}{2}\phi-7A+B} \sin^2 \theta}\right)^\prime}{\e^{3B} \sin^2 \theta}\, \alpha^\prime + \e^{-\tfrac{3}{2}\phi-7A-2B} \alpha^{\prime\prime}  = \alpha \e^{\phi-7A} F_0^2,
\end{equation}
where $\e^A$, $\e^B$ and $\e^\phi$ are non-singular. Thus, $\alpha$ must satisfy
\begin{equation}
\mathrm{sgn}\, \alpha^{\prime\prime} = \mathrm{sgn}\, \alpha \label{top-constraint}
\end{equation}
at every extremum with $\alpha^\prime=0$ such that the sign of $\alpha$ determines whether the extremum is a maximum or a minimum. One can check that this is also true at either of the poles ($\theta=0$, $\theta=\pi$) if there are no sources located at that pole.

\subsubsection{Regularised Sources}
\label{ch:simple-non-bps.localisation.regular}

\begin{figure}[t]
\centering
\includegraphics[trim = 0mm 50mm 0mm 0mm, clip, width=0.5 \textwidth]{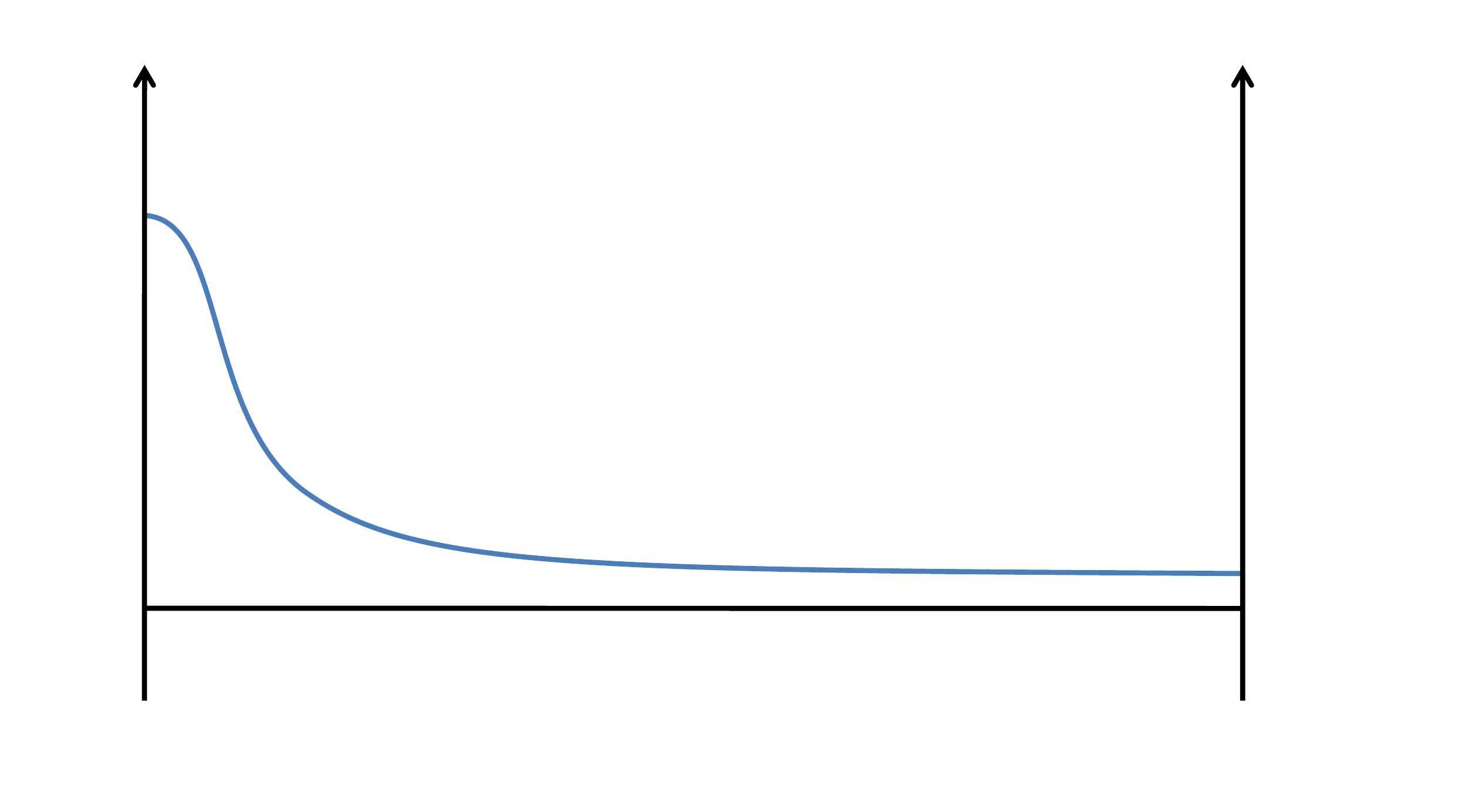}
\put(-200,104){$\scriptstyle z(\theta)$}
\put(-200,20){$\scriptstyle \theta=0$}
\put(-54,20){$\scriptstyle \theta=\pi$}
\caption{A smooth brane profile. \label{fig_regsources1}}
\end{figure}

Before we turn to the case of fully localised sources, it is instructive to study above setup with regularised source profiles of finite extent, where the contribution of the sources to the equations of motion is non-zero at points away from the poles.\footnote{This is consistent with our ansatz if the source profiles respect the assumed $\mathrm{SO}(3)$ symmetry of the solution.} Such profiles can be viewed as intermediate steps between smeared and fully localised sources and can approximate a delta function profile with arbitrary precision. They can therefore be useful to get an intuition for the problems that arise from localisation in our setup. We first consider extremal sources with $|Q| = T > 0$ and a profile of the form
\begin{equation}
T \delta(\Sigma) = z(\theta) \e^{-3B(\theta)}, \label{regularsource}
\end{equation}
where $z(\theta)$ is a smooth function that becomes a constant in the smeared limit (cf. Figure \ref{fig_regsources1}). Note that we took out a factor $\e^{-3B(\theta)}$ on the right-hand side since the delta function $\delta(\Sigma)$ contains an inverse square root of the metric determinant.

We can check whether such a configuration is consistent with the equations of motion \eqref{eoms-theta-bianchi}, \eqref{eoms-theta-dilaton}, \eqref{eoms-theta-exteinstein}, \eqref{eoms-theta-inteinsteintheta} and \eqref{eoms-theta-inteinsteintrans} by performing a Taylor expansion
\begin{gather}
\e^{-A(\theta)} = \sum\limits_n a_n \left({\theta-\theta_0}\right)^n, \qquad \e^{-2B(\theta)} = \sum\limits_n b_n \left({\theta-\theta_0}\right)^n, \notag \\ \e^{-\tfrac{1}{4}\phi(\theta)} = \sum\limits_n f_n \left({\theta-\theta_0}\right)^n, \qquad \lambda(\theta) = \sum\limits_n \lambda_n \left({\theta-\theta_0}\right)^n
\end{gather}
around an arbitrary point $\theta_0 \!\in\,\, ]0,\pi[$ for which $z(\theta_0) \neq 0$.\footnote{Note that our conventions for the coefficients $a_n$, $b_n$ and $f_n$ differ from those used in \cite{Blaback:2011nz}.} Substituting this expansion into the equations of motion, we obtain algebraic equations at every order $n$ that have to be satisfied at $\theta = \theta_0$. For every $n$, we can then use the four equations \eqref{eoms-theta-bianchi}, \eqref{eoms-theta-dilaton}, \eqref{eoms-theta-exteinstein} and \eqref{eoms-theta-inteinsteintheta} to determine $a_{n+2}$, $b_{n+2}$, $f_{n+2}$ and $\lambda_{n+2}$ in terms of the lowest order coefficients $q = \left\{{a_0,b_0,f_0,\lambda_0,a_1,b_1,f_1,\lambda_1}\right\}$. The remaining equation \eqref{eoms-theta-inteinsteintrans} leads to an additional constraint between the elements of $q$ only. At zeroth order ($n=0$), for example, the first four equations determine $a_2$, $b_2$, $f_2$ and $\lambda_2$ in terms of $q$, while the fifth equation relates the coefficients $q$ to each other, and so on. Going to higher orders $n$ then potentially produces an infinite amount of such relations for $q$ and should therefore lead to strong constraints.

As detailed in Appendix \ref{app:computations.regularised}, the equations of the first five orders yield
\begin{equation}
a_1 = f_1 = \lambda_1 = 0, \qquad \lambda_0 = \mp \frac{5}{2}, \qquad F_0^2 = \frac{48}{7} a_0^2f_0^{10},
\end{equation}
where the upper sign is for (regularised) D$6$-branes and the lower sign for \dsix-branes. Since $\theta_0$ is arbitrary, these conditions must be satisfied for all $\theta_0 \!\in\,\, ]0,\pi[$ such that
\begin{equation}
\e^{-A(\theta)} = a = \textrm{const.}, \qquad\!\! \e^{-\tfrac{1}{4}\phi(\theta)} = f = \textrm{const.}, \qquad\!\! \lambda(\theta) = \mp \frac{5}{2}, \qquad\!\! F_0^2 = \frac{48}{7} a^2 f^{10}
\end{equation}
everywhere. We can now substitute these expressions back into the equations of motion \eqref{eoms-theta-bianchi}, \eqref{eoms-theta-dilaton}, \eqref{eoms-theta-exteinstein}, \eqref{eoms-theta-inteinsteintheta} and \eqref{eoms-theta-inteinsteintrans} and solve for $B(\theta)$ and $z(\theta)$ to find
\begin{align}
& \e^{2B(\theta)} = \frac{4}{3} \left[{\xi \cos \theta - \sqrt{16 a^2 + \xi^2}}\right]^{-2}, \label{solution-regularised} \\ & z(\theta) = \frac{320}{7 \sqrt{3}} a^2f^3 \left|{\xi \cos \theta - \sqrt{16 a^2 + \xi^2}}\right|^{-3},
\end{align}
where $\xi$ is a free parameter. It then follows from \eqref{regularsource} that the source terms in the equations of motion are constant for this solution. Furthermore, all fields except for $B(\theta)$ are also constant. This suggests that the solution is just the smeared solution written in a different coordinate system. Indeed, the corresponding coordinate transformation can be found considering
\begin{equation}
\d s_3^2 = r^2 \left({\d \tilde \theta^2 + \sin^2 \tilde \theta\, \d \Omega_2^2}\right) = \e^{2B(\theta)} \left({\d \theta^2 + \sin^2 \theta\, \d \Omega_2^2}\right),
\end{equation}
which yields two equations that determine $\tilde \theta$ in terms of $\theta$ and $B(\theta)$. Eliminating $\tilde \theta$ then leads to an ordinary differential equation for $B(\theta)$ with solutions that exactly agree with \eqref{solution-regularised}. Thus, we have shown that the smeared solution is the only solution to the equations of motion for which the source profile is a smooth function.

\begin{figure}[t]
\centering
\includegraphics[trim = 0mm 50mm 0mm 0mm, clip, width=0.5 \textwidth]{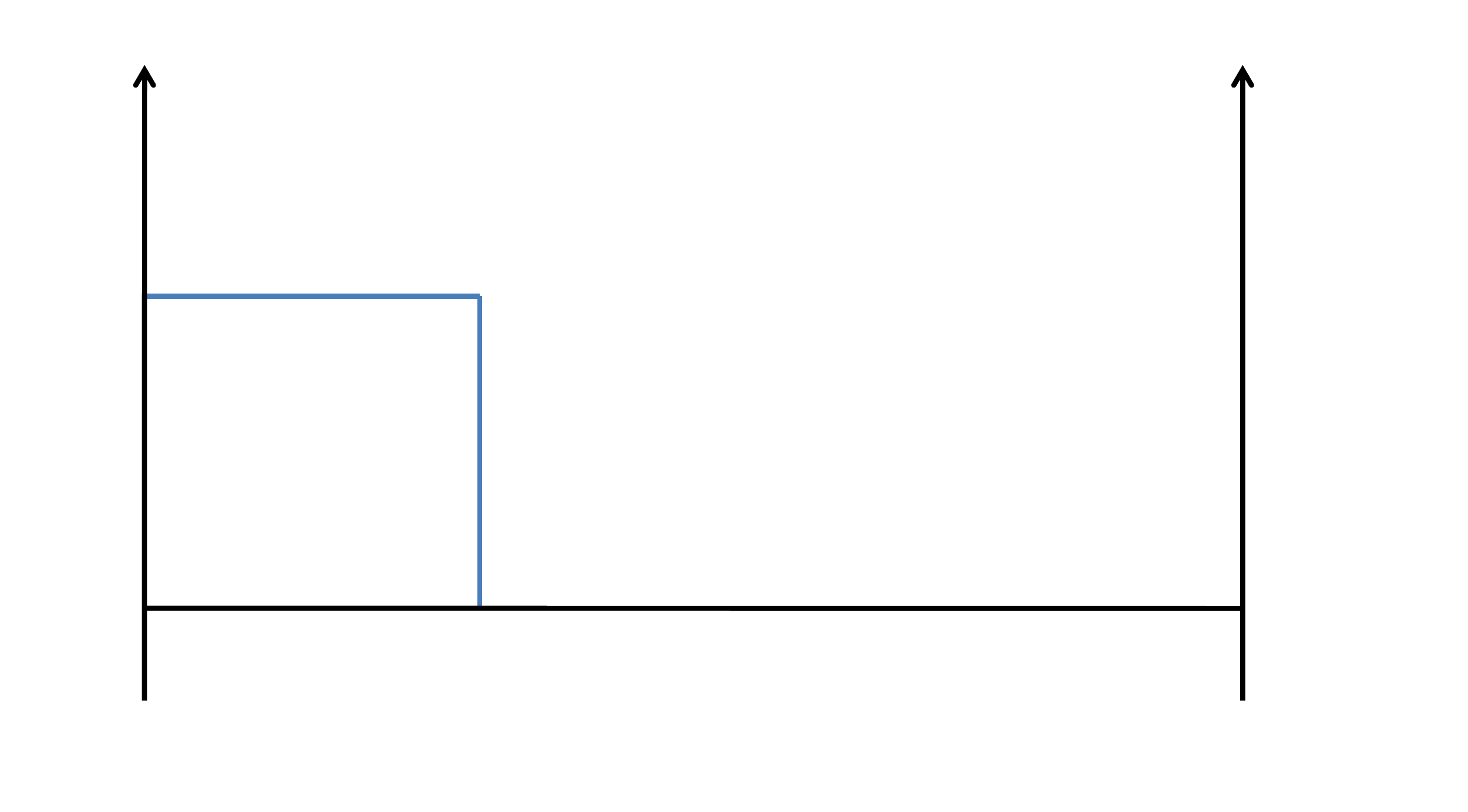}
\put(-200,104){$\scriptstyle z(\theta)$}
\put(-200,20){$\scriptstyle \theta=0$}
\put(-160,20){$\scriptstyle \theta=\epsilon$}
\put(-54,20){$\scriptstyle \theta=\pi$}
\caption{A brane profile in the shape of a step function. \label{fig_regsources2}}
\end{figure}
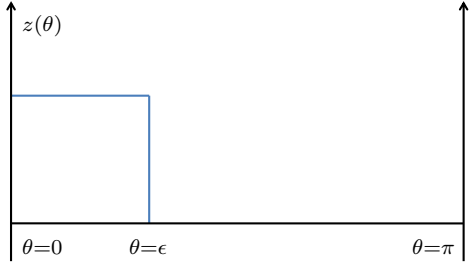

Since this proof is only valid in the support of $z(\theta)$, it does not rule out source profiles in the shape of a step function. Such source profiles can, however, be excluded using the topological constraint \eqref{top-constraint}. In order to show this, let us consider a source that extends from $\theta=0$ to $\theta=\epsilon$ as in Figure \ref{fig_regsources2}. It then follows from above discussion that the solution in the region $\theta \le \epsilon$, where the contribution of the source is non-zero, is the smeared solution. Outside the source, we have to solve the vacuum equations and then match the two solutions at the boundary $\theta=\epsilon$. For the smeared solution inside the source, the $F_2$ Bianchi identity implies that the charge of the source exactly cancels the charge that is induced by the fluxes within that region. The solution outside the source must therefore have zero integrated flux-induced charge in order that the global tadpole is cancelled,
\begin{equation}
\int\limits^\pi_\epsilon F_0 H = F_0^2 \int\limits^\pi_\epsilon \alpha \e^{\phi-7A} \star_3 1 = 0. \label{stepfunction-constraint}
\end{equation}
Since $\e^{A}$ and $\e^\phi$ are positive, this implies that $\alpha$ must vary outside the source and change its sign at some point between $\theta=\epsilon$ and $\theta=\pi$.

\begin{figure}[t]
\centering
\includegraphics[width=0.5 \textwidth]{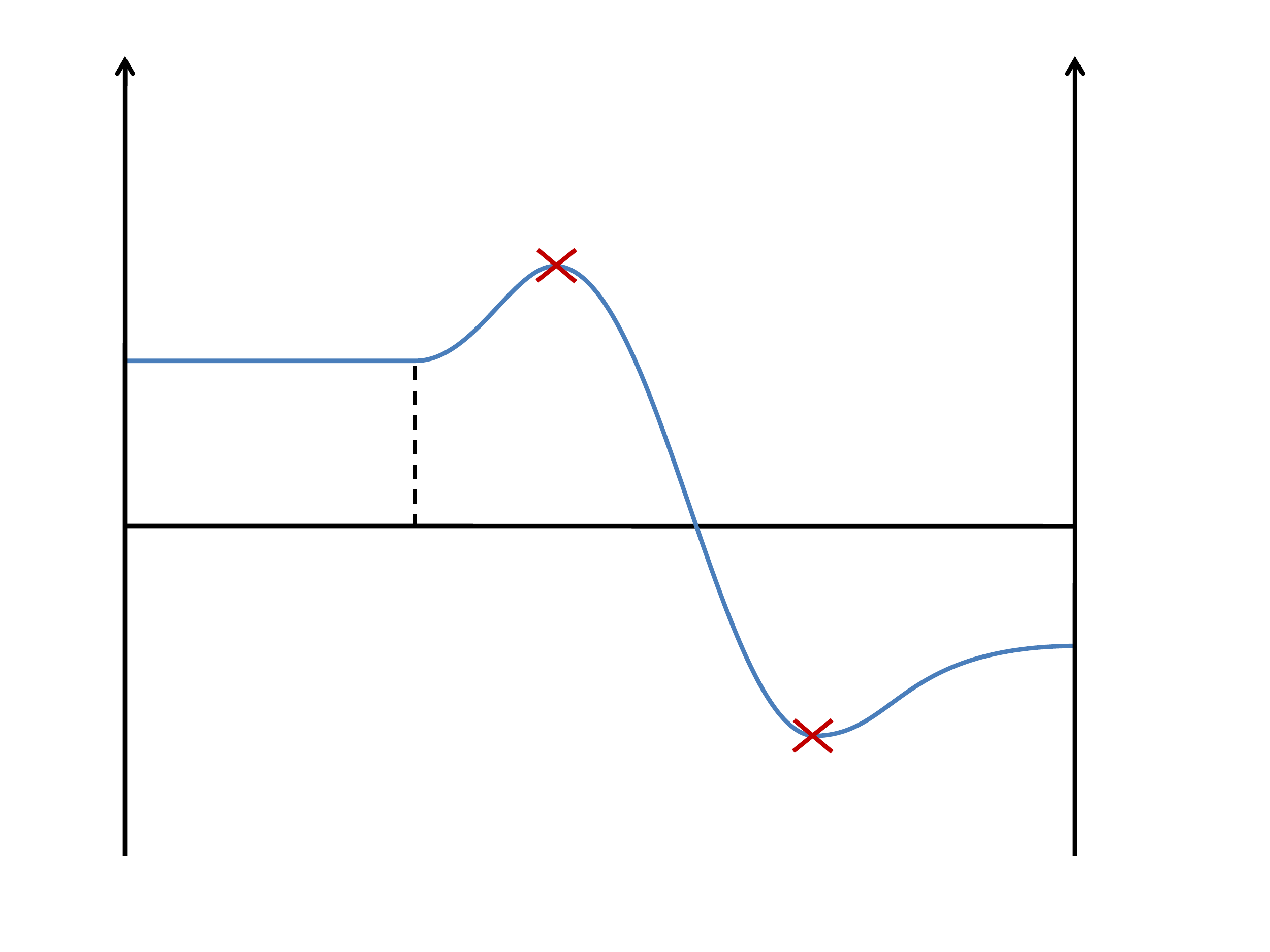}
\put(-200,148){$\scriptstyle \alpha(\theta)$}
\put(-200,20){$\scriptstyle \theta=0$}
\put(-54,20){$\scriptstyle \theta=\pi$}
\caption{An example for a flux configuration in the presence of a step function source with extrema that are required for global tadpole cancellation but forbidden by the constraint $\mathrm{sgn}\, \alpha^{\prime\prime} = \mathrm{sgn}\, \alpha$. \label{fig_excluded-reg}}
\end{figure}

It is now straightforward to show that such a behaviour of $\alpha$ is not consistent with the constraint \eqref{top-constraint}. Let us, for example, consider a solution that has positive $\alpha$ in the region inside the source. The argument for the case of negative $\alpha$ is then analogous. Assuming that the fields and their first derivatives are continuous at the boundary $\theta=\epsilon$, we have to impose $\alpha^\prime=0$ at the step. Since $\alpha$ is positive there by assumption, it then follows from \eqref{top-constraint} that $\alpha^{\prime\prime} > 0$ such that $\alpha$ increases as we move away from the source. Because of \eqref{stepfunction-constraint}, however, $\alpha$ must eventually become negative to cancel the global tadpole and must therefore reach a maximum at some point. This is in conflict with \eqref{top-constraint}, which states that any extremum of $\alpha$ must be a minimum if $\alpha$ is positive at that point. Due to rotational symmetry, we must furthermore impose $\alpha^\prime=0$ at the south-pole $\theta=\pi$, which leads to at least one more extremum forbidden by \eqref{top-constraint} (cf. Figure \ref{fig_excluded-reg}). Regularised source profiles in the shape of a step function can therefore not lead to a globally consistent solution. One can check that the same is true for configurations that have step function sources at both poles.

The results of this section confirm the intuitive argument of Section \ref{ch:smeared-vs-loc.discussion.loc}, where we pointed out that localisation may be problematic in setups with sources that are not mutually BPS with the fluxes since it can lead to non-zero forces that render the solution unstable. It could have been that such instabilities are somehow evaded by a delicate balance between the flux configuration and the sources if one considers the most general ansatz for localisation. However, at least for regularised source profiles, this is not the case, and the only consistent solution with such a profile is the smeared solution.

\subsubsection{Delta Function Sources}
\label{ch:simple-non-bps.localisation.delta}

Let us now consider fully localised D$6$/\dsix-branes with a delta function profile. In order to examine whether consistent solutions can exist with such sources in our model, we have to know the boundary conditions for the supergravity fields in the near-brane region. These can be computed by expanding the vacuum equations around one of the poles, \mbox{e.\ g.} the north-pole with $\theta=0$. A general ansatz for such an expansion is
\begin{align}
& \e^{-A(\theta)} = a_0 \theta^A + a_1 \theta^{A+\zeta} + a_2 \theta^{A+\xi} + \ldots, \notag \\ & \e^{-2B(\theta)} = b_0 \theta^B + b_1 \theta^{B+\zeta} + b_2 \theta^{B+\xi} + \ldots, \notag \\ & \e^{-\tfrac{1}{4}\phi(\theta)} = f_0 \theta^F + f_1 \theta^{F+\zeta} + f_2 \theta^{F+\xi} + \ldots, \notag \\ & \lambda(\theta) = \lambda_0 \theta^L + \lambda_1 \theta^{L+\zeta} + \lambda_2 \theta^{L+\xi} + \ldots, \label{eq:d6-ansatz-bc}
\end{align}
where the exponents $A$, $B$, $F$, $L$ and $\zeta < \xi < \ldots$ on the right-hand sides are unknown (possibly not integral) real numbers. We also assume that $\e^{-A}$, $\e^{-2B}$, $\e^{-\frac{1}{4}\phi}$ and $\lambda$ do not have essential singularities, i.\ e.\ that there is a finite leading power of $\theta$ for each function. The coefficients $a_0$, $b_0$, $f_0$ and $\lambda_0$ are taken to be non-zero such that $A$, $B$, $F$ and $L$ by definition determine the leading order divergences of the fields. Furthermore, $a_0$, $b_0$ and $f_0$ must be non-negative since they are the leading order coefficients in the expansion of the exponential functions $\e^{-A}$, $\e^{-2B}$ and $\e^{-\frac{1}{4}\phi}$. However, we allow all sub-leading order coefficients (such as, for example, $a_1$, $b_5$, $\lambda_2$, etc.) to be zero such that the steps between the different powers of $\theta$ in the various expansion series are not necessarily the same at all orders and for all functions.\footnote{If, for example, $\zeta=\frac{1}{2}$, $\xi=1$ and $a_1 \neq 0$, $f_1=0$, $f_2 \neq 0$, we would obtain $\e^{-A(\theta)} = a_0 \theta^A + a_1 \theta^{A+\frac{1}{2}} + \ldots$ and $\e^{-\frac{1}{4}\phi(\theta)} = f_0 \theta^F + f_2 \theta^{F+1} + \ldots$}

Substituting this ansatz into the equations of motion \eqref{eoms-theta-bianchi}, \eqref{eoms-theta-dilaton}, \eqref{eoms-theta-exteinstein}, \eqref{eoms-theta-inteinsteintheta} and \eqref{eoms-theta-inteinsteintrans}, we can explicitly check which choices for the powers $A$, $B$, $F$, $L$, $\zeta$, $\xi$, $\ldots$ and the coefficients $a_n$, $b_n$, $f_n$ and $\lambda_n$ are consistent. The computation turns out rather lengthy and is detailed in Appendix \ref{app:computations.bc}. The result, however, is surprisingly simple: we find that only five different boundary conditions are locally allowed by the equations of motion. As explained in Appendix \ref{app:computations.bc.sources}, the behaviour of the fields near $\theta=0$ determines the tension and the RR charge of the source that must be present at the pole. This reveals that one of the five boundary conditions leads to inconsistent results for the tension in the different equations of motion and another one has singular fields at the pole without the presence of a localised source there. We therefore discard these two cases and only list the remaining three boundary conditions. These are
\begin{itemize}
\item the smooth solution with no sources sitting at the pole,
\begin{equation}
L = A = B = F = 0,
\end{equation}
\item the standard BPS boundary condition \cite{Janssen:1999sa},
\begin{equation}
L=0, \qquad A= -\frac{1}{16}, \qquad B = \frac{7}{8}, \qquad F= -\frac{3}{16}, \qquad \lambda_0 = \pm 1, \label{simple-bpsbc}
\end{equation}
which yields extremal branes with $|Q|=T$,
\item a previously unknown boundary condition with divergent $\lambda$,
\begin{equation}
L=-1, \qquad A= -\frac{1}{16}, \qquad B = \frac{7}{8}, \qquad F= -\frac{3}{16}, \label{new-bc}
\end{equation}
which also yields extremal branes with $|Q|=T$.
\end{itemize}
The properties of these boundary conditions are summarised in Table \ref{table-bc} and derived in Appendix \ref{app:computations.bc}.\footnote{Our ansatz can even be extended to allow logarithmic divergences of the form $\e^{-A(\theta)} = a_0 \theta^A (\ln \theta)^{\tilde A} + \ldots$, $\e^{-2B(\theta)} = b_0 \theta^B (\ln \theta)^{\tilde B} + \ldots$, etc. One can then show that all powers of logarithms have to be zero in the leading order terms of the fields such that no further boundary conditions arise from this more general ansatz.}

\begin{table}[t]\renewcommand{\arraystretch}{1.4}\setlength{\tabcolsep}{11pt}
\begin{center}
  \begin{tabular}{ |c | c | c | c || c | c || c | c || c | }
    \hline
    $L$ & $A$ & $B$ & $F$ & $\lambda(0)$ & $\alpha(0)$ & source & $\frac{|Q|}{T}$ & valid \\ \hline\hline
    $0$ & $0$ & $0$ & $0$ & finite & finite & none & - & \checkmark \\ \hline
    $0$ & $-\frac{1}{16}$ & $\frac{7}{8}$ & $-\frac{3}{16}$ & $\pm 1$ & $ 0 $ & D$6$/\dsix & $1 $ & only locally \\ \hline
    $-1$ & $-\frac{1}{16}$ & $\frac{7}{8}$ & $-\frac{3}{16}$ & $\pm \infty $ & finite & D$6$/\dsix & $1$  & \checkmark \\ \hline
    \end{tabular}
\caption{The boundary conditions that are locally allowed by the equations of motion.}
\label{table-bc}
\end{center}
\end{table}

Let us now analyse in more detail the two boundary conditions \eqref{simple-bpsbc} and \eqref{new-bc}, which support the presence of localised sources at the pole(s) where they are imposed. The boundary condition given by \eqref{simple-bpsbc} is the standard boundary condition that also appears in the BPS solution for D$6$-branes found in \cite{Janssen:1999sa}, which, like our setup, has $F_0$ and $H$ flux and a non-vanishing $F_2$ field strength. If we impose this boundary condition in our setup, the fluxes are locally BPS with the branes in their vicinity. In the region very close to the branes, we can therefore expect that there are no forces between the different ingredients that could destabilise the configuration and thus destroy a possible solution. Global tadpole cancellation requires, however, that the integrated flux and the branes are of opposite BPS type. Further away from the sources, the flux must therefore change its type, and it is not obvious anymore whether such a configuration can still be stable. Indeed, we can invoke a topological argument similar to the one used in the previous section to show that the BPS boundary condition cannot lead to a globally consistent solution in our model.

This can be seen as follows. Assuming the presence of a source at $\theta=0$, we can use \eqref{lambda-alpha}, \eqref{eq:d6-ansatz-bc} and \eqref{simple-bpsbc} to compute the behaviour of $\alpha$ near the pole,
\begin{equation}
\alpha(\theta) = \pm \frac{1}{a_0^7f_0^3} \theta + \mathcal{O}(\theta^2),
\end{equation}
where $a_0$ and $f_0$ are by definition non-zero and positive. Thus, we find $\alpha(0)=0$. The sign of $\alpha^\prime(0)$ is determined by the charge of the source that sits at the pole. As derived in Appendix \ref{app:computations.bc}, the charge and the tension of the source are given by
\begin{equation}
Q = \pm \frac{f_0^3}{\sqrt{b_0}}, \qquad T= \frac{f_0^3}{\sqrt{b_0}}.
\end{equation}
We thus find that $\alpha^\prime$ is positive near the pole for $Q>0$ (D$6$-branes) and negative for $Q<0$ (\dsix-branes).

\begin{figure}[t]
\centering
\includegraphics[width=0.5 \textwidth]{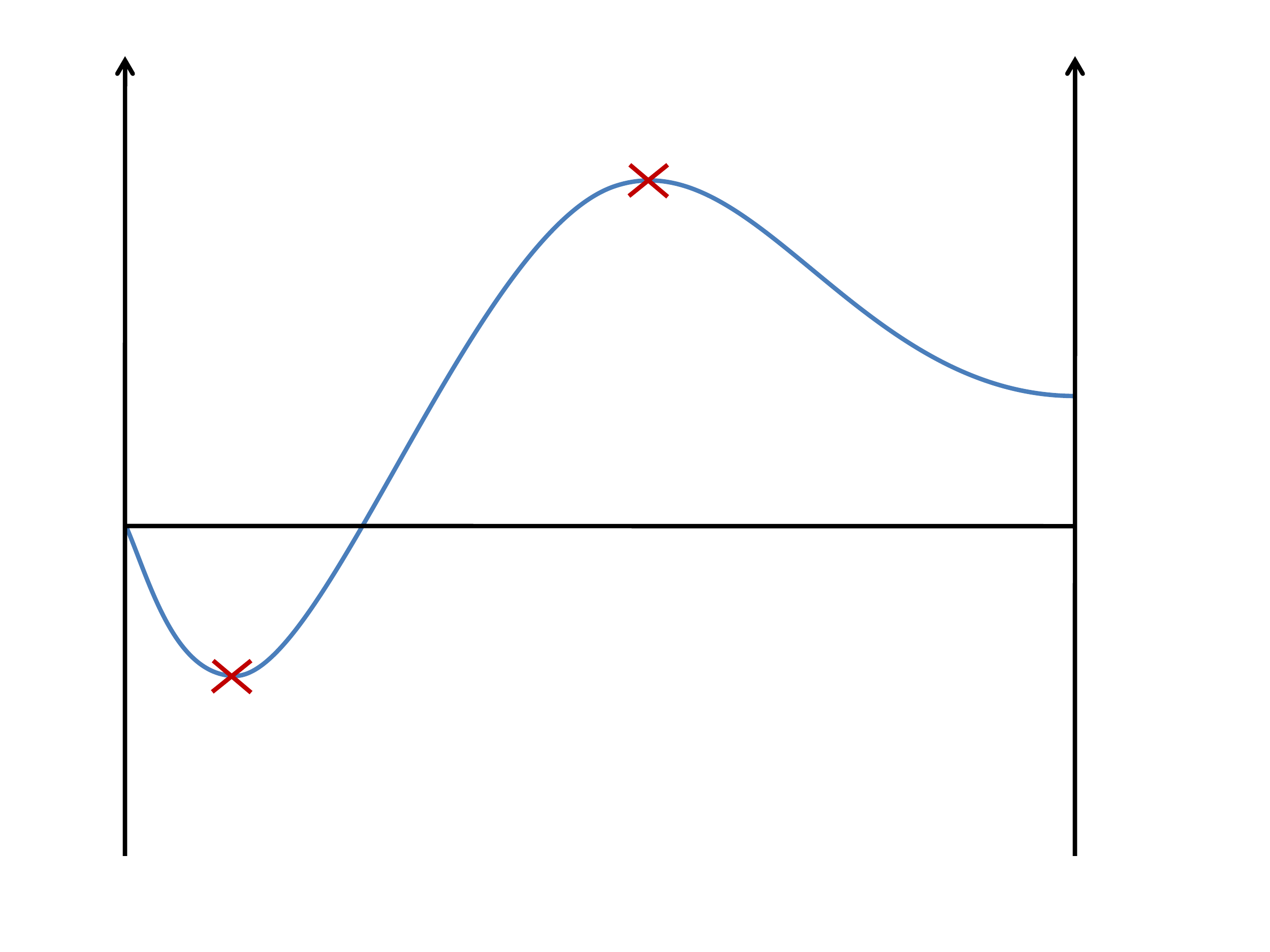}
\put(-200,148){$\scriptstyle \alpha(\theta)$}
\put(-200,20){$\scriptstyle \theta=0$}
\put(-54,20){$\scriptstyle \theta=\pi$}
\put(-200,80){\scriptsize \dsix}
\includegraphics[width=0.5 \textwidth]{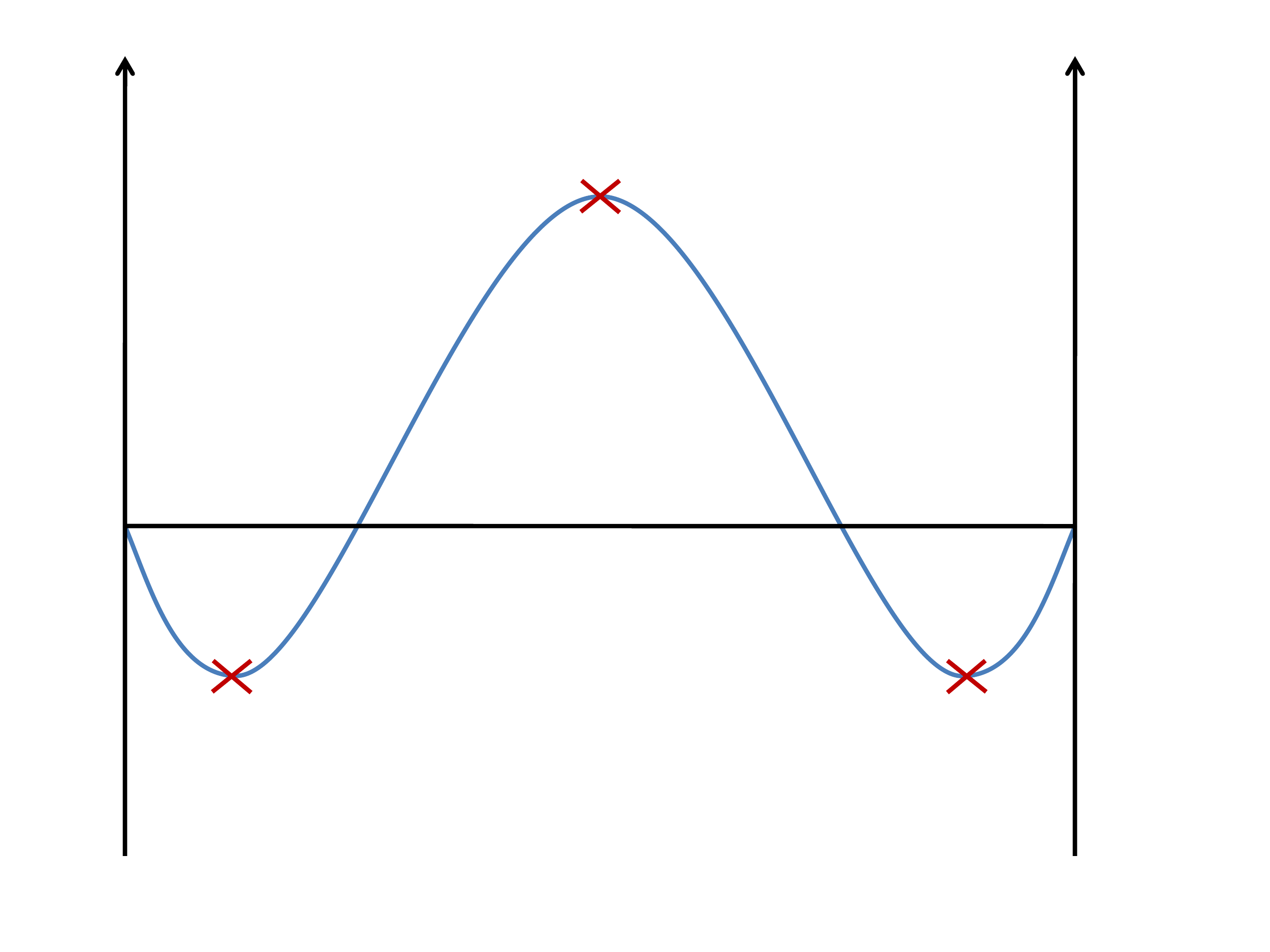}
\put(-200,148){$\scriptstyle \alpha(\theta)$}
\put(-200,20){$\scriptstyle \theta=0$}
\put(-54,20){$\scriptstyle \theta=\pi$}
\put(-200,80){\scriptsize \dsix}
\put(-50,80){\scriptsize \dsix}
\caption{Flux configurations for a setup with \dsix-branes on one pole (left picture) and for a setup with \dsix-branes on both poles (right picture). Both configurations with BPS boundary conditions, which are excluded due to the requirement of global tadpole cancellation and the constraint $\mathrm{sgn}\, \alpha^{\prime\prime} = \mathrm{sgn}\, \alpha$ that must be fulfilled at every extremum away from the sources.\label{fig_excluded}}
\end{figure}
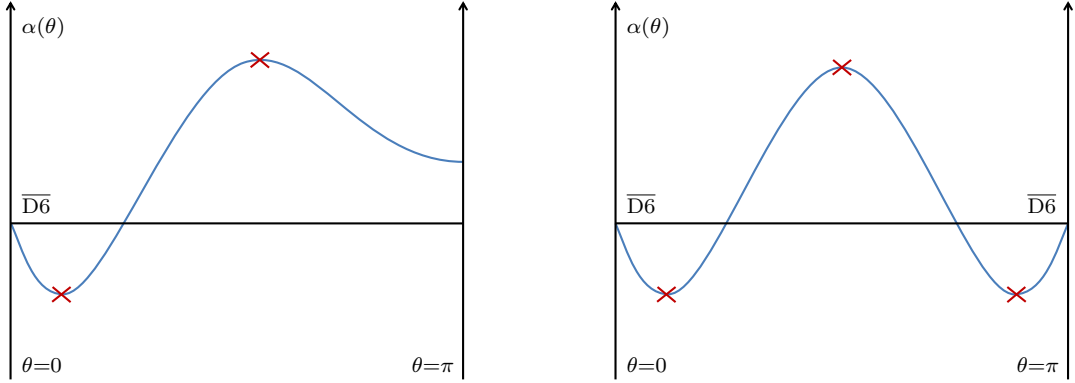

The near-brane behaviour of $\alpha$ can now be used together with the constraint \eqref{top-constraint} to rule out all possible flux configurations involving the BPS boundary condition. Let us, for simplicity, discuss this for a configuration with an \dsix-brane located at $\theta=0$ and no source at the other pole $\theta=\pi$. As argued above, $\alpha$ then starts out zero at the source and becomes negative since $\alpha^\prime(0)<0$. The $F_2$ Bianchi identity requires, however, that the total flux-induced charge is positive in order that the global tadpole is cancelled,
\begin{equation}
\int F_0 H = F_0^2 \int \alpha \e^{\phi-7A} \star_3 1 > 0. \label{nonbps-tadpole}
\end{equation}
Therefore, $\alpha$ must change its sign and become positive somewhere between $\theta=0$ and $\theta=\pi$, which implies that it must reach a minimum before. This is, however, forbidden by \eqref{top-constraint}, which does not allow an extremum with $\alpha^{\prime\prime} > 0$ and $\alpha<0$. Due to rotational symmetry, we must furthermore impose $\alpha^\prime = 0$ at the south-pole $\theta=\pi$, where no source is located, which leads to at least one more extremum forbidden by \eqref{top-constraint} (cf. Figure \ref{fig_excluded}). It is straightforward to show that similar arguments hold if we replace the \dsix-brane by a D$6$-brane or consider a configuration with sources that are located at both poles. The BPS boundary condition, although locally allowed by the equations of motion, can therefore not lead to a globally consistent solution.

The newly found boundary condition \eqref{new-bc}, on the other hand, is not ruled out by such an argument. In order to see this, we can again compute the near-brane behaviour of $\alpha$ for this boundary condition. Substituting \eqref{eq:d6-ansatz-bc} and \eqref{new-bc} into \eqref{lambda-alpha} yields
\begin{equation}
\alpha(\theta) = \frac{\lambda_0}{a_0^7f_0^3} \pm \frac{1}{a_0^7f_0^3} \theta + \mathcal{O}(\theta^2),
\end{equation}
where we have also used that some of the higher order coefficients in the expansion \eqref{eq:d6-ansatz-bc} can be expressed in terms of the leading order coefficients (cf. Appendix \ref{app:computations.bc}). Since $\lambda_0$, $a_0$ and $f_0$ are by definition non-zero, $\alpha$ is finite and non-zero near the pole. The sign of $\lambda_0$, however, is not determined by the equations of motion, and so $\alpha(0)$ can a priori be positive or negative.\footnote{To be precise, there is no \emph{local} constraint from solving the equations of motion in the vicinity of the pole. We will show in Chapter \ref{ch:scaling-symmetries}, however, that there is a global constraint that fixes $\alpha(0)$ in terms of the cosmological constant and the charge of the sources located at $\theta=0$. One can then verify that $\alpha(0)$ must be negative for the case of D$6$-branes and positive for the case of \dsix-branes.}

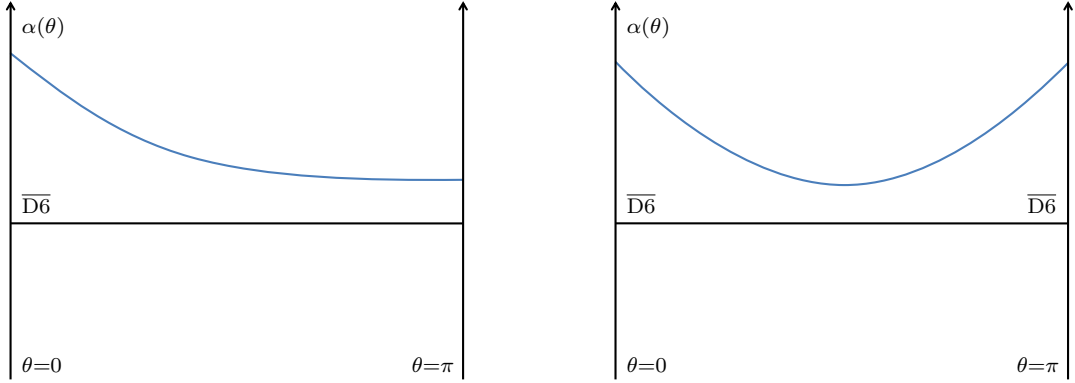
\begin{figure}[t]
\centering
\includegraphics[width=0.5 \textwidth]{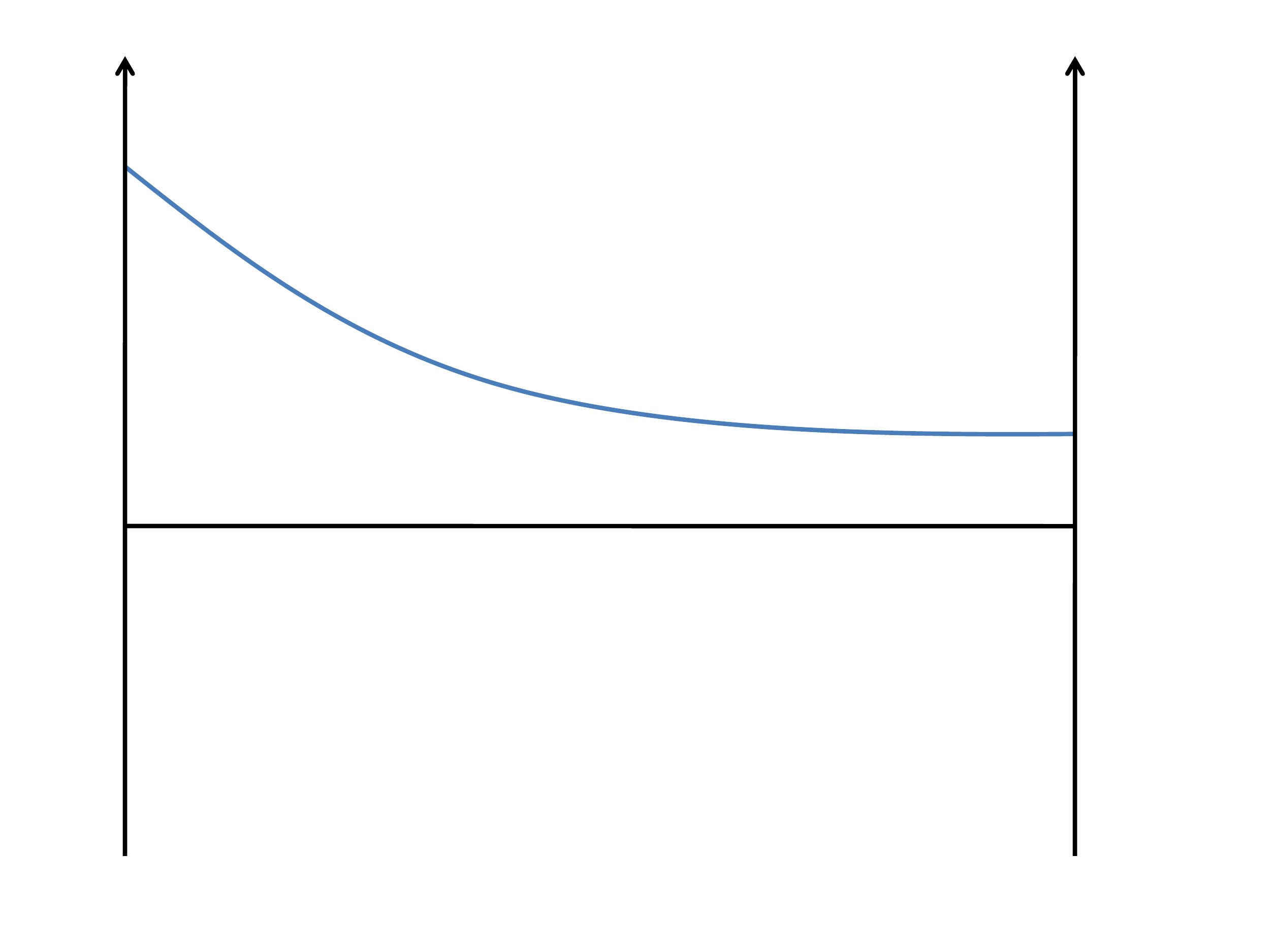}
\put(-200,148){$\scriptstyle \alpha(\theta)$}
\put(-200,20){$\scriptstyle \theta=0$}
\put(-54,20){$\scriptstyle \theta=\pi$}
\put(-200,80){\scriptsize \dsix}
\includegraphics[width=0.5 \textwidth]{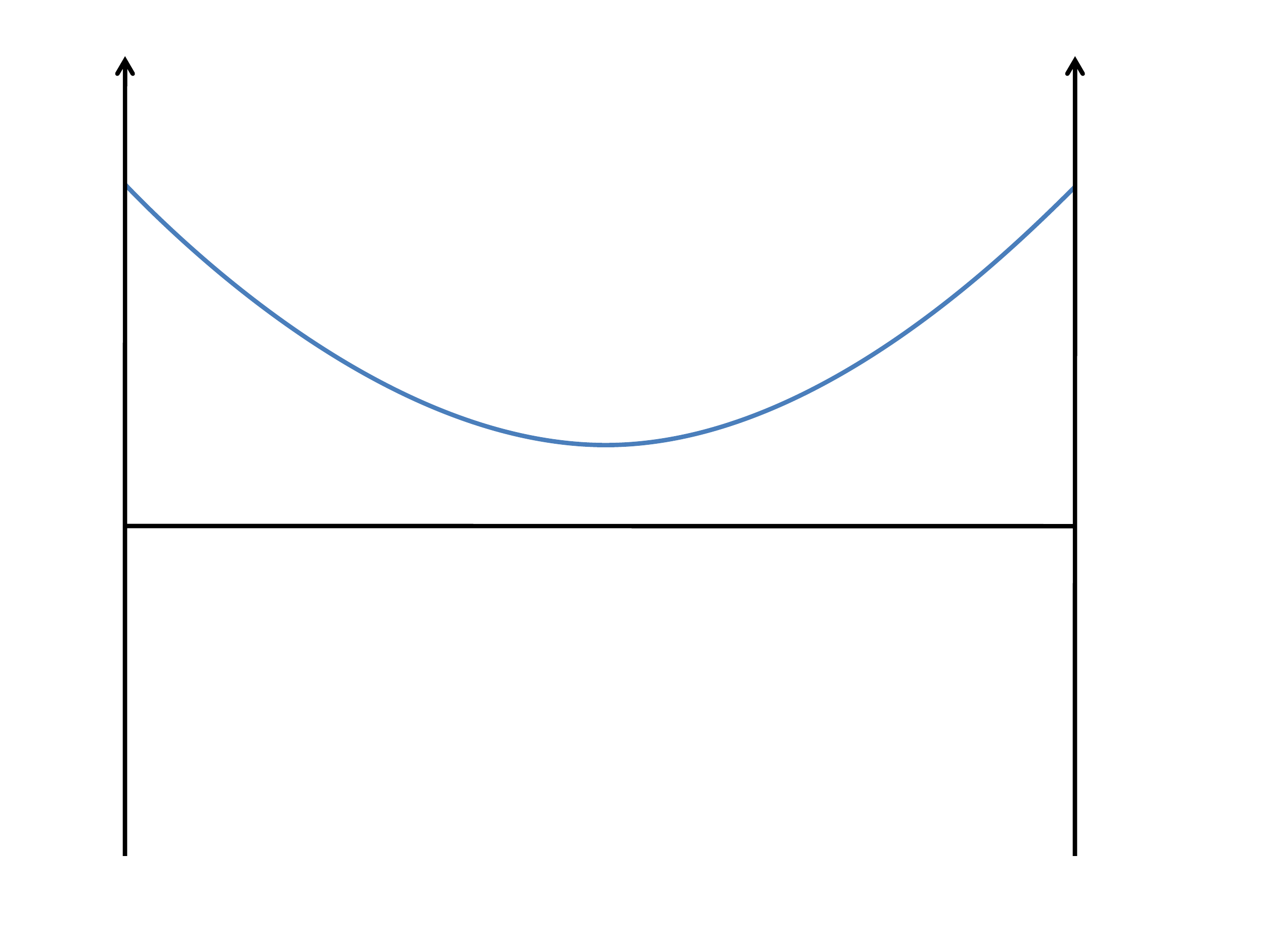}
\put(-200,148){$\scriptstyle \alpha(\theta)$}
\put(-200,20){$\scriptstyle \theta=0$}
\put(-54,20){$\scriptstyle \theta=\pi$}
\put(-200,80){\scriptsize \dsix}
\put(-50,80){\scriptsize \dsix}
\caption{Flux configurations for a setup with \dsix-branes on one pole (left picture) and for a setup with \dsix-branes on both poles (right picture). Both configurations with unusual boundary conditions, which are not excluded by the topological argument.\label{fig_allowed}}
\end{figure}

The sign of $\alpha^\prime(0)$ cannot be chosen freely but is again determined by the charge of the source that sits at the pole. As shown in Appendix \ref{app:computations.bc}, the charge and the tension of the source are given by
\begin{equation}
Q = \pm \frac{f_0^3}{\sqrt{b_0}}, \qquad T= \frac{f_0^3}{\sqrt{b_0}}.
\end{equation}
Thus, $\alpha^\prime$ is positive near the pole for $Q>0$ (D$6$-branes) and negative for $Q<0$ (\dsix-branes). For the case of \dsix-branes, this boundary condition thus evades the topological argument if $\alpha(0) >0$ (see Figure \ref{fig_allowed}). For negative $\alpha(0)$, on the other hand, $\alpha$ cannot become positive without having a minimum that is forbidden by \eqref{top-constraint}. $\alpha$ would therefore have to be negative everywhere, and the tadpole cancellation condition \eqref{nonbps-tadpole} could not be satisfied. Analogously, one can show that, for the case of D$6$-branes, the topological argument is evaded for $\alpha(0) < 0$.

Contrary to the BPS boundary condition, the new boundary condition \eqref{new-bc} is therefore consistent with the existence of a global solution. It is straightforward to verify, however, that the boundary condition yields a singularity in the energy density of the $H$ flux,
\begin{equation}
\e^{-\phi} |H|^2 \propto \theta^{-\tfrac{1}{8}}. \label{h-singularity}
\end{equation}
Although this singularity is integrable,
\begin{equation}
\sqrt{g_{10}}\, \e^{-\phi} |H|^2 \propto \e^{7A+3B} \sin^2 \theta\, \theta^{-\tfrac{1}{8}} \propto \theta,
\end{equation}
it is a priori not clear whether it is acceptable in a possible localised solution of our model since $H$ does not directly couple to the branes. This may indicate that the singularity is not resolved in string theory such that the corresponding solution is not physical and has to be discarded. It was proposed, for example, that the singular solution decays perturbatively via brane-flux annihilation \cite{Blaback:2011pn, Blaback:2012nf, Bena:2012ek, Vanriet:2013}. Another possibility, however, is that the singularity is resolved by taking into account the open string dynamics, similar to how it happens in \cite{Polchinski:2000uf}. Due to the non-trivial flux background, the D$6$-branes (or \dsix-branes), which we took to be located at $\theta=0$, might actually expand into a non-commutative configuration \cite{Myers:1999ps} and thus polarise into a ``fuzzy'' D$8$-brane wrapping a trivial two-cycle with a finite radius. Assumed that our local D$6$-brane solution can be extended to a globally consistent solution, this solution would then only be valid sufficiently far away from the pole, while, in the small $\theta$ region, it would have to be replaced by a solution with a D$8$-brane. If this is indeed the case, one may imagine that the singularity will be resolved in the true solution since a D$8$-brane has a different codimension than a D$6$-brane and therefore induces a different field behaviour in the near-source region. We will elaborate on this possibility in the following section.
\\

\subsection{Brane Polarisation}
\label{ch:simple-non-bps.myers-effect}

In this section, we discuss whether the singularity found in the previous section is resolved by brane polarisation. In order to make contact with the solution for \dthree-branes in the Klebanov-Strassler background, our main focus will be a non-compact version of the setup discussed above, i.\ e.\ we will consider \dsix-branes in flat space with non-zero $F_0$ and $H$ flux. This setup is related by T-duality to partially smeared \dthree-branes on $\mathbb{R}^{1,3} \times T^3 \times \mathbb{R}^3$ with $F_3$ and $H$ flux and was therefore argued to approximate the near-tip solution for partially smeared \dthree-branes in the Klebanov-Strassler background \cite{Massai:2012jn}. In the vicinity of the \dsix-branes, the equations of motion for the non-compact setup are equivalent to those of the compact model up to higher order corrections such that most of the results found in the previous sections of this chapter carry over. Using our local solution in the near-brane region, we analyse the polarisation potential for a probe D$8$-brane and find that, in the regime where our calculation is valid, a polarisation is excluded for all values of the free parameters of the solution. The flux singularity is therefore not resolved by brane polarisation in the non-compact \dsix-brane model. As explained in the introduction to this chapter, this suggests that the singularity generated by \dthree-branes in the Klebanov-Strassler background is not resolved by brane polarisation either. In the compact version of our model, the results are less conclusive and depend on global properties of the solution that are not fixed by solving the equations of motion locally around the branes. A resolution of the flux singularity by brane polarisation is therefore not excluded in the compact model.

\subsubsection{Setup}

We consider a setup with \dsix-branes in flat space that are placed in a flux background with non-zero Romans mass $F_0$ and $H$ flux along the three transverse directions. The flux background is chosen such that it is not mutually BPS with the \dsix-branes but instead carries D$6$-brane charge. This means that we require the solution to approach the BPS D$6$-brane solution \cite{Janssen:1999sa} at a certain distance far enough away from the \dsix-branes.\footnote{Even further away from the anti-branes, the non-compact solution then yields a naked UV singularity. This is, however, not relevant for the present discussion since we are only interested in the IR solution close to the anti-branes to study the near-tip physics of \dthree-branes in the Klebanov-Strassler background.} This is the T-dual analogue of placing partially smeared \dthree-branes in an ISD flux background, i.\ e.\ one that carries D$3$-brane charge.

Assuming that the \dsix-branes are located at one point in transverse space, our setup exhibits an $\mathrm{SO}(3)$ symmetry, and we can make the ansatz
\begin{equation}
\d s_{10}^2 = \e^{2A(r)} \eta_{\mu\nu} \d x^\mu \d x^\nu + \e^{2B(r)} \left({\d r^2 + r^2 \d \Omega_2^2}\right) \label{nc-metricansatz}
\end{equation}
for the metric, where $r$ is the distance to the branes. The most general ansatz for the form fields is
\begin{equation}
H = \lambda F_0 \e^{\tfrac{7}{4}\phi} \star_3 1, \qquad F_2 = \e^{-\tfrac{3}{2}\phi-7A} \star_3 \d \alpha, \label{nc-fluxansatz}
\end{equation}
where $\phi$, $\lambda$ and $\alpha$ are functions depending on $r$ and $F_0$ is a constant. It is straightforward to check that, with this ansatz for the metric and the form fields, the equations of motion are almost the same as those found for the compact model in Section \ref{ch:simple-non-bps.localisation.ansatz}, except that we have to replace $\sin^2 \theta \to r^2$ in all equations and discard the curvature terms in the Einstein equations.

The equations of motion \eqref{nonbps-eom1}, \eqref{nonbps-eom2}, \eqref{nonbps-eom3} and \eqref{nonbps-eom4} thus simplify as follows. The $F_2$ Bianchi identity reads
\begin{equation}
\frac{\left({\e^{-\tfrac{3}{2}\phi-7A+B} r^2 \, \alpha^\prime}\right)^\prime}{\e^{3B} r^2} = \e^{\tfrac{7}{4}\phi} \lambda F_0^2 + Q \delta(\Sigma), \label{eoms-theta-bianchi-nc}
\end{equation}
where primes denote derivatives with respect to $r$. The $H$ equation furthermore implies that $\alpha$ can be eliminated in terms of $\lambda$,
\begin{equation}
\alpha = \e^{\tfrac{3}{4}\phi+7A} \lambda + \alpha_0, \label{lambda-alpha-nc}
\end{equation}
where $\alpha_0$ is an integration constant, which we set to zero in the following. The dilaton equation reads
\begin{equation}
\frac{\left({\e^{7A+B} r^2\, \phi^\prime}\right)^\prime}{\e^{7A+3B} r^2} = \e^{\tfrac{5}{2}\phi} F_0^2 \left({\frac{5}{4} - \frac{\lambda^2}{2}}\right) + \frac{3}{4} \e^{-14A-2B-\tfrac{3}{2}\phi} \left({\alpha^\prime}\right)^2 + \frac{3}{4} \e^{\tfrac{3}{4}\phi} T \delta(\Sigma). \label{eoms-theta-dilaton-nc}
\end{equation}
Taking the trace of the external Einstein equation, we find
\begin{align}
\frac{16}{7} R_7 &= - 16 \e^{-2B} \left[{7 \left({A^\prime}\right)^2 + A^\prime B^\prime + \frac{\left({r^2\, A^\prime}\right)^\prime}{r^2}}\right] \nll = \e^{\tfrac{5}{2}\phi} F_0^2 \left({1 - 2\lambda^2}\right) - \e^{-14A-2B-\tfrac{3}{2}\phi} \left({\alpha^\prime}\right)^2 - \e^{\tfrac{3}{4}\phi} T \delta(\Sigma). \label{eoms-theta-exteinstein-nc}
\end{align}
The internal Einstein equation in radial direction is
\begin{align}
0 &= \frac{\left({r^2 \, B^\prime}\right)^\prime}{r^2} + 7 \left({A^\prime}\right)^2 + B^{\prime\prime} + 7 A^{\prime\prime} - 7 A^\prime B^\prime \nl +\frac{1}{2} \left({\phi^\prime}\right)^2 + \frac{1}{16} \e^{\tfrac{5}{2}\phi+2B} F_0^2 \left({1+6\lambda^2}\right) - \frac{1}{16} \e^{-14A-\tfrac{3}{2}\phi} \left({\alpha^\prime}\right)^2 + \frac{7}{16} \e^{\tfrac{3}{4}\phi+2B}T\delta(\Sigma), \label{eoms-theta-inteinsteintheta-nc}
\end{align}
and the Einstein equations in the remaining directions are
\begin{align}
0 &= \frac{\left({r^2 \, B^\prime}\right)^\prime}{r^2} + \left({B^\prime}\right)^2 + r^{-1} \left({7A+B}\right)^\prime + 7 A^\prime B^\prime \nl + \frac{1}{16} \e^{\tfrac{5}{2}\phi+2B} F_0^2 \left({1+6\lambda^2}\right) + \frac{7}{16} \e^{-14A-\tfrac{3}{2}\phi} \left({\alpha^\prime}\right)^2 + \frac{7}{16} \e^{\tfrac{3}{4}\phi+2B}T\delta(\Sigma). \label{eoms-theta-inteinsteintrans-nc}
\end{align}
All other equations of motion are automatically satisfied with above ansatz for the metric and the form fields. As in the compact model, we can furthermore combine \eqref{eoms-theta-bianchi-nc} and \eqref{lambda-alpha-nc} to obtain the equation
\begin{equation}
\frac{\left({\e^{-\tfrac{3}{2}\phi-7A+B} r^2}\right)^\prime}{\e^{3B} r^2}\, \alpha^\prime + \e^{-\tfrac{3}{2}\phi-7A-2B} \alpha^{\prime\prime}  = \alpha \e^{\phi-7A} F_0^2,
\end{equation}
which leads to the constraint
\begin{equation}
\mathrm{sgn}\, \alpha^{\prime\prime} = \mathrm{sgn}\, \alpha \label{top-constraint-nc}
\end{equation}
at every extremum with $\alpha^\prime=0$. One can also verify that, in an expansion of the fields around the brane position as carried out in Section \ref{ch:simple-non-bps.localisation.delta}, above equations are identical to the ones in Section \ref{ch:simple-non-bps.localisation.ansatz} at leading order. The possible near-brane boundary conditions are therefore the same in our non-compact setting, except that some of the sub-leading order coefficients receive corrections (see Appendix \ref{app:computations.bc.non-compact}).

In order to have a flux background that carries D$6$-brane charge, we demand that it approaches the value of the BPS D$6$-brane solution \cite{Janssen:1999sa} away from the \dsix-branes, i.\ e.\ we have to impose the UV boundary condition
\begin{equation}
\lambda^\textrm{UV} \to 1. \label{lambda-uv}
\end{equation}
As in the compact model, we can then use our topological argument to also fix the IR boundary conditions at the anti-brane position $r=0$. The standard BPS boundary condition \eqref{simple-bpsbc} is again excluded since it has $\alpha(0)=0$ and $\alpha^\prime(0) < 0$ such that $\alpha$ would start out zero and then become negative for small $r$. In order to match the UV boundary condition \eqref{lambda-uv}, however, $\alpha$ would have to change its sign and eventually become positive far away from the anti-branes. This implies that it would first have to reach a minimum at a point where it is still negative, which is forbidden by the constraint \eqref{top-constraint-nc} (see also Section \ref{ch:simple-non-bps.localisation.delta}). Thus, the only possible boundary condition at $r=0$ is the singular boundary condition \eqref{new-bc}. The behaviour of the fields in the vicinity of the \dsix-branes is therefore
\begin{align}
& \e^{-A(r)} = r^{-\tfrac{1}{16}} \left({a_0  + a_1 r + a_2 r^2}\right) + \ldots, \notag \\ & \e^{-2B(r)} = r^{\tfrac{7}{8}} \left({b_0 + b_1 r + b_2 r^2}\right) + \ldots, \notag \\ & \e^{-\tfrac{1}{4}\phi(r)} = r^{-\tfrac{3}{16}} \left({f_0 + f_1 r + f_2 r^2}\right) + \ldots, \notag \\ & \lambda(r) = r^{-1} \left({\lambda_0 + \lambda_1 r + \lambda_2 r^2}\right) + \ldots \label{eq:d6-ansatz-bc-noncompact}
\end{align}
For convenience, all relevant expansion coefficients are explicitly spelled out in Appendix \ref{app:computations.bc.non-compact}.

\subsubsection[The D$8$-brane Potential]{The D$\boldsymbol{8}$-brane Potential}
\label{ch:simple-non-bps.myers-effect.d8}

In order to analyse whether the \dsix-branes polarise into a D$8$-brane in above setup, we now compute the potential for the worldvolume scalar whose vev determines the size of the trivial two-cycle wrapped by the D$8$-brane. To this end, we consider the action of a probe D$8$-brane carrying $n \gg 1$ units of \dsix-brane charge in a background sourced by $N \gg n$ \dsix-branes.\footnote{Note that the probe approximation is justified here since we consider a probe of $n$ anti-branes on top of a background with $N \gg n$ anti-branes such that the probe is a small perturbation \emph{everywhere} on the compact space. This is different from the situation where a probe anti-brane is considered on top of a background without any anti-branes. In that case, the perturbation of the background is typically not small in the near-brane region, where the curvature and/or the string coupling may blow up. Even far away from the anti-branes, it is then not clear whether the probe calculation approximates the fully backreacted solution since, as we discuss in this thesis, the latter might simply not exist.} As explained earlier, the D$8$-brane polarisation channel is T-dual to the D$5$-brane polarisation channel for partially smeared \dthree-branes on $\mathbb{R}^{1,3}\! \times T^3\! \times \mathbb{R}^3$. If brane polarisation does not help to resolve the flux singularity in our model, this would therefore suggest that the same is true for the singularities generated by \dthree-branes in the Klebanov-Strassler background.

The D$8$-brane action in Einstein frame is
\begin{equation}
S^{(8)}_{\textrm{loc}} = S^{(8)}_\textrm{DBI} + S^{(8)}_\textrm{CS} \label{nc-d8brane-action1}
\end{equation}
with
\begin{equation}
S^{(8)}_\textrm{DBI} = - \mu_8 \int \d^9 \xi\, \e^{\tfrac{5}{4}\phi} \sqrt{-\det \big( g_{\alpha\beta} - \e^{-\phi/2} \mathcal{F}_{\alpha\beta}\big)}, \quad S^{(8)}_\textrm{CS} = \mu_8 \int \left({C_9 - \mathcal{F} \w C_7}\right), \label{nc-d8brane-action2}
\end{equation}
where $\mathcal{F} = B + 2\pi F$ in string units and $F$ is the worldvolume gauge field strength. It is fixed by demanding that the CS term of the D$8$-brane action yields $n$ units of induced \dsix-brane charge.\footnote{Note that our conventions for $\mathcal{F}$, $F$ and the \dsix-brane charge differ from those used in \cite{Bena:2012tx} and we work in Einstein frame instead of string frame.} We thus obtain
\begin{equation}
F = \frac{n}{2} \omega_2, \label{nc-fluxansatz2}
\end{equation}
where $\omega_2$ is the volume-form of the unit two-sphere and we have used that the absolute value of the \dsix-brane charge satisfies $\mu_6 = 4\pi^2 \mu_8$. The gauge potentials that appear in the DBI and CS action are given by $\d B = H$, $\d C_7 = - \e^{\frac{3}{2} \phi} \star_{10} F_2$ and $\d C_9 = \e^{\frac{5}{2}\phi} \star_{10} F_0 + H \wedge C_7$ in our conventions (cf. Appendix \ref{app:conventions.sugra}). Substituting \eqref{nc-fluxansatz} and \eqref{lambda-alpha-nc}, we then find
\begin{equation}
B = \beta \omega_2, \qquad C_7 = \alpha \tilde \star_7 1, \qquad C_9 = \gamma \tilde \star_7 1 \w \omega_2, \label{nc-fluxansatz3}
\end{equation}
where
\begin{equation}
\beta^\prime= F_0 \alpha \e^{\phi-7A+3B} r^2, \qquad \gamma^\prime = F_0 \left({ \alpha^2 \e^{\phi-7A+3B} - \e^{\tfrac{5}{2}\phi+7A+3B} }\right) r^2. \label{nc-fluxansatz4}
\end{equation}

In order to calculate the polarisation potential, we consider a static configuration of a probe D$8$-brane that is located at a fixed $r=r_\star$ and extends along the rest of the coordinates. We can then use our metric ansatz \eqref{nc-metricansatz} together with the expressions \eqref{nc-fluxansatz2}, \eqref{nc-fluxansatz3} and \eqref{nc-fluxansatz4} in the D$8$-brane action \eqref{nc-d8brane-action2} and, since we consider a regime where $n$ is large, expand the result in powers of $1/n$. Provided that possible extrema of the potential are located at small $r$, we can furthermore substitute our near-brane expansion \eqref{eq:d6-ansatz-bc-noncompact} together with the corresponding expansion coefficients (cf. Appendix  \ref{app:computations.bc.non-compact}). The potential can then schematically be written as
\begin{equation}
V(\hat r) \sim \pi n c_2 \hat r^2 - c_3 \hat r^3 + \frac{1}{\pi n} c_4 \hat r^4, \label{pol-potential}
\end{equation}
where $c_2$, $c_3$ and $c_4$ are certain coefficients, which are stated below, and $\hat r = 2\pi r$ is the worldvolume scalar whose vev determines the radius of the trivial two-cycle wrapped by the D$8$-brane. All other terms can be shown to scale with higher powers of $\hat r$ and/or $1/n$ in \eqref{pol-potential} and can therefore be neglected in the regime of small $\hat r$ and large $n$. We will verify that this assumption is self-consistent in Section \ref{ch:simple-non-bps.myers-effect.validity}.

Let us now analyse whether the potential \eqref{pol-potential} admits a polarisation of the \dsix-branes into a D$8$-brane. In order for the anti-branes to polarise, the potential must have a minimum at a finite $\hat r$. Depending on the balance between the coefficients $c_2$, $c_3$ and $c_4$, such a minimum may be stable (if the vacuum energy is lower than the vacuum energy at the origin), it may be meta-stable (if the vacuum energy is higher than at the origin), or it may not exist at all. It is straightforward to check that, if the coefficients satisfy
\begin{equation}
c_3^2 < \frac{32}{9} c_2 c_4, \label{pol-condition}
\end{equation}
then the potential does not have a minimum away from the origin, and, consequently, the anti-branes will not polarise.\footnote{The sign of the cubic term in \eqref{pol-potential} can always be flipped by changing the orientation of the D$8$-brane. One can verify, however, that then all three terms are positive such that a polarisation is excluded as well.} The coefficients of the potential \eqref{pol-potential} obtained from above calculation are
\begin{equation}
c_2 = \frac{1}{12} \frac{\lambda_0^2 F_0^2}{a_0^7 b_0 f_0^{13}}, \qquad c_3 = \frac{1}{3} \frac{ \lambda_0 F_0}{a_0^7 b_0^{3/2} f_0^{10}}, \qquad c_4 = \frac{1}{2} \frac{1}{a_0^7 b_0^2 f_0^{7}}. \label{pol-coefficients}
\end{equation}
Thus, the condition \eqref{pol-condition} is satisfied, and the anti-branes do not polarise. Remarkably, this conclusion holds for all values of the free parameters $a_0$, $b_0$, $f_0$, $\lambda_0$ and $F_0$ such that, in the regime where our calculation is valid, our knowledge of the solution in the near-brane region is sufficient to exclude polarisation in our model.

\subsubsection{Regime of Validity}
\label{ch:simple-non-bps.myers-effect.validity}

In order for our calculation to be self-consistent, we have to ensure that several conditions are satisfied.

\begin{itemize}
\item We considered the polarisation of $n$ \dsix-branes into a non-commutative configuration due to the Myers effect \cite{Myers:1999ps}. We studied an effective description of this configuration in terms of a D$8$-brane carrying $n$ units of \dsix-brane charge and treated the brane as a probe in a background sourced by $N$ \dsix-branes. These approximations are justified if
\begin{equation}
1 \ll n \ll N. \label{pol-validity1}
\end{equation}

\item It follows from \eqref{pol-potential} and \eqref{pol-coefficients} that the radius $r=r_\star$, at which a minimum could in principle have been possible, is of the order
\begin{equation}
r_\star \sim \pi n \sqrt{\frac{c_2}{c_4}} \sim n F_0 \lambda_0 \frac{\sqrt{b_0}}{f_0^3}. \label{pol-radius}
\end{equation}
In order to be able to trust the expansion \eqref{eq:d6-ansatz-bc-noncompact}, we have to demand that $r_\star$ is much smaller than $a_0/a_1$, $b_0/b_1$, $f_0/f_1$ and $\lambda_0/\lambda_1$. Some of these expansion coefficients are not fixed by solving the equations of motion locally in the near-brane region, and so their magnitude is a priori not determined. It is possible, however, to nevertheless obtain an estimate by comparing our solution with the BPS D$6$-brane solution of \cite{Janssen:1999sa}. In the near-brane region, the two solutions only differ by the sign and the small $r$ behaviour of the function $\lambda$, while the other functions $\e^{-A}$, $\e^{-2B}$ and $\e^{-\frac{1}{4}\phi}$ diverge in the same way in both solutions. It is therefore reasonable to assume that, for small $r$, we can approximate these functions in our solution by the corresponding expressions of the BPS solution,
\begin{equation}
\e^{-A} \approx g_s^{1/4} h_6^{1/16}, \qquad \e^{-2B} \approx  g_s^{1/2} h_6^{-7/8}, \qquad \e^{-\tfrac{1}{4}\phi} \approx g_s^{-1/4} h_6^{3/16}, \label{gn-scaling}
\end{equation}
where $g_s$ is the string coupling and
\begin{equation}
h_6(r) = 1 + \frac{\pi g_s N}{2r} - \frac{1}{2} F_0^2 r^2 \approx \frac{\pi g_s N}{2r} \label{gn-scaling-wf}
\end{equation}
is the warp function of \cite{Janssen:1999sa} in our conventions. Using \eqref{eq:d6-ansatz-bc-noncompact} and \eqref{gn-scaling-wf} in \eqref{gn-scaling}, we then find that the expansion coefficients $a_0$, $b_0$ and $f_0$ scale like
\begin{equation}
a_0 \sim g_s^{5/16} N^{1/16}, \qquad b_0 \sim g_s^{-3/8} N^{-7/8}, \qquad f_0 \sim g_s^{-1/16} N^{3/16}. \label{gn-scaling2}
\end{equation}
They are related to the other coefficients $a_1$, $b_1$, $f_1$, $\lambda_0$ and $\lambda_1$ by the expressions stated in Appendix \ref{app:computations.bc.non-compact}. Assuming that the different terms in \eqref{bc.non-compact} are of the same order of magnitude, we thus find\footnote{Due to a four-parameter rescaling symmetry of the solution, it is always justified to assume that the expansion coefficients $a_0$, $b_0$, $f_0$, etc. scale like in \eqref{gn-scaling2} and \eqref{gn-scaling3} with respect to $g_s$, $N$ and $F_0$. In addition to two global scaling symmetries exhibited by all solutions of the classical type II supergravity equations (cf. Chapter \ref{ch:scaling-symmetries}), the specific model considered here also allows a rescaling of the radial coordinate $r$ and a rescaling of the warp factor by an arbitrary constant.}
\begin{equation}
\frac{a_0}{a_1} \sim \frac{b_0}{b_1} \sim \frac{f_0}{f_1} \sim \frac{\lambda_0}{\lambda_1} \sim \left({\frac{N}{g_s F_0^2}}\right)^{\tfrac{1}{3}}, \qquad \lambda_1 \sim 1. \label{gn-scaling3}
\end{equation}
It then follows from \eqref{pol-radius} that
\begin{equation}
r_\star \sim n \left({\frac{F_0}{g_s N^2}}\right)^{\tfrac{1}{3}}. \label{pol-radius2}
\end{equation}
In order that $r_\star$ is smaller than $a_0/a_1$, $b_0/b_1$, $f_0/f_1$ and $\lambda_0/\lambda_1$, we therefore require 
\begin{equation}
n \ll \frac{N}{F_0}. \label{pol-validity2}
\end{equation}

\item Our expansion of the square root of the DBI action in powers of $1/n$ is justified if $\det (\e^{\phi/2} g^\perp_{\alpha\beta}) \ll \det \mathcal{F}_{\alpha\beta}$, where $g^\perp_{\alpha\beta}$ is the pullback of the metric along the two-sphere wrapped by the D$8$-brane. This requirement leads to the condition
\begin{equation}
n \gg \frac{r_\star^{3/2}}{f_0^2b_0}.
\end{equation}
Using \eqref{gn-scaling2} and \eqref{pol-radius2}, we find that the condition reduces to \eqref{pol-validity2}.

\item The radius of the two-sphere wrapped by the D$8$-brane should be large in string units. This radius is determined by the value of $\det ( \e^{\phi/2} g^\perp_{\alpha\beta})$ at $r=r_\star$ such that
\begin{equation}
\frac{r_\star^{3/2}}{f_0^2b_0} \gg 1
\end{equation}
or, using again \eqref{gn-scaling2} and \eqref{pol-radius2},
\begin{equation}
n \gg \left({\frac{N}{F_0}}\right)^{\tfrac{1}{3}}. \label{pol-validity3}
\end{equation}
One can verify that this condition also ensures that the background curvature at $r=r_\star$ is small in string units.

\item We finally require the string coupling $\e^\phi$ to be small at $r=r_\star$. Using \eqref{eq:d6-ansatz-bc-noncompact}, \eqref{gn-scaling2} and \eqref{pol-radius2}, this implies
\begin{equation}
f_0 r_\star^{-\tfrac{3}{16}} \gg 1
\end{equation}
and
\begin{equation}
n \ll \left({\frac{N^5}{F_0}}\right)^{\tfrac{1}{3}}. \label{pol-validity4}
\end{equation}
\end{itemize}

It is straightforward to check that one can always fulfill the conditions \eqref{pol-validity1}, \eqref{pol-validity2}, \eqref{pol-validity3} and \eqref{pol-validity4} simultaneously by choosing appropriate values for $n$, $N$ and $F_0$ such that our calculation in Section \ref{ch:simple-non-bps.myers-effect.d8} is self-consistent. One may wonder, however, whether the regime considered here allows us to draw a conclusion about the polarisation of \dthree-branes in the Klebanov-Strassler background, which was our main motivation for analysing the present setup. Because of \eqref{pol-validity2}, our calculation is valid in a regime where $N/F_0 \gg n$ such that the number of \dsix-branes has to be much larger than the units of $F_0$ flux. In the T-dual toy model that captures the near-tip physics in the Klebanov-Strassler background, this would correspond to a regime where the number of \dthree-branes is much larger than the units of $F_3$ flux. We do therefore not directly address the situation considered in \cite{Kachru:2002gs}, where it was shown in a probe calculation that a meta-stable vacuum with an NS$5$-brane only exists in the Klebanov-Strassler background if the number of \dthree-branes divided by the units of $F_3$ flux, $p/M$, is less than $8\%$. It is possible to show, however, that, if one were to consider a polarisation into \emph{multiple} NS$5$-branes instead of a polarisation into \emph{one} NS$5$-brane, then the calculation of \cite{Kachru:2002gs} would yield meta-stable vacua for any number of \dthree-branes as long as $p/M$ divided by the number of NS$5$-branes is less than $8\%$. The result in our toy model, on the other hand, suggests that such a polarisation does not occur if the backreaction of the anti-branes is properly taken into account. Our calculation therefore rules out brane polarisation in a regime where a probe calculation along the lines of \cite{Kachru:2002gs} would find it.

Let us finally comment on brane polarisation in the compact version of our model on AdS$_7 \times S^3$, which we studied in Sections \ref{ch:simple-non-bps.setup} to \ref{ch:simple-non-bps.localisation}. Repeating the steps discussed in this section with the appropriate expansion coefficients of the compact model (cf. Appendix \ref{app:computations.bc}) leads to a polarisation potential of the form
\begin{equation}
V(\hat \theta) \sim \pi n c_2 \hat \theta^2 - c_3 \hat \theta^3 + \frac{1}{\pi n} c_4 \hat \theta^4,
\end{equation}
where $\hat \theta = 2\pi\theta$ is the worldvolume scalar whose vev determines whether a polarisation happens or not. One can check that the potential is identical to the one in the non-compact model except for an additional term at order $\sim \hat \theta^2$. This term is due to the AdS curvature in the compact model and leads to a modified coefficient
\begin{equation}
c_2 = - \frac{7}{a_0^5 b_0 f_0^3} + \frac{1}{12} \frac{\lambda_0^2 F_0^2}{a_0^7 b_0 f_0^{13}}.
\end{equation}
Since the parameters $a_0$, $b_0$, $f_0$, $\lambda_0$ and $F_0$ are not fixed locally, our knowledge of the near-brane solution is not sufficient here to decide whether $c_2$ is positive or negative. Unlike in the non-compact case discussed above, we are therefore not able to exclude brane polarisation in the compact model.
\\

\subsection{Discussion}
\label{ch:simple-non-bps.discussion}

In this chapter, we analysed the backreaction of D$6$/\dsix-branes in a simple non-BPS setup with $F_0$ and $H$ flux. As anticipated by the intuitive arguments of Chapter \ref{ch:smeared-vs-loc}, the presence of localised sources led to several issues in our model. We first focussed on a compact version of our model on AdS$_7 \times S^3$ and studied the backreaction of sources with regularised profiles. The constraints from the equations of motion turned out to be surprisingly strong for such profiles, and we found that all solutions except for the smeared one are excluded. Furthermore, we were able to rule out the existence of a solution with fully localised branes unless it exhibits a singular energy density of the $H$ flux in the near-brane region. We then analysed the possibility that this singularity is resolved by brane polarisation due to the Myers effect. In the compact model, the results were not conclusive since the polarisation potential depends on parameters that are locally not fixed by the equations of motion. It is therefore possible that a fully localised, non-singular solution indeed exists for this model. It would be interesting to further study this possibility in order to settle the initially raised question under which conditions smearing is justified in non-BPS compactifications.

In the non-compact version of our model, we found that, in the regime we studied, our knowledge of the local solution in the near-brane region is sufficient to exclude brane polarisation for all values of the free parameters. Although our calculation is only valid if we consider a small fraction of the \dsix-branes as a probe in the background created by the others, mean-field arguments discussed in \cite{Polchinski:2000uf, Bena:2012tx, Bena:2012vz} actually suggest that our conclusion extends beyond this regime such that brane polarisation should even be excluded in the full polarisation problem for all \dsix-branes. As explained earlier in this chapter, our result indicates that \dthree-branes in the Klebanov-Strassler background do not polarise either, and, indeed, further evidence for this intuition has recently been presented in \cite{Bena:2012vz}. The singularity in the energy densities of $H$ and $F_3$ found in that setup is therefore not explained by brane polarisation. While there may in principle exist another, yet unknown mechanism in string theory that resolves the singularity, it was recently argued in \cite{Bena:2012ek, Vanriet:2013} that this is probably not the case. Instead, it was proposed in \cite{Blaback:2011pn, Blaback:2012nf, Bena:2012ek, Vanriet:2013} that the singularity might indicate a perturbative instability of the solution in the sector of the closed string modes. This instability was argued to lead to an attraction of flux towards the anti-branes, which then in turn triggers a decay of the vacuum via brane-flux annihilation. It would be very interesting to verify this claim explicitly.
\\

\cleardoublepage
\thispagestyle{plain}

\section{Cosmological Constant, Near-brane Behaviour and Singularities}
\label{ch:scaling-symmetries}

As we argued in the preceding chapters, a better understanding of string compactifications involving localised sources is an important task for string phenomenology. Unfortunately, for most scenarios, a full solution to the ten-dimensional equations of motion seems to be out of reach even in the supergravity approximation since the involved differential equations are too complex. On the other hand, commonly used procedures for simplifying this task---such as a smearing of the localised sources over the compact space---might be problematic and do not necessarily capture essential features of the true solution (see also \cite{Douglas:2010rt, Blaback:2010sj, Blaback:2011nz, Blaback:2011pn}). It would therefore be desirable to be able to compute important observables such as the cosmological constant without having to know the full ten-dimensional dynamics or rely on simplifications such as smearing.

In the first part of this chapter, we will show that such a method often exists in type II supergravity \cite{Gautason:2013}, building upon previous work that had already pointed towards this possibility \cite{Aghababaie:2003ar, Burgess:2011rv, Gautason:2012tb}. In particular, we will argue that the cosmological constant $\Lambda$ can often be expressed as a sum of terms that are due to the action of localised sources,
\begin{equation}
\Lambda \propto \sum_{p} c_p \left({S^{(p)}_\textrm{DBI} + S^{(p)}_\textrm{CS}}\right),
\end{equation}
where $S^{(p)}_\textrm{DBI}$ and $S^{(p)}_\textrm{CS}$ are the on-shell evaluated DBI and Chern-Simons actions of the D$p$-branes and/or O$p$-planes present in the corresponding supergravity solution and $c_p$ are $p$-dependent constants. Thus, in compactification scenarios where our reasoning holds, $\Lambda$ is entirely specified by the classical boundary conditions of some of the bulk fields at the positions of the sources and independent of the details of the ten-dimensional bulk dynamics.

Such a property was noticed before in \cite{Aghababaie:2003ar, Burgess:2011rv} for the special case of compactifications with spacetime-filling codimension $2$ sources, however, without explicitly considering the possible effects of topologically non-trivial fluxes.\footnote{In \cite{Aghababaie:2003ar}, the same property was found for two models in six-dimensional supergravity that involve $3$-branes, $4$-branes and magnetic flux. The expression for the cosmological constant derived in \cite{Burgess:2011rv} for type II supergravity can implicitly also contain contributions from Dirac strings such that it is in principle also valid in the presence of non-trivial flux. As will be explained below, however, our work in \cite{Gautason:2013} goes beyond this result in that it makes this hidden flux contribution explicit so that our expression can be used to compute the cosmological constant in concrete examples. Furthermore, we show that the flux contribution can in many cases be gauged away such that, even in the presence of flux, the cosmological constant is often fully determined by the on-shell actions of the D-branes and O-planes present in the corresponding solution.} Using a scaling symmetry of the actions of different supergravity theories, the authors were able to relate $\Lambda$ to boundary terms involving the supergravity fields that have to be evaluated in the near-source region. From a somewhat different angle, the results of \cite{Gautason:2012tb} suggested that such a behaviour might in fact be quite generic. There, it was shown that the cosmological constant in solutions of perturbative heterotic string theory is zero to all orders in $\alpha^\prime$, unless one introduces spacetime-filling fluxes or considers string loop or non-perturbative corrections. Since the argument only used the scaling properties of the effective potential with respect to the dilaton, it was then conjectured that a similar reasoning should also be applicable for the type II string, with the exception that then also localised sources such as D-branes and O-planes should contribute to $\Lambda$. For classical solutions of type II supergravity, this suggests that, in absence of spacetime-filling flux, any non-zero contribution to $\Lambda$ must be generated by terms that are due to localised sources.

It turns out, however, that the intuitive scaling argument of \cite{Gautason:2012tb} is complicated in the type II string by a subtlety related to the RR fields: in a frame where the bulk action scales uniformly with the dilaton, non-trivial couplings of the RR potentials with derivatives of the dilaton of the form $\d \phi \w C_{9-n} \w F_n$ arise. These couplings are only present in the type II string but not in the heterotic string. In the presence of background fluxes, they can be shown to yield non-zero contributions to $\Lambda$, thus spoiling the argument sketched above.

We will argue below, however, that it is still true in many cases that $\Lambda$ is completely determined by a sum of source terms. The reason is that classical type II (and also heterotic) supergravity exhibits a two-parameter scaling symmetry, which is related to the dilaton scaling and the mass scaling of the classical action \cite{Witten:1985xb, Burgess:1985zz}. Both the scaling symmetry exploited in \cite{Aghababaie:2003ar, Burgess:2011rv} and the one implicitly used in \cite{Gautason:2012tb} are special cases of this more general symmetry. As we will show below, it ensures that one can often find a particular combination of the equations of motion such that all bulk terms are eliminated from the equation determining $\Lambda$, leaving a contribution entirely from localised sources. The cosmological constant is then indeed given by a sum of source terms as initially claimed. More precisely, this can be shown to hold for maximally symmetric compactifications of type II supergravity involving sources of arbitrary dimension and at most NSNS $H$ flux and one type of RR flux.

In the second part of this chapter, we apply our results to the idea of placing \dthree-branes at the tip of the Klebanov-Strassler solution \cite{Klebanov:2000hb, Kachru:2002gs}, which was proposed as a mechanism to construct meta-stable de Sitter vacua in string theory \cite{Kachru:2003aw}. As we discussed in Chapter \ref{ch:intro}, the backreaction of \dthree-branes on the Klebanov-Strassler geometry has been heavily studied in the literature \cite{DeWolfe:2008zy, McGuirk:2009xx, Bena:2009xk, Bena:2011hz, Bena:2011wh, Dymarsky:2011pm, Blaback:2011nz, Blaback:2011pn, Blaback:2012nf, Massai:2012jn, Bena:2012tx, Bena:2012bk, Bena:2012vz, Bena:2012ek, Gautason:2013, Vanriet:2013}, and increasing evidence suggests the presence of an unusual singularity in the energy densities of the NSNS and RR three-form field strengths $H$ and $F_3$, which do not directly couple to the anti-branes. Using our expression for the cosmological constant, we give a simple global argument \cite{Gautason:2013} showing that this observation is not an artifact of the approximations used previously in the literature, such as a partial smearing of the anti-branes or a linearisation of the supergravity equations around the BPS background. Under some general assumptions that we discuss in detail, we find that the singularity is also generated by fully localised anti-branes that backreact on the full non-linear equations of motion.

This chapter is based on \cite{Gautason:2013} and organised as follows. In Section \ref{ch:scaling-symmetries.conventions}, we state a number of conventions that will be important for the arguments presented in this chapter. In Section \ref{ch:scaling-symmetries.cc}, we discuss the two scaling symmetries of classical type II supergravity. We then show that the cosmological constant can be written as a sum of source terms and a term involving topological background fluxes, which can in many cases be gauged away by exploiting a combination of the symmetries. In Section \ref{ch:scaling-symmetries.examples}, we present several explicit examples of compactifications of type II supergravity and show how our framework can be applied to them in order to obtain an expression for the cosmological constant in terms of the actions of localised sources. In Section \ref{ch:scaling-symmetries.kklt}, we consider the backreaction of \dthree-branes on the Klebanov-Strassler throat glued to a compact space in type IIB string theory. We then use our previous results to show that, under some general assumptions, the backreaction yields a singularity in the energy densities of $H$ and $F_3$. We conclude with some comments in Section \ref{ch:scaling-symmetries.discussion}.
\\

\subsection{Conventions}
\label{ch:scaling-symmetries.conventions}

In this section, we state a number of conventions that will be important for the arguments presented in this chapter. A more complete account of the notation and conventions used in this thesis can be found in Appendix \ref{app:conventions}.

In the tree-level supergravity approximation, the low-energy effective action of type II string theory in Einstein frame can be written as
\begin{equation}
S = S_\textrm{bulk} + S_\textrm{loc} \label{cc:action}
\end{equation}
with
\begin{equation}
S_\textrm{bulk} = S_\textrm{NSNS} + S_\textrm{RR} = \int\star_{10}\bigg[ R -\frac{1}{2} \left({\partial \phi}\right)^2 - \frac{1}{2}\e^{-\phi}|H|^2 - \frac{1}{4}
\sum_n \e^{\tfrac{5-n}{2}\phi}|F_{n}|^2\bigg], \label{cc:bulkaction}
\end{equation}
where we have set $2\kappa_{10}^2=1$. As in the previous chapters, $R$ denotes the curvature scalar of the metric $g_{MN}$, $\star_{10}$ is the ten-dimensional Hodge operator associated with $g_{MN}$, $\phi$ is the dilaton, $H$ is the NSNS three-form field strength, and $F_n$ are the RR field strengths, which are doubled in the democratic formulation such that the sum in above equation also contains the dual fields with $n > 5$. The RR field strengths are related to one another by the duality relations
\begin{equation}
\e^{\tfrac{5-n}{2}\phi}F_n  = \star_{10}\, \sigma(F_{10-n}), \label{rrduality}
\end{equation}
which have to be imposed at the level of the equations of motion. The operator $\sigma$ here acts on an $n$-form $\omega_n$ like
\begin{equation}
\sigma(\omega_n) = (-1)^{\tfrac{n(n-1)}{2}} \omega_n. \label{reversal}
\end{equation}

The term $S_\textrm{loc}$ denotes the action of the localised sources, which we take to be either D$p$-branes or O$p$-planes.\footnote{In all examples discussed in this chapter, the D-branes are pointlike in the internal space such that a $B$ field along the worldvolume cannot occur and we can neglect the $B$ field term in the DBI action. Also note that we do not consider configurations with a non-zero worldvolume gauge field strength.} It reads
\begin{equation}
S_\textrm{loc} = \sum_p S_\textrm{loc}^{(p)} = \sum_p\left(S_\textrm{DBI}^{(p)} + S_\textrm{CS}^{(p)}\right) \label{cc:locaction}
\end{equation}
with
\begin{align}
S_\textrm{DBI}^{(p)} = \mp \mu_p \int \star_{p+1} \e^{\tfrac{p-3}{4}\phi} \w \sigma(\delta_{9-p}), \qquad S_\textrm{CS}^{(p)} = \Bigg\{ \begin{split} & +\mu_p \int \langle C\wedge \e^{-B} \rangle_{p+1} \w \sigma(\delta_{9-p}) \\ & - \mu_p \int C_{p+1} \w \sigma(\delta_{9-p}) \end{split}, \label{cc:locaction2}
\end{align}
where the upper line is for D$p$-branes and the lower line for O$p$-planes and $\mu_p > 0$ is the absolute value of the D$p$-brane/O$p$-plane charge. For $\overline{\textrm{D}p}$-branes and $\overline{\textrm{O}p}$-planes, the CS action would have the opposite sign. Also note that $\star_{p+1} 1$ is the volume form on the $(p+1)$-dimensional worldvolume $\Sigma$ of the corresponding source and $\star_{9-p} 1$ is the $(9-p)$-dimensional volume form transverse to the source, which we normalise such that $\star_{10} 1 = \star_{p+1} 1 \w \star_{9-p} 1$. We furthermore define $\delta_{9-p}=\delta(\Sigma) \,\sigma(\star_{9-p} 1)$, where $\delta(\Sigma)$ is the delta distribution with support on $\Sigma$. For readability, we will often also use the polyform notation in this chapter. In \eqref{cc:locaction2}, the polyform $C= \sum_n C_{n-1}$ denotes the sum of all electric and magnetic RR potentials that appear in type IIA or type IIB supergravity, and the polyform $\e^{-B}$ is defined as a power series of wedge products. The symbol $\langle\cdots \rangle_{p+1}$ denotes a projection to the form degree $p+1$,
\begin{equation}
\langle C\w \e^{-B}\rangle_{p+1} = C_{p+1} - C_{p-1}\w B + \frac{1}{2}C_{p-3}\w B\w B - \ldots
\end{equation}

Throughout this chapter, we will restrict ourselves to warped compactifications to $d \ge 4$ dimensions that preserve maximal symmetry in the non-compact $d$-dimensional spacetime. Accordingly, we only consider spacetime-filling sources extending in $p+1 \ge d$ dimensions. Furthermore, all fields are assumed to depend only on the internal coordinates $x^m$. The form fields are allowed to have legs in external directions only if they are spacetime-filling, i.\ e.\ they have to be of degree $d$ or higher. All other form fields are purely internal. We assume a warped metric of the form
\begin{equation}
\d s_{10}^2 = g_{\mu\nu} \d x^\mu \d x^\nu + g_{mn} \d x^m \d x^n, \qquad g_{\mu\nu} = \e^{2A} \tilde g_{\mu\nu}, \label{metric}
\end{equation}
where $A$ is the warp factor and $\tilde g_{\mu\nu}$ is the unwarped $d$-dimensional metric of the Minkowski or (A)dS spacetime. We will also put a tilde on quantities such as Hodge operators, covariant derivatives or contractions of tensors if they are constructed using the unwarped metric instead of the warped one.

Let us now list the relevant equations of motion. The trace of the external Einstein equation reads
\begin{equation}
R_d = \frac{d}{2} \Big({\mathcal{L} - \sum_p\mathcal{L}^{(p)}_\textrm{CS}}\Big) + \frac{d}{4} \sum_n \e^{\tfrac{5-n}{2}\phi} |F_n^\textrm{ext}|^2, \label{einsteinx}
\end{equation}
where $R_d = R_{\mu\nu} g^{\mu\nu}$ is the $d$-dimensional Ricci scalar and we denote the spacetime-filling RR field strengths by $F_n^\textrm{ext}$. For the warped metric \eqref{metric}, one finds
\begin{equation}
R_d = \frac{2d}{d-2} \e^{-2A} \Lambda - \e^{-dA} \tilde \nabla^2 \e^{dA}, \label{ricci-lambda}
\end{equation}
where $\Lambda$ is the $d$-dimensional cosmological constant (cf. Appendix \ref{app:computations.curvature}). Substituting this into \eqref{einsteinx} and integrating over the ten-dimensional spacetime then yields
\begin{equation}
\frac{8 v \mathcal{V}}{d-2} \Lambda = 2 \Big(S - \sum_p S^{(p)}_\textrm{CS}\Big) + \sum_n \int \star_{10}\, \e^{\tfrac{5-n}{2}\phi} |F_n^\textrm{ext}|^2, \label{einstein-int}
\end{equation}
where we have introduced the volume factors
\begin{equation}
v = \int\tilde \star_d 1, \qquad \mathcal{V} = \int \star_{10-d}\, \e^{(d-2)A}. \label{volumes}
\end{equation}
The Bianchi identities for the RR fields in polyform notation are
\begin{equation}
\d_{-H} F + j = 0, \label{bianchix}
\end{equation}
where $F= \sum_n F_n$ is the polyform containing the sum over all RR field strengths, $\d_{-H} = \d - H \w$ is the twisted exterior derivative, and $j$ is the polyform containing the sum over all source contributions of the different Bianchi identities, where $j = \sum_p \mu_p \langle\delta \w \e^{B}\rangle_{9-p}$ for D-branes and $j = - \sum_p \mu_p \delta_{9-p}$ for O-planes. Finally, we state the equation of motion and Bianchi identity for $H$ in polyform notation,
\begin{equation}
\d\left(\e^{-\phi}\star_{10} H\right) - \frac{1}{2} \left\langle F \w \sigma(F) \right\rangle_{8} = 0, \qquad \d H = 0. \label{h-eoms}
\end{equation}
\\[-0.7cm]

\subsection{The Cosmological Constant as a Sum of Source Terms}
\label{ch:scaling-symmetries.cc}

In this section, we will introduce two scaling symmetries satisfied by the action \eqref{cc:action} and use them to derive an expression for the cosmological constant $\Lambda$ in terms of the (on-shell evaluated) action of localised sources.

\subsubsection{Two Scaling Symmetries}
\label{ch:scaling-symmetries.cc.scaling}

It is well-known that, in the absence of localised sources, the low-energy effective action of string theory is classically scale invariant \cite{Witten:1985xb}. This property is due to the coupling of the dilaton to the worldsheet curvature and manifest in the ten-dimensional action of the bulk fields in string frame, where all terms scale uniformly like $\e^{-2\phi}$ at tree-level since they all derive from a string worldsheet with the same Euler characteristic. The lowest order terms in the action for D-branes or O-planes, on the other hand, scale like $\e^{-\phi}$ since the corresponding worldsheet contains a boundary or a cross-cap, respectively. In Einstein frame, this simple scaling property is less apparent due to the field redefinition. In type II string theory, the corresponding scaling transformation then involves the dilaton, the metric and the RR gauge potentials,
\begin{equation}
\e^{-\phi} \to s\e^{-\phi}, \qquad g_{MN} \to \sqrt{s} g_{MN}, \qquad C_{n-1} \to s C_{n-1}, \label{jhgskg}
\end{equation}
where $s$ is a scaling parameter. This leads to
\begin{equation}
S^{(\chi)} \to s^\chi S^{(\chi)},
\end{equation}
where $\chi$ is the Euler characteristic of the worldsheet from which the contribution $S^{(\chi)}$ to the effective action is derived. For the usual low-energy effective action of type II string theory, which consists of the classical two-derivative action for the bulk supergravity fields and the leading order action for D-branes and O-planes, we obtain
\begin{equation}
S = S_{\textrm{bulk}} + S_{\textrm{loc}} \to s^2 S_{\textrm{bulk}} + s S_{\textrm{loc}}, \label{dilatonscaling}
\end{equation}
as expected from the form of the string frame action. This can be verified using \eqref{jhgskg} in \eqref{cc:bulkaction} and \eqref{cc:locaction2}. Thus, in the absence of localised sources, the effect of \eqref{jhgskg} is to rescale the tree-level supergravity action by an overall factor $s^2$. The transformations \eqref{jhgskg} are then a symmetry of the theory since they leave the equations of motion invariant.

A second scaling symmetry \cite{Burgess:1985zz} can be obtained from the mass dimension of the fields, which can be determined from the fact that the effective action is a derivative expansion and has mass dimension zero. Using that the mass dimension of the coordinates is $-1$ and the mass dimension of a derivative is $+1$, one can count the number of derivatives of a given term in the action and the number of dimensions that are integrated over to determine the mass dimension of the fields. If one then scales the fields in the effective action according to their mass dimension but leaves the coordinates unscaled, one obtains a non-trivial scaling of the terms in the action. The corresponding scaling of the bosonic fields in type II string theory is\footnote{This symmetry is sometimes also referred to as ``trombone'' symmetry \cite{Cremmer:1997xj}. Also note that, in our conventions, the powers of $t$ in \eqref{gshogg} correspond to the \emph{inverse} mass dimensions of the corresponding fields.}
\begin{equation}
g_{MN} \to t^{-2} g_{MN}, \qquad C_{n-1} \to t^{-(n-1)} C_{n-1}, \qquad B\to t^{-2} B, \label{gshogg}
\end{equation}
where $t$ is another scaling parameter. The terms in the low-energy action then scale like
\begin{equation}
S_i^D \to t^{i-D} S_i^D,
\end{equation}
where $D$ denotes the number of dimensions that are integrated over (which is typically less than ten for source terms) and $i$ is the number of derivatives in the corresponding term. For a two-derivative bulk action and zero-derivative source terms with $(p+1)$-dimensional worldvolume, we thus get
\begin{equation}
S = S_{\textrm{bulk}} + S_{\textrm{loc}} \to t^{-8} S_{\textrm{bulk}} + \sum_p t^{-p-1} S^{(p)}_{\textrm{loc}}, \label{massscaling}
\end{equation}
as can be verified using \eqref{gshogg} in \eqref{cc:bulkaction} and \eqref{cc:locaction2}. In the absence of localised sources, the transformations \eqref{gshogg} are a symmetry since they rescale the bulk action by an overall factor $t^{-8}$ and thus leave the equations of motion invariant. Together with \eqref{dilatonscaling}, this implies that the type II supergravity action at tree-level has two global scaling symmetries, which are explicitly broken by terms that are due to the presence of localised sources.

\subsubsection{The Method}
\label{ch:scaling-symmetries.cc.method}

As mentioned in the introduction to this chapter, the above scaling symmetries can often be used to derive an expression for the cosmological constant $\Lambda$ in terms of the on-shell action of localised sources. In those cases where this is true, $\Lambda$ is thus determined by the boundary conditions of some of the bulk supergravity fields at the positions of the sources and independent of the details of the dynamics in the bulk. We will argue below that this is possible for compactifications that involve at most NSNS $H$ flux and not more than one type of RR flux. Our argument is an extension of \cite{Burgess:2011rv}, where a similar result was obtained for compactifications with codimension $2$ sources but the effect of topologically non-trivial background fluxes was not made explicit. In the following, we generalise this work in that we consider setups with spacetime-filling sources of arbitrary codimension and explicitly take into account the effect of topological fluxes. In contrast to \cite{Burgess:2011rv}, where it was sufficient to use one of the two scaling symmetries discussed above, our generalisation requires to exploit both of the symmetries.

The strategy for deriving our expression for $\Lambda$ is as follows. At first, the scaling symmetries are used to derive an expression for the action \eqref{cc:action} that holds on-shell. This on-shell expression can then be substituted into the integrated Einstein equation \eqref{einstein-int}, which, as we will show, eliminates the dependence of the equation on the bulk fields up to certain flux terms and yields the desired result for $\Lambda$. Before we discuss how to derive the on-shell action in the general case, let us at first review the basic principle \cite{Aghababaie:2003ar, Burgess:2011rv}  using a simple example. Consider an action $S[\psi_i]$ that depends on a number of fields $\psi_i$ and satisfies a scaling symmetry,
\begin{equation}
S[\tau^{k_i}\psi_i] = \tau^k S[\psi_i], \label{scalingex}
\end{equation}
where the scaling parameter $\tau$ is a real number and $k$ is assumed to be non-vanishing. We can then take the $\tau$ derivative of \eqref{scalingex} to obtain
\begin{equation}
\int \sum_i k_i\tau^{k_{i}-1}\psi_i\de{S[\tau^{k_i}\psi_i]}{(\tau^{k_{i}}\psi_i)} = k\tau^{k-1}S[\psi_i], \label{scalingex2}
\end{equation}
where we have written the result in terms of a functional derivative (which, for derivative terms in $S[\tau^{k_i}\psi_i]$, implicitly involves partial integrations). Evaluating the equation at $\tau=1$ and using the fact that the fields satisfy the equations of motion $\delta S[\psi_i]/ \delta\psi_i = 0$, we then find that the left-hand side of \eqref{scalingex2} vanishes and
\begin{equation}
S[\psi_i] = 0 \label{onshellaction}
\end{equation}
on-shell.

In deriving \eqref{onshellaction}, however, we made two simplifications that do in general not hold in the context of string compactifications. The right-hand side of the equation is therefore often more complicated than in this simple example. First, we assumed that all terms in the action $S[\psi_i]$ scale uniformly with $\tau$. However, this is not true in the presence of localised sources, as follows from \eqref{dilatonscaling} for $\tau=s$ and from \eqref{massscaling} for $\tau=t$.\footnote{The assumption also breaks down if one includes, for example, $\alpha^{\prime}$ or loop corrections.} Second, when we evaluated $\d S[\tau^{k_i}\psi_i]/\d \tau$ to arrive at \eqref{scalingex2}, we had to integrate by parts all those terms in $S[\tau^{k_i}\psi_i]$ that involve derivatives of $\psi_i$. In string theory, however, many compactifications involve the presence of non-trivial background fluxes. The corresponding NSNS and/or RR field strength(s) then have a non-exact part such that, globally, they cannot be written in terms of a gauge potential. Instead, their gauge potentials are only locally defined. Thus, total derivatives involving the NSNS or RR gauge potentials do not necessarily integrate to zero anymore but may involve non-trivial contributions from patches of different gauge charts, which would yield an extra contribution when one integrates by parts. When we repeat the  above calculation for the general action \eqref{cc:action}, we therefore expect that the right-hand side of \eqref{onshellaction} receives two contributions: one contribution due to the presence of localised sources and another one due to non-trivial background fluxes.

In order to account for the possibility of flux, we explicitly divide the NSNS and RR field strengths into a flux part, which is closed but not exact, and a fluctuation, which is exact and given in terms of a globally defined gauge potential. For $H$, we thus write
\begin{equation}
H= \d B + H^\textrm{b}, \label{def-nsns}
\end{equation}
where $H^\textrm{b}$ denotes the background flux and $B$ is the fluctuating globally defined NSNS potential. Since $H^\textrm{b}$ is closed, the Bianchi identity $\d H = 0$ is satisfied such that our definition is consistent.\footnote{We do not consider compactifications involving NS$5$-branes such that the Bianchi identity for $H$ does not contain a source term.}

For the RR field strengths, separating off the non-exact part is more subtle. This is related to the fact that their Bianchi identities are more complicated and, in particular, that some of them receive contributions from localised sources. Since we only consider spacetime-filling sources in this chapter, they enter the Bianchi identities as delta forms whose legs are always in some of the internal directions. Thus, a source term can only show up in the Bianchi identity for the purely internal part of the corresponding RR field strength. It is therefore convenient to split the polyform $F = \sum_n F_n$ into a part $F^\textrm{int} = \sum_n F^\textrm{int}_n$, which contains all RR field strengths that are purely internal and may have a source term in their Bianchi identity, and a part $F^\textrm{ext} = \sum_n F^\textrm{ext}_n$, which contains all RR field strengths that are spacetime-filling (with possible additional legs in the internal space) and, accordingly, do not have a source term in their Bianchi identity,
\begin{equation}
F = F^\textrm{int} + F^\textrm{ext}. \label{split-rrfields}
\end{equation}
For $F^\textrm{ext}$, the Bianchi identities \eqref{bianchix} then simplify to
\begin{equation}
\d_{-H}F^\textrm{ext} =0. \label{bi-rrfields}
\end{equation}
This allows us to make the ansatz
\begin{equation}
F^\textrm{ext} = \d_{-H} C^\textrm{ext} + \e^B\w F^\textrm{b}, \label{def-rrfields}
\end{equation}
where $F^\textrm{b}$ is a $\d_{-H^\textrm{b}}$-closed but non-exact polyform containing the sum over the spacetime-filling background fluxes and $C^\textrm{ext}$ is a polyform containing the sum over the spacetime-filling RR potentials. In a (maximally symmetric) type IIB compactification to four dimensions, for example, we would have $F^\textrm{b} = F_5^\textrm{b} + F_7^\textrm{b} + F_9^\textrm{b}$ and $C^\textrm{ext} = C_4^\textrm{ext} + C_6^\textrm{ext} + C_8^\textrm{ext}$ since only forms of degree $4$ or higher would be allowed to be spacetime-filling. One can verify that \eqref{def-rrfields} solves the Bianchi identities \eqref{bi-rrfields} and is therefore a consistent ansatz for the field strengths $F^\textrm{ext}$.

The Bianchi identities of the internal field strengths $F^\textrm{int}$, however, may contain source terms such that these field strengths can in general not be written in a way similar to \eqref{def-rrfields} everywhere on the compact space. We will circumvent this problem by simply expressing $F^\textrm{int}$ in terms of their dual field strengths $F^\textrm{ext}$ on-shell, which then in turn can be expressed in terms of \eqref{def-rrfields}. If, for example, $F_3=F_3^\textrm{int}$ is internal, we can express it in terms of the spacetime-filling $F_7=F_7^\textrm{ext}$ via the duality relation $F_3^\textrm{int}=- \e^{-\phi} \star_{10} F_7^\textrm{ext}$ and then use \eqref{def-rrfields} to split $F_7^\textrm{ext}$ into an exact and a non-exact part.\footnote{A subtlety occurs for $F_5$, which is self-dual, and $F_4$, which can have both internal and spacetime-filling components in compactifications to four dimensions. In these cases, only the internal components $F_5^\textrm{int}$ and $F_4^\textrm{int}$ can have a source term in the Bianchi identity. We therefore express those in terms of their duals $F_5^\textrm{ext}$ and $F_6^\textrm{ext}$, which can in turn be written in terms of \eqref{def-rrfields}.}

Let us finally note that, since we put the non-exact parts of the NSNS and RR field strengths into $H^\textrm{b}$ and $F^\textrm{b}$, we can assume that the gauge potentials $B$ and $C^\textrm{ext}$ are globally defined. This implies that total derivatives involving $B$ and $C^\textrm{ext}$ integrate to zero on a compact space, which will be used below. It should also be mentioned that, under the scalings \eqref{jhgskg} and \eqref{gshogg}, the flux terms $H^\textrm{b}$ and $F^\textrm{b}$ behave in the same way as the corresponding gauge potentials do. This follows from the fact that the mass dimension and the coupling to the dilaton are the same for the exact and the non-exact parts of the NSNS and RR field strengths.

\subsubsection{On-shell Action and Cosmological Constant}
\label{ch:scaling-symmetries.cc.onshellrel}

Let us now discuss how to derive the on-shell expression for the action \eqref{cc:action} that will later be used in the integrated Einstein equation \eqref{einstein-int} to obtain our result for $\Lambda$. Contrary to the simple example sketched in the previous section, the calculation is rather involved if one considers the general case including sources and fluxes. Let us therefore note that there is an alternative way to obtain our result, which only uses the equations of motion instead of exploiting the scaling symmetries. This second derivation may serve as a double-check of our results and is detailed in Appendix \ref{app:ccandsources}. In the following, we will continue to discuss the first method, using the scaling symmetries. The reader who is less interested in the technical details of the derivation may also jump directly to \eqref{y} and the subsequent discussion, where we present our result for $\Lambda$.

Let $\tau$ denote the scaling parameter, where $\tau$ equals $s$ if we consider the dilaton scaling \eqref{jhgskg} and $t$ in case of the mass scaling \eqref{gshogg}. Moreover, we will use primes to denote the $\tau$-transformed fields and the corresponding $\tau$-transformed action. Thus, if $\tau=s$, we have, for example, $g'_{MN} = \sqrt{s}g_{MN}$, and if $\tau=t$, we have $g'_{MN} = t^{-2}g_{MN}$. According to \eqref{dilatonscaling} and \eqref{massscaling}, the action \eqref{cc:action} then scales as
\begin{equation}
S^\prime =  S^\prime_\textrm{bulk} + S^\prime_\textrm{loc} = \tau^k S_\textrm{bulk} + \sum_p \tau^{l_{p}} S^{(p)}_{\textrm{loc}}, \label{scalingex3}
\end{equation}
where $k=2$, $l_{p}=1$ for $\tau=s$ and $k=-8$, $l_{p}=-p-1$ for $\tau=t$. Taking the $\tau$ derivative and evaluating the equation at $\tau=1$, we find
\begin{equation}
\left.{\frac{\d S^\prime_\textrm{bulk}}{\d \tau}}\right|_{\tau=1} + \left.{\frac{\d S^\prime_\textrm{loc}}{\d \tau}}\right|_{\tau=1} = k S_\textrm{bulk} + \sum_p l_{p} S^{(p)}_{\textrm{loc}}. \label{scalingex4}
\end{equation}
We now proceed as in the simple example discussed in Section \ref{ch:scaling-symmetries.cc.method}: we first evaluate the terms on the left-hand side of the equation and integrate by parts to express them in terms of a functional derivative of the action with respect to the fields. We then substitute the equations of motion to simplify the expressions.

The first term on the left-hand side of \eqref{scalingex4} yields\footnote{We define functional derivatives with respect to form fields $\omega_n$ such that $\delta S = \int \delta S/ \delta \omega_n \wedge \delta \omega_n$.}
\begin{align}
\left.{\diff{S_\textrm{bulk}^\prime}{\tau}}\right|_{\tau=1} = \int \bigg[ & \de{S_\textrm{bulk}}{g_{MN}}\diff{g_{MN}^\prime}{\tau} + \de{S_\textrm{bulk}}{\phi}\diff{\phi^\prime}{\tau} + \de{S_\textrm{NSNS}}{H}\w \diff{H^\prime}{\tau} \notag \\ & + \left.{ \left\langle\de{S_\textrm{RR}}{F}\w\diff{F^\prime}{\tau}\right\rangle_{10}\bigg] }\right|_{\tau=1}, \label{sbulkdiff}
\end{align}
where we have implicitly used partial integration to write the first two terms in the integrand as variations of $S_\textrm{bulk}$ with respect to the metric and the dilaton. Evaluating the remaining two terms is more involved since $H$ and $F$ may contain flux (cf. \eqref{def-nsns} and  \eqref{def-rrfields}), and so we will consider them separately later. Let us at first evaluate the $\d S^\prime_\textrm{loc}/ \d \tau$ term in \eqref{scalingex4},
\begin{align}
\left.{\diff{S^\prime_\textrm{loc}}{\tau}}\right|_{\tau=1} &= \int \left.{\left[\de{S_\textrm{loc}}{g_{MN}}\diff{g^\prime_{MN}}{\tau} + \de{S_\textrm{loc}}{\phi}\diff{\phi^\prime}{\tau} + \left\langle \de{S_\textrm{loc}}{C}\w\diff{C^\prime}{\tau}\right\rangle_{10} + \de{S_\textrm{loc}}{B}\w\diff{B^\prime}{\tau}\right]}\right|_{\tau=1} \nll = \int \left.{\left[\de{S_\textrm{loc}}{g_{MN}}\diff{g^\prime_{MN}}{\tau} + \de{S_\textrm{loc}}{\phi}\diff{\phi^\prime}{\tau}\right]}\right|_{\tau=1} + \sum_p \diff{S_\textrm{CS}^{\prime (p)}}{\tau} \bigg|_{\tau=1}. \label{locscaling}
\end{align}
Since $S_\textrm{loc}$ does not depend on any field derivatives but only on the fields themselves, we did not have to integrate by parts here. We can now combine \eqref{sbulkdiff} and \eqref{locscaling} and use the equations of motion $\delta S / \delta g_{MN} = \delta S / \delta \phi = 0$ to obtain
\begin{align}
\left.{\diff{S_\textrm{bulk}^\prime}{\tau}}\right|_{\tau=1} + \left.{\diff{S^\prime_\textrm{loc}}{\tau}}\right|_{\tau=1} &= \int \left.{\left[\de{S_\textrm{NSNS}}{H}\w \diff{H^\prime}{\tau} + \left\langle\de{S_\textrm{RR}}{F}\w\diff{F^\prime}{\tau} \right\rangle_{10}\right] }\right|_{\tau=1} \nl + \sum_p \diff{S_\textrm{CS}^{\prime (p)}}{\tau} \bigg|_{\tau=1}. \label{sscaling1}
\end{align}

The two terms involving $\delta H$ and $\delta F$ are evaluated as follows. Substituting \eqref{def-nsns} into the $\delta S_\textrm{NSNS}/\delta H$ term in \eqref{sscaling1}, we can integrate by parts to obtain 
\begin{align}
\int \left.{ \de{S_\textrm{NSNS}}{H}\w \diff{H^\prime}{\tau}}\right|_{\tau=1}
 &= \int \left.{\left[\d \de{S_\textrm{NSNS}}{H}\w \diff{B^\prime}{\tau} + \de{S_\textrm{NSNS}}{H}\w\diff{H^{\prime \textrm{b}}}{\tau}\right]}\right|_{\tau=1} \nll =  \int \left.{\left[\de{S_\textrm{NSNS}}{B}\w\diff{B^\prime}{\tau} + \de{S_\textrm{NSNS}}{H}\w\diff{H^{\prime \textrm{b}}}{\tau}\right]}\right|_{\tau=1}. \label{NSNSscaling}
\end{align}
The $\delta S_\textrm{RR}/\delta F$ term in \eqref{sscaling1} can be computed in a similar fashion but is more complicated due to the subtleties explained in Section \ref{ch:scaling-symmetries.cc.method}. We first use \eqref{split-rrfields} and write
\begin{equation}
\int \left.{\left\langle\de{S_\textrm{RR}}{F}\w\diff{F^\prime}{\tau}\right\rangle_{10}}\right|_{\tau=1} = \int \left.{\left\langle\de{S_\textrm{RR}}{F^\textrm{ext}}\w\diff{F^{\prime \textrm{ext}}}{\tau} + \de{S_\textrm{RR}}{F^\textrm{int}}\w\diff{F^{\prime \textrm{int}}}{\tau}\right\rangle_{10}}\right|_{\tau=1}. \label{frewrite0}
\end{equation}
We now have to replace all RR field strengths $F_n^\textrm{int}$ by their dual field strengths $F_{10-n}^\textrm{ext}$ in order to be able to write them in terms of the globally defined gauge potentials $C^\textrm{ext}$ using \eqref{def-rrfields}, which in turn will allow us to integrate by parts in \eqref{frewrite0}. Using the duality relations \eqref{rrduality} as well as the scalings \eqref{jhgskg} and \eqref{gshogg}, we find for the two cases $\tau=s$ and $\tau=t$:
\begin{align}
\int \left.{\left\langle\de{S_\textrm{RR}}{F}\w\diff{F^\prime}{s}\right\rangle_{10}}\right|_{s=1} &= \sum_n \int \left(\de{S_\textrm{RR}} {F_n^\textrm{ext}}\w F_n^\textrm{ext} + \de{S_\textrm{RR}}{F_n^\textrm{int}}\w F_n^\textrm{int} \right) \nll = \sum_n \int \left( \de{S_\textrm{RR}} {F_n^\textrm{ext}}\w F_n^\textrm{ext} - \de{S_\textrm{RR}}{F_{10-n}^\textrm{ext}}\w F_{10-n}^\textrm{ext} \right) \nll = 0, \displaybreak[3] \label{frewrite1} \\
\int \left.{\left\langle\de{S_\textrm{RR}}{F}\w\diff{F^\prime}{t}\right\rangle_{10}}\right|_{t=1} &= \sum_n (1-n) \int \left(\de{S_\textrm{RR}} {F_n^\textrm{ext}}\w F_n^\textrm{ext} + \de{S_\textrm{RR}}{F_n^\textrm{int}}\w F_n^\textrm{int} \right) \nll = \sum_n (1-n) \int \left( \de{S_\textrm{RR}} {F_n^\textrm{ext}}\w F_n^\textrm{ext} - \de{S_\textrm{RR}}{F_{10-n}^\textrm{ext}}\w F_{10-n}^\textrm{ext} \right) \nll = \sum_n (10-2n) \int \de{S_\textrm{RR}}{F_n^\textrm{ext}}\w F_n^\textrm{ext}. \label{frewrite2}
\end{align}
These two expressions can now be rewritten in a way that will become convenient further below. In order to do so, we again exploit the scalings \eqref{jhgskg} and \eqref{gshogg} and make use of the identity $ \delta S_\textrm{RR} / \delta F_n^\textrm{ext} \w F_n^\textrm{ext} = - \frac{1}{2}\star_{10}\, \e^{\frac{5-n}{2}\phi} |F_n^\textrm{ext}|^2$, which can be derived from \eqref{cc:bulkaction}. We thus find
\begin{align}
\int \left.{\left\langle\de{S_\textrm{RR}}{F}\w\diff{F^\prime}{\tau}\right\rangle_{10}}\right|_{\tau=1} &= 2 \int \left.{\left\langle \de{S_\textrm{RR}}{F^\textrm{ext}}\w \diff{F^{\prime \textrm{ext}}}{\tau} \right\rangle_{10}}\right|_{\tau=1} - 2 k \int \left\langle \de{S_\textrm{RR}}{F^\textrm{ext}}\w F^\textrm{ext} \right\rangle_{10} \nl - \frac{k}{2} \sum_n \int \star_{10}\, \e^{\tfrac{5-n}{2}\phi} |F_n^\textrm{ext}|^2, \label{frewrite5}
\end{align}
where $k=2$ for $\tau=s$ and $k=-8$ for $\tau=t$, as in \eqref{scalingex3}.

We now integrate by parts on the right-hand side of \eqref{frewrite5}. Taking into account \eqref{def-nsns} and \eqref{def-rrfields},
this yields\footnote{The factor $\frac{1}{2}$ that appears when rewriting $\delta S_\text{RR}/\delta C^\textrm{ext}$ in terms of $\delta S/\delta C^\textrm{ext}$ and $\delta S_\textrm{loc}/\delta C^\textrm{ext}$ is related to a subtlety regarding the variation of the CS action of the RR fields. One only obtains the correct equations of motion if one takes the coupling of the RR fields to the sources to be half the coupling that one would get from the ``naive'' variation of the action. One can think of this as being due to the fact that one half of $\sum_p S^{(p)}_\textrm{CS}$ represents an electric coupling of the RR fields to the sources, whereas the other half is due to a magnetic coupling of the dual RR fields to the sources. This subtlety is known in the literature and has, for example, been discussed in footnote 6 of \cite{Giddings:2001yu} and also in \cite{Koerber:2007hd}.}
\begin{align}
& \int \left.{ \left\langle\de{S_\textrm{RR}}{F^\textrm{ext}} \w\diff{F^{\prime \textrm{ext}}}{\tau}\right\rangle_{10} }\right|_{\tau=1} \nll
= \int \left\langle\de{S_\textrm{RR}}{F^\textrm{ext}}\w\left(\d_{-H}\diff{C^{\prime \textrm{ext}}}{\tau} + \e^B\w\diff{F^{\prime \textrm{b}}}{\tau} - \diff{(\d B^\prime + H^{\prime \textrm{b}})}{\tau}\w C^\textrm{ext}\right.\right. \notag \\
&\qquad\quad \left.{\left.\left. +\, \diff{B^\prime}{\tau}\w\e^B\w F^\textrm{b}\right)\right\rangle_{10} }\right|_{\tau=1} \nll
=  \left. \int\left\langle \de{S_\textrm{RR}}{C^\textrm{ext}}\w\diff{C^{\prime \textrm{ext}}}{\tau} + \de{S_\textrm{RR}}{F^\textrm{ext}}\w \left({\e^B \w \diff{F^{\prime \textrm{b}}}{\tau} + \de{F^\textrm{ext}}{B} \w \diff{B^\prime}{\tau}  + \de{F^\textrm{ext}}{H} \w \diff{H^{\prime \textrm{b}}}{\tau}}\right)\right\rangle_{10}\right|_{\tau=1} \nll
= \int \left. \left\langle \left(\de{S}{C^\textrm{ext}}  - \frac{1}{2}\de{S_\textrm{loc}}{C^\textrm{ext}}\right)\w\diff{C^{\prime \textrm{ext}}}{\tau} + \frac{1}{2}\de{S_\textrm{RR}}{B}\w\diff{B^\prime}{\tau} + \de{S_\textrm{RR}}{F^\textrm{ext}}\w \left({\e^B \w \diff{F^{\prime \textrm{b}}}{\tau} }\right.\right.\right. \notag \\
&\qquad\quad \left.\left.\left.{ +\, \de{F^\textrm{ext}}{H} \w \diff{H^{\prime \textrm{b}}}{\tau}}\right)\right\rangle_{10}\right|_{\tau=1} \nll
= \int \left. \left\langle \left(\de{S}{C^\textrm{ext}} - \frac{1}{2}\de{S_\textrm{loc}}{C^\textrm{ext}}\right)\w\diff{C^{\prime \textrm{ext}}}{\tau} + \frac{1}{2}\left(\de{S}{B} - \de{S_\textrm{NSNS}}{B} - \de{S_\textrm{loc}}{B}\right)\w\diff{B^\prime}{\tau}\right.\right. \notag \\
& \qquad\quad \left.\left. +\, \de{S_\textrm{RR}}{F^\textrm{ext}}\w\left(\e^B \w \diff{F^{\prime \textrm{b}}}{\tau} + \de{F^\textrm{ext}}{H}\w\diff{H^{\prime \textrm{b}}}{\tau}\right)\right\rangle_{10}\right|_{\tau=1},
\end{align}
where we also used the identity
\begin{equation}
2 \left\langle\de{S_\textrm{RR}}{F^\textrm{ext}}\w\de{F^\textrm{ext}}{B}\right\rangle_8 = \left\langle F^\textrm{ext} \w \sigma(F^\textrm{int})\right\rangle_8 - \de{S_\textrm{loc}}{B} = \de{S}{B} - \de{S_\textrm{NSNS}}{B} - \de{S_\textrm{loc}}{B} = \de{S_\textrm{RR}}{B},
\end{equation}
which can be derived using \eqref{cc:bulkaction}, \eqref{h-eoms}, \eqref{def-nsns} and \eqref{def-rrfields}. With the equations of motion $\delta S/\delta C^\textrm{ext} = \delta S/\delta B = 0$, we finally obtain
\begin{align}
\int \left. \left\langle\de{S_\textrm{RR}}{F^\textrm{ext}} \w\diff{F^{\prime \textrm{ext}}}{\tau}\right\rangle_{10} \right|_{\tau=1} &= \int \left.{\left\langle \de{S_\textrm{RR}}{F^\textrm{ext}}\w\left(\e^B \w \diff{F^{\prime \textrm{b}}}{\tau} + \de{F^\textrm{ext}}{H}\w\diff{H^{\prime \textrm{b}}}{\tau}\right)\right\rangle_{10}}\right|_{\tau=1} \nl -\frac{1}{2} \sum_p \diff{S^{\prime (p)}_\textrm{CS}}{\tau}\bigg|_{\tau=1} - \frac{1}{2} \int \left.{\de{S_\textrm{NSNS}}{B}\w\diff{B^\prime}{\tau}}\right|_{\tau=1} \label{identity1}
\end{align}
and, evaluating this equation for $\tau=s$ using \eqref{jhgskg},
\begin{equation}
\int \left\langle \de{S_\textrm{RR}}{F^\textrm{ext}}\w F^\textrm{ext} \right\rangle_{10} = \int \left\langle \de{S_\textrm{RR}}{F^\textrm{ext}}\w \e^B \w F^\textrm{b}\right\rangle_{10} -\frac{1}{2} \sum_p S^{(p)}_\textrm{CS}. \label{identity2}
\end{equation}
Substituting \eqref{identity1} and \eqref{identity2} into \eqref{frewrite5} then leads to
\begin{align}
\int \left.{\left\langle\de{S_\textrm{RR}}{F}\w\diff{F^\prime}{\tau}\right\rangle_{10}}\right|_{\tau=1} &=
 2 \int \left.{\left\langle \de{S_\textrm{RR}}{F^\textrm{ext}}\w\left(\e^B \w \diff{F^{\prime \textrm{b}}}{\tau} + \de{F^\textrm{ext}}{H}\w\diff{H^{\prime \textrm{b}}}{\tau}\right)\right\rangle_{10}}\right|_{\tau=1} \nl - 2k \int \left.{\left\langle \de{S_\textrm{RR}}{F^\textrm{ext}}\w \e^B\w F^\textrm{b} \right\rangle_{10}}\right|_{\tau=1} -\sum_p \diff{S^{\prime (p)}_\textrm{CS}}{\tau}\bigg|_{\tau=1} \nl + k \sum_p S^{(p)}_\textrm{CS} - \int \left.\de{S_\textrm{NSNS}}{B}\w\diff{B^\prime}{\tau}\right|_{\tau=1} \nl - \frac{k}{2} \sum_n \int \star_{10}\, \e^{\tfrac{5-n}{2}\phi} |F_n^\textrm{ext}|^2. \label{frewrite6}
\end{align}

We can now put everything together by using \eqref{frewrite6} and \eqref{NSNSscaling} in \eqref{sscaling1} such that we arrive at
\begin{align}
\frac{\d S^\prime_\textrm{bulk}}{\d \tau}\bigg|_{\tau=1} + \frac{\d S^\prime_\textrm{loc}}{\d \tau}\bigg|_{\tau=1} &= 2\int \left.{\left\langle \de{S_\textrm{RR}}{F^\textrm{ext}}\w\left(\e^B \w \diff{F^{\prime \textrm{b}}}{\tau} + \de{F^\textrm{ext}}{H}\w\diff{H^{\prime \textrm{b}}}{\tau}\right)\right\rangle_{10}}\right|_{\tau=1} \nl -2k \int \left.{\left\langle \de{S_\textrm{RR}}{F^\textrm{ext}}\w \e^B \w F^\textrm{b}\right\rangle_{10}}\right|_{\tau=1} + k \sum_p S^{(p)}_\textrm{CS} \nl + \int\left.\de{S_\textrm{NSNS}}{H}\w\diff{H^{\prime \textrm{b}}}{\tau}\right|_{\tau=1} - \frac{k}{2} \sum_n \int \star_{10}\, \e^{\tfrac{5-n}{2}\phi} |F_n^\textrm{ext}|^2. \label{scalingresult}
\end{align}
Using \eqref{scalingex4} on the left-hand side and the two scaling symmetries \eqref{jhgskg} and \eqref{gshogg} on the right-hand side and evaluating the functional derivatives then leads to the two equations 
\begin{align}
2 S_\textrm{bulk} + S_\textrm{loc} &= 2\sum_p S_\textrm{CS}^{(p)}  - \sum_n \int \star_{10}\, \e^{\tfrac{5-n}{2}\phi} |F_n^\textrm{ext}|^2 \nl - \sum_n \int F^\textrm{b}_n \w \left\langle \e^B\w\sigma(F^\textrm{int})\right\rangle_{10-n}, \label{onshell1} \\
- 8 S_\textrm{bulk} - \sum_p (p+1) S_\textrm{loc}^{(p)} &= -8\sum_p S_\textrm{CS}^{(p)} + 4 \sum_n \int \star_{10} \, \e^{\tfrac{5-n}{2}\phi} |F_n^\textrm{ext}|^2 \nl + \sum_n (9-n) \int F^\textrm{b}_n\w\left\langle \e^B \w \sigma(F^\textrm{int})\right\rangle_{10-n} \nl - 2\int H^\textrm{b}\w \left( \e^{-\phi}\star_{10} H - \left\langle \sigma(F^\textrm{int})\w C^\textrm{ext}\right\rangle_7\right), \label{onshell2}
\end{align}
where $\sigma$ is the operator defined in \eqref{reversal}. We can now linearly combine \eqref{onshell1} and \eqref{onshell2} introducing a free parameter $c$ and rearrange the source terms using $S = S_\textrm{bulk} + S_\textrm{loc}$ and $S_\textrm{loc}^{(p)} = S_\textrm{DBI}^{(p)} + S_\textrm{CS}^{(p)}$, which yields
\begin{align}
& 2S-2\sum_pS^{(p)}_\textrm{CS} + \sum_n \int \star_{10} \,
\e^{\tfrac{5-n}{2}\phi} |F_n^\textrm{ext}|^2 \notag \\ & \qquad = \sum_p  \left(1+\frac{p-3}{2}c\right) \left[S_{\textrm{DBI}}^{(p)}+S_\textrm{CS}^{(p)}\right] - \sum_n \left(1+\frac{n-5}{2}c\right) \int F^\textrm{b}_n \w\left\langle \e^B\w\sigma(F^\textrm{int})\right\rangle_{10-n} \notag \\ & \qquad\quad - c \int H^\textrm{b}\w \left({ \e^{-\phi}\star_{10} H - \left\langle \sigma(F^\textrm{int})\w C^\textrm{ext}\right\rangle_7}\right). \label{long}
\end{align}
Substituting this into the integrated Einstein equation \eqref{einstein-int} and collecting all contributions from background fluxes into a single term $\mathcal{F}(c)$, we find the result
\begin{equation}
\frac{8 v \mathcal{V}}{d-2} \Lambda = \sum_{p} \left(1+\frac{p-3}{2}c\right) \left[S^{(p)}_\textrm{DBI} + S_\textrm{CS}^{(p)}\right] + \int \mathcal{F}(c) \label{y}
\end{equation}
with the volume factors $v$ and $\mathcal{V}$ defined as in \eqref{volumes}. Note that all terms on the right-hand side of \eqref{y} contain an implicit factor of the external ``volume'' $v$ such that it cancels out in the equation and $\Lambda$ does not depend on it. The flux term $\mathcal{F}(c)$ takes the form
\begin{align}
\mathcal{F} (c) &= - \sum_{n \ge d} \left(1+\frac{n-5}{2}c\right)F^\textrm{b}_{n}\w\left\langle\e^B\w \sigma(F^\textrm{int})\right\rangle_{10-n} \nl - c\, H^\textrm{b}\w\left( \e^{-\phi}\star_{10} H - \left\langle \sigma(F^\textrm{int})\w C^\textrm{ext}\right\rangle_7\right), \label{sigma}
\end{align}
where the summation range is determined by the fact that the background fluxes $F_n^\textrm{b}$ are spacetime-filling by definition and must therefore be of degree $d$ or higher (cf. the discussion in Section \ref{ch:scaling-symmetries.cc.method}).

As stated earlier, the contribution of the flux term $\mathcal{F}(c)$ can often be gauged away in \eqref{y} by choosing an appropriate numerical value for the free parameter $c$. Up to an overall volume factor $\mathcal{V}$ (whose sign is known to be positive), $\Lambda$ is then completely determined by the on-shell actions of the localised sources that appear in the corresponding solution. If only one of the fluxes in \eqref{sigma} is non-zero, it is straightforward to see that $\mathcal{F}(c)$ can be set to zero since one can then simply choose $c$ such that the $c$-dependent prefactor of the corresponding term vanishes in \eqref{sigma}.\footnote{$F_5$ flux is an exception since it does not have a $c$-dependent prefactor in $\mathcal{F}(c)$ and can therefore not be gauged away in \eqref{y}. This is the reason for the existence of the Freund-Rubin solutions of type IIB supergravity on $\mathrm{AdS}_5 \times S^5$ \cite{Freund:1980xh}.} For a compactification with non-zero $H^\textrm{b}$, for example, one would choose $c=0$, and, for a compactification with non-zero $F_7^\textrm{b}$, one would choose $c=-1$.

Even if the NSNS flux $H^\textrm{b}$ and one of the RR fluxes with $n \neq 5$ are both non-zero, it is still often possible to find a $c$ such that $\mathcal{F}(c)$ vanishes. The reason is that the term multiplying $H^\textrm{b}$ in \eqref{sigma} is proportional to
\begin{equation}
\de{S_\textrm{NSNS}}{H}+2\left\langle\de{S_\textrm{RR}}{F^\textrm{ext}} \w \de{F^\textrm{ext}}{H}\right\rangle_7 = - \e^{-\phi}\star_{10} H + \left\langle \sigma (F^\textrm{int})\w C^\textrm{ext}\right\rangle_7. \label{gauge1}
\end{equation}
If the $H$ equation of motion implies that $\d \left[{\e^{-\phi}\star_{10} H - \left\langle \sigma (F^\textrm{int})\w C^\textrm{ext}\right\rangle_7}\right] = 0$, which is the case in many interesting examples, we can write
\begin{equation}
- \e^{-\phi}\star_{10} H + \left\langle \sigma (F^\textrm{int})\w C^\textrm{ext}\right\rangle_7 = \omega_7, \label{gauge2}
\end{equation}
where $\omega_7$ is a closed (but not necessarily exact) seven-form. If it is furthermore possible to make a gauge transformation of the RR potentials such that $\omega_7$ is cancelled in \eqref{gauge2}, the term multiplying $H^\textrm{b}$ in \eqref{sigma} vanishes for any $c$, and we can choose the value for $c$ such that also the RR flux term in \eqref{sigma} vanishes. Consider, for example, a compactification of type IIA supergravity with non-zero $H^\textrm{b}$ and $F_0$. The non-trivial background fluxes appearing in \eqref{sigma} are then $H^\textrm{b}$ and $F^\textrm{b}_{10}$,
\begin{equation}
\mathcal{F} (c) =  - \left(1+\frac{5}{2}c\right)F^\textrm{b}_{10}\w F_0 - c\,H^\textrm{b}\w\left( \e^{-\phi}\star_{10} H - \left\langle \sigma(F^\textrm{int})\w C^\textrm{ext}\right\rangle_7\right). \label{gauge4}
\end{equation}
Assuming that $\d \left[{\e^{-\phi}\star_{10} H - \left\langle \sigma (F^\textrm{int})\w C^\textrm{ext}\right\rangle_7}\right] = 0$ by the $H$ equation, \eqref{gauge1} and \eqref{gauge2} imply that the term multiplying $H^\textrm{b}$ can be cancelled by a gauge transformation $C_7 \to C_7 - \omega_7/F_0$. This is a valid gauge transformation that leaves all RR field strengths unchanged. In the new gauge, we then have $\e^{-\phi}\star_{10} H - \left\langle \sigma (F^\textrm{int})\w C^\textrm{ext}\right\rangle_7 =0$ such that \eqref{gauge4} reduces to $\mathcal{F}(c) = - \left(1+\frac{5}{2}c\right)F^\textrm{b}_{10} \w F_0$. We can therefore choose $c=-\frac{2}{5}$ so that $\mathcal{F}\left(-\frac{2}{5}\right) = 0$.\footnote{Note that, even though $\mathcal{F}(c)$ is not gauge invariant, one can convince oneself that the full expression for $\Lambda$ in \eqref{y} is gauge invariant.}

In the presence of more than one type of RR flux, this reasoning does not work anymore since it is then not possible to choose an appropriate $c$ such that each term in $\mathcal{F}(c)$ is set to zero individually. We may still be able to find a $c=c_0$ that solves the equation $\int \mathcal{F}(c_0) = 0$ such that $\int \mathcal{F}(c_0)$ vanishes as a whole, but the numerical value of $c_0$ then depends on the bulk fields that appear in \eqref{sigma}. This will in general not be useful since it just has the effect of trading the explicit dependence of $\Lambda$ on the bulk dynamics for an implicit dependence hidden in the value of $c_0$. We will explain this in more detail in Section \ref{ch:scaling-symmetries.examples}, where we discuss several examples for string compactifications in which $\mathcal{F}(c)$ can be set to zero and one counterexample in which it cannot be set to zero.

\subsubsection{Validity of the Supergravity Approximation}

Before we proceed by applying the above results to some explicit examples, a comment on their regime of validity is in order. In the vicinity of localised sources, field derivatives and the string coupling often blow up such that $\alpha^\prime$ and loop corrections can become large, making the reliability of the supergravity approximation questionable. Given that the right-hand side of \eqref{y} is evaluated directly at the positions of the sources, one might therefore wonder about the self-consistency of our expression for $\Lambda$.

In order to interpret our result \eqref{y} correctly, it is important to appreciate the fact that it was obtained from the action given in \eqref{cc:bulkaction} and \eqref{cc:locaction2} and therefore relates the cosmological constant in the supergravity approximation to the near-source boundary conditions of the fields in the supergravity approximation. Since \eqref{y} is a consequence of the supergravity equations of motion, it is an \emph{exact} expression within this theory and as good as any other method to compute the cosmological constant. Let us, for simplicity, discuss this for the case where only one type of sources is present in the compactification. We can then schematically write $\Lambda^{\textrm{class}}=\kappa S_\textrm{loc}^\textrm{class}$, where the superscript $^\text{class}$ denotes the values of $\Lambda$ and $S_\textrm{loc}$ in the supergravity approximation and $\kappa$ is a constant. If corrections to the classical supergravity calculation are negligible in the lower-dimensional effective theory (as in the usual regime of large volume and small string coupling), the full cosmological constant $\Lambda^{\textrm{full}}$ is well-approximated by the supergravity result such that $\Lambda^{\textrm{full}}\approx \Lambda^\text{class}$. It then follows that also $\Lambda^{\textrm{full}}\approx \kappa S_\textrm{loc}^\textrm{class}$. Note that this is true even when $S_\textrm{loc}^\textrm{class}$ is not a good approximation to $S_\textrm{loc}^{\textrm{full}}$.

As we will show in Section \ref{ch:scaling-symmetries.examples} for several examples, \eqref{y} can therefore often be used to compute the cosmological constant from the classical boundary conditions of the supergravity fields at the source positions. In Section \ref{ch:scaling-symmetries.kklt}, on the other hand, we will use \eqref{y} in the opposite direction, i.\ e.\ we will use our knowledge of the cosmological constant in the KKLT scenario to draw conclusions about the near-brane behaviour of the supergravity fields and thus find a singularity in the energy densities of $H$ and $F_3$. It is then again important to emphasise that the existence of this singularity is inferred from the near-brane behaviour in the supergravity approximation, like it was also done in earlier analyses of the anti-brane backreaction \cite{McGuirk:2009xx, Bena:2009xk, Bena:2011hz, Bena:2011wh, Bena:2012bk}. By construction, our argument does not make any statements about whether or not the singularity is resolved by stringy effects. It indicates, however, that the singularity is not just an artifact of a partial smearing of the \dthree-branes or a linearisation around the BPS background but instead also present in the full supergravity analysis.

\subsection{Examples}
\label{ch:scaling-symmetries.examples}

In this section, we discuss different solutions of type IIA and IIB supergravity that have appeared in the literature and show
how \eqref{y} can be evaluated in our framework to obtain an explicit expression for the cosmological constant.

\subsubsection{The GKP Solutions}
\label{ch:scaling-symmetries.examples.gkp}

Here, we consider warped compactifications of type IIB supergravity to four-dimensional Minkowski space with $H$ and $F_3$ flux along the lines of \cite{Giddings:2001yu} (see also Chapter \ref{ch:smeared-vs-loc}). In order to cancel the tadpole generated by the fluxes, the solutions also require the presence of localised sources. For simplicity, we specialise to models involving only O$3$-planes. In \cite{Giddings:2001yu}, the authors also consider orientifold limits \cite{Sen:1997gv} of F-theory compactifications involving D$7$-branes and O$7$-planes. The discussion of such models in our framework is analogous albeit more lengthy.

Following \cite{Giddings:2001yu}, we find that the non-vanishing fields must satisfy
\begin{align}
F_3 = - \e^{-\phi} \star_6 H, \quad F_5 = -(1+\star_{10}) \e^{-4A} \star_6 \d \alpha, \quad C_4^\textrm{ext} = \tilde \star_4(\alpha + a), \quad \alpha = \e^{4A}, \label{gkp_fields}
\end{align}
where the warp factor $A$ and the dilaton $\phi$ are functions on the compact space and $a$ is an integration constant corresponding to a gauge transformation. Also note that $F_5 =\star_{10} F_5 = F_5^\textrm{int} + F_5^\textrm{ext}$ with $F_5^\textrm{ext} = \d C_4^\textrm{ext}$. The topologically non-trivial fluxes are $F_3$ and $H$ such that the relevant fluxes appearing in the definition of $\mathcal{F}(c)$, which is given by \eqref{sigma}, are
\begin{equation}
H^\textrm{b}, \qquad F_7^\textrm{b},
\end{equation}
while all other terms in \eqref{sigma} vanish. Thus, \eqref{sigma} reduces to
\begin{equation}
\mathcal{F}(c) = -c\,H^\textrm{b}\w\left[\e^{-\phi} \star_{10} H + F_3\w C_4^\textrm{ext}\right] + \left({1+c}\right) F_7^\textrm{b}\w F_3. \label{gkp_sigma}
\end{equation}
Using \eqref{gkp_fields}, we find that the first term can be set to zero by gauge fixing $a=0$.\footnote{Note that, although $\mathcal{F}(c)$ is not gauge invariant, the full expression for the cosmological constant $\Lambda$ is gauge invariant since it contains a term $C_4 \w \mu_3 \delta_6$, which changes under a gauge transformation such that the total $a$-dependence of $\Lambda$ cancels out as it should.} Furthermore, $F_3$ and $H$ are related by a special condition, which is given in \eqref{gkp_fields}. This condition can be shown to saturate a BPS bound and is equivalent to the ISD condition of the complex three-form field strength in the notation of \cite{Giddings:2001yu}. It follows from this condition that also the second term in \eqref{gkp_sigma} is zero,
\begin{align}
\int F_7^\textrm{b} \w F_3  &= \int  \left(F_7 - \d C_6^\textrm{ext} + H\w C_4^\textrm{ext}\right) \w F_3 \nll = \int\left(F_7 \w F_3 + \e^{\phi}\star_6 F_3 \w\left(\tilde \star_4 \e^{4A}\right)\w F_3\right) = 0,
\end{align}
where, in the last step, we used $F_7 = -\e^{\phi}\star_{10} F_3 = -\e^{\phi}\star_6 F_3\w \tilde \star_4 \e^{4 A}$. Thus, $\mathcal{F}(c)$ reduces to zero for any choice of $c$. This is expected in this model since also the contribution of the localised sources to $\Lambda$ is independent of $c$ for sources with $p=3$.

We thus find that \eqref{y} yields
\begin{equation}
\Lambda = \frac{1}{4v\mathcal{V}} \left({S^{(3)}_\textrm{DBI} + S^{(3)}_\textrm{CS}}\right).
\end{equation}
Spelling out the contributions from the O$3$-planes and using \eqref{gkp_fields}, we arrive at
\begin{equation}
\Lambda = \frac{1}{4v\mathcal{V}}\, \mu_3\! \int \left(\tilde \star_4\e^{4A} - C_4^\textrm{ext}\right) \w \sigma(\delta_6) = \frac{1}{4\mathcal{V}}\, N_{\textrm{O}3}\, \mu_3\! \left({\e^{4A_0} - \alpha_0}\right),
\end{equation}
where $A_0$ and $\alpha_0$ denote the values of $A$ and $\alpha$ at the position(s) of the O$3$-plane(s) and $\mu_3 > 0$ is the absolute value of the O$3$-plane charge. Since $\alpha=\e^{4A}$, the DBI and Chern-Simons parts of the source action cancel out such that
\begin{equation}
\Lambda = 0
\end{equation}
as expected.

\subsubsection[$\overline{\textrm{D}6}$-branes on $\mathrm{AdS}_7 \times S^3$]{$\overline{\textrm{D}\boldsymbol{6}}$-branes on $\boldsymbol{\mathrm{AdS}_7 \times S^3}$}
\label{ch:scaling-symmetries.examples.d6}

Let us now consider type IIA supergravity with \dsix-branes on AdS$_7 \times S^3$, i.\ e.\ the setup studied in Chapter \ref{ch:simple-non-bps} (see also \cite{Blaback:2011nz,Blaback:2011pn,Bena:2012tx}). While a smeared solution can be constructed explicitly for this setup, we found that a supergravity solution with fully localised branes, if existent at all, necessarily yields a singularity in the energy density of the $H$ flux. As we will see below, it is rather straightforward to reproduce this result in the present framework.

As follows from the discussion in Section \ref{ch:simple-non-bps.localisation.ansatz}, the non-vanishing fields in this setup must satisfy the ansatz
\begin{align}
F_0 = \textrm{const.}, \quad H = \alpha F_0 \e^{\phi-7A} \star_3 1, \quad F_2 = \e^{-3/2 \phi - 7A} \star_3 \d \alpha, \quad C_7^\textrm{ext} = \tilde \star_7 (\alpha + a), \label{ads7_fields}
\end{align}
where $A$, $\phi$ and $\alpha$ are functions on the internal space and $a$ is an integration constant related to a gauge freedom. The tadpole for the \dsix-branes is cancelled by a non-zero $H$ flux on the three-sphere and a non-zero Romans mass, i.\ e.\ $F_0$ ``flux''. The relevant fluxes appearing in $\mathcal{F}(c)$ are therefore
\begin{equation}
H^\textrm{b}, \qquad F^\textrm{b}_{10},
\end{equation}
and \eqref{sigma} reduces to
\begin{equation}
\mathcal{F} (c) =  -c\,H^\textrm{b}\w\left[\e^{-\phi} \star_{10} H - F_0\w C_7^\textrm{ext}\right] - \left({1+\frac{5}{2}c}\right) F_{10}^\textrm{b}\w F_0. \label{ads7_sigma}
\end{equation}
Using \eqref{ads7_fields}, we find that the first term vanishes by a convenient gauge choice, $a=0$. The second term can be set to zero by choosing $c=-\frac{2}{5}$.

We can now substitute this into \eqref{y} to find
\begin{equation}
\Lambda = \frac{1}{4v\mathcal{V}} \left({S^{(6)}_\textrm{DBI} + S^{(6)}_\textrm{CS}}\right).
\end{equation}
Spelling out the contributions of the \dsix-branes and using \eqref{ads7_fields} then yields
\begin{equation}
\Lambda = \frac{1}{4v\mathcal{V}}\, \mu_6\! \int \left(- \tilde\star_7\e^{3/4 \phi+7A} - C_7^\textrm{ext} \right) \w \sigma(\delta_3) =  
- \frac{1}{4\mathcal{V}}\, N_{\overline{\textrm{D}6}}\, \mu_6\! \left({ \e^{3/4 \phi_0+7A_0} + \alpha_0}\right), \label{d6-lambda}
\end{equation}
where $A_0$, $\alpha_0$ and $\phi_0$ denote the values of $A$, $\alpha$ and $\phi$ at the brane position and $\mu_6>0$ is the absolute value of the \dsix-brane charge. Assuming that, at leading order in the distance $\theta$ to the branes, the dilaton and the warp factor diverge as they would in flat space \cite{Janssen:1999sa}, 
\begin{equation}
\e^{2A} \sim \theta^{1/8}, \qquad \e^{\phi} \sim \theta^{3/4}, \label{wddiv}
\end{equation}
it is straightforward to show that the first term in \eqref{d6-lambda} (which comes from the DBI part of the brane action) is actually zero. That this assumption is correct was explicitly proven in the analysis carried out in Section \ref{ch:simple-non-bps.localisation.delta}.

The cosmological constant is therefore exclusively determined by $\alpha_0$,
\begin{equation}
\Lambda \sim - \mu_6 \alpha_0.
\end{equation}
Since $\Lambda$ is negative, it then follows that $\alpha$ has to be non-zero and positive at the source. Together with \eqref{wddiv}, this implies that, near the source, the energy density of the $H$ flux diverges like the inverse of the warp factor,
\begin{equation}
\e^{-\phi} |H|^2 = \alpha^2\e^{-14A} \e^{\phi} F_0^2 \sim \e^{-2A}.
\end{equation}
This is consistent with the result found in Section \ref{ch:simple-non-bps.localisation.delta}, where we also argued that finite $\alpha_0$ implies a singular energy density of the $H$ flux. As we will show in Section \ref{ch:scaling-symmetries.kklt}, a similar argument holds for meta-stable de Sitter vacua that are obtained by placing \dthree-branes at the tip of the Klebanov-Strassler background. Under a few assumptions that we will discuss in detail, one then finds a singularity similar to the one observed in the \dsix-brane model.

\subsubsection[$\mathrm{SU}(3)$-structure Manifolds with O$6$-planes]{$\boldsymbol{\mathrm{SU}(3)}$-structure Manifolds with O$\boldsymbol{6}$-planes}

Here, we discuss a particular model of compactifications of type IIA supergravity on $\mathrm{SU}(3)$-structure manifolds that was studied in \cite{Caviezel:2008tf}, namely O$6$-planes on $\mathrm{dS}_4 \times \mathrm{SU}(2) \times \mathrm{SU}(2)$ (see also \cite{Danielsson:2011au} for more examples of this type). This setup allows (unstable) critical points with positive $\Lambda$.

According to \cite{Caviezel:2008tf}, the form fields satisfy
\begin{equation}
F_0 = m, \qquad F_2 = m^i Y_i^{(2-)}, \qquad H = p \left({Y_1^{(3-)}+Y_2^{(3-)}-Y_3^{(3-)}+Y_4^{(3-)}}\right),
\end{equation}
where $Y_i^{(2-)}$ and $Y_i^{(3-)}$ are certain two-forms and three-forms, respectively, and $m$, $m^i$ and $p$ are constant coefficients that are not relevant for the following discussion. The tadpole generated by the O$6$-planes is cancelled by non-zero $H$ and $F_0$ flux. However, while there is a non-trivial field strength $F_2$ (induced by the presence of the O$6$-planes), there is no topological $F_2$ flux since it is not allowed by the cohomology of $\mathrm{SU}(2) \times \mathrm{SU}(2)$. For the same reason, one finds $F_8^\textrm{b}=0$ such that the non-zero background fluxes appearing in $\mathcal{F}(c)$ are
\begin{equation}
H^\textrm{b}, \qquad F^\textrm{b}_{10}.
\end{equation}
Considering \eqref{sigma} for this setup, we thus find
\begin{equation}
\mathcal{F}(c) = -c\,H^\textrm{b}\w\left[\e^{-\phi} \star_{10} H - F_0\w C_7^\textrm{ext}\right] - \left({1+\frac{5}{2}c}\right) F^\textrm{b}_{10} \w F_0. \label{su3_sigma}
\end{equation}
As discussed in Section \ref{ch:scaling-symmetries.cc.onshellrel}, the $H$ equation of motion
\begin{equation}
\d \left[\e^{-\phi} \star_{10} H - F_0\w C_7^\textrm{ext}\right] = 0
\end{equation}
implies that we can choose a gauge for $C_7^\textrm{ext}$ such that the first term on the right-hand side of \eqref{su3_sigma} vanishes. The second term can be set to zero by choosing $c=-\frac{2}{5}$.

Evaluating \eqref{y}, we therefore find that the cosmological constant is given by
\begin{equation}
\Lambda = \frac{1}{10 v \mathcal{V}} \left({S^{(6)}_\textrm{DBI} + S^{(6)}_\textrm{CS}}\right) = \frac{1}{10 v \mathcal{V}}\, \mu_6\! \int \left({ \e^{3/4 \phi} \, \star_4 1 \w \star_3 1 - C_7^\textrm{ext}}\right) \w \sigma(\delta_3), \label{lambda-su3}
\end{equation}
where the right-hand side should be understood as a sum over the various O$6$-plane terms and $\mu_6 > 0$ is the absolute value of the O$6$-plane charge. In \cite{Caviezel:2008tf}, the setup was considered in the smeared limit, where the delta forms $\delta_3$ are replaced by volume forms of the space transverse to the corresponding sources. In a (hypothetical) localised version of this solution, \eqref{lambda-su3} would relate the value of the cosmological constant to the classical boundary conditions of the supergravity fields at the O-planes.

\subsubsection{The DGKT Solutions}

Finally, we consider type IIA supergravity compactified on $T^6/\mathbb{Z}^2_3$, which is an explicit example for the type IIA flux compactifications considered in \cite{DeWolfe:2005uu, Acharya:2006ne}. In order to stabilise the moduli, the model requires the presence of NSNS flux as well as several RR fluxes of different form degrees. As discussed in Section \ref{ch:scaling-symmetries.cc.onshellrel}, it is therefore a counterexample, where it is in general not possible to set the flux-dependent terms in \eqref{y} to zero and write $\Lambda$ as a sum of localised source terms only.

The NSNS and RR field strengths in this model are given by 
\begin{equation}
H = -p \beta_0,\qquad F_0 = m_0, \qquad F_2=0, \qquad F_4 = F_4^\textrm{int} + F_4^\textrm{ext} = e_i \tilde\omega^i + \star_{4}\, e_0,
\end{equation}
where $p$, $m_0$, $e_0$ and $e_i$ are numbers, $\beta_0$ is an odd three-form and $\tilde\omega^i$ are even four-forms under the orientifold involution.\footnote{Note that the spacetime-filling part of $F_4$, which is given by $F_4^\textrm{ext}$, is treated as internal $F_6$ in the conventions of \cite{DeWolfe:2005uu}.} The non-trivial fluxes appearing in \eqref{sigma} are thus
\begin{equation}
H^\textrm{b}, \qquad F^\textrm{b}_{10}, \qquad F^\textrm{b}_{6}, \qquad F^\textrm{b}_{4}
\end{equation}
such that
\begin{align}
\mathcal{F}(c) &= - c\,H^\textrm{b}\w \left[ \e^{-\phi}\star_{10} H - F_0\w C_7^\textrm{ext}\right] - \left(1 + \frac{5}{2}c\right) F_{10}^\textrm{b}\w F_0 \nl - \left(1+\frac{1}{2}c\right)F_6^\textrm{b}\w F_4^\textrm{int} + \left(1 - \frac{1}{2}c\right) F_4^\textrm{b} \w F_6^\textrm{int}, \label{sigma_dgkt}
\end{align}
where we have used that the fluctuation $B$ is zero on-shell. The first term on the right-hand side can be set to zero by choosing an appropriate gauge for $C_7^\textrm{ext}$. Since the other terms do in general not vanish, however, we cannot choose $c$ such that all of them are set to zero simultaneously.

As pointed out in Section \ref{ch:scaling-symmetries.cc.onshellrel}, we can still solve the equation $\int \mathcal{F}(c) = 0$ for some $c=c_0$ (unless its $c$-dependence coincidentally cancels out on-shell) and use it in \eqref{y} to arrive at an expression for $\Lambda$ that formally only depends on source terms,
\begin{equation}
\Lambda = \frac{2+3c_0}{8v \mathcal{V}} \left({S^{(6)}_\textrm{DBI} + S^{(6)}_\textrm{CS}}\right).
\end{equation}
However, the resulting numerical value for $c_0$ then implicitly depends on the bulk fields appearing in $\mathcal{F}(c)$. It is therefore hard to approximate its numerical value or even its sign in compactification scenarios with more than one type of RR flux, unless the full solution is already known (as in the present example). This is contrary to the previous examples, where $c$ could be fixed to a known number such that, up to a volume factor, $\Lambda$ was completely determined by the boundary conditions of the fields in the near-source region.
\\

\subsection[Singular $\overline{\textrm{D}3}$-branes in the Klebanov-Strassler Background]{Singular $\overline{\textrm{D}\boldsymbol{3}}$-branes in the Klebanov-Strassler Background}
\label{ch:scaling-symmetries.kklt}

In this section, we apply our previous results to meta-stable de Sitter vacua in type IIB string theory that are obtained by placing \dthree-branes at the tip of a warped throat geometry along the lines of \cite{Kachru:2003aw}. Under a few assumptions that we discuss in detail, we give a simple topological argument stating that the backreaction of fully localised \dthree-branes yields a singularity in the energy densities of $H$ and $F_3$.

\subsubsection{Ansatz}

Following \cite{Kachru:2003aw}, we consider type IIB no-scale Minkowski solutions obtained by embedding the Klebanov-Strassler solution \cite{Klebanov:2000hb} into a compact setting \cite{Giddings:2001yu}. In order to stabilise the geometric moduli, we also include non-perturbative effects, which may come from Euclidean D$3$-brane instantons \cite{Witten:1996bn} or gaugino condensation \cite{Kachru:2003aw}. The resulting supersymmetric AdS vacuum is then uplifted to a meta-stable de Sitter vacuum by adding a small number of \dthree-branes \cite{Kachru:2002gs,Kachru:2003aw}.

In order to apply the results of Section \ref{ch:scaling-symmetries.cc} to this scenario, we split the total cosmological constant into a part $\Lambda^\textrm{class}$, which is due to the classical equations of motion and given by evaluating \eqref{y} at the solution, and the rest $\Lambda^\textrm{np}$, which contains all corrections from non-perturbative effects that are not captured by the classical computation,
\begin{equation}
\Lambda = \Lambda^\textrm{class} + \Lambda^\textrm{np}. \label{lambda_split}
\end{equation}
Let us now discuss the explicit form of $\Lambda^\textrm{class}$ in the present setup. For simplicity, we will restrict ourselves to the case where the no-scale solutions of \cite{Giddings:2001yu} are realised in a model with O$3$-planes and the non-perturbative effects come from Euclidean D$3$-brane instantons. In \cite{Giddings:2001yu}, also orientifold limits of F-theory compactifications involving D$7$-branes and O$7$-planes are discussed. We checked that it is also possible to study such models in our framework, but the discussion becomes more involved since the presence of these sources induces a non-trivial $F_1$ field strength.

Our ansatz for the different fields thus reads\footnote{Note that, unlike in the solution discussed in Section \ref{ch:scaling-symmetries.examples.gkp}, $H$ and $F_3$ need in general not satisfy an ISD condition and $\alpha$ is not necessarily related to the warp factor.}
\begin{gather}
C_4^\textrm{ext} = \tilde \star_4 (\alpha + a), \qquad F_5 = - (1+\star_{10}) \e^{-4A} \star_6 \d \alpha, \notag \\  H = \e^{\phi-4A} \star_6 \left({ \alpha F_3 + X_3 }\right), \qquad F_1 = 0, \label{kklt_fields}
\end{gather}
where $A$, $\alpha$ and $\phi$ are functions on the internal space, $a$ is an integration constant corresponding to a gauge freedom and $X_3$ is an a priori unknown three-form satisfying $\d X_3=0$. One can check that this ansatz follows from the form equations of motion and the requirement that the non-compact spacetime be maximally symmetric if only sources with $p=3$ are present.

As in the examples discussed in Section \ref{ch:scaling-symmetries.examples}, the flux-dependent terms $\mathcal{F}(c)$ in \eqref{y} can now be simplified by a convenient choice of the parameter $c$. Since the relevant fluxes in the present case are
\begin{equation}
H^\textrm{b}, \qquad F_7^\textrm{b},
\end{equation}
\eqref{sigma} reduces to
\begin{equation}
\mathcal{F} (c) = - c\, H^\textrm{b}\w\left[\e^{-\phi} \star_{10} H + F_3\w C_4^\textrm{ext}\right] + \left({1+c}\right) F^\textrm{b}_7 \w F_3. \label{sigma-kklt}
\end{equation}
Using \eqref{kklt_fields}, we find that the first expression on the right-hand side of \eqref{sigma-kklt} cancels out for $a=0$ except for a term $\sim\! X_3$. The second term in \eqref{sigma-kklt} can be set to zero by the choice $c=-1$, yielding\footnote{To be precise, one finds that the integrated dilaton equation implies $ - \int H^\textrm{b}\w\left[\e^{-\phi} \star_{10} H + F_3\w C_4^\textrm{ext}\right] + \int F^\textrm{b}_7 \w F_3 = 0$ in absence of sources with $p \neq 3$ such that $\int \mathcal{F} (c) = - \int \tilde \star_4 1 \w H^\textrm{b} \w X_3$ actually holds for any choice of $c$. This is consistent with the fact that also the source part of \eqref{y} is independent of $c$ for $p=3$. Thus, the value of $\Lambda^\textrm{class}$ is uniquely determined by \eqref{y} as it should be.}
\begin{equation}
\mathcal{F} (-1) = - \tilde \star_4 1 \w H^\textrm{b}\w X_3. \label{sigma-kklt2}
\end{equation}
We will argue below that a certain choice for the UV boundary conditions of the three-form field strengths actually implies that $H^\textrm{b} \w X_3 \approx 0$ everywhere on the compact space such that the contribution of the flux term to \eqref{y} vanishes.

Keeping the flux term for the moment, we can substitute \eqref{sigma-kklt2} into \eqref{y} and  write
\begin{align}
\Lambda^\textrm{class} &= \frac{1}{4v\mathcal{V}} \left({S^{(3)}_\textrm{DBI} + S^{(3)}_\textrm{CS}}\right) + \frac{1}{4v\mathcal{V}} \int \mathcal{F} (-1) \nll = \frac{1}{4v\mathcal{V}}\, \mu_3\! \int \left( -\tilde\star_4\e^{4A} - C_4^\textrm{ext}\right) \w \sigma\big(\delta^{(\overline{\textrm{D}3})}_6\big) + \frac{1}{16v\mathcal{V}}\, \mu_3\! \int \left(\tilde\star_4\e^{4A} - C_4^\textrm{ext} \right) \w \sigma\big(\delta^{(\textrm{O}3)}_6\big) \nl - \frac{1}{4v\mathcal{V}} \int \tilde \star_4 1 \w H^\textrm{b} \w X_3,
\end{align}
where we have spelled out the contributions of the localised sources. Note that the O$3$-plane charge is $\frac{1}{4}$ of the \dthree-brane charge $\mu_3$, where $\mu_3 >0$ in our conventions. Evaluating the above equation, we find that the total cosmological constant \eqref{lambda_split} is given by
\begin{equation}
\Lambda = - \frac{1}{4\mathcal{V}}\, N_{\overline{\textrm{D}3}}\, \mu_3\! \left({\e^{4A_0} + \alpha_0}\right) + \frac{1}{16\mathcal{V}}\, N_{\textrm{O}3}\, \mu_3\! \left({ \e^{4A_*} - \alpha_*}\right) - \frac{1}{4\mathcal{V}} \int_{\mathcal{M}_6} H^\textrm{b} \w X_3 + \Lambda^\textrm{np}, \label{lambda_kklt}
\end{equation}
where $A_0$, $\alpha_0$ and $A_*$, $\alpha_*$ denote the values of $A$, $\alpha$ at the positions of the \dthree-branes and O$3$-planes, respectively.

\subsubsection{The Argument}

Our goal is now to evaluate \eqref{lambda_kklt} and relate it to the near-tip behaviour of the energy density of $H$. In order to do so, we make the following assumptions.

\begin{enumerate}
\item {\bf Topological flux.}
In the region of the conifold, $F_3$ carries a non-trivial topological flux along the directions of a three-cycle called the A cycle, $H$ carries a topological flux along the directions of the dual three-cycle called the B cycle, and all other components of $H$ and $F_3$ are exact.  This assumption is due to the fact that the deformed conifold is topologically a cone over $S^2 \times S^3$, where the deformation has the effect of replacing the singular apex of the conifold by a finite $S^3$ (see e.\ g.\ \cite{Candelas:1989js, Minasian:1999tt}). The deformed conifold therefore has a non-trivial compact three-cycle along the $S^3$ (the A cycle) and a dual, non-compact three-cycle (the B cycle). We will assume that, also in our compact setting, the relevant cycles threaded by topological flux are the A cycle and the B cycle, at least in the region of the conifold. Following the literature \cite{Klebanov:2000hb}, we then place $F_3$ flux along the A cycle and $H$ flux along the B cycle. On general compact manifolds, there may exist additional cycles that are threaded by flux. We will assume, however, that such additional topologically non-trivial terms in $F_3$ and $H$ only become relevant deep in the UV, i.\ e.\ far away from the \dthree-branes.

\item {\bf IR boundary conditions.} The \dthree-branes locally deform the geometry as they would do in flat space. This implies in particular that the warp factor goes to zero in the vicinity of the \dthree-branes as it usually does,
\begin{equation}
\e^{2A} \to 0. \label{wftozero}
\end{equation}
It also implies that we can locally approximate the internal geometry by 
\begin{equation}
g_{mn} \approx \e^{-2A} \tilde g_{mn}
\end{equation}
at leading order in an expansion around the distance $r$ to the brane, with $\tilde g_{mn}$  regular (in suitable coordinates). This is a standard assumption, which is, for example, discussed in \cite{Bena:2012bk, Bena:2012vz} for the case of partially smeared \dthree-branes in the non-compact Klebanov-Strassler solution. In an analogous setting, we explicitly verified it in Section \ref{ch:simple-non-bps.localisation.delta} for our toy model with \dsix-branes, where both the warp factor and the internal metric indeed diverge exactly as they would do in the corresponding flat space solution \cite{Janssen:1999sa} at leading order in the distance parameter $\theta$. It would be interesting to carry out a similar derivation also for the \dthree-branes considered here, but this is beyond the scope of this work.

In order that the unperturbed deformed conifold metric $\tilde g_{mn}$ shrinks smoothly at the tip, we furthermore expect that the energy density of $F_3$ along the A cycle contracted with $\tilde g_{mn}$ does not vanish at the tip,
\begin{equation}
\e^\phi |\tilde F^A_3|^2 \neq 0, \label{finitef3}
\end{equation}
where the superscript denotes the component of $F_3$ along the A cycle.\footnote{This is not to be confused with the notation of \cite{Massai:2012jn, Bena:2012vz}, where the superscript in $F_3^A$ is an index running over all components of $F_3$.} This is motivated by the fact that the non-vanishing energy density of $F_3^A$ prevents the A cycle from collapsing at the tip of the deformed conifold before the perturbation by the \dthree-branes \cite{Klebanov:2000hb}. Using the results of \cite{Bena:2012vz}, one can verify that \eqref{finitef3} indeed holds for the case of partially smeared \dthree-branes.

\item {\bf UV boundary conditions.} The boundary conditions for the O$3$-planes in the UV far away from the \dthree-branes are approximately the standard BPS boundary conditions,
\begin{equation}
\alpha_* \approx \e^{4A_*},\label{distortedrelation}
\end{equation}
such that the O$3$-plane term in \eqref{lambda_kklt} is negligible compared to the other terms. Thus, far away from the anti-branes, the function $\alpha$ approaches the BPS behaviour of the unperturbed GKP solution (cf. Section \ref{ch:scaling-symmetries.examples.gkp}). This assumption is reasonable when we consider a large flux background with a large number of O-planes that is perturbed by a small number of anti-branes at the tip of a warped throat. The anti-branes are then strongly redshifted and give a small direct contribution to the cosmological constant due to the classical brane action. Their effect on the field behaviour far away in the UV, on the other hand, is expected to be of higher order such that a possible deviation from \eqref{distortedrelation} due to the anti-brane backreaction should be negligible in \eqref{lambda_kklt}. This is analogous to the usual assumption of BPS asymptotics in the UV imposed in non-compact treatments of anti-brane backreaction (see \mbox{e.\ g.} \cite{Bena:2012tx, Bena:2012bk}). It would be an interesting extension of our work to explicitly compute the boundary conditions at the O-planes, e.\ g.\ following the analysis carried out in Section \ref{ch:simple-non-bps.localisation.delta}.

Similarly, we also assume that the three-form field strengths approach their unperturbed values and thus become ISD in the UV far away from the \dthree-branes, which implies
\begin{equation}
X_3^\textrm{UV} \approx 0. \label{uv-isd}
\end{equation}
One might again wonder whether a small deviation from the ISD condition in the UV due to the anti-brane backreaction could be relevant for the value of the cosmological constant. As discussed above, however, it would be very surprising if the effect of such a deviation far away from the anti-branes would not be negligible compared to their direct effect in the IR, and so we will adopt \eqref{uv-isd} as a reasonable assumption.

\item {\bf Non-perturbative corrections.} Non-perturbative corrections to the effective potential (due to Euclidean D$3$-brane instantons \cite{Witten:1996bn} or gaugino condensation on D$7$-branes \cite{Kachru:2003aw}) are captured by adding a \emph{negative} term to the overall cosmological constant, i.\ e.\
\begin{equation}
\Lambda = \Lambda^\textrm{class} - |\Lambda^\textrm{np}|. 
\end{equation}
This assumption consists of two parts. The first part is that the non-perturbative effect itself yields a negative contribution to the cosmological constant, and the second one is that it does not significantly change the classical contribution. This is implicitly also assumed in \cite{Kachru:2003aw}, where it is argued that the non-perturbative corrections stabilise the K{\"{a}}hler moduli in a supersymmetric AdS vacuum without significantly changing the vevs and the masses of the dilaton and the complex structure moduli, which are already stabilised by fluxes at the classical level. The uplift to de Sitter is furthermore assumed to be only due to the classical action of the \dthree-branes and argued not to modify the shape of the potential for the moduli such that the moduli vevs and masses approximately remain at the values before the uplift.

There has also been some progress in understanding non-perturbative effects explicitly from a ten-dimensional point of view \cite{Koerber:2007xk, Baumann:2010sx, Heidenreich:2010ad, Dymarsky:2010mf}. In \cite{Heidenreich:2010ad}, it was argued that a non-vanishing gaugino bilinear $\langle\overline{\lambda}\lambda\rangle$ on D$7$-branes leads to a negative contribution $\sim\! |\langle\overline{\lambda}\lambda\rangle|^2$ to the cosmological constant. The backreaction of the gaugino condensation on the classical contribution $\Lambda^{\textrm{class}}$, on the other hand, is expected to be a higher order effect and should therefore be negligible. Similar properties are expected for non-perturbative corrections due to Euclidean D$3$-brane instantons.

\item {\bf Cosmological constant.} The presence of the \dthree-branes uplifts the solution to a meta-stable de Sitter vacuum such that the total cosmological constant of the solution is positive,
\begin{equation}
\Lambda > 0,
\end{equation}
as proposed in \cite{Kachru:2003aw}.
\end{enumerate}

Under assumptions 1 to 5, our ansatz \eqref{lambda_kklt} for the cosmological constant drastically simplifies. Let us at first discuss the flux term in \eqref{lambda_kklt}. Since $X_3$ is closed by definition, we can make the ansatz
\begin{equation}
X_3 = \beta \omega_3^A + \d \omega_2 \label{bla}
\end{equation}
in the conifold region, where $\beta$ is an unknown function of the internal coordinates, $\omega_2$ is a two-form and $\omega_3^A$ is the harmonic three-form along the A cycle satisfying $\d \omega_3^A = 0$. We have split $X_3$ into a part along the A cycle, which can in general be non-exact, and a part that is not necessarily along the A cycle and has to be exact.\footnote{Note that, assuming the presence of $F_3$ flux along the A cycle, $X_3$ is not allowed to have a non-exact component along the B cycle, as follows from the $F_1$ equation $\e^{-\phi} H \w \star_{10} F_3 = 0$ and the ansatz for $H$ stated in \eqref{kklt_fields}.} Using $\d X_3 = \d \omega_3^A = 0$, we find from \eqref{bla} that
\begin{equation}
\d \beta \w \omega_3^A = 0,
\end{equation}
which implies that $\beta$ is only a function of the coordinates parametrising the $S^3$ but constant over the remaining directions. We can therefore set $\beta = \beta^\textrm{UV}=0$ without loss of generality, where $\beta^\textrm{UV}$ denotes the value of $\beta$ in the UV region of the warped throat far away from the \dthree-branes.

The flux term in \eqref{lambda_kklt} then simplifies as follows. Since, under assumption 1, $H$ only carries a flux along the B cycle in the conifold region, we find $H^\textrm{b} \w X_3 = H^\textrm{b} \w (\beta \omega_3^A + \d \omega_2) = H^\textrm{b} \w \beta^\textrm{UV} \omega_3^A - \d (H^\textrm{b} \w \omega_2)$.
We can therefore write
\begin{equation}
\int_{\mathcal{M}_6} H^\textrm{b} \w X_3 = \int_{\mathcal{M}_6} H^\textrm{b} \w X^\textrm{UV}_3 =0 \label{bla2}
\end{equation}
such that the integral is completely determined by the units of $H$ flux present in the compactification and the UV boundary conditions for the three-form field strengths but independent of the IR physics close to the \dthree-branes.

Using \eqref{bla2} together with assumptions 2 to 4, we find that \eqref{lambda_kklt} reduces to 
\begin{equation}
\Lambda \approx - \frac{1}{4\mathcal{V}}\, N_{\overline{\textrm{D}3}}\, \mu_3\, \alpha_0 - |\Lambda^\textrm{np}|.
\end{equation}
From assumption 5, it then follows that
\begin{equation}
- \frac{1}{4\mathcal{V}}\, N_{\overline{\textrm{D}3}}\, \mu_3\, \alpha_0 > |\Lambda^\textrm{np}|,
\end{equation}
which implies that $\alpha_0$ must be finite and negative.\footnote{Note that $\alpha$ must change its sign somewhere in between the BPS region around the O$3$-planes (where $\alpha \approx \e^{4A}$) and the tip of the throat (where $\alpha < 0$). In the toy model discussed in Chapter \ref{ch:simple-non-bps}, we used a similar constraint to formulate a topological no-go theorem, which is rederived in the framework of the present chapter in Section \ref{ch:scaling-symmetries.examples.d6}.}

It is straightforward to see that this yields a singular energy density of the $H$ flux in the region near the \dthree-branes. As argued above, we can locally approximate the internal metric as $g_{mn} \approx \e^{-2A} \tilde g_{mn}$, where $\tilde g_{mn}$ is regular. Using \eqref{kklt_fields}, we can then write
\begin{equation}
\e^{-\phi} |H|^2 = \e^{\phi-8A} |\alpha F_3 + X_3|^2 \ge \alpha^2\e^{-8A} \e^{\phi} |F^A_3|^2 \approx \alpha^2\e^{-2A} \e^{\phi} |\tilde F^A_3|^2
\end{equation}
in the near-brane region, where we have used that the component of $X_3$ along $F^A_3$ vanishes. Since $\e^{\phi}|\tilde F_3^A|^2$ is expected to be non-zero at the tip of the conifold, it then follows from \eqref{wftozero} and $\alpha_0 \neq 0$ that the energy density of the $H$ flux at least diverges like the inverse of the warp factor,
\begin{equation}
\e^{-\phi} |H|^2 \sim \e^{-2A}.
\end{equation}
Assuming a regular dilaton\footnote{If the dilaton is singular at the anti-branes even though it does not directly couple to them, $\e^{-\phi} |H|^2$ still diverges, but the dilaton equation does not necessarily imply that also $\e^{\phi} |F_3|^2$ diverges.}, the dilaton equation $\nabla^2 \phi = - \frac{1}{2} \e^{-\phi} |H|^2 + \frac{1}{2} \e^{\phi} |F_3|^2$ furthermore implies that the divergence in the energy density of $H$ must be cancelled by a divergent term in the energy density of $F_3$. We thus find that the energy densities of $H$ and $F_3$ diverge at least like\footnote{Evaluating this equation for the case of partially smeared \dthree-branes, we recover the result of \cite{Bena:2012vz}, where it was shown that $\e^{2A} \sim \tau^{1/2}$ and $\e^{-\phi} |H|^2 \sim \e^\phi |F_3|^2 \sim \tau^{-1/2}$ near the tip of the conifold (with $\tau$ being the radial coordinate transverse to the branes in the conventions of \cite{Bena:2012vz}).}
\begin{equation}
\e^{-\phi} |H|^2 \sim \e^{-2A}, \qquad \e^\phi |F_3|^2 \sim \e^{-2A}.
\end{equation}
Note that, due to its global nature, the argument is independent of most details of the bulk dynamics and does therefore not require simplifications such as a partial smearing of the anti-branes or a linearisation around the BPS background. Under the assumptions discussed above, it holds for fully localised anti-branes that backreact on the full non-linear equations of motion.
\\

\subsection{Discussion}
\label{ch:scaling-symmetries.discussion}

In this chapter, we showed that the classical cosmological constant in type II flux compactifications can be written as a sum of terms due to the action of localised sources and a contribution due to topologically non-trivial background fluxes. We then argued that the flux contribution can be set to zero in many interesting examples such that the cosmological constant is fully determined by the boundary conditions of the supergravity fields in the near-source region. This generalises and makes more explicit previous work in the literature \cite{Aghababaie:2003ar, Burgess:2011rv}. Our formalism should be a useful tool to compute the cosmological constant in compactifications with fully backreacting sources, and it would be interesting to use it in other examples than those discussed in Section \ref{ch:scaling-symmetries.examples}. We also applied our result to the KKLT scenario and found a simple global argument showing that, under a few general assumptions, the backreaction of fully localised \dthree-branes generates a singularity in the energy densities of $H$ and $F_3$. Prior to our work, this singularity had already been noticed for partially smeared \dthree-branes in the Klebanov-Strassler solution. It was found in \cite{McGuirk:2009xx, Bena:2009xk, Bena:2011hz, Bena:2011wh} at the level of linearised perturbations around the BPS background and recently also in \cite{Bena:2012bk} taking into account the full non-linear supergravity equations. The argument discussed in this chapter indicates that the singularity is not an artifact of these approximations but also present for fully localised anti-branes. As will be discussed below, this raises a number of interesting questions that could be investigated in future work.

\cleardoublepage
\thispagestyle{plain}

\section{Conclusions and Outlook}
\label{ch:conclusions}

In this thesis, we addressed a number of problems related to the backreaction of localised sources in string compactifications \cite{Blaback:2010sj, Blaback:2011nz, Blaback:2011pn, Bena:2012tx, Gautason:2013}. As discussed in the introductory chapter, localised sources like D-branes and O-planes are important ingredients in many solutions of string theory that are appealing from a phenomenological point of view. However, the presence of such objects typically leads to complicated dynamics in the compact dimensions such that, even in the supergravity approximation, it is often not possible to find the full solution to the ten-dimensional equations of motion. In many solutions in the literature, the sources are therefore taken to be smeared over the transverse space such that the equations of motion are only solved in an integrated sense, while the backreaction of the sources on the internal fields is neglected. Our aim in this thesis was to investigate to what extent this simplification is justified.

In Chapter \ref{ch:smeared-vs-loc}, we considered a class of flux compactifications of type II supergravity with O-planes and an internal space that, in the smeared limit, is either Ricci-flat or an everywhere negatively curved twisted torus. Our solutions saturate a BPS bound and are related to the well-known GKP solution \cite{Giddings:2001yu} by a chain of T-dualities. We found that all solutions can be localised using a relatively simple localisation prescription, and we argued that the complex structure moduli do not shift in the localised solutions due to the BPS condition. We furthermore gave an intuitive argument for why localisation is expected to be more difficult in non-BPS solutions. We finally discussed the Douglas-Kallosh problem and explicitly showed that it is evaded in our localised solutions due to large warping everywhere on the compact space.

In Chapter \ref{ch:simple-non-bps}, we considered a simple setup on AdS$_7 \times S^3$ with $F_0$ and $H$ flux and D$6$/\dsix-branes that are not mutually BPS with the flux background. We showed that this setup has a smeared solution, which is stable in the closed string sector. Contrary to the BPS case, however, we found that the localisation of our solution is problematic. Considering branes with a regularised source profile, we were able to show that there are no solutions except for the smeared one. For fully localised sources, we computed the allowed boundary conditions in the near-brane region by means of an expansion of the fields around the sources. Using a simple topological argument, we then showed that only one boundary condition is consistent with the equations of motion both locally and globally. This boundary condition yields a singularity in the energy density of $H$. We then focussed on a non-compact version of our model, which yields the same singularity and can be argued to capture the near-tip physics of \dthree-branes in the Klebanov-Strassler background \cite{Massai:2012jn}. In this model, we demonstrated that the singularity is not resolved by brane polarisation due to the Myers effect \cite{Myers:1999ps}. In the compact model, our results were less conclusive, and it remains unclear whether brane polarisation helps to resolve the singularity there.

In Chapter \ref{ch:scaling-symmetries}, we discussed two global scaling symmetries of the classical type II supergravity action and showed that they imply that the classical cosmological constant is given by a sum of terms due to the action of localised sources and a contribution from non-trivial background fluxes. We furthermore argued that, in many interesting examples, the flux term can be set to zero by a convenient gauge choice such that the cosmological constant is fully determined by the classical field behaviour at the source positions. We finally used this result to analyse the backreaction of fully localised \dthree-branes in the KKLT scenario and argued that it yields a singularity in the energy densities of $H$ and $F_3$.

The results discussed in this thesis suggest several interesting directions for future research. First of all, it would certainly be desirable to understand more generally under which conditions smeared solutions can be localised and how smearing affects moduli values. These questions were only partially answered in this thesis. While we showed for the BPS solutions of Chapter \ref{ch:smeared-vs-loc} that they are localisable and that the moduli values do not shift upon localisation, this need not be the case for BPS solutions in general. On the other hand, we found in Chapters \ref{ch:simple-non-bps} and \ref{ch:scaling-symmetries} that anti-branes in certain non-BPS flux backgrounds yield an unexpected singularity, but we were not able to fully settle the question of how this singularity should be interpreted. It is therefore neither clear whether the saturation of a BPS bound is sufficient to justify smearing, nor is it clear whether it is necessary. It would be very interesting to come back to this issue in future work. A first step in this direction could be to explicitly construct classes of localised solutions that are more general than those described in Chapter \ref{ch:smeared-vs-loc}, possibly with intersecting sources. On a related note, it would also be important to better understand the warped effective field theory that describes the low-energy limit of string compactifications with localised sources. This would then allow to study the four-dimensional low-energy physics of string theory without resorting to the potentially unreliable simplification of smearing. It is possible that the formula developed in Chapter \ref{ch:scaling-symmetries}, which in many cases relates the on-shell effective potential directly to the action of localised sources, also turns out helpful for determining the off-shell potential.

Finally, it would be interesting to understand the origin of the singularity that is generated by the backreaction of the \dthree-branes on the Klebanov-Strassler background. Whether this singularity shows that the backreacted solution is unphysical or whether there is a mechanism for its resolution in string theory is currently still debated. There are, however, arguments suggesting that the singularity is neither resolved by brane polarisation \cite{Bena:2012tx, Bena:2012vz} (see also Section \ref{ch:simple-non-bps.myers-effect}) nor stringy effects \cite{Bena:2012ek, Vanriet:2013}. It was recently proposed that the singularity rather indicates the existence of a perturbative decay channel via brane-flux annihilation, triggered by an instability in the closed string modes \cite{Blaback:2011pn, Blaback:2012nf, Bena:2012ek, Vanriet:2013}. If this proposal is correct, \dthree-branes in warped throat geometries do probably not lead to meta-stable de Sitter vacua. In that case, an important question would be whether such issues are specific to \dthree-brane uplifting or point towards a more general problem in string theory. It would be interesting to understand, for example, whether alternative proposals like the K{\"{a}}hler uplifting scenario \cite{Balasubramanian:2004uy, Westphal:2006tn, Rummel:2011cd, Louis:2012nb} also suffer from instabilities or other problems. Given the apparent difficulty of constructing de Sitter solutions in purely geometric compactifications, another interesting direction for future research are non-geometric flux compactifications \cite{Shelton:2005cf}, which have recently gained renewed attention because they might allow for stable de Sitter solutions \cite{Danielsson:2012by, Danielsson:2013}. However, since the supergravity approximation is in general not reliable in such setups, new techniques may be required in order to fully understand their properties.

\cleardoublepage
\thispagestyle{plain}

\section*{Acknowledgements}

I would like to thank my supervisor Marco Zagermann for giving me the chance to work on this exciting topic and for his continuous support over the last three years. I also thank my collaborators Iosif Bena, Johan Bl{\r{a}}b{\"{a}}ck, Ulf Danielsson, Fri\fontencoding{T1}\selectfont \dh \fontencoding{OT1}\selectfont rik Freyr Gautason, Stanislav Kuperstein, Thomas Van Riet and Timm Wrase for many interesting and helpful discussions. Finally, I am grateful to Olaf Lechtenfeld and Ulf Danielsson for kindly accepting to co-referee my thesis and to Herbert Pfn{\"{u}}r for agreeing to be the chairman of the defense.

\cleardoublepage

\appendix

\thispagestyle{plain}

\section{Notation and Conventions}
\label{app:conventions}

Here, we establish our notation and conventions and collect a number of useful formulae for tensors and differential forms. We furthermore present the type II supergravity action and state the equations of motion that are used in the main text.

\subsection{Conventions for Tensors and Differential Forms}
\label{app:conventions.tensors}

Our conventions for tensors and differential forms are adopted from \cite{Danielsson:2009ff}. Throughout this thesis, we will often consider compactifications where the ten-dimensional spacetime is a warped product of a $d$-dimensional external spacetime and a $(10-d)$-dimensional internal space, $\mathcal{M}_{10}=\mathcal{M}_d \times_{\textrm{w}} \mathcal{M}_{10-d}$. We then use Greek indices $\mu, \nu, \ldots$ to label the external directions $0, \ldots, d-1$, Latin indices $m, n, \ldots$ to label the internal directions $d, \ldots, 9$ as well as capital Latin indices $M, N, \ldots$, which run over all ten directions $0, \ldots, 9$. In the present section, we will furthermore use early Latin indices $a, b, \ldots$ for definitions that are valid for all types of indices. Indices in brackets indicate an \mbox{(anti-)}symmetrisation with unit weight, i.\ e.\ $A_{(ab)} = \frac{1}{2}\left({A_{ab}+A_{ba}}\right)$, $A_{[ab]} = \frac{1}{2}\left({A_{ab}-A_{ba}}\right)$ and accordingly for tensors with more than two indices.

We define an $n$-form $A_n$ as
\begin{equation}
A_n = \frac{1}{n!} A_{\left[{a_1 \ldots a_n}\right]} \d x^{a_1} \wedge \ldots \wedge \d x^{a_n},
\end{equation}
where the wedge product of an $n$-form $A_n$ and an $m$-form $B_m$ has the property
\begin{equation}
A_n \wedge B_m = \left({-1}\right)^{nm} B_m \wedge A_n
\end{equation}
and is given by
\begin{equation}
A_n \wedge B_m = \frac{1}{n!m!} A_{\left[{a_1 \ldots a_n}\right.} B_{\left.{b_1 \ldots b_m}\right]} \d x^{a_1} \wedge \ldots \wedge \d x^{a_n} \wedge \d x^{b_1} \wedge \ldots \wedge \d x^{b_m}.
\end{equation}
The exterior derivative is defined as
\begin{equation}
\d A_n = \frac{1}{n!} \partial_{\left[{a}\right.} A_{\left.{b_1 \ldots b_n}\right]} \d x^{a} \wedge \d x^{b_1} \wedge \ldots \wedge \d x^{b_n}
\end{equation}
and obeys the Leibniz rule
\begin{equation}
\d \left({A_n \wedge B_m}\right) = \d A_n \wedge B_m + \left({-1}\right)^n A_n \wedge \d B_m.
\end{equation}

The $d$-dimensional Levi-Civita symbol $\varepsilon_{a_1 \ldots a_d}$ is totally anti-symmetric and normalised such that $\varepsilon_{01 \ldots d-1}=1$ and $\varepsilon_{a_1 \ldots a_d} \varepsilon_{a_{d+1} \ldots a_{10}} = \varepsilon_{a_1 \ldots a_{10}}$. One can then construct a $d$-dimensional Levi-Civita tensor
\begin{equation}
\epsilon_{a_1 \ldots a_d} = \sqrt{|g|} \varepsilon_{a_1 \ldots a_d},
\end{equation}
where $g$ is the determinant of the $d$-dimensional metric $g_{ab}$ and the indices of $\epsilon_{a_1 \ldots a_d}$ can be raised and lowered with (inverse) metrics. Contractions of the Levi-Civita tensor yield
\begin{equation}
\epsilon_{a_1 \ldots a_na_{n+1} \ldots a_d}\epsilon^{a_1 \ldots a_nb_{n+1} \ldots b_d} = \left({-1}\right)^t n! \left({d-n}\right)! \, \delta^{[b_{n+1}}_{[a_{n+1}} \ldots \delta^{b_d]}_{a_d]},
\end{equation}
where $t$ is the number of timelike dimensions of the $d$-dimensional space.

The $d$-dimensional Hodge operator $\star_d$ maps $n$-forms to $\left({d-n}\right)$-forms via
\begin{equation}
\star_d A_n = \frac{1}{n!\left({d-n}\right)!} \epsilon_{a_1 \ldots a_{d-n}}\vphantom{}^{b_1 \ldots b_n}A_{b_1 \ldots b_n} \d x^{a_1} \wedge \ldots \wedge \d x^{a_{d-n}} \label{eq:hodge_definition}
\end{equation}
and has the useful property
\begin{equation}
\star_d \star_d A_n = \left({-1}\right)^{n\left({d-n}\right)+t} A_n. \label{eq:doubled_hodge}
\end{equation}
On a warped product space $\mathcal{M}_{10}=\mathcal{M}_d \times_{\textrm{w}} \mathcal{M}_{10-d}$, one also has the identity
\begin{equation}
\star_{10} \left({ A_n \wedge B_m }\right) = \left({-1}\right)^{n\left({10-d-m}\right)} \star_d A_n \wedge \star_{10-d} B_m, \label{eq:split_hodge}
\end{equation}
where $\star_d$ and $\star_{10-d}$ denote the Hodge operators associated with the corresponding metric factors, $A_n$ is an $n$-form on the $d$-dimensional external spacetime and $B_m$ is an $m$-form on the $(10-d)$-dimensional internal space. 

We furthermore define
\begin{equation}
\left|{A_n \cdot B_n}\right| = \frac{1}{n!} A_{a_1 \ldots a_n} B^{a_1 \ldots a_n}, \qquad \left|{A_n \cdot B_n}\right|_{ab} = \frac{1}{\left({n-1}\right)!} A_{aa_2 \ldots a_n} B_b^{a_2 \ldots a_n}, \label{eq:tensor_contr}
\end{equation}
which yields the useful relations
\begin{align}
& \star_d\! A_n \wedge B_n = \star_d B_n \wedge A_n = |A_n  \cdot B_n| \star_d 1, \label{eq:forms_vs_tensors1} \\
& |\star_d\! A_n \cdot \star_d B_n| = \left({-1}\right)^t |A_n \cdot B_n|, \label{eq:forms_vs_tensors2} \\
& |\star_d\! A_n \cdot \star_d B_n|_{ab} = g_{ab} \left({-1}\right)^t |A_n \cdot B_n| - \left({-1}\right)^t |A_n \cdot B_n|_{ab}, \label{eq:forms_vs_tensors3}
\end{align}
where $g_{ab}$ is the $d$-dimensional metric. One can also verify that
\begin{equation}
\d \star_d \d \alpha = - (-1)^d \,\nabla^2 \alpha \star_d 1
\end{equation}
for an arbitary function $\alpha$.
\\

\subsection{Type II Supergravity}
\label{app:conventions.sugra}

Throughout this thesis, we use the democratic formulation of type II supergravity \cite{Bergshoeff:2001pv}. Our conventions are those of \cite{Koerber:2010bx}, except that we work in (ten-dimensional) Einstein frame and flip the signs of the $B$ field and the worldvolume gauge field strength $F$.

\subsubsection{Action in the Democratic Formulation}

In the classical supergravity approximation, the bosonic part of the low-energy effective action of type II string theory in Einstein frame can be written as
\begin{equation}
S = S_\textrm{bulk} + S_\textrm{loc} \label{app:conv.action}
\end{equation}
in terms of a bulk action and an action of localised sources, where the bulk action is given by
\begin{equation}
S_\textrm{bulk} = S_\textrm{NSNS} + S_\textrm{RR} = \int\star_{10}\bigg[ R -\frac{1}{2} \left({\partial \phi}\right)^2 - \frac{1}{2}\e^{-\phi}|H|^2 - \frac{1}{4}
\sum_n \e^{\tfrac{5-n}{2}\phi}|F_{n}|^2\bigg] \label{app:conv.bulkaction}
\end{equation}
and we have set $2\kappa_{10}^2=1$. Here, $R$ is the curvature scalar of the metric $g_{MN}$, $\star_{10}$ denotes the ten-dimensional Hodge operator associated with $g_{MN}$, $\phi$ is the dilaton, $H$ is the NSNS three-form field strength and $F_n$ are the RR field strengths. In the democratic formulation, the RR fields are doubled such that the sum in \eqref{app:conv.bulkaction} also includes the dual fields with $n>5$. The index $n$ then runs over $0,2,4,6,8,10$ in type IIA string theory and over $1,3,5,7,9$ in type IIB string theory. The RR field strengths are related to one another by the duality relations
\begin{equation}
\e^{\tfrac{5-n}{2}\phi}F_n  = \star_{10}\, \sigma(F_{10-n}), \label{app:conv.rrduality}
\end{equation}
which have to be imposed at the level of the equations of motion. The operator $\sigma$ here acts on an $n$-form $\omega_n$ like
\begin{equation}
\sigma(\omega_n) = (-1)^{\tfrac{n(n-1)}{2}} \omega_n. \label{app:conv.reversal}
\end{equation}
Away from the localised sources, the NSNS field strength $H$ and the RR field strengths $F_n$ with $n \ge 1$ are defined in terms of the Kalb-Ramond field $B$ and the RR gauge potentials $C_{n-1}$ via
\begin{equation}
H = \d B, \qquad F_n = \d C_{n-1} - H \w C_{n-3} + F_0 \e^B. \label{app:def-fieldstrengths}
\end{equation}
$F_0$ is the Romans mass in type IIA string theory and does not have any propagating degrees of freedom. Note that, in the presence of topologically non-trivial background fluxes, the gauge potentials $B$ and $C_{n-1}$ are only defined on local gauge patches. It is therefore sometimes convenient to choose a different definition of the field strengths, where their non-exact part is explicitly separated off and only the exact part is defined in terms of a gauge potential (cf. Section \ref{ch:scaling-symmetries.cc.method}).

$S_\textrm{loc}$ denotes the action of localised sources, which can, for example, include D-branes, O-planes, NS$5$-branes, fundamental strings or KK monopoles in type II string theory. In this thesis, we only consider compactifications with D-branes and O-planes. Their action consists of a Dirac-Born-Infeld (DBI) part and a Chern-Simons (CS) part,
\begin{equation}
S_\textrm{loc} = \sum_p S_\textrm{loc}^{(p)} = \sum_p\left(S_\textrm{DBI}^{(p)} + S_\textrm{CS}^{(p)}\right), \label{app:conv.locaction}
\end{equation}
and reads
\begin{equation}
S_\textrm{loc}^{(p)} = - \mu_p \int_\Sigma \d^{p+1} \xi\, \e^{\tfrac{p-3}{4}\phi} \sqrt{-\det (g_{\alpha\beta} - \e^{-\phi/2} \mathcal{F}_{\alpha\beta})}\, + \mu_p \int_\Sigma \left\langle C \w \e^{-\mathcal{F}} \right\rangle_{p+1} \label{app:conv.d-brane}
\end{equation}
for a D$p$-brane and
\begin{equation}
S_\textrm{loc}^{(p)} = \mu_p \int_\Sigma \d^{p+1} \xi\, \e^{\tfrac{p-3}{4}\phi} \sqrt{-\det (g_{\alpha\beta})}\, - \mu_p \int_\Sigma C_{p+1} \label{app:conv.o-plane}
\end{equation}
for an O$p$-plane. Here, $\mu_p > 0$ is the absolute value of the D$p$-brane/O$p$-plane charge, and the O$p$-plane charge equals $-2^{p-5}$ times the charge of a D$p$-brane. For $\overline{\textrm{D}p}$-branes and $\overline{\textrm{O}p}$-planes, the CS terms in \eqref{app:conv.d-brane} and \eqref{app:conv.o-plane} would have the opposite sign. We have denoted the pullback of $g_{MN}$ to the worldvolume by $g_{\alpha\beta}$ and written $\mathcal{F}_{\alpha\beta} = B_{\alpha\beta} + 2\pi\alpha^\prime F_{\alpha\beta}$, where $B_{\alpha\beta}$ is the pullback of $B_{MN}$ to the worldvolume and $F_{\alpha\beta}$ is the worldvolume gauge field strength.\footnote{In the DBI part of the D-brane action, we suppressed the kinetic terms for the worldvolume scalars that determine the position of the branes in transverse space. There is also a subtlety related to the CS action in massive type IIA supergravity \cite{Green:1996bh}, which we disregard here since it does not affect the equations of motion that are stated below \cite{Koerber:2010bx}.} In \eqref{app:conv.d-brane}, we have also used the polyform notation, i.\ e.\ $C= \sum_n C_{n-1}$ denotes the sum over all electric and magnetic RR potentials that appear in type IIA or type IIB supergravity, and $\e^{-\mathcal{F}}$ is defined as a power series of wedge products. The symbol $\langle\cdots \rangle_{p+1}$ denotes a projection to the form degree $p+1$,
\begin{equation}
\left\langle C \w \e^{-\mathcal{F}} \right\rangle_{p+1} = C_{p+1} - C_{p-1} \w \mathcal{F} + \frac{1}{2} C_{p-3} \w \mathcal{F} \w \mathcal{F} - \ldots
\end{equation}

To derive the equations of motion, it is convenient to express $S_\textrm{loc}$ in terms of a ten-dimensional integral involving delta distributions. Omitting the contribution of $\mathcal{F}$ in the DBI action for simplicity, we can write\footnote{For the most part of this thesis, we only consider configurations with vanishing worldvolume gauge field strength and no $B$ field along the brane worldvolume such that the general form of the DBI action is not required.}
\begin{align}
S_\textrm{DBI}^{(p)} = \mp \mu_p \int \star_{p+1} \e^{\tfrac{p-3}{4}\phi} \w \sigma(\delta_{9-p}), \quad S_\textrm{CS}^{(p)} = \Bigg\{ \begin{split} & +\mu_p \int \langle C\wedge \e^{-\mathcal{F}} \rangle_{p+1} \w \sigma(\delta_{9-p}) \\ & - \mu_p \int C_{p+1} \w \sigma(\delta_{9-p}) \end{split}, \label{app:conv.locaction2}
\end{align}
where the upper line is for D$p$-branes and the lower line for O$p$-planes. Here, $\star_{p+1} 1$ is the volume form on the $(p+1)$-dimensional worldvolume $\Sigma$ of the corresponding source, and $\star_{9-p} 1$ is the $(9-p)$-dimensional volume form transverse to the source, which we normalise such that $\star_{10} 1 = \star_{p+1} 1 \w \star_{9-p} 1$. We also define $\delta_{9-p}=\delta(\Sigma) \,\sigma(\star_{9-p} 1)$, where $\delta(\Sigma)$ is the delta distribution with support on $\Sigma$.

\subsubsection{Equations of Motion}

Let us now state the equations of motion that follow from the above action. The ten-dimensional Einstein equation reads
\begin{align}
R_{MN} - \frac{1}{2} g_{MN} R &= \frac{1}{2} \left({\partial_M \phi}\right) \left({\partial_N \phi}\right) - \frac{1}{4} g_{MN} \left({\partial \phi}\right)^2 + \frac{1}{2}\, \e^{-\phi} \left({ |H|^2_{MN} - \frac{1}{2} g_{MN} |H|^2}\right) \nl + \frac{1}{4} \sum\limits_{n} \e^{\tfrac{5-n}{2}\phi} \left({ |F_n|^2_{MN} - \frac{1}{2}\, g_{MN} |F_n|^2}\right) + \frac{1}{2} T^\textrm{loc}_{MN}.
\end{align}
It is often convenient to consider the trace-reversed equation,
\begin{align}
R_{MN} &= \frac{1}{2} \left({\partial_M \phi}\right) \left({\partial_N \phi}\right) + \frac{1}{2}\, \e^{-\phi} \left({ |H|^2_{MN} - \frac{1}{4} g_{MN} |H|^2}\right) \nl + \frac{1}{4} \sum\limits_{n} \e^{\tfrac{5-n}{2}\phi} \left({ |F_n|^2_{MN} - \frac{n-1}{8}\, g_{MN} |F_n|^2}\right) + \frac{1}{2} \left({T^\textrm{loc}_{MN}-\frac{1}{8}g_{MN}T^\textrm{loc}}\right).
\end{align}
For localised sources that extend along the external spacetime and for which $\mathcal{F}$ vanishes in the DBI action, the energy-momentum tensor takes the simple form
\begin{equation}
T^\textrm{loc}_{\mu\nu} = \mp \mu_p \,\e^{\tfrac{p-3}{4}\phi} g_{\mu\nu} \delta(\Sigma), \qquad T^\textrm{loc}_{mn} = \mp \mu_p \,\e^{\tfrac{p-3}{4}\phi} \Pi^\Sigma_{mn} \delta(\Sigma),
\end{equation}
where the upper sign is for D$p$-branes and the lower sign for O$p$-planes. $\Pi^\Sigma_{mn}$ denotes the projector to the cycle $\Sigma$ wrapped by the corresponding source and is given by $\Pi^\Sigma_{mn} = g_{mn}$ in the parallel directions and $\Pi^\Sigma_{mn}=0$ in the transverse directions.

The equation of motion for the dilaton reads
\begin{equation}
\nabla^2 \phi = - \frac{1}{2} \e^{-\phi} |H|^2 + \sum_n \frac{5-n}{8} \e^{\tfrac{5-n}{2}\phi} |F_n|^2 - \frac{1}{\sqrt{-g}} \frac{\delta S_\textrm{DBI}}{\delta \phi}. \label{app:conv.dilaton}
\end{equation}
Assuming that $\mathcal{F}$ does not contribute in the DBI action, the source term reduces to
\begin{equation}
\frac{1}{\sqrt{-g}} \frac{\delta S_\textrm{DBI}}{\delta \phi} = \mp \sum_p \frac{p-3}{4} \mu_p \e^{\tfrac{p-3}{4}\phi} \delta(\Sigma).
\end{equation}
The equations of motion for the RR form fields are
\begin{equation}
\d \left({ \e^{\tfrac{5-n}{2}\phi} \star_{10} F_n }\right) + \e^{\tfrac{3-n}{2}\phi} H \w \star_{10} F_{n+2} - \sigma(j_{11-n}) = 0,
\end{equation}
and their Bianchi identities read
\begin{equation}
\d F_n - H \w F_{n-2} + j_{n+1} = 0, \label{app:conv.bianchix}
\end{equation}
where $j_{9-p} = \mu_p \langle\delta \w \e^{\mathcal{F}}\rangle_{9-p}$ for D$p$-branes and $j_{9-p} = - \mu_p \delta_{9-p}$ for O$p$-planes. The equation of motion and the Bianchi identity for $H$ are
\begin{equation}
\d \left({\e^{-\phi} \star_{10} H}\right) + \frac{1}{2} \sum\limits_n \e^{\tfrac{5-n}{2}\phi} \star_{10} F_n \w F_{n-2} - \frac{\delta S_\textrm{DBI}}{\delta B} = 0, \qquad \d H = 0. \label{app:conv.h-eoms}
\end{equation}
The source term in the $H$ equation is discussed in more detail in \cite{Koerber:2007hd}. It is not relevant in this thesis since it vanishes for all cases we consider.
\\

\cleardoublepage
\thispagestyle{plain}

\section{Curvature in Warped Compactifications}
\label{app:computations.curvature}

In this appendix, we compute curvature tensors for different metrics with non-trivial warping.

\subsection{Block Diagonal Metric}
\label{app:computations.curvature.block}

Let us consider a warped metric of the form
\begin{equation}
\d s^2 = \e^{2A} \d \tilde s_{p+1}^2 + \e^{2B} \d \tilde s_{9-p}^2 = \e^{2A} \tilde g_{\mu\nu} \d x^\mu \d x^\nu + \e^{2B} \tilde g_{ik} \d x^i \d x^k, \label{eq:app-warpedmetric}
\end{equation}
where $A$ and $B$ are functions of the internal coordinates $x^i$. Following standard textbooks such as \cite{Wald:1984rg}, we can compute the curvature of this metric by introducing a derivative operator $\tilde \nabla_M$ associated with the unwarped metric $\tilde g_{MN}$. We can then define the connection
\begin{equation}
\Gamma^M_{NP} = \frac{1}{2} g^{MR} \left({ \tilde \nabla_P g_{RN} + \tilde \nabla_N g_{RP} - \tilde \nabla_R g_{NP}}\right)
\end{equation}
and evaluate the right-hand side for the metric \eqref{eq:app-warpedmetric}. The non-trivial components are
\begin{gather}
\Gamma^\mu_{\nu l} = \frac{1}{2} g^{\mu \rho} \tilde \nabla_l g_{\rho \nu} = \delta^\mu_\nu \partial_l A, \qquad \Gamma^i_{\mu \nu} = - \frac{1}{2} g^{im} \tilde \nabla_m g_{\mu \nu} = - \e^{2A-2B} \tilde g_{\mu \nu} \tilde \partial^i A, \notag \\
\Gamma^i_{kl} = \frac{1}{2} g^{im} \left({ \tilde \nabla_l g_{mk} + \tilde \nabla_k g_{ml} - \tilde \nabla_m g_{kl}}\right) = \delta^i_k \partial_l B + \delta^i_l \partial_k B - \tilde g_{kl} \tilde \partial^i B,
\end{gather}
where all other components are either zero or can be obtained using the symmetry $\Gamma_{NP}^M=\Gamma_{PN}^M$.

The Riemann tensor is given by
\begin{equation}
R_{MRN} \vphantom{}^P = - \tilde \nabla_M \Gamma^P_{RN} + \tilde \nabla_R \Gamma^P_{MN} + \Gamma^S_{NM} \Gamma^P_{RS} - \Gamma^S_{NR} \Gamma^P_{MS} + \tilde R_{MRN} \vphantom{}^P,
\end{equation}
and its non-trivial components are
\begin{align}
R_{ijk} \vphantom{}^l &= 2 \delta^l_{[i} \tilde \nabla_{j]} \partial_k B - 2 \tilde g_{k[i} \tilde \nabla_{j]} \tilde \partial^l B  - 2 \delta^l_{[i} (\partial_{j]} B) (\partial_k B) - 2 \tilde g_{k[i} \delta^l_{j]} (\partial_m B) (\tilde \partial^m B) \nl + 2 \tilde g_{k[i} (\partial_{j]} B) (\partial^l B) + \tilde R_{ijk} \vphantom{}^l, \\
R_{\mu\nu\lambda} \vphantom{}^\rho &= - 2\e^{2A-2B} \tilde g_{\lambda[\mu} \delta_{\nu]}^\rho (\partial_m A) (\tilde \partial^m A) + \tilde R_{\mu\nu\lambda} \vphantom{}^\rho, \\
R_{\mu j \lambda} \vphantom{}^l &= - \e^{2A-2B} \tilde g_{\mu\lambda} \tilde \nabla_j \tilde \partial^l A - \e^{2A-2B} \tilde g_{\mu\lambda} (\partial_j A) (\tilde \partial^l A) + \e^{2A-2B} \tilde g_{\mu\lambda} (\partial_j B) (\tilde \partial^l A) \nl + \e^{2A-2B} \tilde g_{\mu\lambda} (\partial_j A) (\tilde \partial^l B) - \e^{2A-2B} \delta_j^l \tilde g_{\mu\lambda} (\partial_m A) (\tilde \partial^m B), \\
R_{i \nu k} \vphantom{}^\rho &= - \delta^\rho_\nu \tilde \nabla_i \partial_k A + 2 \delta^\rho_\nu (\partial_{(i} A) (\partial_{k)} B) - \delta^\rho_\nu (\partial_i A) (\partial_k A) - \delta^\rho_\nu \tilde g_{ik} (\partial_m A) (\tilde \partial^m B),
\end{align}
where all other components are zero or implied by the antisymmetry $R_{MNP}\vphantom{}^R = - R_{NMP}\vphantom{}^R$. The components of the Ricci tensor $R_{MN}= R_{MPN} \vphantom{}^P$ are then
\begin{align}
R_{\mu\nu} &= \e^{2A-2B} \tilde g_{\mu\nu} \left[{- \left({p+1}\right) (\partial_m A)(\tilde \partial^m A) - \tilde \nabla^2 A + \left({p-7}\right) (\partial_m A) (\tilde \partial^m B)}\right] \nl + \tilde R_{\mu\nu}, \\
R_{ik} &= \left({p-7}\right)\tilde \nabla_i \partial_k B - \tilde g_{ik} \tilde \nabla^2 B + \left({7-p}\right) (\partial_i B) (\partial_k B) + \left({p-7}\right)\tilde g_{ik} (\partial_m B) (\tilde \partial^m B) \nl - \left({p+1}\right) \tilde \nabla_i \partial_k A - \left({p+1}\right) (\partial_i A) (\partial_k A) + 2 \left({p+1}\right) (\partial_{(i} A) (\partial_{k)} B) \nl - \left({p+1}\right) \tilde g_{ik} (\partial_m A) (\tilde \partial^m B) + \tilde R_{ik}.
\end{align}
For the special case $B = \frac{p+1}{p-7}A$ used in Section \ref{ch:smeared-vs-loc.bps-ricciflat}, this reduces to
\begin{align}
R_{\mu\nu} &= - \e^{\tfrac{16}{7-p}A} \tilde g_{\mu\nu} \tilde \nabla^2 A + \tilde R_{\mu\nu}, \\
R_{ik} &= - \frac{p+1}{p-7} \tilde g_{ik} \tilde \nabla^2 A + 8 \frac{p+1}{p-7} (\partial_i A) (\partial_k A) + \tilde R_{ik}.
\end{align}
\\[-0.7cm]

\subsection{Twisted Tori}
\label{app:computations.curvature.twisted}

Here, we consider a warped metric of the form
\begin{equation}
\d s^2 = \e^{2A} \tilde g_{\mu\nu} \d x^\mu \d x^\nu + \e^{2B} \tilde g_{ik} \d x^i \d x^k + \e^{2A} \tilde g_{99} e^9 e^9 \label{eq:metric_e_basis}
\end{equation}
with basis forms
\begin{equation}
\d x^\mu, \qquad \d x^i, \qquad e^9 = \d x^9 + \frac{1}{2} f^9_{ik} x^i \d x^k, \label{eq:e_basis}
\end{equation}
where $\mu = 0,\ldots,p-1$ and $i =p,\ldots,8$. Furthermore, we take $A$ and $B$ to be functions of the internal coordinates $x^i$. Since $e^9$ is not closed,
\begin{equation}
\d e^9 = \frac{1}{2} f^9_{ik} \d x^i \wedge \d x^k, \label{eq:de9}
\end{equation}
the structure constants $f^9_{ik}$ contribute to the curvature, which can be taken into account by switching to a coordinate basis spanned by the one-forms $\d x^M$. Using \eqref{eq:e_basis} in \eqref{eq:metric_e_basis}, we find
\begin{equation}
\d s^2 = g_{\mu\nu} \d x^\mu \d x^\nu + g_{ik} \d x^i \d x^k + 2 g_{9k} \d x^9 \d x^k + g_{99} \d x^9 \d x^9 \label{eq:metric-dx_basis}
\end{equation}
with the metric components
\begin{gather}
g_{\mu\nu} = \e^{2A} \tilde g_{\mu\nu}, \qquad g_{ik} = \e^{2B} \tilde g_{ik} + \frac{1}{4} \e^{2A} \tilde g_{99} f^9_{mi} x^m f^9_{nk} x^n, \notag \\ g_{9k} = \frac{1}{2} \e^{2A} \tilde g_{99} f^9_{mk} x^m, \qquad g_{99} = \e^{2A} \tilde g_{99}. \label{eq:inv_metrics_dx_basis}
\end{gather}
For convenience, we also state the inverse components,
\begin{gather}
g^{\mu\nu} = \e^{-2A} \tilde g^{\mu\nu}, \qquad g^{ik} = \e^{-2B} \tilde g^{ik}, \qquad g^{9k} = -\frac{1}{2} \e^{-2B} \tilde g^{ik} f^9_{mi} x^m, \notag \\ g^{99} = \e^{-2A} \tilde g^{99} + \frac{1}{4} \e^{-2B} \tilde g^{ik} f^9_{mi} x^m f^9_{nk} x^n. \label{eq:inv_metrics_dx_basis2}
\end{gather}
Note that the metrics with tildes are the unwarped metrics in the ``vielbein'' basis \eqref{eq:metric_e_basis}, whereas metrics without tildes are with respect to the coordinate basis \eqref{eq:metric-dx_basis}.

We now introduce a derivative operator $\tilde \nabla_M$ associated with the unwarped metric $\tilde g_{MN}$ in \eqref{eq:metric_e_basis} and define the connection
\begin{equation}
\Gamma^M_{NP} = \frac{1}{2} g^{MR} \left({ \tilde \nabla_P g_{RN} + \tilde \nabla_N g_{RP} - \tilde \nabla_R g_{NP}}\right). \label{eq:app-connection-twisted}
\end{equation}
Substituting \eqref{eq:inv_metrics_dx_basis} and \eqref{eq:inv_metrics_dx_basis2}, we find
\begin{align}
\Gamma^i_{kl} &= 2 \delta^i_{(k} \partial_{l)} B - \tilde g_{kl} \tilde \partial^i B - \frac{1}{4} \e^{2A-2B} \tilde g_{99} f^9_{mk} x^m f^9_{jl} x^j \tilde \partial^i A + \frac{1}{2} \e^{2A-2B} \tilde g_{99} \tilde g^{im} f^9_{m(l} f^9_{k)j} x^j \nl + \hat \Gamma^i_{kl}, \\
\Gamma^9_{kl} &= f^9_{i(k} x^i \partial_{l)} A - f^9_{i(k}x^i \partial_{l)} B + \frac{1}{2} \tilde g_{kl} f^9_{im}x^i \tilde \partial^m B + \frac{1}{8} \e^{2A-2B} \tilde g_{99} f^9_{ik}x^i f^9_{jl}x^j f^9_{mn}x^m \tilde \partial^n A \nl - \frac{1}{4} \e^{2A-2B} \tilde g_{99} \tilde g^{mn} f^9_{im} x^i f^9_{n(l} f^9_{k)j} x^j + \hat \Gamma^9_{kl}, \\
\Gamma^i_{9l} &= -\frac{1}{2} \e^{2A-2B} \tilde g_{99} f^9_{jl}x^j \tilde \partial^i A + \frac{1}{2} \e^{2A-2B} \tilde g_{99} \tilde g^{ij}f^9_{lj} + \hat \Gamma^i_{9l}, \\
\Gamma^i_{99} &= - \e^{2A-2B} \tilde g_{99} \tilde \partial^i A, \\
\Gamma^9_{9l} &= \partial_l A + \frac{1}{4} \e^{2A-2B} \tilde g_{99} f^9_{il} x^i f^9_{jm}x^j \tilde \partial^m A - \frac{1}{4} \e^{2A-2B} \tilde g_{99} \tilde g^{mn} f^9_{im}x^i f^9_{ln} + \hat \Gamma^9_{9l}, \\
\Gamma^9_{99} &= \frac{1}{2} \e^{2A-2B} \tilde g_{99} f^9_{im}x^i \tilde \partial^m A, \\
\Gamma^i_{\mu\nu} &= - \e^{2A-2B} \tilde g_{\mu\nu} \tilde \partial^i A, \\
\Gamma^\lambda_{\mu l} &= \delta^\lambda_\mu \partial_l A, \\
\Gamma^9_{\mu\nu} &= \frac{1}{2} \e^{2A-2B} \tilde g_{\mu\nu} f^9_{im}x^i \tilde \partial^m A,
\end{align}
where all other components are zero or implied by the symmetry $\Gamma^M_{NP}=\Gamma^M_{PN}$. Note that some of the components of the connection can be non-zero for $A=B=0$ since \eqref{eq:app-connection-twisted} is defined with respect to the unwarped metric in \eqref{eq:metric_e_basis}, which is not equal to the unwarped metric in \eqref{eq:metric-dx_basis}. The terms $\hat \Gamma^M_{NP}$ in above expressions contain warp factors, structure constants and factors $\sim\! \tilde \Gamma_{kl}^i x^l$ and are non-zero if $\tilde g_{MN}$ is not flat. We have omitted to spell them out explicitly since they do not contribute to the Ricci tensor in the vielbein basis and are therefore not relevant for the following discussion.

We now compute the Ricci tensor
\begin{equation}
R_{MN} = R_{MPN} \vphantom{}^P = - \tilde \nabla_M \Gamma^P_{PN} + \tilde \nabla_P \Gamma^P_{MN} + \Gamma^R_{NM} \Gamma^P_{PR} - \Gamma^R_{NP} \Gamma^P_{MR} + \tilde R_{MN},
\end{equation}
which yields
\begin{align}
R_{\mu\nu} &= \e^{2A-2B} \tilde g_{\mu\nu} \left[{- \tilde \nabla^2 A - \left({p+1}\right) (\partial_m A)(\tilde \partial^m A) + \left({p-7}\right) (\partial_m A) (\tilde \partial^m B)}\right] + \tilde R_{\mu\nu}, \\
R_{ik} &= - \left({p+1}\right) \tilde \nabla_i \partial_k A + \left({p-7}\right)\tilde \nabla_i \partial_k B - \tilde g_{ik} \tilde \nabla^2 B - \left({p+1}\right) (\partial_i A) (\partial_k A) \nl + \left({7-p}\right) (\partial_i B) (\partial_k B) + \left({p-7}\right) \tilde g_{ik} (\partial_m B)(\tilde \partial^m B) + 2 \left({p+1}\right) (\partial_{(i} A) (\partial_{k)} B) \nl - \left({p+1}\right) \tilde g_{ik} (\partial_m A) (\tilde \partial^m B)
+ \frac{1}{2} \e^{2A-2B} \tilde g_{99} \tilde g^{mn} f^9_{m(k}f^9_{i)n} - \frac{1}{4} \e^{2A-2B} \tilde g_{99} f^9_{mi}x^m \nl \cdot f^9_{nk}x^n \tilde \nabla^2 A - \frac{p+1}{4} \e^{2A-2B} \tilde g_{99} f^9_{mi}x^m f^9_{nk}x^n (\partial_l A) (\tilde \partial^l A) +\frac{p-7}{4} \e^{2A-2B} \tilde g_{99} \nl \cdot f^9_{mi}x^m f^9_{nk}x^n (\partial_l A) (\tilde \partial^l B) + \frac{p+3}{2} \e^{2A-2B} \tilde g_{99} f^9_{m(i}x^m f^9_{k)n} \tilde \partial^n A - \frac{p-5}{2} \e^{2A-2B} \nl \cdot \tilde g_{99} f^9_{m(i}x^m f^9_{k)n} \tilde \partial^n B - \frac{1}{16} \e^{4A-4B} \tilde g_{99} \tilde g_{99} \tilde g^{jn} \tilde g^{lm} f^9_{jl} f^9_{mn} f^9_{pi}x^p f^9_{qk}x^q + \hat R_{ik} + \tilde R_{ik}, \\
R_{9k} &= - \frac{1}{2} \e^{2A-2B} \tilde g_{99} f^9_{ik}x^i \tilde \nabla^2 A - \frac{p+1}{2} \e^{2A-2B} \tilde g_{99} f^9_{ik}x^i (\partial_m A) (\tilde \partial^m A) \nl + \frac{p-7}{2} \e^{2A-2B} \tilde g_{99} f^9_{ik}x^i (\partial_m A) (\tilde \partial^m B) - \frac{p+3}{2} \e^{2A-2B} \tilde g_{99} f^9_{ik} \tilde \partial^i A + \frac{p-5}{2} \e^{2A-2B} \nl \cdot \tilde g_{99} f^9_{ik} \tilde \partial^i B - \frac{1}{8} \e^{4A-4B} \tilde g_{99} \tilde g_{99} \tilde g^{jn} \tilde g^{lm}  f^9_{jl} f^9_{mn} f^9_{ik}x^i + \hat R_{9k} + \tilde R_{9k}, \\
R_{99} &= - \e^{2A-2B} \tilde g_{99} \tilde \nabla^2 A - \left({p+1}\right) \e^{2A-2B} \tilde g_{99} (\partial_m A) (\tilde \partial^m A) + \left({p-7}\right) \e^{2A-2B} \tilde g_{99} \nl \cdot (\partial_m A) (\tilde \partial^m B) - \frac{1}{4} \e^{4A-4B} \tilde g_{99} \tilde g_{99} \tilde g^{jn} \tilde g^{lm} f^9_{jl} f^9_{mn} + \hat R_{99} + \tilde R_{99},
\end{align}
where all other components are zero or implied by the symmetry $R_{MN} = R_{NM}$. Furthermore, we wrote $\hat R_{MN}$ for all terms that include factors of $\hat \Gamma^M_{NP}$. One can show that these terms do not appear in the Ricci tensor in the vielbein basis, and so we omit to spell them out explicitly.

In order to switch back from the coordinate basis to the vielbein basis \eqref{eq:e_basis}, we now transform the Ricci tensor using
\begin{equation}
R_{AB} = R_{MRN} \vphantom{}^{P} e^M_{A} e^R_{C} e^N_{B} e^{C}_P,
\end{equation}
where
\begin{equation}
e^{\alpha}_M = \delta^\alpha_M, \qquad e^{a}_M = \delta^a_M, \qquad e^{9}_M = \delta^9_M + \frac{1}{2} \delta^m_M f^9_{im} x^i. \label{eq:app-twisted-transform}
\end{equation}
In order to distinguish the two types of indices in above two equations, we have used letters from the beginning of the alphabet (such as $A,B,\ldots$, $a,b,\ldots$ and $\alpha,\beta,\ldots$) for indices in the vielbein basis and letters from the middle of the alphabet (such as $M,N,\ldots$, $m,n,\ldots$ and $\mu,\nu,\ldots$) for indices in the coordinate basis. The index ``$9$'' in \eqref{eq:app-twisted-transform} is also a vielbein index. The vielbein indices are raised and lowered with the metric components given in \eqref{eq:metric_e_basis} including the corresponding warp factors, while the coordinate indices are raised and lowered using \eqref{eq:inv_metrics_dx_basis} and \eqref{eq:inv_metrics_dx_basis2}. The components of the Ricci tensor in the vielbein basis \eqref{eq:e_basis} are thus given by
\begin{align}
R_{\mu\nu} &= \e^{2A-2B} \tilde g_{\mu\nu} \left[{- \tilde \nabla^2 A - \left({p+1}\right) (\partial_m A)(\tilde \partial^m A) + \left({p-7}\right) (\partial_m A) (\tilde \partial^m B)}\right] + \tilde R_{\mu\nu}, \\
R_{ik} &= - \left({p+1}\right) \tilde \nabla_i \partial_k A + \left({p-7}\right)\tilde \nabla_i \partial_k B - \tilde g_{ik} \tilde \nabla^2 B - \left({p+1}\right) (\partial_i A) (\partial_k A) \nl + \left({7-p}\right) (\partial_i B) (\partial_k B) + \left({p-7}\right) \tilde g_{ik} (\partial_m B)(\tilde \partial^m B) + 2 \left({p+1}\right) (\partial_{(i} A) (\partial_{k)} B) \nl - \left({p+1}\right) \tilde g_{ik} (\partial_m A) (\tilde \partial^m B) + \frac{1}{2} \e^{2A-2B} \tilde g_{99} \tilde g^{mn} f^9_{m(k}f^9_{i)n} + \tilde R_{ik}, \\
R_{9k} &= - \frac{p+3}{2} \e^{2A-2B} \tilde g_{99} f^9_{ik} \tilde \partial^i A + \frac{p-5}{2} \e^{2A-2B} \tilde g_{99} f^9_{ik} \tilde \partial^i B + \tilde R_{9k}, \\
R_{99} &= - \e^{2A-2B} \tilde g_{99} \tilde \nabla^2 A - \left({p+1}\right) \e^{2A-2B} \tilde g_{99} (\partial_m A) (\tilde \partial^m A) + \left({p-7}\right) \e^{2A-2B} \tilde g_{99} \nl \cdot (\partial_m A) (\tilde \partial^m B) - \frac{1}{4} \e^{4A-4B} \tilde g_{99} \tilde g_{99} \tilde g^{jn} \tilde g^{lm} f^9_{jl} f^9_{mn} + \tilde R_{99}.
\end{align}
Note that, in the limit $f^9_{ik} = 0$ and taking the $9$-direction to be external, the expressions reduce to those computed in Section \ref{app:computations.curvature.block} as expected.

Finally, we state the components of the Ricci tensor for the special case $B = \frac{p+1}{p-7}A$ used in Section \ref{ch:smeared-vs-loc.bps-twisted}. We find
\begin{align}
R_{\mu\nu} &= - \e^{\tfrac{16}{7-p}A} \tilde g_{\mu\nu} \tilde \nabla^2 A + \tilde R_{\mu\nu}, \\
R_{ik} &= - \frac{p+1}{p-7} \tilde g_{ik} \tilde \nabla^2 A + 8 \frac{p+1}{p-7} (\partial_i A) (\partial_k A) - \frac{1}{2} \e^{\tfrac{16}{7-p}A} \tilde g_{99} |\d \tilde e^9|^2_{ik} + \tilde R_{ik}, \\
R_{9k} &= - \frac{8}{p-7} \e^{\tfrac{16}{7-p}A} \tilde g_{99} |e^9 \wedge \d A \cdot \d \tilde e^9|_{9k} + \tilde R_{9k}, \\
R_{99} &= - \e^{\tfrac{16}{7-p}A} \tilde g_{99} \tilde \nabla^2 A + \frac{1}{2} \e^{\tfrac{32}{7-p}A} \tilde g_{99} \tilde g_{99} |\d \tilde e^9|^2 + \tilde R_{99},
\end{align}
where we have used \eqref{eq:de9} to rewrite the structure constants $f_{ik}^9$ in terms of $\d e^9$.
\\

\cleardoublepage
\thispagestyle{plain}

\section{Regularised Sources in the Non-BPS Model}
\label{app:computations.regularised}

Here, we present the details of the computation discussed in Section \ref{ch:simple-non-bps.localisation.regular}, where we argue that the simple non-BPS setup with D$6$/\dsix-branes on AdS$_7 \times S^3$ does not have consistent solutions with a regular source profile other than the smeared one. Solving \eqref{eoms-theta-bianchi}, \eqref{eoms-theta-dilaton}, \eqref{eoms-theta-exteinstein} and \eqref{eoms-theta-inteinsteintheta} in a Taylor expansion around $\theta=\theta_0$ and substituting the results into \eqref{eoms-theta-inteinsteintrans}, we find the following constraints at the first five orders in the expansion parameter $(\theta-\theta_0)$:
\begin{align}
F_0^2 &= \frac{f_0^8}{a_0^2 b_0\left({\lambda_0^2-1}\right)} \left({a_0^2b_1^2f_0^2 + 84 a_0^4b_0f_0^2 + 84 a_1^2b_0^2f_0^2 + 49a_1^2b_0^2f_0^2\lambda_0^2 - 16 a_0^2 b_0^2 f_1^2}\right. \nl + a_0^2b_0^2f_0^2\lambda_1^2 - 4 a_0^2b_0b_1f_0^2 \cot \theta_0 - 56a_0a_1b_0^2f_0^2\cot\theta_0 + 28 a_0 a_1 b_0 b_1f_0^2 - 4 a_0^2b_0^2f_0^2 \nl + \left.{9a_0^2b_0^2f_1^2\lambda_0^2 - 14 a_0a_1b_0^2f_0^2\lambda_0\lambda_1 - 6 a_0^2b_0^2f_0f_1\lambda_0\lambda_1 + 42a_0a_1b_0^2f_0f_1\lambda_0^2 }\right), \label{constraints-regular1} \\
a_1 &= \frac{1}{7} \frac{a_0\left({3f_1-3f_1\lambda_0+f_0\lambda_1}\right)}{f_0\left({\lambda_0-1}\right)}, \\
b_1 &= -\frac{14}{3} \frac{b_0f_1}{f_0} + 2 b_0 \cot \theta_0 \pm \frac{2 \sqrt{\left({b_0^2\lambda_0-b_0^2+42a_0^2b_0\sin^2 \theta_0}\right)\left({\lambda_0-1}\right)}}{\left({\lambda_0-1}\right)\sin \theta_0}, \\
f_1 &= \frac{3}{16} \frac{f_0\lambda_1}{\lambda_0-1}, \label{constraints-regular4} \\
\lambda_0 &= -\frac{5}{2}, \label{constraints-regular5}
\end{align}
where we simplified each constraint using the previously found constraints of the lower orders. One can verify that choosing $\lambda_0 = \pm 1$ is not consistent with the equations of motion, and so it is justified to use the first four constraints in order to arrive at the last one. Since \eqref{constraints-regular5} holds everywhere on the three-sphere except for the poles, $\lambda(\theta)$ is constant, which implies $\lambda_1 = 0$. Substituting this back into the constraints \eqref{constraints-regular4} to \eqref{constraints-regular1} in reverse order then yields
\begin{equation}
a_1 = f_1 = \lambda_1 = 0, \qquad \lambda_0 = -\frac{5}{2}, \qquad F_0^2 = \frac{48}{7} a_0^2f_0^{10}.
\end{equation}
Note that we only spelled out the case of D$6$-brane sources here. The expressions for the case of \dsix-branes are obtained by replacing $\lambda_0 \to -\lambda_0$ and $\lambda_1 \to -\lambda_1$.
\\

\cleardoublepage
\thispagestyle{plain}

\section{Near-source Boundary Conditions in the Non-BPS Model}
\label{app:computations.bc}

In this appendix, we compute the boundary conditions of localised D$6$/\dsix-branes in the simple non-BPS setup on AdS$_7 \times S^3$, which are used in Section \ref{ch:simple-non-bps.localisation.delta}. In order to derive the boundary conditions, it is convenient to introduce the notation
\begin{equation}
a(\theta) = \e^{-A(\theta)}, \qquad b(\theta) = \e^{-2B(\theta)}, \qquad f(\theta) = \e^{-\tfrac{1}{4} \phi(\theta)}
\end{equation}
and take advantage of different combinations of the equations of motion for which certain terms cancel out. In particular, we will use a combination of the external Einstein equation \eqref{eoms-theta-exteinstein} and the dilaton equation \eqref{eoms-theta-dilaton} such that the source term cancels out,
\begin{align}
0 &= 4\frac{f^{\prime\prime}}{f} - 4 \frac{f^{\prime 2}}{f^2} + 8 \cot(\theta) \frac{f^\prime}{f} - 28 \frac{a^\prime f^\prime}{af} - 2\frac{b^\prime f^\prime}{bf} - 12 \frac{a^{\prime\prime}}{a} + 96 \frac{a^{\prime 2}}{a^2} - 24 \cot(\theta) \frac{a^\prime}{a} + 6\frac{a^\prime b^\prime}{ab} \nl + 72 \frac{a^2}{b} - 2 \frac{\lambda^2 F_0^2}{bf^{10}} + 2 \frac{F_0^2}{bf^{10}}, \label{eq:e1}
\end{align}
the dilaton equation \eqref{eoms-theta-dilaton},
\begin{align}
0 &= 4\frac{f^{\prime\prime}}{f} - 4 \frac{f^{\prime 2}}{f^2} + 8 \cot(\theta) \frac{f^\prime}{f} - 28 \frac{a^\prime f^\prime }{af} - 2 \frac{b^\prime f^\prime}{bf} - \frac{1}{2} \frac{\lambda^2 F_0^2}{bf^{10}} + \frac{5}{4} \frac{F_0^2}{bf^{10}} + \frac{147}{4} \frac{\lambda^2 a^{\prime 2}}{a^2} \nl + \frac{63}{2} \frac{\lambda^2 a^\prime f^\prime}{af} - \frac{21}{2} \frac{\lambda a^\prime \lambda^\prime}{a} + \frac{27}{4} \frac{\lambda^2 f^{\prime 2}}{f^2} - \frac{9}{2} \frac{\lambda f^\prime \lambda^\prime}{f} + \frac{3}{4} \lambda^{\prime 2} + \frac{3}{4} T \frac{\sqrt{b}}{f^3} \frac{\delta(\theta)}{\sin^2(\theta)},
\label{eq:e2}
\end{align}
the Bianchi identity \eqref{eoms-theta-bianchi},
\begin{align}
0 &= 7\frac{\lambda a^{\prime\prime}}{a} - 7\frac{\lambda a^{\prime 2}}{a^2} + 14 \cot(\theta) \frac{\lambda a^\prime}{a} + 3\frac{\lambda f^{\prime\prime}}{f} + 6 \frac{\lambda f^{\prime 2}}{f^2} + 6\cot(\theta) \frac{\lambda f^\prime}{f} - \lambda^{\prime\prime} - 2 \cot(\theta) \lambda^\prime \nl + \frac{\lambda F_0^2}{bf^{10}} - \frac{7}{2} \frac{\lambda a^\prime b^\prime}{ab} - \frac{3}{2} \frac{\lambda b^\prime f^\prime}{bf} + \frac{1}{2} \frac{b^\prime \lambda^\prime}{b} + 21 \frac{\lambda a^\prime f^\prime}{af} + 7 \frac{a^\prime \lambda^\prime}{a} + Q \frac{\sqrt{b}}{f^3} \frac{\delta(\theta)}{\sin^2(\theta)},
\label{eq:e3}
\end{align}
a combination of the internal Einstein equations \eqref{eoms-theta-inteinsteintheta} and \eqref{eoms-theta-inteinsteintrans} such that all $F_0^2$-dependent terms cancel out,
\begin{align}
0 &= - 7 \cot(\theta) \frac{a^\prime}{a} - \frac{1}{2} \cot(\theta) \frac{b^\prime}{b} + 7 \frac{a^{\prime\prime}}{a} - 14 \frac{a^{\prime 2}}{a^2} + \frac{1}{2} \frac{b^{\prime\prime}}{b} - \frac{1}{4} \frac{b^{\prime 2}}{b^2} + 7 \frac{a^\prime b^\prime}{ab} - 8 \frac{f^{\prime 2}}{f^2} + \frac{49}{2} \frac{\lambda^2 a^{\prime 2}}{a^2} \nl + 21 \frac{\lambda^2 a^\prime f^\prime}{af} - 7 \frac{\lambda a^\prime \lambda^\prime}{a} + \frac{9}{2} \frac{\lambda^2 f^{\prime 2}}{f^2} - 3\frac{\lambda f^\prime \lambda^\prime}{f} + \frac{1}{2} \lambda^{\prime 2},
\label{eq:i1}
\end{align}
the internal Einstein equations transverse to the $(\theta\theta)$-direction \eqref{eoms-theta-inteinsteintrans},
\begin{align}
0 &= 7 \cot(\theta) \frac{a^\prime}{a} + \frac{3}{2} \cot(\theta) \frac{b^\prime}{b} + \frac{1}{2} \frac{b^{\prime\prime}}{b} - \frac{3}{4} \frac{b^{\prime 2}}{b^2} - \frac{7}{2} \frac{a^\prime b^\prime}{ab} + 2 - \frac{1}{16} \frac{F_0^2}{bf^{10}} - \frac{343}{16} \frac{\lambda^2 a^{\prime 2}}{a^2} \nl - \frac{147}{8} \frac{\lambda^2 a^\prime f^\prime}{af} + \frac{49}{8} \frac{\lambda a^\prime \lambda^\prime}{a} - \frac{63}{16} \frac{\lambda^2 f^{\prime 2}}{f^2} + \frac{21}{8} \frac{\lambda f^\prime \lambda^\prime}{f} - \frac{7}{16} \lambda^{\prime 2} - \frac{3}{8} \frac{\lambda^2 F_0^2}{bf^{10}} \nl - \frac{7}{16} T \frac{\sqrt{b}}{f^3} \frac{\delta(\theta)}{\sin^2(\theta)},
\label{eq:i2}
\end{align}
a combination of the dilaton equation \eqref{eoms-theta-dilaton}, the Bianchi identity \eqref{eoms-theta-bianchi} and the transverse internal Einstein equations \eqref{eoms-theta-inteinsteintrans} such that all $F_0^2$-dependent terms cancel out,
\begin{align}
0 &= 2 - \frac{7}{5} \frac{a^\prime f^\prime}{af} - \frac{1}{10} \frac{b^\prime f^\prime}{bf}- \frac{7}{2} \frac{a^\prime b^\prime}{ab} + \frac{14}{5} \frac{\lambda^2 a^{\prime\prime}}{a} + \frac{6}{5} \frac{\lambda^2 f^{\prime\prime}}{f} - \frac{1}{5} \frac{f^{\prime 2}}{f^2} + \frac{2}{5} \cot(\theta) \frac{f^\prime}{f} \nl + 7 \cot(\theta) \frac{a^\prime}{a} - \frac{112}{5} \frac{\lambda^2 a^{\prime 2}}{a^2} - \frac{6}{5} \frac{\lambda^2 f^{\prime 2}}{f^2} + \frac{42}{5} \frac{\lambda a^\prime \lambda^\prime}{a} + \frac{12}{5} \frac{\lambda f^\prime \lambda^\prime}{f} - \frac{2}{5} \lambda^{\prime 2} - \frac{42}{5} \frac{\lambda^2 a^\prime f^\prime}{af} \nl - \frac{3}{4} \frac{b^{\prime 2}}{b^2} + \frac{1}{5} \frac{f^{\prime\prime}}{f} + \frac{3}{2} \cot(\theta) \frac{b^\prime}{b} + \frac{28}{5} \cot(\theta) \frac{\lambda^2 a^\prime}{a} + \frac{12}{5} \cot(\theta) \frac{\lambda^2 f^\prime}{f} + \frac{1}{5} \frac{\lambda b^\prime \lambda^\prime}{b} \nl - \frac{4}{5} \cot(\theta) \lambda \lambda^\prime - \frac{7}{5} \frac{\lambda^2 a^\prime b^\prime}{ab} - \frac{3}{5} \frac{\lambda^2 b^\prime f^\prime}{bf} - \frac{2}{5} \lambda \lambda^{\prime\prime} + \frac{1}{2} \frac{b^{\prime\prime}}{b} - \frac{2}{5} T \frac{\sqrt{b}}{f^3} \frac{\delta(\theta)}{\sin^2(\theta)} \nl + \frac{2}{5} Q \frac{\lambda \sqrt{b}}{f^3} \frac{\delta(\theta)}{\sin^2(\theta)},
\label{eq:i3}
\end{align}
a combination of the dilaton equation \eqref{eoms-theta-dilaton} and the internal Einstein equations \eqref{eoms-theta-inteinsteintheta} and \eqref{eoms-theta-inteinsteintrans} such that all $\lambda$-dependent terms cancel out,
\begin{align}
0 &= 2 + 21\frac{a^\prime f^\prime}{af} + \frac{3}{2} \frac{b^\prime f^\prime}{bf} + \frac{21}{2} \frac{a^\prime b^\prime}{ab} - 28 \frac{a^{\prime 2}}{a^2} - 13 \frac{f^{\prime 2}}{f^2} - 6 \cot(\theta) \frac{f^\prime}{f} - 7 \cot(\theta) \frac{a^\prime}{a} - \frac{F_0^2}{bf^{10}} \nl - 3 \frac{f^{\prime\prime}}{f} + 14 \frac{a^{\prime\prime}}{a} - \frac{5}{4} \frac{b^{\prime 2}}{b^2} + \frac{3}{2} \frac{b^{\prime\prime}}{b} + \frac{1}{2} \cot(\theta) \frac{b^\prime}{b} - T \frac{\sqrt{b}}{f^3} \frac{\delta(\theta)}{\sin^2(\theta)},
\label{eq:i4}
\end{align}
and finally a combination of the dilaton equation \eqref{eoms-theta-dilaton} and the Einstein equations \eqref{eoms-theta-exteinstein}, \eqref{eoms-theta-inteinsteintheta} and \eqref{eoms-theta-inteinsteintrans} such that all $\lambda$-dependent terms cancel out,
\begin{align}
0 &= 2 - 7\frac{a^\prime f^\prime}{af} - \frac{1}{2} \frac{b^\prime f^\prime}{bf} - \frac{11}{2} \frac{a^\prime b^\prime}{ab} - 32 \frac{a^{\prime 2}}{a^2} - \frac{f^{\prime 2}}{f^2} + 2 \cot(\theta) \frac{f^\prime}{f} + 15 \cot(\theta) \frac{a^\prime}{a} + \frac{f^{\prime\prime}}{f} \nl + 4\frac{a^{\prime\prime}}{a} - 24 \frac{a^2}{b} - \frac{3}{4} \frac{b^{\prime 2}}{b^2} + \frac{1}{2} \frac{b^{\prime\prime}}{b} + \frac{3}{2} \cot(\theta) \frac{b^\prime}{b}. \label{eq:i5}
\end{align}
\\[-0.7cm]

\subsection{Leading Order Behaviour}

Let us now substitute the ansatz \eqref{eq:d6-ansatz-bc} into equations \eqref{eq:e1} to \eqref{eq:i5}. Since the different scalings $A$, $B$, $F$ and $L$ are a priori unknown, we do not know by how many powers of $\theta$ two terms in the equations differ unless they are of the same order. We can therefore only trust the leading order (LO) terms in the equations, whereas the sub-leading order (SLO) parts may secretly contain more terms than we have written down, regardless of how many orders we take into account in the expansion \eqref{eq:d6-ansatz-bc}. It is convenient to consider the three different cases $L>0$, $L=0$ and $L<0$ separately.
\\

{\noindent\bf Case 1 ($\boldsymbol{L > 0}$).} Substituting our ansatz into \eqref{eq:e2}, we find that the possible LO terms of the equation are
\begin{equation}
0 = -2 F \left({B-2+14A}\right) \theta^{-2} + \frac{5}{4} \frac{F_0^2}{b_0 f_0^{10}}\, \theta^{-B-10F}. \label{eq:b}
\end{equation}
We therefore need to consider three possibilities: $B>2-10F$, $B=2-10F$ and $B<2-10F$. If $B>2-10F$, we find a contradiction since then only the last term on the right-hand side of \eqref{eq:b} is of LO, and, thus, the equation is not solved. If $B=2-10F$, all terms in the equation are of LO, and we find
\begin{equation}
F_0^2 = -\frac{4}{25}\left({B-2}\right)\left({B-2+14A}\right) b_0 f_0^{10}. \label{eq:a}
\end{equation}
The LO of \eqref{eq:i3} then gives
\begin{equation}
0 = \frac{1}{25} \left({6 B^2+84AB-168A-24B-1}\right) \theta^{-2},
\end{equation}
which yields $A = -\frac{1}{84}\frac{6B^2-24B-1}{B-2}$. Note that $B \neq 2$ since otherwise \eqref{eq:i3} yields $1=0$. The expression for $A$ is therefore well-defined. Substituting $A$ into \eqref{eq:a}, we find
\begin{equation}
F_0^2 = -\frac{2}{3} b_0 f_0^{10},
\end{equation}
which is not allowed since $F_0^2$ has to be positive. The possibility $B=2-10F$ is therefore excluded as well.

The last possibility we need to check is $B<2-10F$. The LO of \eqref{eq:b} is then solved for $A = \frac{1}{7} - \frac{1}{14}B$ or $ F=0$. Substituting the first option into \eqref{eq:i3} leads to
\begin{equation}
\theta^{-2} = 0
\end{equation}
at LO and is therefore excluded. Substituting the second option into \eqref{eq:i3} yields
\begin{equation}
-\frac{1}{4} \left({-28A+14AB-4B+B^2}\right) \theta^{-2} = 0
\end{equation}
at LO, which implies $ A = -\frac{1}{14} \frac{B\left({B-4}\right)}{B-2} $. Note that, by assumption, $B < 2-10F$ and $F=0$, and so the expression for $A$ is well-defined. Using our results in \eqref{eq:i1} and considering the LO terms, we find
\begin{equation}
-\frac{1}{7} \frac{B\left({B-4}\right)\left({2B^2-8B+7}\right)}{\left({B-2}\right)^2}\, \theta^{-2} = 0.
\end{equation}
This yields the four solutions $0$, $2-\frac{1}{2} \sqrt{2}$, $2 + \frac{1}{2} \sqrt{2}$ and $4$ for $B$, where the last two solutions can be discarded because they violate our assumption $ 0 = 10 F < 2 - B $. The other two solutions are excluded due to \eqref{eq:e3}. Evaluating its LO contribution for either $B=0$ or $B=2-\frac{1}{2} \sqrt{2}$ yields
\begin{equation}
L \left({2+2L}\right) \theta^{L-2} = 0, \qquad \left({4L+\sqrt{2}}\right)\left({4L+7\sqrt{2}}\right) \theta^{L-2} = 0,
\end{equation}
respectively, which cannot be fulfilled for $L>0$. Thus, we have shown that all possible values for $A$, $B$ and $F$ are excluded such that there are no consistent boundary conditions with $L > 0$.
\\

{\noindent\bf Case 2 ($\boldsymbol{L = 0}$).} Substituting our ansatz into \eqref{eq:e2}, we find that the possible LO terms are
\begin{align}
0 &= \left({4F -28AF -2BF +\frac{27}{4}\lambda_0^2 F^2+\frac{147}{4}\lambda_0^2A^2+\frac{63}{2}\lambda_0^2AF}\right) \theta^{-2} \nl + \left({\frac{5}{4} - \frac{1}{2} \lambda_0^2}\right) \frac{F_0^2}{b_0 f_0^{10}}\, \theta^{-B-10F}.
\end{align}
Therefore, we again need to consider the three possibilities $B>2-10F$, $B=2-10F$ and $B<2-10F$. The first possibility $B>2-10F$ is excluded by \eqref{eq:e3}, which yields
\begin{equation}
\frac{F_0^2}{b_0 f_0^{10}}\, \theta^{-B-10F} = 0
\end{equation}
at LO and is therefore not solved. The second possibility $B=2-10F$ can be excluded by combining the LO of \eqref{eq:e3} and \eqref{eq:i4} such that a linear equation for $B$ remains. This equation yields the condition $A \neq 0$ and can be solved for $B$,
\begin{equation}
B=2 + 2A + \frac{1}{7A} + \frac{3}{28} \frac{F_0^2}{b_0f_0^{10}A}.
\end{equation}
We then use this in the LO of \eqref{eq:i1} and \eqref{eq:i3} and combine the equations such that we obtain a quadratic equation for $A$. This equation yields the condition $\lambda_0^2 \neq \frac{3}{2}$ and can be solved for $A$,
\begin{equation}
A = \pm \frac{\sqrt{14}}{112} \frac{\sqrt{b_0 \left({-3+2\lambda_0^2}\right)\left({9\lambda_0^2 F_0^2 -6 F_0^2-8b_0f_0^{10}+12\lambda_0^2b_0f_0^{10}}\right)}}{b_0f_0^5\left({-3+2\lambda_0^2}\right)}.
\end{equation}
We can substitute this into \eqref{eq:e3} and \eqref{eq:i3} and solve both equations for $F_0^2$ to find
\begin{equation}
F_0^2 = \frac{2b_0f_0^{10}\left({7\lambda_0^2-10}\right)}{36-56\lambda_0^2+21\lambda_0^4}, \qquad F_0^2 = -\frac{2}{3}\frac{b_0f_0^{10}\left({54-73\lambda_0^2+28\lambda_0^4}\right)}{48-82\lambda_0^2+35\lambda_0^4}, \label{eq:app-d6-f0equations}
\end{equation}
where it can be checked that the denominators are not zero since \eqref{eq:e3} or \eqref{eq:i3} would not be satisfied for the corresponding values of $\lambda_0$. We now combine the two equations in \eqref{eq:app-d6-f0equations} such that $F_0^2$ cancels out and solve for $\lambda_0$. Substituting the solutions back into \eqref{eq:app-d6-f0equations}, we then find that $F_0^2$ is negative for all possible $\lambda_0$. Therefore, the initial assumption $B=2-10F$ is not consistent with the equations of motion.

The last possibility to check is $B<2-10F$. Let us first assume that $F=0$. We then find $A=0$ from the LO of \eqref{eq:e2} and $B=0$ or $B=4$ from the LO of \eqref{eq:i1}. Since $B=4$ violates our assumption $B<2-10F$, only $B=0$ is consistent, and we find
\begin{equation}
A=B=F=L=0.
\end{equation}
One can check that this solves all equations at LO such that we have found a consistent boundary condition, which we will refer to as \emph{boundary condition 1}. Since there are no divergent fields, this is the boundary condition for the case where no sources are present at the pole.

Let us now assume $F \neq 0$. We can then solve the LO of \eqref{eq:e2} for $B$, which yields
\begin{equation}
B = \frac{1}{8F} \left({16F-112AF+27\lambda_0^2F^2+147\lambda_0^2A^2+126\lambda_0^2AF}\right). \label{eq:app-d6-b}
\end{equation}
Substituting this expression into the LO of \eqref{eq:e3} and solving for $A$ gives the solutions
\begin{equation}
A = -\frac{3}{7} F, \qquad A = -\frac{1}{21} \frac{F\left({-16+9\lambda_0^2}\right)}{\lambda_0^2}. \label{eq:app-d6-a}
\end{equation}
An exception is the special case $\lambda_0^2 = 0$, where we only find $A = -\frac{3}{7} F$. Substituting the first solution for $A$ into the LO of \eqref{eq:i1} leads to a contradiction. Using the second solution in \eqref{eq:i2}, we can solve for $F$ to obtain
\begin{equation}
F = \pm \frac{3}{16} \lambda_0.
\end{equation}
The LO of \eqref{eq:i1} then yields an equation for $\lambda_0$. Solving this equation and testing the solutions for $\lambda_0$ in the LO of \eqref{eq:e1}, we find that only two solutions do not lead to a contradiction. This yields
\begin{equation}
\lambda_0 = \pm 1, \qquad F = - \frac{3}{16},
\end{equation}
where the sign of $F$ is determined by our initial assumption $B<2-10F$. Using these expressions in \eqref{eq:app-d6-b} and \eqref{eq:app-d6-a}, we find the boundary condition
\begin{equation}
L=0, \qquad A = -\frac{1}{16}, \qquad B = \frac{7}{8}, \qquad F = -\frac{3}{16}, \qquad \lambda_0 = \pm 1,
\end{equation}
which we will refer to as \emph{boundary condition 2}. For convenience, we also state the expansion of $\alpha(\theta)$ for this boundary condition,
\begin{equation}
\alpha(\theta) = \frac{\lambda(\theta)}{a(\theta)^7 f(\theta)^3} = \pm \frac{1}{a_0^7 f_0^3} \theta + \mathcal{O}(\theta^2).
\end{equation}
Thus, $\alpha^\prime(0)>0$ for the solution with the upper sign and $\alpha^\prime(0)<0$ for the solution with the lower sign.
\\

{\noindent\bf Case 3 ($\boldsymbol{L < 0}$).} Considering the LO of \eqref{eq:i1}, we find
\begin{equation}
A = \frac{1}{7}\left({L - 3 F}\right). \label{eq:app-d6-aaaa}
\end{equation}
The LO of \eqref{eq:e2}, \eqref{eq:e3} and \eqref{eq:i2} are then also solved if the condition $ B < 2 - 10F$ is satisfied. The possible LO contributions of \eqref{eq:i4} and \eqref{eq:i5} are
\begin{align}
0 &= -\frac{1}{28}\left({-7B^2+772F^2-168F+8L^2-132FL+84L+84BF-42BL}\right. \nl +\left.{\vphantom{B^2} 28B}\right) \theta^{-2} \label{eq:x1}
\end{align}
and
\begin{align}
0 &= \frac{1}{28}\left({28B- 104F-16L^2+68FL-60F^2+44L+52BF-22BL-7B^2}\right) \theta^{-2} \nl - 24\frac{a_0^2}{b_0}\, \theta^{\tfrac{2}{7}L-\tfrac{6}{7}F-B}. \label{eq:x2}
\end{align}
There are two possibilities to solve \eqref{eq:x2} at LO, $\,\frac{2}{7}L-\frac{6}{7}F-B = -2\,$ or $\,\frac{2}{7}L-\frac{6}{7}F-B > -2 $. The first possibility can be solved for $B$ and used in \eqref{eq:x1} and \eqref{eq:x2} to find
\begin{equation}
L =  \frac{1}{4} \frac{32b_0F^2 + b_0 + 6a_0^2}{b_0 F},
\end{equation}
where one can check that $F \neq 0$. Substituting the expressions for $B$ and $L$ into \eqref{eq:x1}, we then find that its right-hand side is necessarily negative. Thus, $\frac{2}{7}L-\frac{6}{7}F-B = -2 $ is excluded.

Now considering the second possibility $\frac{2}{7}L-\frac{6}{7}F-B > -2 $, we first combine \eqref{eq:x1} and \eqref{eq:x2} such that the $B^2$ terms cancel out. The resulting equation can then be solved for $B$ to find
\begin{equation}
B = \frac{2\left({8F-3L^2+25FL-104F^2-5L}\right)}{8F-5L}, \label{eq:app-d6-aaab}
\end{equation}
where it can be checked that $F \neq \frac{5}{8}L$ such that the denominator is not zero. Using \eqref{eq:app-d6-aaaa} and \eqref{eq:app-d6-aaab}, we find that the possible LO contributions of \eqref{eq:e1} are
\begin{equation}
0 = -2\frac{\lambda_0^2 F_0^2}{b_0f_0^{10}}\, \theta^{\tfrac{4\left({4F+L}\right)\left({-8F+L}\right)}{-8F+5L}-2} + \frac{32}{7} \frac{\left({-16F+3L}\right)\left({16F^2-6FL+L^2}\right)}{-8F+5L}\, \theta^{-2}. \label{eq:x3}
\end{equation}
This again yields two possibilities: either only the second term is of LO or both terms are. For the first possibility, \eqref{eq:x3} implies $\left({-16F+3L}\right)\left({16F^2-6FL+L^2}\right)=0$, which gives one real and two complex solutions for $L$. Choosing the real solution $L= \frac {16}{3}F$, we can evaluate \eqref{eq:x1} to find $F= \pm \frac{3}{16}$. The sign of $F$ is then fixed to be negative by demanding that the previously found conditions $B < 2 - 10F$ and $\frac{2}{7}L-\frac{6}{7}F-B > -2$ hold. We thus find the boundary condition
\begin{equation}
L = -1, \qquad A=-\frac{1}{16}, \qquad B=\frac{7}{8}, \qquad F=-\frac{3}{16}, \label{eq:bc3}
\end{equation}
which we will refer to as \emph{boundary condition 3}.

The second possibility is realised if $\left({4F+L}\right)\left({-8F+L}\right) = 0$ in \eqref{eq:x3}, which yields the solutions $F=-\frac{1}{4}L$ and $F=\frac{1}{8}L$. Substituting this into \eqref{eq:x1} and \eqref{eq:x3}, we find the two boundary conditions
\begin{equation}
L = -\frac{\sqrt{42}}{6}, \quad A=-\frac{5\sqrt{42}}{336}, \quad B=2-\frac{\sqrt{42}}{8}, \quad F=-\frac{\sqrt{42}}{48}, \quad \lambda_0 = \pm  \frac{\sqrt{b_0}f_0^5}{\sqrt{3}F_0} \label{eq:bc1}
\end{equation}
and
\begin{equation}
L = -\frac{\sqrt{3}}{6}, \quad A=-\frac{\sqrt{3}}{24}, \quad B=2-\frac{3\sqrt{3}}{4}, \quad F=\frac{\sqrt{3}}{24}, \quad \lambda_0 = \pm \frac{\sqrt{2 b_0}f_0^5}{\sqrt{3}F_0}, \label{eq:bc2}
\end{equation}
which we will refer to as \emph{boundary condition 4} and \emph{boundary condition 5}. One can verify that both boundary conditions are consistent with the conditions $B < 2 - 10F$ and $\frac{2}{7}L-\frac{6}{7}F-B > -2$.
\\

\subsection{Sub-leading Order Behaviour}

For two of the boundary conditions found above, we need to compute the SLO behaviour of the fields in order to be able to determine the source types they support. This is the subject of the present section.
\\

{\noindent\bf Boundary condition 3.} We now consider the ansatz \eqref{eq:d6-ansatz-bc} including next-to-leading order (NLO) and next-to-next-to-leading order (NNLO) terms, i.\ e.\
\begin{equation}
\e^{-A(\theta)} = a_0 \theta^A + a_1 \theta^{A+\zeta} + a_2 \theta^{A+\xi}
\end{equation}
and accordingly for the other functions $\e^{-2B(\theta)}$, $\e^{-\frac{1}{4}\phi(\theta)}$ and $\lambda(\theta)$. We then substitute this ansatz into the equations of motion and use our knowledge of the LO scalings,
\begin{equation}
L = -1, \qquad A=-\frac{1}{16}, \qquad B=\frac{7}{8}, \qquad F=-\frac{3}{16},
\end{equation}
in order to determine the NLO and NNLO terms. The possible NLO contributions of \eqref{eq:i5} are
\begin{align}
0 &= \left({2a_0b_1f_0 + a_0b_1f_0 \zeta + 22a_1b_0f_0 + 8a_1b_0f_0 \zeta + 2a_0b_0f_1 + 2a_0b_0f_1 \zeta}\right) \zeta\, \theta^{-3+\zeta} \nl + 48a_0^3f_0\, \theta^{-2},
\end{align}
which cannot be solved unless $\zeta \le 1$. The NLO contributions of \eqref{eq:e1}, \eqref{eq:e3} and \eqref{eq:i4} are
\begin{align}
0 &= 2 b_0 f_0^9 \left({3a_1f_0-a_0f_1}\right) \zeta \left({\zeta+1}\right) \theta^{-3+\zeta} + a_0 \lambda_0^2 F_0^2\, \theta^{-2} - 36 a_0^3 f_0^{10}\, \theta^{-2}, \\
0 &= \left({-a_0 f_0 \lambda_1 + 7a_1 f_0 \lambda_0 + 3a_0 f_1 \lambda_0}\right) \zeta \left({\zeta-1}\right) \theta^{-4+\zeta}, \\
0 &= \left({-6a_0b_0f_1 + 3a_0b_1f_0 + 28a_1b_0f_0}\right) \zeta \left({\zeta-1}\right) \theta^{-3+\zeta}.
\end{align}
Solving the equations for $\zeta < 1$ yields $a_1 = b_1 = f_1 = \lambda_1 = 0$, and so we find that non-trivial NLO terms have $\zeta = 1$. Considering \eqref{eq:e1}, \eqref{eq:e3}, \eqref{eq:i4} and \eqref{eq:i5} at NNLO and assuming $\xi < 2$ yields analogous constraints and leads to $a_2 = b_2 = f_2 = \lambda_2 = 0$. We can therefore also set $\xi=2$.

Using $ \zeta=1 $, the NLO contributions of \eqref{eq:e1} and \eqref{eq:i5} become
\begin{align}
0 &= 36 a_0^3 f_0^{10} + 4 a_0 b_0 f_0^9 f_1 - F_0^2 a_0 \lambda_0^2 - 12 a_1 b_0 f_0^{10}, \\
0 &= - 30 a_1 b_0 f_0 - 3 a_0 b_1 f_0 + 48 a_0^3 f_0 - 4 a_0 b_0 f_1.
\end{align}
All other equations are identically satisfied at NLO and do therefore not give additional constraints at that order. The NNLO contributions of \eqref{eq:e1}, \eqref{eq:e3}, \eqref{eq:i4} and \eqref{eq:i5} are
\begin{align}
0 &= 120 a_1^2 b_0^2 f_0^{11} - 72 a_0 a_2 b_0^2 f_0^{11} + 24 a_0^2 b_0^2 f_0^{10} f_2 - 100 a_0 a_1 b_0^2 f_0^{10} f_1 + 12 a_0^2 b_0^2 f_0^9 f_1^2 \nl + 18 a_0 a_1 b_0 b_1 f_0^{11} + 144 a_0^3 a_1 b_0 f_0^{11} + 216 a_0^4 b_0 f_0^{10} f_1 - 6 a_0^2 b_0 b_1 f_0^{10} f_1 \nl - 4 F_0^2 a_0^2 b_0 f_0 \lambda_0 \lambda_1 + 14 F_0^2 a_0^2 b_0 f_1 \lambda_0^2 - 108 a_0^4 b_1 f_0^{11} + 3 F_0^2 a_0^2 b_1 f_0 \lambda_0^2, \\
0 &= 4 a_0^2 b_0 f_0^2 \lambda_2 - 14 a_0 a_1 b_0 f_0^2 \lambda_1 + 14 a_1^2 b_0 f_0^2 \lambda_0 - 28 a_0 a_2 b_0 f_0^2 \lambda_0 - 12 a_0^2 b_0 f_0 f_2 \lambda_0 \nl - 42 a_0 a_1 b_0 f_0 f_1 \lambda_0 - 12 a_0^2 b_0 f_1^2 \lambda_0 - a_0^2 b_1 f_0^2 \lambda_1 + 7 a_0 a_1 b_1 f_0^2 \lambda_0 + 3 a_0^2 b_1 f_0 f_1 \lambda_0, \\
0 &= 336 a_1^2 b_0^2 f_0^2 - 336 a_0 a_2 b_0^2 f_0^2 - 16 a_0^2 b_0^2 f_0^2 + 72 a_0^2 b_0^2 f_0 f_2 - 252 a_0 a_1 b_0^2 f_0 f_1 \nl + 156 a_0^2 b_0^2 f_1^2 - 126 a_0 a_1 b_0 b_1 f_0^2 - 36 a_0^2 b_0 b_2 f_0^2 - 18 a_0^2 b_0 b_1 f_0 f_1 + 15 a_0^2 b_1^2 f_0^2, \\
0 &= - 6 a_0^2 b_0^2 f_0 f_2 - 38 a_0 a_2 b_0^2 f_0^2 - 38 a_0 a_1 b_0^2 f_0 f_1 + 47 a_1^2 b_0^2 f_0^2 - 3 a_0^2 b_0^2 f_1^2 - 2 a_0^2 b_0^2 f_0^2 \nl + 72 a_0^4 b_0 f_0 f_1 - 3 a_0^2 b_0 b_1 f_0 f_1 - 4 a_0^2 b_0 b_2 f_0^2  + 48 a_0^3 a_1 b_0 f_0^2 + 13 a_0 a_1 b_0 b_1 f_0^2 \nl - 36 a_0^4 b_1 f_0^2 + 3 a_0^2 b_1^2 f_0^2.
\end{align}
The NNLO of \eqref{eq:e2}, \eqref{eq:i1}, \eqref{eq:i2} and \eqref{eq:i3} each yield
\begin{equation}
\left({a_0 f_0 \lambda_1 - 7a_1f_0\lambda_0 - 3a_0 f_1\lambda_0}\right)^2 = a_0^2 f_0^2. \label{eq:bc3constraint}
\end{equation}
The above seven equations determine seven of the eight coefficients $a_1$, $b_1$, $f_1$, $\lambda_1$, $a_2$, $b_2$, $f_2$ and $\lambda_2$, which is sufficient to later determine the source terms. We find two solutions, which only differ in the signs of some of the terms,
\begin{align}
a_1 &= \frac{27}{16} \frac{a_0^3}{b_0} + \frac{1}{16} \frac{a_0}{\lambda_0}\left({\lambda_1\mp1}\right) - \frac{3}{64} \frac{F_0^2 a_0 \lambda_0^2}{b_0f_0^{10}}, \\
b_1 &= \frac{35}{8} a_0^2 - \frac{7}{8} \frac{b_0}{\lambda_0}\left({\lambda_1\mp1}\right) + \frac{31}{96} \frac{F_0^2 \lambda_0^2}{f_0^{10}}, \\
f_1 &= -\frac{63}{16} \frac{a_0^2f_0}{b_0} + \frac{3}{16} \frac{f_0}{\lambda_0}\left({\lambda_1\mp1}\right) + \frac{7}{64} \frac{F_0^2 \lambda_0^2}{b_0f_0^9}, \\
a_2 &= -\frac{1}{192}a_0 -\frac{1631}{512}\frac{a_0^5}{b_0^2} -\frac{109}{3072}\frac{F_0^2a_0\lambda_0\lambda_1}{b_0f_0^{10}} \mp \frac{59}{3072}\frac{F_0^2a_0\lambda_0}{b_0f_0^{10}} - \frac{3403}{3072}\frac{F_0^2a_0^3\lambda_0^2}{b_0^2f_0^{10}} \nl - \frac{15}{512}\frac{a_0}{\lambda_0^2}\left({\lambda_1\mp1}\right)^2 + \frac{535}{256}\frac{a_0^3}{b_0\lambda_0}\left({\lambda_1\mp1}\right) + \frac{2137}{73728}\frac{F_0^4a_0\lambda_0^4}{b_0^2f_0^{20}}, \\
b_2 &= - \frac{41}{96}b_0 + \frac{14413}{128}\frac{a_0^4}{b_0} + \frac{37}{768}\frac{F_0^2 \lambda_0\lambda_1}{f_0^{10}} \pm \frac{359}{768}\frac{F_0^2\lambda_0}{f_0^{10}} + \frac{4529}{768}\frac{F_0^2a_0^2\lambda_0^2}{b_0f_0^{10}} \nl -\frac{1351}{64}\frac{a_0^2}{\lambda_0}\left({\lambda_1 \mp 1}\right) + \frac{105}{128}\frac{b_0}{\lambda_0^2}\left({\lambda_1 \mp 1}\right)^2 - \frac{2843}{18432} \frac{F_0^4 \lambda_0^4}{b_0f_0^{20}}, \\
f_2 &= -\frac{1}{64}f_0 - \frac{12327}{512}\frac{a_0^4f_0}{b_0^2} + \frac{19}{3072}\frac{F_0^2\lambda_0\lambda_1}{b_0f_0^9} \mp \frac{11}{3072}\frac{F_0^2\lambda_0^2}{b_0f_0^9} + \frac{1519}{1024}\frac{F_0^2a_0^2\lambda_0^2}{b_0^2f_0^9}
 \nl +\frac{567}{256}\frac{a_0^2f_0}{b_0\lambda_0}\left({\lambda_1 \mp 1}\right) - \frac{39}{512}\frac{f_0}{\lambda_0^2}\left({\lambda_1 \mp 1}\right)^2 - \frac{869}{24576} \frac{F_0^4 \lambda_0^4}{b_0^2f_0^{19}}, \\
\lambda_2 &= -\frac{1}{12}\lambda_0 -\frac{511}{4} \frac{a_0^4\lambda_0}{b_0^2} - \frac{13}{48}\frac{F_0^2\lambda_0^2\lambda_1}{b_0f_0^{10}} \mp \frac{3}{16}\frac{F_0^2\lambda_0^2}{b_0f_0^{10}} - \frac{35}{24}\frac{F_0^2a_0^2\lambda_0^3}{b_0^2f_0^{10}} + \frac{91}{4} \frac{a_0^2\lambda_1}{b_0} \nl \mp \frac{63}{4}\frac{a_0^2}{b_0} + \frac{41}{576}\frac{F_0^4\lambda_0^5}{b_0^2f_0^{20}}.
\end{align}
We are thus left with the five free parameters $a_0$, $b_0$, $f_0$, $\lambda_0$ and $\lambda_1$, one of which will be fixed later by the charge of the source. One can check that the equations of motion are solved at higher orders for suitable choices of the coefficients $a_n$, $b_n$, $f_n$ and $\lambda_n$ with $n>2$ but do not yield additional constraints relating the free parameters $a_0$, $b_0$, $f_0$, $\lambda_0$ and $\lambda_1$. For convenience, we also state the expansion of $\alpha(\theta)$ for the present boundary condition,
\begin{equation}
\alpha(\theta) = \frac{\lambda(\theta)}{a(\theta)^7 f(\theta)^3} = \frac{\lambda_0}{a_0^7 f_0^3} \pm \frac{1}{a_0^7 f_0^3} \theta + \mathcal{O}(\theta^2). \label{eq:alphabc3}
\end{equation}
Thus, $\alpha^\prime(0)>0$ for the solution with the upper sign and $\alpha^\prime(0)<0$ for the solution with the lower sign.
\\

{\noindent\bf Boundary condition 4.} We start by showing that, for this boundary condition, the SLO terms in \eqref{eq:d6-ansatz-bc} have a structure involving four different scalings,
\begin{gather}
\zeta = \frac{1}{16} \sqrt{42} + \frac{11}{336} \sqrt{42} \approx 0.617, \qquad \xi=-\frac{1}{12}\sqrt{42}+\frac{1}{12}\sqrt{330} \approx 0.974, \notag \\ \eta=\frac{1}{6} \sqrt{42} \approx 1.080, \qquad \kappa = 2, \label{eq:abc}
\end{gather}
and integer multiples thereof such that
\begin{align}
\e^{-A(\theta)} &= a_0 \theta^A + a_1 \theta^{A+\zeta} + a_2 \theta^{A+\xi} + a_3 \theta^{A+\eta} + a_4 \theta^{A+2\zeta} + a_5 \theta^{A+\xi+\zeta} + a_6 \theta^{A+\eta+\zeta} \nl + a_7\theta^{A+3\zeta} + a_8\theta^{A+2\xi}+ a_9\theta^{A+\kappa}+ a_{10}\theta^{A+\eta+\xi}+ a_{11}\theta^{A+2\eta} + \ldots \label{eq:abcd}
\end{align}
and accordingly for the other functions $\e^{-2B(\theta)}$, $\e^{-\frac{1}{4}\phi(\theta)}$ and $\lambda(\theta)$. In order to determine the source terms in the next section, we will only need the coefficients $a_3$, $b_3$, $f_3$, $\lambda_3$, $a_{11}$, $b_{11}$, $f_{11}$ and $\lambda_{11}$.

Let us now suppose that the expansion of $\e^{-A(\theta)}$ yields an additional term $a_\chi \theta^{A+\chi}$ with $0 < \chi < 2\eta$ somewhere between the terms $a_0 \theta^A$ and $a_{11} \theta^{A+2\eta}$ such that $\theta^{A+\chi}$ does not coincide with one of the orders that have already been written out. Introducing analogous terms for the other functions $\e^{-2B(\theta)}$, $\e^{-\frac{1}{4}\phi(\theta)}$ and $\lambda(\theta)$ and using the previously found scalings
\begin{equation}
L = -\frac{\sqrt{42}}{6}, \qquad A=-\frac{5\sqrt{42}}{336}, \qquad B=2-\frac{\sqrt{42}}{8}, \qquad F=-\frac{\sqrt{42}}{48},
\end{equation}
we can evaluate \eqref{eq:e1}, \eqref{eq:e3}, \eqref{eq:i4} and \eqref{eq:i5} at order $\sim \theta^\chi$, which yields
\begin{align}
0 &= 63 F_0^2 a_0 b_\chi f_0 \lambda_0^2 - 84 F_0^2 a_0 b_0 f_0 \lambda_0 \lambda_\chi + 294 F_0^2 a_0 b_0 f_\chi \lambda_0^2 - 56 \sqrt{42} \chi a_\chi b_0^2 f_0^{11} \nl + 14 \sqrt{42} \chi a_0 b_0^2 f_0^{10} f_\chi + 42 a_0 b_0^2 f_0^{10} f_\chi - 7 a_0 b_0 b_\chi f_0^{11} -\sqrt{42} \chi a_0 b_0 b_\chi f_0^{11} \nl + 84 \chi^2 a_0 b_0^2 f_0^{10} f_\chi - 252 \chi^2 a_\chi b_0^2 f_0^{11}, \\
0 &= \left({- 7 a_\chi f_0 \lambda_0 - 3 a_0 f_\chi \lambda_0 + a_0 f_0 \lambda_\chi}\right)\left({6\chi-\sqrt{42}}\right), \\
0 &= 16 \sqrt{42} a_\chi b_0 f_0 - 168 \chi a_\chi b_0 f_0 + 36 \chi a_0 b_0 f_\chi - 2 \sqrt{42} a_0 b_0 f_\chi + 3 \sqrt{42} a_0 b_\chi f_0 \nl - 18 \chi a_0 b_\chi f_0, \\
0 &= 336 \chi a_\chi b_0 f_0 + 140 \sqrt{42} a_\chi b_0 f_0 + 14 \sqrt{42} a_0 b_0 f_\chi + 84 \chi a_0 b_0 f_\chi + 42 \chi a_0 b_\chi f_0 \nl + 13 \sqrt{42} a_0 b_\chi f_0.
\end{align}
For $\chi \neq -\frac{1}{12}\sqrt{42}+\frac{1}{12}\sqrt{330}$ and $\chi \neq \frac{1}{6} \sqrt{42}$ (which is true by assumption), this yields the solution $a_\chi = b_\chi = f_\chi = \lambda_\chi = 0$. Thus, we have shown that there are no additional terms between $a_0 \theta^A$ and $a_{11} \theta^{A+2\eta}$ other than those written out in \eqref{eq:abcd} (and accordingly for $\e^{-2B(\theta)}$, $\e^{-\frac{1}{4}\phi(\theta)}$ and $\lambda(\theta)$).

We can now compute the coefficients $a_3$, $b_3$, $f_3$, $\lambda_3$, $a_{11}$, $b_{11}$, $f_{11}$ and $\lambda_{11}$. The contributions of \eqref{eq:e1}, \eqref{eq:i4} and \eqref{eq:i5} at order $\sim \theta^\eta$ give
\begin{align}
0 &= 42 F_0^2 a_0 b_0 f_3 \lambda_0^2 + 34 a_0 b_0^2 f_0^{10} f_3 - 2 a_0 b_0 b_3 f_0^{11} - 98 a_3 b_0^2 f_0^{11} + 9 F_0^2 a_0 b_3 f_0 \lambda_0^2 \nl - 12 F_0^2 a_0 b_0 f_0 \lambda_0 \lambda_3, \\
0 &= - 3 a_3 f_0 + a_0 f_3, \\
0 &= 7 a_0 b_0 f_3 + 49 a_3 b_0 f_0 + 5 a_0 b_3 f_0,
\end{align}
while the other equations are identically satisfied. At order $\sim \theta^{2\eta}$, \eqref{eq:e1}, \eqref{eq:e3}, \eqref{eq:i4} and \eqref{eq:i5} yield
\begin{align}
0 &= 280 a_0 a_3 b_0^2 b_3 f_0^{12} + 9 a_0^2 b_0 b_3^2 f_0^{12} - 12 a_0^2 b_0^2 b_{11} f_0^{12} - 108 a_0^2 b_0^2 b_3 f_0^{11} f_3 \nl + 192 a_0^2 b_0^3 f_0^{10} f_3^2 - 45 F_0^2 a_0^2 b_3^2 f_0^2 \lambda_0^2 + 336 F_0^2 a_0^2 b_0^2 f_0 f_3 \lambda_0 \lambda_3 + 1736 a_3^2 b_0^3 f_0^{12} \nl - 48 F_0^2 a_0^2 b_0^2 f_0^2 \lambda_0 \lambda_{11} - 1568 a_0 a_3 b_0^3 f_0^{11} f_3 - 252 F_0^2 a_0^2 b_0 b_3 f_0 f_3 \lambda_0^2 \nl + 168 F_0^2 a_0^2 b_0^2 f_0 f_{11} \lambda_0^2 + 24 F_0^2 a_0^2 b_0^2 f_0^2 + 36 F_0^2 a_0^2 b_0 b_{11} f_0^2 \lambda_0^2 - 672 F_0^2 a_0^2 b_0^2 f_3^2 \lambda_0^2 \nl - 24 F_0^2 a_0^2 b_0^2 f_0^2 \lambda_3^2 + 72 F_0^2 a_0^2 b_0 b_3 f_0^2 \lambda_0 \lambda_3 - 1120 a_0 a_{11} b_0^3 f_0^{12} + 360 a_0^2 b_0^3 f_0^{11} f_{11}, \\
0 &= - 98 a_0 a_3 b_0 f_0^{10} \lambda_3 + 28 a_0^2 b_0 f_0^{10} \lambda_{11} - 196 a_0 a_{11} b_0 f_0^{10} \lambda_0  + 98 a_3^2 b_0 f_0^{10} \lambda_0 \nl + 49 a_0 a_3 b_3 f_0^{10} \lambda_0 - 7 a_0^2 b_3 f_0^{10} \lambda_3 - 294 a_0 a_3 b_0 f_0^9 f_3 \lambda_0 - 84 a_0^2 b_0 f_0^9 f_{11} \lambda_0 \nl + 21 a_0^2 b_3 f_0^9 f_3 \lambda_0 - 84 a_0^2 b_0 f_0^8 f_3^2 \lambda_0 - 12 F_0^2 a_0^2 \lambda_0, \\
0 &= - 1120 a_0 a_{11} b_0^2 f_0^{10} + 952 a_3^2 b_0^2 f_0^{10} - 210 a_0 a_3 b_0 b_3 f_0^{10} - 84 a_0^2 b_0 b_{11} f_0^{10} \nl + 35 a_0^2 b_3^2 f_0^{10} + 280 a_0^2 b_0^2 f_0^9 f_{11}  - 1092 a_0 a_3 b_0^2 f_0^9 f_3 - 70 a_0^2 b_0 b_3 f_0^9 f_3 \nl + 476 a_0^2 b_0^2 f_0^8 f_3^2 + 24 F_0^2 a_0^2 b_0, \\
0 &= - 78 a_0^2 b_0 b_3 f_0 f_3 - 168 a_0^2 b_0^2 f_0 f_{11} - 1008 a_0 a_{11} b_0^2 f_0^2 + 81 a_0^2 b_3^2 f_0^2 - 84 a_0^2 b_0^2 f_3^2 \nl + 1288 a_3^2 b_0^2 f_0^2 - 980 a_0 a_3 b_0^2 f_0 f_3 + 350 a_0 a_3 b_0 b_3 f_0^2 - 108 a_0^2 b_0 b_{11} f_0^2,
\end{align}
and \eqref{eq:i1} gives
\begin{equation}
7 \left({7 a_3 f_0 \lambda_0 + 3a_0 f_3 \lambda_0 - a_0 f_0 \lambda_3}\right)^2 =  6 a_0^2 b_0^2 f_0^2.
\end{equation}
One can check that, if these equations are solved, all other equations are also satisfied at this order. Note that there cannot be any interference with terms from other orders in any of the above equations since this is not admitted by the numerical values \eqref{eq:abc} (\mbox{e.\ g.} $2\zeta \neq \eta$, $\zeta+\xi \neq \eta$, etc.). Solving the equations, we find
\begin{gather}
a_3 = \frac{3}{8} \frac{F_0^2 a_0 \lambda_0 \lambda_3}{b_0 f_0^{10}}, \qquad b_3 = - \frac{21}{4} \frac{F_0^2 \lambda_0 \lambda_3}{f_0^{10}}, \qquad f_3 = \frac{9}{8} \frac{F_0^2 \lambda_0 \lambda_3}{b_0 f_0^9}, \qquad \lambda_3 = \pm \frac{\sqrt{42}}{7}, \qquad\quad \notag \\
a_{11} = - \frac{21}{64} \frac{F_0^2 a_0}{b_0 f_0^{10}}, \qquad b_{11} = \frac{141}{16} \frac{F_0^2}{f_0^{10}}, \qquad f_{11} = - \frac{387}{448} \frac{F_0^2}{b_0 f_0^9}, \qquad \lambda_{11} = 0,
\end{gather}
where $\lambda_0 = \pm \frac{\sqrt{b_0} f_0^5}{\sqrt{3} F_0}$ as derived in the previous section.
\\

\subsection{Determining the Sources}
\label{app:computations.bc.sources}

In order to determine the sources that are supported by the boundary conditions, we have to consider the second derivative terms in the equations of motion. D$6$/\dsix-branes couple to the RR potential $C_7$, the dilaton and the metric. We therefore expect the term $\d F_2$ in the $F_2$ Bianchi identity (which is proportional to the second derivative of $\alpha$) to give the D$6$/\dsix-brane charge. The tension of the branes should be provided by the second derivative of the dilaton $\phi$ in the dilaton equation and by second derivatives of the warp factor $A$ and the conformal factor $B$ in the Einstein equations. In order to simplify the calculation, it is convenient to rewrite the equations of motion \eqref{eoms-theta-bianchi}, \eqref{eoms-theta-dilaton}, \eqref{eoms-theta-exteinstein}, \eqref{eoms-theta-inteinsteintheta} and \eqref{eoms-theta-inteinsteintrans} such that they have total derivative terms on the left-hand side and a delta function with a constant coefficient on the right-hand side. This yields, up to irrelevant terms,
\begin{align}
\left({\e^{-\tfrac{3}{2}\phi-7A+B} \sin^2 \theta \Big( \e^{\tfrac{3}{4}\phi+7A} \lambda \Big)^\prime}\right)^\prime &= \ldots + Q \delta(\theta),\\
\left({\e^{-\tfrac{7}{4}\phi+B}\sin^2\theta \big(\e^\phi\big)^\prime}\right)^\prime &= \ldots + \frac{3}{4}T \delta(\theta),\\
\left({\e^{-\tfrac{3}{4}\phi-16A+B}\sin^2\theta \left({\e^{16A}}\right)^\prime}\right)^\prime &= \ldots + T \delta(\theta),\\
\left({\e^{-\tfrac{3}{4}\phi-7A-B}\sin^2\theta \left({\e^{7A+2B}}\right)^\prime}\right)^\prime &= \ldots - \frac{7}{16} T \delta(\theta),\\
\left({\e^{-\tfrac{3}{4}\phi}\sin^2\theta \left({\e^B}\right)^\prime}\right)^\prime &= \ldots - \frac{7}{16} T \delta(\theta).
\end{align}
Integrating the equations, we find
\begin{align}
\e^{-\tfrac{3}{2}\phi-7A+B}\sin^2 \theta \Big(\e^{\tfrac{3}{4}\phi+7A} \lambda\Big)^\prime &= \int \ldots \d \theta + Q, \label{eq:app-d6-int1} \\ 
\e^{-\tfrac{7}{4}\phi+B}\sin^2\theta \big(\e^\phi\big)^\prime &= \int \ldots \d \theta + \frac{3}{4}T,\\
\e^{-\tfrac{3}{4}\phi-16A+B}\sin^2\theta \left({\e^{16A}}\right)^\prime &= \int \ldots \d \theta + T,\\
\e^{-\tfrac{3}{4}\phi-7A-B}\sin^2\theta \left({\e^{7A+2B}}\right)^\prime &= \int \ldots \d \theta - \frac{7}{16} T,\\
\e^{-\tfrac{3}{4}\phi}\sin^2\theta \left({\e^B}\right)^\prime &= \int \ldots \d \theta - \frac{7}{16} T. \label{eq:app-d6-int5}
\end{align}
We can now substitute the expansion \eqref{eq:d6-ansatz-bc} and the results of the previous sections into the left-hand sides of the equations and read off the constant terms in order to determine $Q$ and $T$. This yields the following results.
\begin{itemize}
\item {\noindent\bf Boundary condition 1.} This boundary condition has $L=A=B=F=0$ and does not give rise to any sources,
\begin{equation}
Q = T = 0.
\end{equation}
\item {\noindent\bf Boundary condition 2.} This boundary condition has $L = 0$, $A=-\frac{1}{16}$, $B=\frac{7}{8}$ and $F=-\frac{3}{16}$ and yields extremal sources,
\begin{equation}
Q = \frac{\lambda_0 f_0^3}{\sqrt{b_0}} = \pm \frac{f_0^3}{\sqrt{b_0}}, \qquad T = \frac{f_0^3}{\sqrt{b_0}}.
\end{equation}
\item {\noindent\bf Boundary condition 3.} This boundary condition has $L = -1$, $A=-\frac{1}{16}$, $B=\frac{7}{8}$ and $F=-\frac{3}{16}$ and yields extremal sources at NLO. Using the NLO coefficients derived in the previous section, we find
\begin{equation}
Q = \pm \frac{f_0^3}{\sqrt{b_0}}, \qquad T = \frac{f_0^3}{\sqrt{b_0}}.
\end{equation}
\item {\noindent\bf Boundary condition 4.} This boundary condition has $L = -\frac{\sqrt{42}}{6}$, $A=-\frac{5\sqrt{42}}{336}$, $B=2-\frac{\sqrt{42}}{8}$ and $F=-\frac{\sqrt{42}}{48}$ and yields inconsistent sources. Using the SLO coefficients derived in the previous section, we find
\begin{equation}
Q = \pm \frac{f_0^3}{\sqrt{b_0}},\qquad T = \left\{{ \begin{array}{l} \frac{1}{3}\sqrt{\frac{14}{3}} \frac{f_0^3}{\sqrt{b_0}}\\ 5 \sqrt{\frac{2}{21}}\frac{f_0^3}{\sqrt{b_0}}\\ \frac{96-11\sqrt{42}}{21}\frac{f_0^3}{\sqrt{b_0}}\\ \frac{16-\sqrt{42}}{7}\frac{f_0^3}{\sqrt{b_0}} \end{array}. }\right.
\end{equation}
\item {\noindent\bf Boundary condition 5.} This boundary condition has $L = -\frac{\sqrt{3}}{6}$, $A=-\frac{\sqrt{3}}{24}$, $B=2-\frac{3\sqrt{3}}{4}$ and $F=\frac{\sqrt{3}}{24}$. Since the LO terms in \eqref{eq:app-d6-int1} to \eqref{eq:app-d6-int5} have a positive power of $\theta$ for this boundary condition, there cannot be any SLO terms that are constant, and so
\begin{equation}
Q = T = 0.
\end{equation}
\\[-0.7cm]
\end{itemize}

\subsection{Non-compact Setup}
\label{app:computations.bc.non-compact}

Here, we state the NLO and NNLO coefficients of boundary condition 3 that are required in the discussion of the non-compact setup in Section \ref{ch:simple-non-bps.myers-effect}. The calculation of the boundary conditions in the non-compact case is analogous to the one detailed in the previous sections and yields the same results at LO. The only difference is that the curvature terms have to be discarded and $\sin^2 \theta$ has to be replaced by $r^2$ in the equations of motion \eqref{eoms-theta-bianchi}, \eqref{eoms-theta-dilaton}, \eqref{eoms-theta-exteinstein}, \eqref{eoms-theta-inteinsteintheta} and \eqref{eoms-theta-inteinsteintrans}, which modifies some of the SLO coefficients. We therefore only state the result here. For the NLO coefficients, we obtain
\begin{gather}
a_1 = \frac{1}{16} \frac{a_0}{\lambda_0}\left({\lambda_1\mp1}\right) - \frac{3}{64} \frac{F_0^2 a_0 \lambda_0^2}{b_0f_0^{10}}, \qquad b_1 = - \frac{7}{8} \frac{b_0}{\lambda_0}\left({\lambda_1\mp1}\right) - \frac{31}{96} \frac{F_0^2 \lambda_0^2}{f_0^{10}}, \notag \\
f_1 = \frac{3}{16} \frac{f_0}{\lambda_0}\left({\lambda_1\mp1}\right) + \frac{7}{64} \frac{F_0^2 \lambda_0^2}{b_0f_0^9}, \label{bc.non-compact}
\end{gather}
where the solution with the upper sign has $\alpha^\prime(0)>0$ and the one with the lower sign has $\alpha^\prime(0)<0$. We also need one particular combination of the NNLO coefficients,
\begin{equation}
\frac{\lambda_2}{\lambda_0} - 7 \frac{a_2}{a_0} - 3\frac{f_2}{f_0} = \pm \frac{1}{4}\frac{b_1}{b_0\lambda_0} + \frac{7}{2}\frac{a_1\lambda_1}{a_0\lambda_0} - \frac{7}{2}\frac{a_1^2}{a_0^2} + 3\frac{f_1^2}{f_0^2} + \frac{21}{2} \frac{a_1f_1}{a_0f_0},
\end{equation}
which is not modified as compared to the compact case. The other higher order coefficients are not relevant for the computation carried out in Section \ref{ch:simple-non-bps.myers-effect}.
\\

\cleardoublepage
\thispagestyle{plain}

\section{The Cosmological Constant as a Sum of Source Terms}
\label{app:ccandsources}

Here, we present an alternative derivation of the main result \eqref{y} of Chapter \ref{ch:scaling-symmetries}, which only uses the equations of motion instead of directly exploiting the scaling symmetries. We first consider the Bianchi identity \eqref{bianchix} for the internal RR field strength $F^\textrm{int}_{8-p}$ and multiply by $\sigma(C^\textrm{ext}_{p+1})$,
\begin{align}
0 &= -(-1)^p\,\sigma(C^\textrm{ext}_{p+1})\w \left( \d F^\textrm{int}_{8-p} - H \w F^\textrm{int}_{6-p} + j_{9-p} \right) \nll
= \d\left[\sigma(C^\textrm{ext}_{p+1})\w F^\textrm{int}_{8-p}\right] + (-1)^p\,\sigma(\d C^\textrm{ext}_{p+1} - H \w C^\textrm{ext}_{p-1}) \w F^\textrm{int}_{8-p} \nl + (-1)^p\,\sigma(H\w C^\textrm{ext}_{p-1})\w F^\textrm{int}_{8-p}
+(-1)^p\,\sigma(C^\textrm{ext}_{p+1})\w H\w F_{6-p}^\textrm{int} - (-1)^p\,\sigma(C^\textrm{ext}_{p+1}) \w j_{9-p} \nll
= \d\left[\sigma(C^\textrm{ext}_{p+1})\w F^\textrm{int}_{8-p}\right] + (-1)^p\,\sigma\left\langle F^\textrm{ext} - \e^B \w F^\textrm{b} \right\rangle_{p+2}\w F^\textrm{int}_{8-p} \nl - (-1)^p\,\sigma(F_{8-p}^\textrm{int})\w H\w C^\textrm{ext}_{p-1}
+ (-1)^p\,\sigma(F_{6-p}^\textrm{int})\w H\w C^\textrm{ext}_{p+1}  + (-1)^p\,\sigma(j_{9-p})\w C^\textrm{ext}_{p+1} \nll
= \d\left[\sigma(C^\textrm{ext}_{p+1})\w F^\textrm{int}_{8-p}\right] - \star_{10}\, \e^{\tfrac{p-3}{2}\phi}|F_{8-p}^\textrm{int}|^2 + \left\langle \e^B \w F^\textrm{b} \right\rangle_{p+2}\w \sigma(F^\textrm{int}_{8-p}) \nl
- H\w C^\textrm{ext}_{p-1}\w\sigma(F^\textrm{int}_{8-p}) + H\w C^\textrm{ext}_{p+1} \w \sigma(F^\textrm{int}_{6-p}) - C^\textrm{ext}_{p+1}\w \sigma(j_{9-p}). \label{bi}
\end{align}
Multiplying the $H$ equation \eqref{h-eoms} by $B$ and using the duality relations \eqref{rrduality} yields
\begin{align}
0 &= 2B \w \d\left(\e^{-\phi}\star_{10} H\right) - (-1)^p\left\langle B\w \sigma(F) \w F\right\rangle_{10}\nll
= 2B \w \d\left(\e^{-\phi}\star_{10} H\right) - 2(-1)^p\left\langle B\w \sigma(F^\textrm{int})\w F^\textrm{ext}\right\rangle_{10} \nll
= 2\d\left\langle\e^{-\phi}B\w\star_{10} H - B\w\sigma(F^\textrm{int})\w C^\textrm{ext}\right\rangle_9 - 2(H-H^\textrm{b})\w \left(\e^{-\phi}\star_{10} H\right) \nl
+ 2\left\langle (\d + H \w )(B\w\sigma(F^\textrm{int}))\w C^\textrm{ext} - (-1)^p B\w\sigma(F^\textrm{int})\w \e^B \w F^\textrm{b}\right\rangle_{10} \nll
= 2\d\left\langle\e^{-\phi}B\w\star_{10} H - B\w\sigma(F^\textrm{int})\w C^\textrm{ext}\right\rangle_9 - 2(H-H^\textrm{b})\w \big(\e^{-\phi}\star_{10} H \nl - \left\langle\sigma(F^\textrm{int})\w C^\textrm{ext}\right\rangle_7\big) - 2\left\langle -B \w C^\textrm{ext}\w\sigma(j) + \e^B \w F^\textrm{b} \w B\w\sigma(F^\textrm{int})\right\rangle_{10}. \label{eom_h}
\end{align}
Note that $F_6^\textrm{int}$ does not contribute to any term in above equation since $F^\textrm{int}$ is everywhere multiplied by either $B$ or $H$, which must both be purely internal in a maximally symmetric compactification to $d \ge 4$ dimensions. We now take the combination $\left(1+\frac{p-3}{2}c\right)$ times \eqref{bi} plus $\frac{c}{2}$ times \eqref{eom_h} and sum over $p$. Substituting the definition of $j$ from Section \ref{ch:scaling-symmetries.conventions}, this yields
\begin{align}
0 &= \sum_{3\le p}\left(1+\frac{p-3}{2}c\right)\left(- \star_{10} \e^{\tfrac{p-3}{2}\phi} |F^\textrm{int}_{8-p}|^2  - C^\textrm{ext}_{p+1}\w\sigma(j_{9-p})\right) \nl + c\left\langle \star_{10}\, \e^{-\phi} |H|^2 + B\w C^\textrm{ext}\w\sigma(j) \right\rangle_{10} - \Sigma(c) + \textrm{total derivatives} \nll = \sum_{3\le p} \left(1+\frac{p-3}{2}c\right)\left(- \star_{10} \e^{\tfrac{p-3}{2}\phi}|F^\textrm{int}_{8-p}|^2 - S_\textrm{CS}^{(p)}\right) \nl + c \star_{10} \e^{-\phi} |H|^2 - \Sigma(c) + \textrm{total derivatives}, \label{all_bi}
\end{align}
where $c$ is a free parameter. We also introduced the shorthand
\begin{align}
\Sigma (c) &= -\sum_{2\le p} \left(1+\frac{p-3}{2}c\right)F^\textrm{b}_{p+2}\w\langle\e^B\w \sigma(F^\textrm{int})\rangle_{8-p} + \left(1-\frac{1}{2}c\right)F^\textrm{b}_4\w \sigma(F^\textrm{int}_6) \nl - c\, H^\textrm{b}\w\left( \e^{-\phi} \star_{10} H - \langle \sigma( F^\textrm{int})\w C^\textrm{ext}\rangle_7\right), \label{sigma2}
\end{align}
where we have combined all terms that depend on background fluxes to simplify our notation.

The trace of the external components of the (trace-reversed) Einstein equation reads
\begin{equation}
\frac{4}{d} R_d =  - \frac{1}{2} \e^{-\phi} |H|^2 + \sum_{3\le p} \frac{p-7}{4}\left(\e^{\tfrac{p-3}{2}\phi} |F_{8-p}^\textrm{int}|^2  \pm \mu_p \e^{\tfrac{p-3}{4}\phi} \delta (\Sigma)\right) + \frac{5}{4} \e^{\phi/2}|F_4^\textrm{ext}|^2\text, \label{einstein-app}
\end{equation}
where the upper sign is for D-branes and the lower sign for O-planes and we have used $|F_5^\textrm{ext}|^2 = -|F_5^\textrm{int}|^2$ to rewrite the spacetime-filling part of $|F_5|^2$. Note that spacetime-filling $F_4$ flux can only be present for $d=4$ in type IIA supergravity, while $F_5$ flux can be present for $d=4$ or $d=5$ in type IIB supergravity.

The dilaton equation reads
\begin{equation}
0 = -\nabla^2 \phi - \frac{1}{2} \e^{-\phi} |H|^2 + \sum_{3\le p} \frac{p-3}{4}\left( \e^{\tfrac{p-3}{2}\phi} |F^\textrm{int}_{8-p}|^2 \pm \mu_p \e^{\tfrac{p-3}{4}\phi} \delta(\Sigma)\right) + \frac{1}{4}\e^{\phi/2}|F_4^\textrm{ext}|^2. \label{dilaton-app}
\end{equation}
Combining \eqref{einstein-app} and \eqref{dilaton-app}, we find
\begin{align}
\frac{4}{d} R_d &=  c\, \e^{-\phi} |H|^2 + \sum_{3\le p} \left({1+\frac{p-3}{2}c}\right)\left(-\e^{\tfrac{p-3}{2}\phi} |F^\textrm{int}_{8-p}|^2  \mp \mu_p \e^{\tfrac{p-3}{4}\phi} \delta(\Sigma)\right) \nl + \left({1-\frac{c}{2}}\right) \e^{\phi/2}|F_4^\textrm{ext}|^2 +\textrm{total derivatives}. \label{es_dil}
\end{align}
Finally, we can combine \eqref{es_dil} with \eqref{all_bi} to get
\begin{align}
\frac{4}{d} \star_{10} R_d &=  \sum_{3\le p}\left({1+\frac{p-3}{2}c}\right)\left(\mp \star_{10}\mu_p \, \e^{\tfrac{p-3}{4}\phi} \delta(\Sigma) + S_\textrm{CS}^{(p)} \right) +\mathcal{F}(c) \nl +\textrm{total derivatives}, \label{es_dil_IIa}
\end{align}
where we defined
\begin{equation}
\mathcal{F}(c) = \Sigma(c) - \left(1 - \frac{c}{2}\right) F_4^\textrm{b} \w \sigma( F^\textrm{int}_6) \label{sigma3}
\end{equation}
and used $\e^{\phi/2} \star_{10} F_4^\textrm{ext} = - \sigma( F^\textrm{int}_6)$, which follows from the duality relations \eqref{rrduality}.
Integrating over the ten-dimensional spacetime and using \eqref{ricci-lambda}, we obtain
\begin{equation}
\frac{8v\mathcal{V}}{d-2} \Lambda = \sum_p\left({1+\frac{p-3}{2}c}\right) \left[S^{(p)}_\textrm{DBI} + S_\textrm{CS}^{(p)}\right] + \int\mathcal{F}(c),
\end{equation}
with the volume factors $v$ and $\mathcal{V}$ defined as in \eqref{volumes}.
\\

\cleardoublepage
\thispagestyle{plain}

\phantomsection

\addcontentsline{toc}{section}{Bibliography}
\bibliographystyle{utphys}
\bibliography{groups-arxiv}

\end{document}